\documentclass[]{aa}
\usepackage{graphicx}
\usepackage{multirow}
\usepackage{txfonts}
\usepackage{color}
\usepackage{booktabs}      

\newcommand{\kms}{km\,s$^{-1}$}

\begin{document}
\title{Organic molecules, ions, and rare isotopologues in the remnant  of the stellar-merger candidate, CK Vulpeculae (Nova 1670)\thanks{Reduced IRAM and APEX spectra are available at the CDS.}}

\author{T. Kami\'nski\inst{\ref{inst1},\ref{inst2}}, K.M. Menten\inst{\ref{inst3}}, R. Tylenda\inst{\ref{inst4}}, 
        A. Karakas\inst{\ref{inst5}}, A. Belloche\inst{\ref{inst3}} 
        and N.A. Patel\inst{\ref{inst1}} 
       }

\institute{\centering 
       Harvard-Smithsonian Center for Astrophysics, 60 Garden Street, Cambridge, MA, USA\label{inst1} 
       \and Submillimeter Array fellow, \email{tkaminsk@cfa.harvard.edu}\label{inst2},
       \and Max-Planck-Institut f\"ur Radioastronomie, Auf dem H\"ugel 69, 53121 Bonn, Germany\label{inst3} 
       \and Nicolaus Copernicus Astronomical Center, Polish Academy of Sciences, Rabia{\'n}ska 8, 87-100 Toru\'n\label{inst4}
       \and Monash Centre for Astrophysics, School of Physics and Astronomy, Monash University, VIC 3800, Australia\label{inst5}
       }

\date{Received; accepted}
\abstract {CK\,Vul is an enigmatic star whose outburst was observed in 1670--72. A stellar-merger event was proposed to explain its ancient eruption.} 
{We aim to investigate the composition of the molecular gas recently discovered in the remnant of CK\,Vul. Deriving the chemical, elemental, and isotopic composition is crucial for identifying the nature of the object and obtaining clues on its progenitor(s).} 
{We observed millimeter and submillimeter-wave spectra of CK\,Vul using the IRAM 30\,m and APEX telescopes. Radiative-transfer modeling of the observed molecular features was performed to yield isotopic ratios for various elements.} 
{The spectra of CK\,Vul reveal a very rich molecular environment of low excitation ($T_{\rm ex} \lesssim 12$\,K). Atomic carbon and twenty seven different molecules, including two ions, were identified. They range from simple diatomic to complex polyatomic species of up to seven atoms large. The chemical composition of the molecular gas is indicative of carbon and nitrogen-driven chemistry but oxides are also present. Additionally, the abundance of fluorine may be enhanced. The spectra are rich in isotopologues that are very rare in most known sources. All stable isotopes of C, N, O, Si, and S are observed and their isotopic ratios are derived.} 
{The composition of the remnant's molecular gas is most peculiar and gives rise to a very unique millimeter and submillimeter spectrum. The observation of ions and complex molecules suggests the presence of a photoionizing source but its nature (a central star or shocks) remains unknown. The elemental and isotopic composition of the gas cannot be easily reconciled with standard stellar nucleosynthesis  but processing in hot CNO cycles and partial helium burning can explain most of the chemical peculiarities. The isotopic ratios of CK\,Vul are remarkably close to those of presolar "nova grains" but the link of Nova\,1670 to objects responsible for these grains is unclear. The accuracy of isotopic ratios can be improved by future observations at higher angular resolutions and with realistic models of the kinematical structure of the remnant.}
        
\titlerunning{Molecules in CK Vul}
\authorrunning{Kami\'nski et al.}
\maketitle
\section{Introduction}\label{sec-intro}
CK\,Vul was observed in outburst in 1670--72 as a very bright object and is the oldest known nova-like variable with a well documented  light curve \citep{hevelius,shara85}. Its maximum visual brightness was close to 2.6$^{mag}$. Light variations at a time scale of months were apparent to the naked eye and were followed by contemporary observers, including V. Anthelme, J. Hevelius, and G. Cassini. The object had not been recovered until 1981, when a remnant in the form of a bipolar optical nebula was found \citep{shara82,shara85,hajduk2007}. To date, no stellar object has been identified within the remnant, although a weak radio source is located at its center. The outburst of CK\,Vul was unusual: it lasted more than two years; the light curve displayed three peaks; and the object appeared reddish in color. Such observational characteristics do not resemble classical novae in outburst. Several other scenarios have been proposed to explain the outburst of Nova\,1670: a slow nova \citep{shara85}, a diffusion-induced nova \citep{DIN}, a late thermal pulse (or a born-again object) \citep{harrison96,evans2002}, and a stellar-merger event \citep{kato,tyl-blg360}. Although all of the proposed hypotheses have problems explaining observational constraints in hand \citep{evans,hajduk2007}, the merger event seems to be the most promising one and CK\,Vul is now considered to be a red nova. Red novae (or intermediate luminosity optical transients (ILOTs) or red transients) are a group of eruptive stars similar to V838\,Mon \citep{tyl838} which were demonstrated to erupt in stellar-merger events \citep{ST2003,TS2006,v1309}. A recent discovery of molecular gas of extraordinary chemical and elemental composition in CK\,Vul strongly supports the stellar-merger scenario for this ancient transient \citep{kamiNat}. Here, we investigate deeper this highly unusual composition of the CK\,Vul remnant.  

Since the re-discovery of CK\,Vul by Shara et al., the nature of the remnant has been extensively investigated in search for clues that would help explaining the ancient eruption. Identifying the central object and determining the chemical composition of the remnant are particularly important.

The nebula of CK\,Vul has a spectrum similar to Herbig-Haro objects and in the optical it is dominated by atomic lines of \ion{H}{I}, [\ion{N}{II}], [\ion{O}{III}], and [\ion{S}{II}] \citep{shara82,cohen85,naylor}. The (hydrogen) recombination time-scale of the gas was estimated to be $\gtrsim$300\,yr. This time seemed to be consistent with the gas originating in the 1670--72 events and no ionization source was necessary to explain the observations. Later observations of \citet{hajduk2007} revealed that the nebula is much more extended (71\arcsec) than the earlier observations had suggested and has an hourglass shape reminiscent of some planetary nebulae (PNe) and pre-PNe. At a distance of 700$\pm$150\,pc \citep{hajduk2013}, the large structure has a size of 0.24\,pc. The brightest knots of the nebula are located closer (within $\lesssim$8\arcsec) to the radio source which presumably is the location of the stellar remnant. Hajduk et al. proposed that the knots are excited directly by the stellar remnant. They calculated a recombination timescale for nitrogen of 75\,yr, which is shorter than that of hydrogen derived by \citet{shara85} and which suggests an ionization mechanism that has been active after the ancient eruption. Subsequent observations indicated that the large-scale nebula is quickly fading, with a 20--30\% decrease between 1991 and 2009 \citep{hajduk2013}. The tips of the large-scale nebula are moving with a de-projected velocity of 900\,\kms \citep{hajduk2013} and the rest of the nebula expands at similar or smaller velocities. To explain the atomic emission in the inner optical nebula, an active ionization mechanism is necessary and \citet{hajduk2007} proposed that shocks of velocities >100\,\kms\ can explain the observed intensity ratios of the recombination lines. The radio source discovered in the central part of the nebula is compact and its emission is thought to be produced by the free-free mechanism. According to \citet{hajduk2007}, the radio source requires a central photoionizing source with a luminosity of $\sim$1\,L$_{\sun}$. Additionally, \citet{evans} observed in the mid-infrared (MIR) emission lines of H$_2$ and of atoms at relatively high ionization degree, i.e. [\ion{Si}{II}], [\ion{S}{III--IV}], and [\ion{O}{IV}], which they explain by postulating the presence of a photoionization stellar source of a temperature  >50\,000\,K. In summary, the optical and MIR observations suggest an ionizing source affecting the remnant's material at different spatial scales, but it is not clear if it is caused by an unidentified hot central source or by the fast shocks. 

The first interpretations of the CK Vul optical spectra suggested that the atomic gas is enhanced in nitrogen, with roughly three times higher nitrogen to oxygen ratio than in the standard cosmic composition \citep{shara82}. In later observations, the radio continuum source was identified but no hydrogen recombination lines were observed at its location. This led \citet{hajduk2007} to suggest that "the core" and some other nebular clumps are hydrogen poor. We note, however, that the central source is heavily obscured by circumstellar dust \citep{kamiNat} and optical photons arising in the central parts of the remnant may be completely trapped. In a subsequent paper presenting deeper images of the nebula, the claim of hydrogen-poor regions was retracted \citep{hajduk2013}. The observed spatial differences in the intensity ratios of the H$\alpha$ and [\ion{N}{II}] lines are not a consequence of varying elemental abundances but rather reflect differences in densities and recombination time scales. There is no conclusive observational evidence for N enhancement or H deficiency in CK\,Vul. More recent observations show however, that the gas in the remnant exhibits certain chemical anomalies. By studying spectra of two backgrounds stars, \citet{hajduk2013} found that the outflow is rich in \ion{Li}{I}. 

Using the APEX telescope in 2014, \citet{kamiNat} discovered bright and chemically complex molecular gas in emission surrounding the central parts of the remnant. Interferometric maps with the Submillimeter Array (SMA) showed that the emission arises from a bipolar nebula of 13\arcsec\ size located at the center of the larger hourglass-shaped optical nebula. The molecular lines have broad profiles with a full width of up to $\sim$400\,\kms. Observations within the millimeter (mm) and submillimeter-wave (submm) range revealed a plethora of molecules: CO, CN, HCN, HNC, H$_2$CO, HCO$^+$, N$_2$H$^+$, CS, CCH, SiO, and SiS. Additionally, with the Effelsberg telescope, lines of NH$_3$ were observed. The unusual inventory of observed molecules suggested that the gas is enriched in N, while O and C seem to have similar relative abundances.  However, a quantitative study of elemental abundances was not possible with these data and the N enhancement was only a tentative result. The molecular spectrum revealed  an extraordinary isotopic composition of the molecular gas, with a very low ratios of $^{12}$C/$^{13}$C=4$\pm$2 (solar ratio is 89), $^{16}$O/$^{18}$O$\approx$26 (solar 499), $^{14}$N/$^{15}$N$\approx$23 (solar 441; terrestrial 272). The isotopic abundances rule out the possibility that the gas is a product of nuclear runaway in a classical nova. They cannot be explained by pure non-explosive CNO cycles, either. It was speculated, however, that the non-explosive CNO cycles with a subsequent incomplete He-burning and with an admixture of H might explain the isotopic ratios of CK\,Vul. A violent merger could be responsible for disrupting a star and dispersing the gas enriched in nuclear burning into the currently observed remnant.

In this paper, we present extensive mm and submm wavelength observations of the molecular gas of the CK\,Vul remnant. The spectral coverage of the observations is substantially wider than in \citet{kamiNat} and the number of detected species is doubled. The observations are presented in Sect.\,\ref{sec-obs} and analyzed in Sect.\,\ref{sec-analyzis}. The revised identification is described in Sect.\,\ref{sec-ident}. The kinematics and excitation of the molecular gas are analyzed in Sects.\,\ref{sec-profiles} and \ref{sec-excit}. In Sect.\,\ref{sec-results}, we discuss the origin and implication of the chemical, elemental, and isotopic compositions of the molecular gas. The main conclusions are summarized in Sect.\,\ref{sec-conclusion}.    

\section{Observations}\label{sec-obs}
\subsection{APEX}\label{sect-obs-APEX}
CK\,Vul was observed with the Atacama Pathfinder Experiment (APEX) 12-m telescope \citep{apex} in May 2014, July 2014, June 2015, November 2015, April 2016, and 18--19 July 2017. The frequency setups observed between 169 and 909\,GHz are listed in Table\,\ref{tab-log-apex}. A fraction of the observations has been already reported in \citet{kamiNat} and they are included here for completeness. For observations up to 200\,GHz, we used the SEPIA Band-5 receiver \citep{sepia} which produces separate spectra for each of the two sidebands. At higher frequencies up to 270\,GHz, we used the SHeFI-1 receiver \citep{shfi} which produces one single-sideband spectrum. For still higher frequencies, between 278 and 492\,GHz, we used the FLASH$^+$ receiver \citep{flash} which operates simultaneously in the atmospheric windows at about 345 and 460\,GHz. FLASH$^+$ separates the two heterodyne sidebands in each of the two 345/460 channels, giving four spectra simultaneously, each 4\,GHz wide. SEPIA, SHeFI-1, and FLASH$^+$ are single-receptor receivers allowing for observations of one position on the sky at the time. Above 690\,GHz, we used the CHAMP$^+$ receiver which consists of two arrays operating in the atmospheric windows at about 660 and 850\,GHz. Each CHAMP$^+$ array has seven receptors whose spatial arrangement is given in \citet{champ}. Each of the fourteen receptors of the CHAMP$^+$ array produced a single-sideband spectrum. The image-band rejection factors for all the instruments used are all above 10\,dB and at most frequencies the rejection is higher than 20\,dB. For SEPIA, SHeFi-1, and FLASH$^+$, we found that the strongest spectral features indeed do not contaminate the image side band. In all the APEX observations, we used the eXtended Fast Fourier Transform Spectrometer \citep[XFFTS;][]{ffts} which typically provided us with resolutions better than 0.1\,\kms. For presentation purposes, we rebinned the spectra to a much lower resolution. 

At the location of APEX, CK\,Vul can only be observed at low elevations of 25\degr--40\degr. Most of our APEX observations were executed in excellent weather condition to compensate for high atmospheric opacity at these low elevations. The typical system temperatures ($T_{\rm sys}$) and rms noise levels reached are given in Table\,\ref{tab-log-apex}. The full width at half maximum (FWHM) beam sizes and the main-beam efficiencies ($\eta_{\rm mb}$) of the APEX antenna at each observed frequency are also listed in the table. We use both the antenna ($T_A^*$) and main-beam  ($T_{\rm mb}$) brightness temperature scales, as indicated accordingly in each instance. Velocities are expressed in the local standard of rest (LSR) frame. 

First APEX observations of CK\,Vul were made with the band centered on the CO(2--1) line and using wobbler switching. In addition to a very broad emission feature (FWHM=156\,\kms) originating from the target itself, the spectra showed a considerable contamination from narrow (FWHM=0.3--0.9\,\kms) emission features present at the source position and at wobbler off-source positions. The emission features are at velocities  between --0.4 and 9.0\,\kms\ and in 16.8--25.9\,\kms. By obtaining a map which covered 3\farcm0$\times$2\farcm5 with sampling of 45\arcsec, we found that this emission extends nearly across the entire field covered by the map. Taking the extent of the CO emission and the small widths of the lines, the emission is most likely interstellar in origin. Because we are interested here in the circumstellar material, we blanked all channels affected by these interstellar features before smoothing the spectra to a coarser resolution. 

By examining continuum maps obtained with {\it Herschel} \citep[cf.][]{kamiNat}, which show patchy emission in a big field around CK\,Vul, we found a nearby position at an offset $\Delta$RA=--180\arcsec and $\Delta$Dec=--100\arcsec\ where the dust emission was weakest and which turned out to have none or very weak emission in CO transitions. For most of the spectra which cover the CO $J$=2--1, 3--2, and 4--3  transitions and which were obtained later, we applied the total-power switching method using this position as a reference (OFF). Higher rotational transitions of CO, all lines of $^{13}$CO, and all spectra that do not cover CO lines were observed with symmetric wobbler switching and with throws of $\pm$60, $\pm$100, or $\pm$120\arcsec; they  are not contaminated by the interstellar component. 

The data processing included spectra averaging and baseline subtraction. Some APEX spectra contain spurious absorption features caused by telluric lines and the most obvious ones were blanked before final averaging. 

By comparing overlapping spectra from different dates and different instrumental setups, we found that the relative calibration is better than 5\%. The absolute calibration uncertainty, including that of the beam efficiency, is of about 15\% at frequencies up to 400\,GHz and worse at higher frequencies. 

\subsection{IRAM 30\,m}
Observations with the IRAM 30\,m telescope and the EMIR dual-sideband receiver \citep{emir} were obtained in February, May, June, August 2015, and April and May 2016. Spectra of each linear polarization covered simultaneously the same range. The Fast Fourier Transform Spectrometer \citep[FTS;][]{ffts} and WILMA correlator\footnote{\url{http://www.iram.fr/IRAMFR/TA/backend/veleta/wilma/index.htm}} were used in parallel as backends. The FTS was used as the primary spectrometer and covered 7.8\,GHz in each sideband at a resolution of 195\,kHz (typically 0.4\,\kms). WILMA, operating at a resolution of 2\,MHz, covered a fraction of the available band and was used only to identify spurious instrumental effects in the FTS spectra, in particular in the regions where three different FTS units are stitched. Technical details of all the setups observed with IRAM/EMIR are given in Table\,\ref{tab-log-iram}.

The spectra produced by the FTS required special processing in which a baseline of low order was subtracted individually from each of the spectrometer units. This removed the ``platforming'' effect. A few spikes introduced by the spectrometer were blanked before spectra were averaged and rebinned to a lower resolution. All IRAM observations were obtained using wobbler switching with OFF positions at 110\arcsec\ from the central position. Similarly to observations with APEX, some molecular transitions were affected by the extended interstellar emission, including the $J$=1--0 line of $^{12}$CO, $^{13}$CO, and HCN. Those features were blanked in the data reduction procedure.   

The collected spectra are considerably affected by imperfect sideband rejection of EMIR. The E0 and E2 units of EMIR operating at the 90\,GHz and 230\,GHz atmospheric windows have image band rejection factors of over 10\,dB. At our signal-to-noise ratio (S/N) reaching 200 for some lines, these rejection factors are too low to block the signal from the image side band and strong lines sink into the symmetric side band. The rejection is even worse in the case of the E1 unit operating in the 150\,GHz window for which the rejection factors were of about 5\,dB and changed strongly with frequency \citep{rejection}. In addition, ghost features were reported for E1 \citep{ghosts}. In order to distinguish real spectral features from those arising in the image side band and to be able to identify other instrumental artifacts, most spectral ranges were observed with at least two different local oscillator frequencies. In cases when a feature was undoubtedly identified as residual emission from the image side band, the feature was blanked before spectra averaging. However, not all spectral ranges were observed more than once and some of those spurious features remain in the spectra. We refer to them as OSB (other-sideband) or "instr." features. 

Spectra obtained for orthogonal polarizations show differences of the order of 10\% within the profiles of strongest lines (most pronounced in lines of HCN). We attribute these differences primarily to the receiver performance. Observations of spectral features obtained on different dates and with different central frequencies (both polarizations averaged) are consistent within 15\%. The overall absolute calibration accuracy of the IRAM spectra is better than 20\%. 

\begin{figure}
\centering
\fbox{\includegraphics[angle=0, width=0.9\columnwidth]{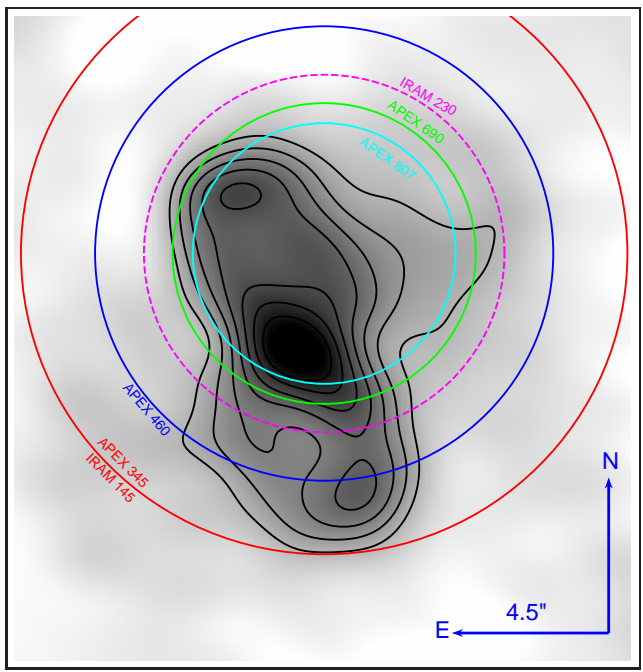}}
\caption{Representative telescope beams overplotted on a map obtained with the SMA and APEX in CO(3--2) \citep{kamiNat}. The diameters of the circles correspond to FWHMs of the beams. From largest to smallest the circles mark beams at: 345\,GHz of APEX and 145\,GHz of IRAM (red); 460\,GHz at APEX (blue); 230\,GHz at IRAM (magenta, dashed); 690\,GHz at APEX (green); and 810\,GHz at APEX (cyan). The beams corresponding to APEX observations below 300\,GHz and to IRAM observations below 100\,GHz are larger than the map and are not shown. The centering of the beams corresponds to perfect antenna pointing. In reality, the observations could have been performed with pointing errors leading to offsets of up to 3\arcsec.}\label{fig-beams}
\end{figure}

\subsection{Spatial characteristics of APEX and IRAM 30m data}\label{sec-spatial}
Although we obtained extra APEX observations aiming to probe the extent of the molecular emission, we confine this paper to observations obtained toward the central source. 

Most of our APEX and IRAM observations were obtained towards CK\,Vul's catalog position from SIMBAD\footnote{\url{http://simbad.u-strasbg.fr/simbad/}} \citep[originally from][]{SimbadPosition}, i.e. at RA=19:47:38.0, Dec=+27:18:48.0 (J2000). It is not the morphological center of the molecular emission \citep{kamiNat}. It is 3\farcs0 off from the radio continuum source observed with the Very Large Array (VLA) \citep{hajduk2007}. This offset is comparable to the pointing accuracy of both telescopes. 

After the morphological details of the molecular emission were revealed by the interferometric maps obtained with the Submillimeter Array \citep[SMA;][]{kamiNat}, it was realized that the single-position APEX and IRAM observations probe slightly different emission regions at different frequencies, as is illustrated in Fig.\,\ref{fig-beams}. While the low-frequency APEX and IRAM observations encompass nearly the entire molecular region, the beams of APEX at frequencies above about 400\,GHz and these of IRAM above 230\,GHz were most sensitive to the central part and northern molecular lobe of the remnant. This spatial coupling is also responsible for systematic shifts in line velocities at different frequencies.


Interferometric observations of molecular gas in CK\,Vul were reported in \citet{kamiNat}. These observations were performed in July 2014 with the SMA in the 230 and 345\,GHz bands and with an angular resolution of 1\farcs9 and 6\farcs4. Since then, we have mapped the remnant using the NOEMA and ALMA interferometers at chosen frequency ranges between about 85 and 235\,GHz and at angular resolutions of 0\farcs6--2\farcs8. Although the maps helped us to better understand the single-dish data and at instances we refer to them in this paper, these interferometric observations will be presented in a dedicated study.    

\begin{figure*}
\centering
\includegraphics[angle=270, width=0.99\textwidth]{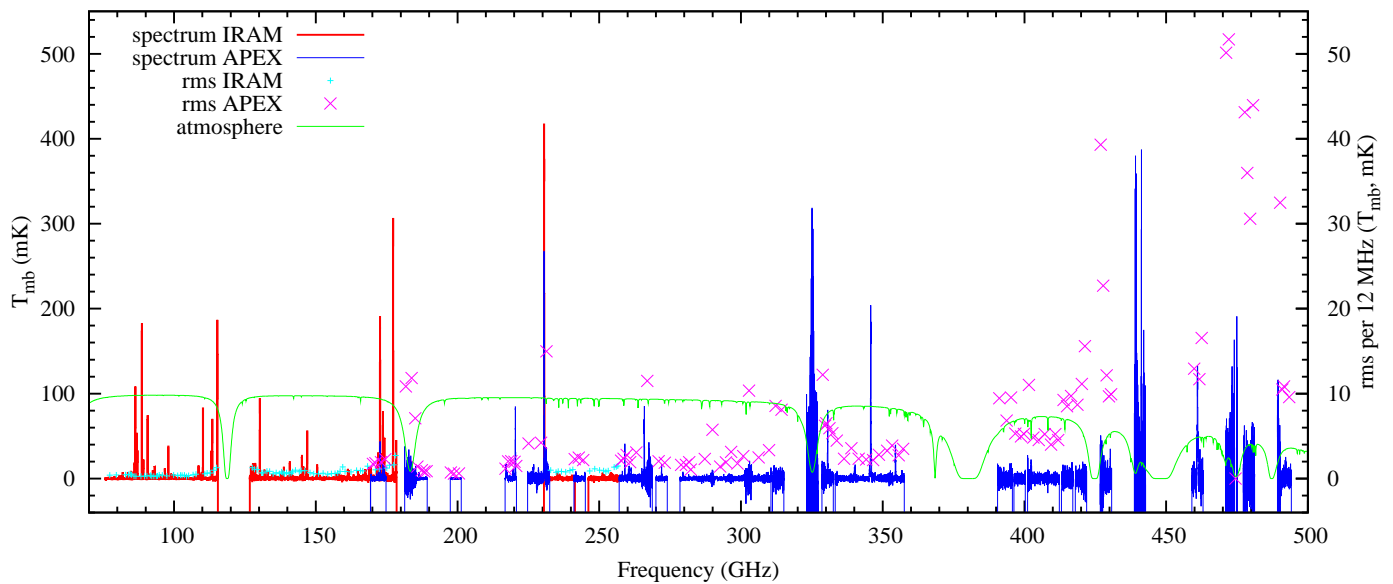}
\caption{Spectral covarage of the data collected with IRAM (red) and APEX (blue). Data up to 500\,GHz are shown but three more spectral ranges were covered at higher frequencies (Table\,\ref{tab-log-apex}). Spectral binning is of 12\,MHz. The spectra are in the $T_{\rm mb}$ scale shown in the left axis. The atmospheric transparency at the APEX site and at 1\,mm of precipitable water vapour is shown (green) to illustrate the location of atmospheric windows \citep{atm}. Zero and full transparency (in \%) correspond to values of 0 and 100 on the left axis. Note that the width and depth of the telluric absorption bands depend on the actual weather conditions and some observations were perfomed within the telluric bands at excellent weather. Plus symbols show the rms levels reached by APEX (bigger, purple) and IRAM (smaller, cyan). They are displayed in the $T_{\rm mb}$ scale shown in the right axis.}\label{fig-fullSpec}
\end{figure*}

\subsection{Spectral characteristics of APEX and IRAM 30m data}
A graphical illustration of the spectral coverage and sensitivities reached is shown in Fig.\,\ref{fig-fullSpec}. The rms values were measured in data combined from different settings and in ranges free of obvious emission lines. However, some spectral ranges appear confusion limited and the rms values may overestimate the noise levels. Our spectral coverage and sensitivities are a result of a search for particular molecular transitions making the spectral survey inhomogeneous and somewhat biased. We covered, however, a very big fraction of the submm and mm-wave spectrum within the available atmospheric windows and the resulting composite spectrum provides a good representation of CK\,Vul's emission spectrum. Also, the observed ranges give a good view on species that are most commonly observed in sources specific for submm and mm astronomy, such as interstellar medium (ISM), and  envelopes of evolved and young  stars. 

The spectra acquired with APEX and IRAM overlap in 169.3--174.7\,GHz and 230.5--232.5\,GHz. In the former range, the telescope beams are much larger than the source size and the intensity of emission lines agrees very well in both spectra when they are compared in the $T_{\rm mb}$  scale corrected by the ratio of the respective beam solid angles. In the latter range, the IRAM beam is smaller than the source size while the APEX beam still encompasses the entire source. These differences in spatial coverage cause a mismatch in intensities of emission features even when spectra are corrected for the beam dilution. The full spectrum of CK\,Vul obtained with the two telescopes is shown in a form of an atlas in Figs.\,\ref{atlas-iram} and \ref{atlas-apex}. 

Single-dish observations of CK\,Vul were also obtained at selected radio frequencies between 1.6 and 25\,GHz using the Effelsberg telescope, including lines of ammonia, OH, and SiO, and they were described in \citet{kamiNat}. Here we focus only on the submm and mm spectra.

\section{Analysis}\label{sec-analyzis}
\subsection{Line identification and inventory of species}\label{sec-ident}
We detected over 320 unique features in the APEX and IRAM spectra. Molecular transitions could be assigned to most of these features. No likely candidate carrier was found for about forty features and they remain unidentified (U). One line is of atomic origin, i.e. of [\ion{C}{I}]. All lines are listed in Tables\,\ref{tab-ID-iram} and \ref{tab-ID-apex}. Some of the listed features are very weak and their presence or identification is uncertain. The identification procedure was based primarily on: line lists of Cologne Database for Molecular Spectroscopy \citep[CDMS;][]{cdms1,cdms2} and Jet Propulsion Laboratory \citep[JPL;][]{jpl}; original papers, e.g. \citet{13ch3nh2} for $^{13}$CH$_3$NH$_2$ and \citet{h15n13c} for H$^{15}$N$^{13}$C; we also used CASSIS\footnote{\url{http://cassis.irap.omp.eu}} and Splatalogue\footnote{\url{http://www.cv.nrao.edu/php/splat}}. For most molecules, we created in CASSIS a simulated spectrum over the full observed spectral range and under the assumption of local thermodynamic equilibrium (LTE; Sect.\,\ref{sec-excit}). The simulation helped us to identify blended lines.   

\begin{table*}\small
\caption{Identified molecules.}\label{tab-mols}\centering
\begin{tabular}{lll}
\hline\hline
2 atoms	& 3 atoms	& 4 atoms\\
\hline\hline
AlF                            & CCH ($^{13}$CCH, C$^{13}$CH)	                 & NH$_3$\tablefootmark{a}  \\
CN ($^{13}$CN, C$^{15}$N, $^{13}$C$^{15}$N)& H$_2$S (H$_2\!^{34}$S?)                         & H$_2$CO (H$_2\!^{13}$CO) \\	
CO ($^{13}$CO, C$^{18}$O, $^{13}$C$^{18}$O, C$^{17}$O?) & HCN (H$^{13}$CN, HC$^{15}$N, H$^{13}$C$^{15}$N) & HNCO\tablefootmark{b} \\			
CS ($^{13}$CS, C$^{34}$S, C$^{33}$S?)	   & HCO$^+$ (H$^{13}$CO$^+$, HC$^{18}$O$^+$?)       & H$_2$CS\\				
NO                                         & HNC (HN$^{13}$C, H$^{15}$NC, H$^{15}$N$^{13}$C?)& \\		
NS                                         & N$_2$H$^+$ ($^{15}$NNH$^+$, N$^{15}$NH$^+$?)    & \\		
PN	                                       & SO$_2$                                          &\\				
SO ($^{34}$SO)		                                 &&\\                        
SiO ($^{29}$SiO, $^{30}$SiO, Si$^{18}$O, Si$^{17}$O?)&&\\  
SiN?\tablefootmark{c,d} &&\\ 				
SiS                  &&\\
\hline\hline
5 atoms	& 6 atoms	& 7 atoms\\
\hline\hline
CH$_2$NH ($^{13}$CH$_2$NH)                                                       & CH$_3$CN ($^{13}$CH$_3$CN)                  & CH$_3$NH$_2$ ($^{13}$CH$_3$NH$_2$?)\tablefootmark{d}\\
HC$_3$N (H$^{13}$CCCN, HC$^{13}$CCN, HCC$^{13}$CN) & CH$_3$OH ($^{13}$CH$_3$OH, CH$_3^{18}$OH?)& \\
\hline\hline
\end{tabular}
\tablefoot{
In parenthesis are rare isotopologes which were identified in addition to the main one. The presence of species followed by ? is questionable.
\tablefoottext{a}{Detected at radio wavelengths \citep{kamiNat}.}
\tablefoottext{b}{Most features of HN$^{13}$CO and HNCO overlap and the presence of the former could not be verified.}
\tablefoottext{c}{Only two features of SiN were detected.}
\tablefoottext{d}{No spectroscopic data available for isotopologues containing $^{15}$N.}
}
\end{table*}



Carbon is the only element observed in atomic form in the submillimeter spectra. All the molecular carriers are listed in Table\,\ref{tab-mols}. The identified transitions belong to 27 different molecules, not counting their rare isotopologues. The list of detected molecules includes many simple diatomic and triatomic species, but more complex polyatomic compounds, including complex organic molecules, are also observed. The largest molecule identified is CH$_3$NH$_2$ (methylamine). The polyatomic molecules have a great number of transitions which however are usually weak (owing to the large partition functions) and, despite the omnipresence of their spectral features, their flux contribution to the observed spectrum is modest. Transitions of simpler molecules containing H and CNO elements dominate the spectrum in terms of intensity; these include lines of: CO, CN, HCO$^+$, HCN, HNC, H$_2$CO, and N$_2$H$^+$. Lines of CO and HCN are by far the most intense ones. Molecules containing heavier elements are also abundant, including AlF, NS, PN, SO, SiO, SiN, SiS, H$_2$CS and SO$_2$. All of the molecules are built out of nine elements: H, C, N, O, F, Al, Si, P, and S. 

Perhaps the most striking spectral features are numerous lines of rare isotopologues, especially those of the CNO elements. Their presence makes the observed spectra very rich and indicates enhanced contribution of isotopes that are very rare in circumstellar envelopes and in the interstellar medium. 

Only two ionic species have been firmly identified, HCO$^+$ and N$_2$H$^+$. Both give rise to very strong emission. 

Assigning transitions to the complex polyatomic species has proven to be most challenging and it is possible that we misidentified a few of their lines. Our basic identification procedure relied on line positions and relative line intensities as predicted for isothermal gas in LTE (Sect.\,\ref{sec-excit}). Considering that both of these simplifying assumptions may be inadequate for the molecular gas in CK\,Vul, in a few cases we relaxed the requirement that the line ratios have to closely follow the LTE prediction. This applies mainly to CH$_3$CN, CH$_3$OH, CH$_2$NH, CH$_3$NH$_2$ and their isotopologues. We assigned transitions of these complex species to spectral features which are considerably stronger in the observed spectrum than in the LTE simulation but their velocity and line profiles match exactly other lines of the given species. These lines are marked with "??" in Tables\,\ref{tab-ID-iram} and \ref{tab-ID-apex} and in figures presented in the appendices.  

As can be seen in Fig.\,\ref{fig-fullSpec} (and figures in the appendices), intense lines is observed mainly in the mm range, in particular at frequencies below about 180\,GHz. In Sect.\,\ref{sec-excit}, we show that the gas excitation temperatures are very low, i.e. $\lesssim$12\,K. At these temperatures, the molecular gas radiates most effectively in the lowest rotational transitions which for a majority of molecules are located in the mm range. In \citet{kamiNat}, we reported and analyzed molecular inventory that was based primarily on submm spectra. That limitation introduced a bias to the earlier analysis. Some species with very weak emission, for instance these of the polyatomic compounds and sulfur oxides, although covered by some of the older submm spectra, were too weak to be detected or too few for unambiguous identification. With the much broader range of frequencies and energy levels covered by the combined data from the IRAM and APEX telescopes, the identifications proposed here are much more reliable. Still, the list of detected transitions is somewhat biased by our choices of observed spectral ranges and achieved sensitivities. 


Many targeted transitions have not been detected despite the high sensitivity. These non-detections are significant for our understanding of CK\,Vul's chemistry. Among the most important searched but non-detected species are aluminum-bearing molecules, i.e. AlO, AlH, AlH$^+$, AlOH, AlS, and oxides, i.e. TiO, TiO$_2$, PO, NO$_2$, N$_2$O, HNO, and H$_2$O. The latter was covered in two transitions, 3(1,3)--2(2,0) at 183.3\,GHz and 5(1,5)--4(2,2) at 325.1\,GHz, which however arise from relatively high-lying energy levels of $E_u$=205\,K and 470\,K and should be weak at the anticipated excitation temperature of $<12$\,K. Multiple strong transitions of the simplest cyclic hydrocarbon $c$-C$_3$H$_2$ and of one relatively well known ion HCS$^+$ are not seen in the spectra. Also, several hydrogen recombination lines were serendipitously covered by our most sensitive observations but are not detected. No deuterated species is detected, although numerous lines easily observable in the ISM were covered.

\begin{figure*}
\centering
\includegraphics[angle=0, width=0.29\textwidth]{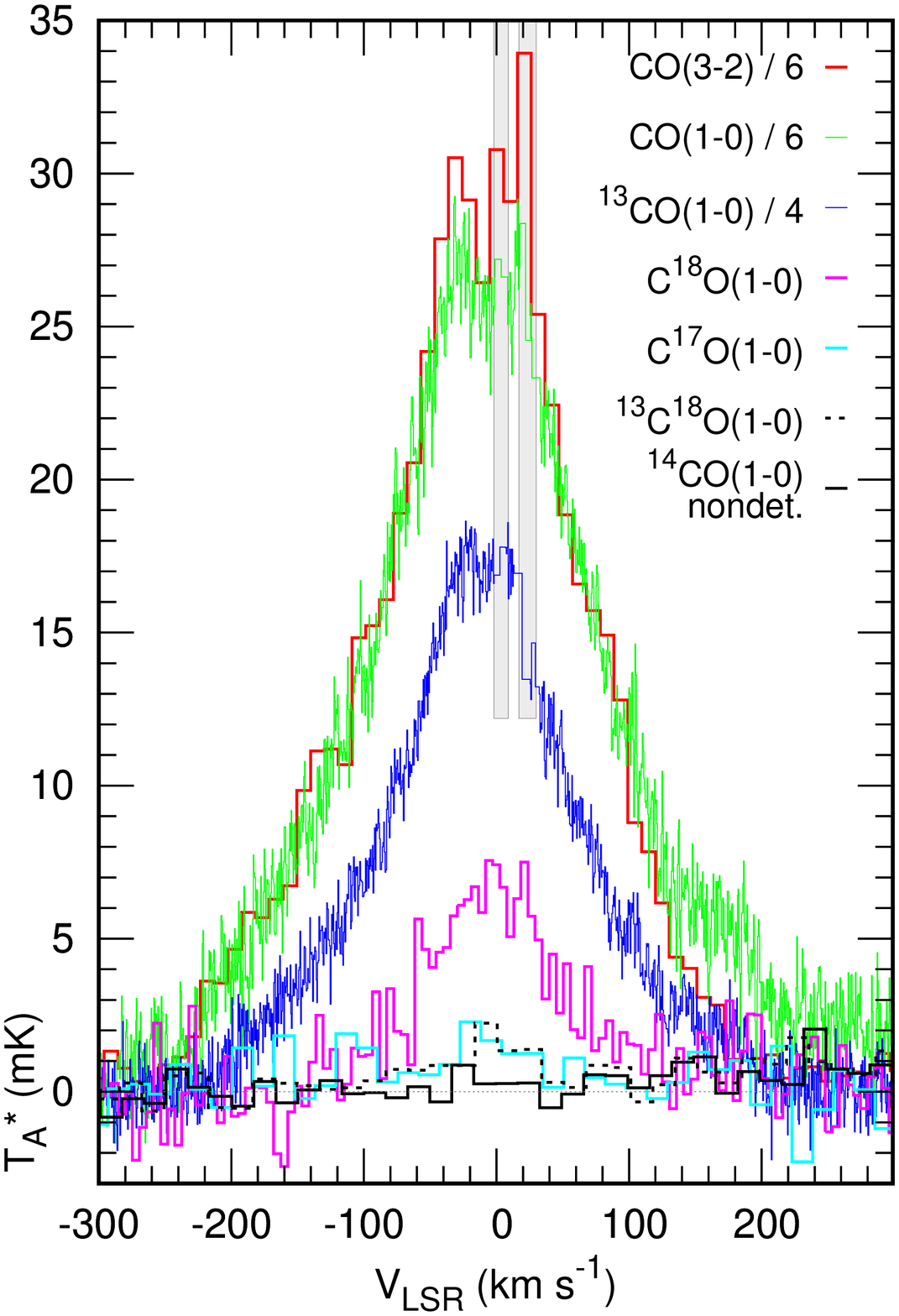}
\includegraphics[angle=0, width=0.29\textwidth]{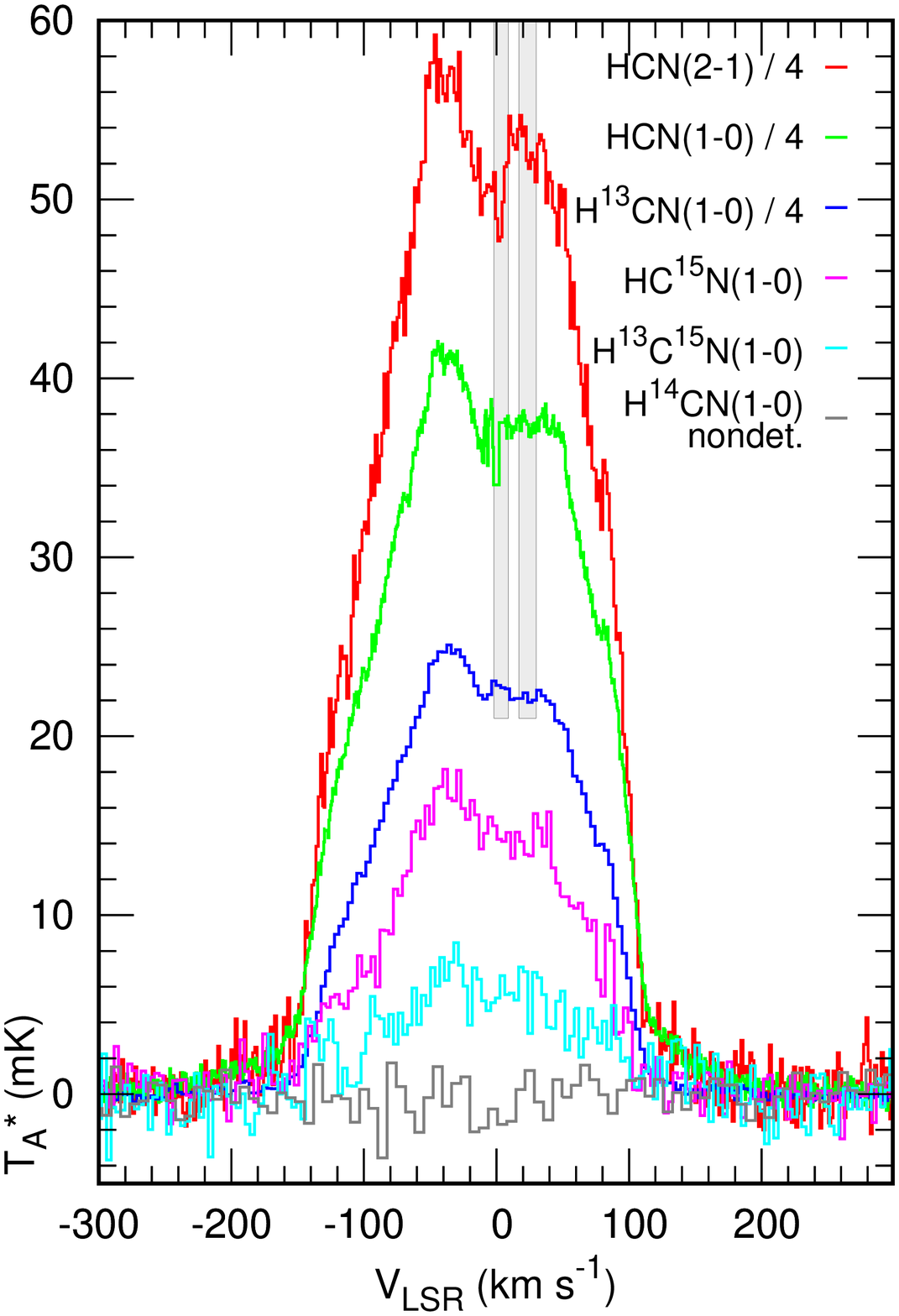}\\
\includegraphics[angle=0, width=0.29\textwidth]{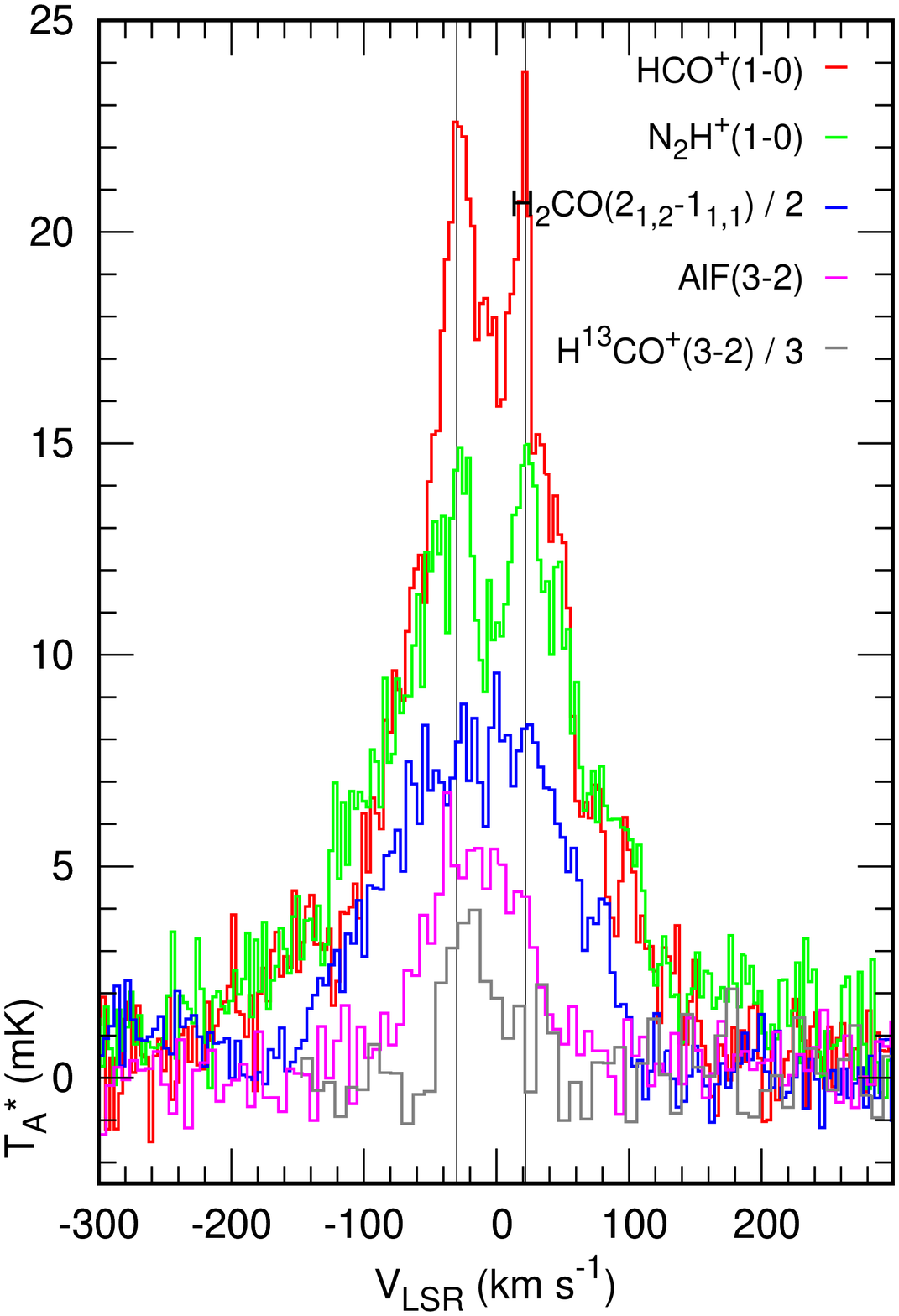}
\includegraphics[angle=0, width=0.29\textwidth]{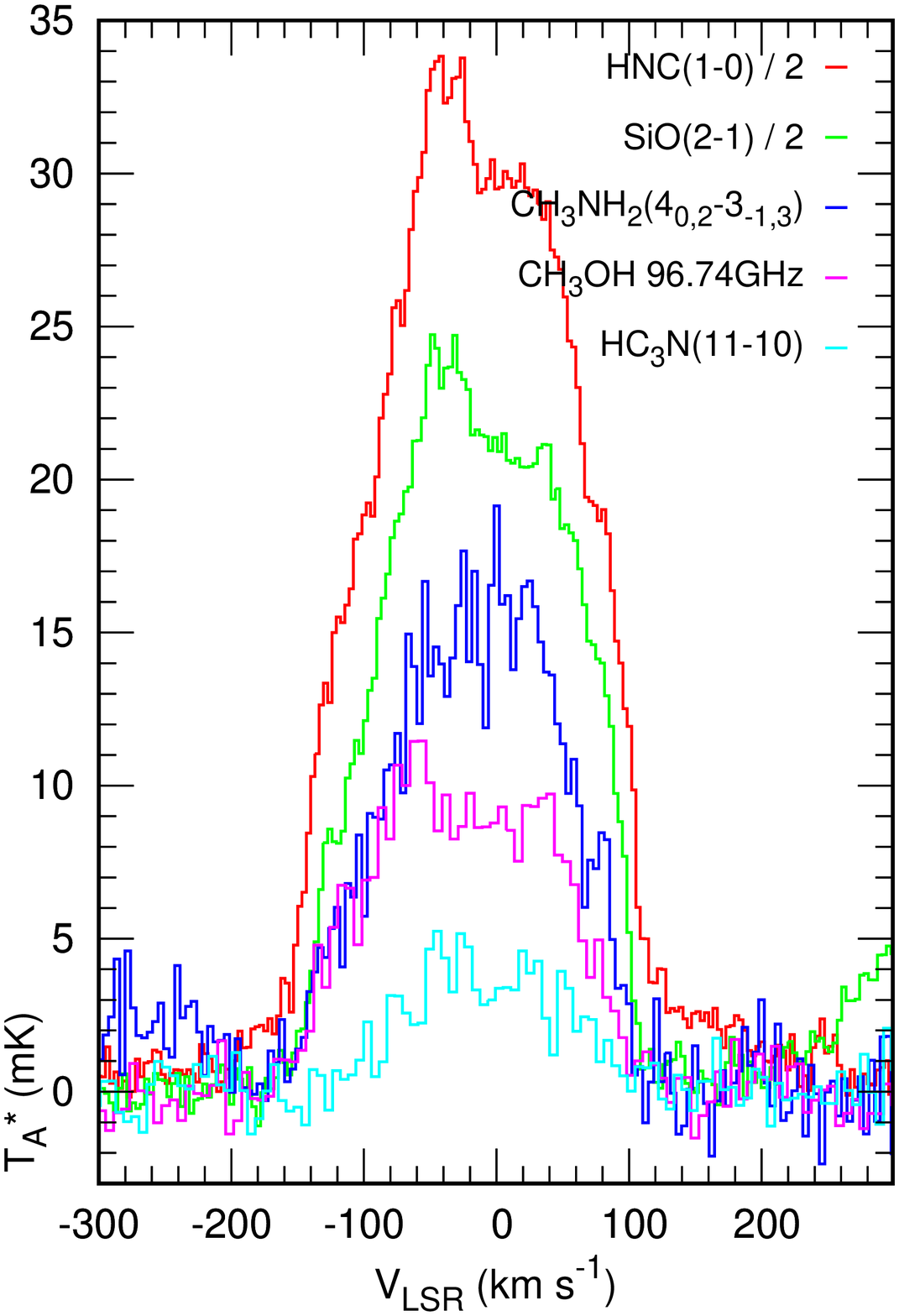}
\includegraphics[angle=0, width=0.29\textwidth]{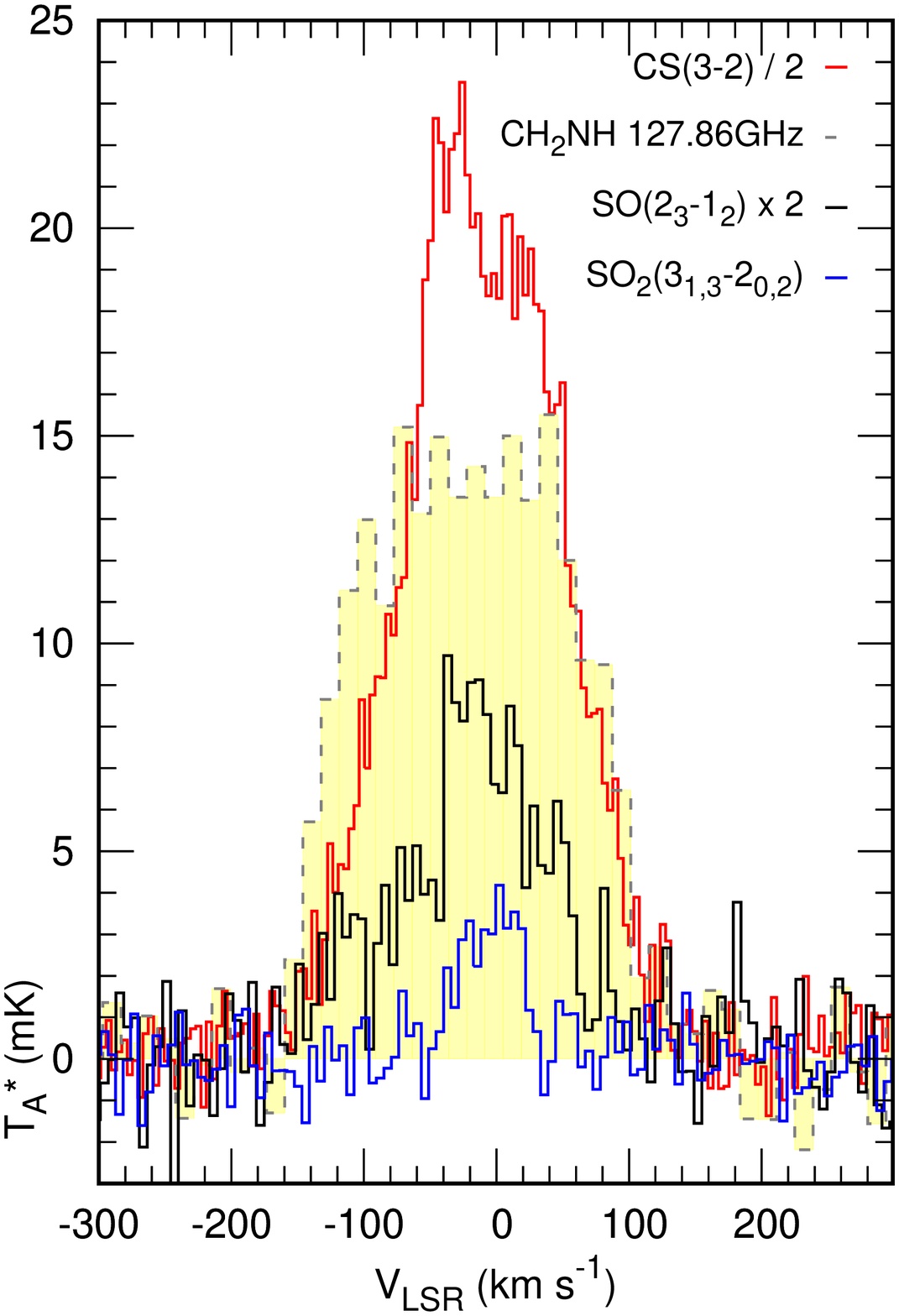}
\caption{Sample emission profiles. The species and transitions shown are indicated in the legends. For CH$_3$OH and CH$_2$NH, the approximate frequencies are given instead of quantum numbers. We could not assign unique quantum numbers to these features because they are blends of components that arise from transitions of different energy levels or species of different symmetry (e.g. $E$ and $A$-type methanol). The shaded areas shown in the top panels mark the positions of interstellar features for which HCN and CO spectra were corrected. Some transitions have hyperfine splitting which, however, is insignificant compared to the total line width. For better presentation, some spectra were smoothed and scaled in intensity, as indicated in the legends.}\label{fig-profiles}
\end{figure*}

\subsection{Diversity of line profiles}\label{sec-profiles}
Figure\,\ref{fig-profiles} illustrates the variety of spectral profiles. 
Among the shown profiles are strong lines of CO which appear triangular, fairly symmetric, and have very broad wings that extend over a range of $\sim$500\,\kms. The profiles of CO features are rather unique in the spectrum  of CK\,Vul. The majority of other species display narrower lines and those observed at a high S/N have typically a FWHM of $\sim$190\,\kms\ and a full width of about 300\,\kms. These widths are comparable to those of optical recombination lines of hydrogen and [\ion{N}{II}] measured by \citet{hajduk2007}. The narrowest molecular profiles, with a FWHM of 50\,\kms, are observed in all lines of SO$_2$ and in H$^{13}$CO$^+$(3--2). The ground-state lines of the two ions, HCO$^+$ and N$_2$H$^+$, show double peaks in the central parts of their profiles, at about --30 and 22\,\kms. The biggest group of lines, e.g. of HCN, HNC, HC$_3$N, SiO and their isotopologues, have an overall parabolic shape with multiple irregularities or overlapping discrete components. These discrete features are consistently present in lines of different species. This group of profiles shows also a small asymmetry -- their blueshifted (negative velocity) part is slightly stronger. 

The variety of profiles observed for different species reflects the differences in kinametical characteristics of the molecular gas. Some differences in profiles of different isotopologues can be partially explained by opacity effects. Most lines at lowest $E_u$ are broad and rounded, suggesting saturation, while transitions from higher energy levels or the same transition of a rarer isotopologue appear narrower or sharper. Excitation conditions are responsible for the observed variations as well. It is not straightforward to distinguish the different effects, especially when one does not know a priori the exact isotopic ratios.

The strongest effect determining the line shape is the spatio-kinematical distribution of gas. From our interferometric maps, to be reported in detail in a forthcoming paper, we can distinguish three basic types of morphologies of the molecular gas: some species are seen predominantly within a radius of $\sim$2\arcsec\ around the central source (e.g. AlF, SO, SO$_2$, CS, HCO$^+$, CH$_3$CN); other species are seen only in the molecular lobes (e.g. CH$_3$OH); and there are species whose emission is seen in the lobes \emph{and} around the central source as in Fig.\,\ref{fig-beams} (e.g. H$_2$CO, HCN, HNC, and CO). However each of the spatial components is often characterized by multiple kinematical components indicating a very complex kinematical and chemical structure within the molecular nebula. There is  no unique link between the appearance of the source-integrated single-dish spectral profiles presented here and the morphology of the molecular emission. The kinematics of the remnant will be studied in detail in an upcoming paper.        


Based on the IRAM and APEX spectra, it is not straightforward to define one `central' velocity that would represent the entire remnant. Our modeling of molecular emission features with CASSIS allowed us to search for central velocity but the best fit results varied from --18 to +8\,\kms\ for different species. Emission of [\ion{C}{I}] (1--0), the only atomic line observed, is centered at a unique velocity of --50$\pm$13\,\kms. However, our spectra show systematic frequency-dependent differences related to how differently-sized beams  couple to the source structure, as discussed in Sect.\,\ref{sec-spatial}. We use the weighted-mean value of $-10$\,\kms\ as a rough representation of the LSR systemic velocity of CK\,Vul.

\subsection{Excitation of the molecular gas}\label{sec-excit}
In order to improve our identification procedure and obtain first constraints on the excitation conditions in the molecular gas, we performed radiative-transfer simulations of the observed molecular spectra. The derived physical parameters are listed in Table\,\ref{tab-cassis} and the final simulation of all species is compared to observations in Figs.\,\ref{fig-model-iram} and \ref{fig-model-apex}. 


All simulations have been performed using the LTE module of CASSIS \citep{vastel}. We used scripted routines\footnote{\url{http://cassis.irap.omp.eu/docs/CassisScriptingDoc/las_index.html}}, in which observations of a given species are compared to a grid of models and the best fit is determined by a $\chi^2$ minimization procedure. A single isothermal component with a Gaussian velocity profile was used to reproduce all observed transitions of a given molecule. The central velocity and the line width were first measured in the observed spectra and usually fixed in the models. In some cases, profile parameters were later iteratively optimized or set as free fit parameters to better reproduce the observations. Of course, the observed profiles are not Gaussian (Sect.\,\ref{sec-profiles}) and the observed single-dish profiles result from a very complex kinematics. Our simplified treatment of line profiles in CASSIS leads to higher line saturation than if a true source structure were taken into account. Additionally, the gas is not isothermal and lines often appear narrower with increasing $E_u$. The varying size of the beams at different frequencies (Sect.\,\ref{sec-spatial}) affects the analysis of a few species, as well. 

In the most basic procedure, the best model was searched in a dense grid of various excitation temperatures ($T_{\rm ex}$) and column densities ($N$), i.e. with two free parameters. The simulated spectra account for opacity effects and for beam dilution. Both effects are a function of the source size ($\theta_S$). For species for which we have obtained interferometric observations, the size was often fixed at a value that was measured directly in maps. For some of the species, however, the size is unknown and was another free parameter. The column density and source size are strongly coupled and their degeneracy led to unrealistically large source sizes for a few species. In such cases, we searched for models at smaller sizes which, although they did not correspond to the absolute $\chi^2$ minimum, still satisfactorily reproduced the observations. In a few cases, we had to use larger source sizes than those measured, in order to avoid complete saturation of the ground-state transitions. It is another limiting consequence of ignoring the real source structure in our model.

The excitation temperature is derived in the LTE models from relative line intensities corrected for the opacity effects. Therefore, $T_{\rm ex}$ can also depend on the source size and can be affected by the $N$--$\theta_S$ degeneracy. Many species were detected in transitions from their lowest energy levels and in a very limited range of $E_u$. This could result in underestimated $T_{\rm ex}$. To reduce this bias, we included in the analysis observations of transitions that were not detected. The final models are still far from satisfactory chiefly because some species may be excited in non-LTE conditions (Sect.\,\ref{sec-ident}) and for many molecules a single temperature is insufficient to explain the observations. Our radiative-transfer simulations included the cosmic background radiation of 2.73\,K.

CASSIS allows us to simultaneously analyze two versions of a given species and we used this mode to determine the column-density ratio of different isotopologues. These simulations benefit from a better coverage of $E_u$ and allow us for a more reliable opacity correction. Although simulations with combined isotopologue data introduce one extra free parameter, their results have proven to be more reliable than single-isotopologue results. The isotopic ratios are discussed in Sect.\,\ref{sec-isot}. The same CASSIS mode was used to yield the abundance ratio of the different spin or symmetry forms of a given molecule, such as the {\it ortho} and {\it para} versions of H$_2$CO, or $A$ and $E$ type CH$_3$OH. In these cases, we used the VASTEL database where the different-symmetry or spin species have separate line lists and partition functions. However, the ratio could be only determined for H$_2$CO at {\it ortho}/{\it para} of 2.5$\pm$0.8. 

The physical parameters derived here are more accurate than those reported in \citet{kamiNat} owing to more transitions covered and the automated minimization procedure. However, they are still subject to many systematic uncertainties and should be treated with caution.



\section{Results and discussions}\label{sec-results}
\subsection{Why so cool?}
The derived excitation temperatures are very low, between 3 and 12.5\,K (Table\,\ref{tab-cassis}). Because, in general, gas may be excited subthermally \citep[see e.g.][]{RD}, these excitation temperatures do not necessarily indicate directly the gas kinetic temperatures. Observed transitions of CO, which have lowest critical densities among all covered transitions, indicate the highest excitation temperature of 10--12\,K, while many transitions of known high-density tracers, e.g. H$_2$CO, HCO$^+$, CCH, HNC, yield temperatures of 3--3.5\,K. This would suggest subthermal excitation of the latter species but there are also high-density tracers, e.g. transitions of SiO and HCN, which yield higher temperatures of 8--12\,K. It is therefore unclear how low the kinetic temperatures are. The relatively large scatter in the derived excitation temperatures may be partially caused by physical variations of kinetic temperature within the remnant and we can conclude that the overall kinetic temperatures in the regions dominating the observed molecular emission are likely below $\sim$12 K.

The 1670--72 eruption probably dispersed material of a relatively high temperature of a few 10$^3$\,K (as these are typical temperatures of material ejected by red novae in outburst) or even higher, i.e. $\gtrsim 10\,000$\,K, since part of the hydrogen gas was ionized in the eruption (Sect.\,\ref{sec-intro}). We assume that the molecular remnant was created in the same event, i.e. after the gas recombined and formed molecules. Radiative cooling may be insufficient to cool down the hot plasma to these low gas temperatures in less than 350\,yr. Adiabatic expansion into a pre-existing medium is another mechanism that can be responsible for efficient cooling of the gas \citep{adiabatic}. Many pre-PNe of dynamical ages of a few hundred years contain molecular gas at similarly low excitation temperatures, e.g. 10--20\,K in OH231.8 \citep{contreras2015} and $\sim$10\,K in Frosty Leo \citep{FrostyLeo}. This suggests that the cool gas in CK\,Vul indeed could have been dispersed only 350\,yr ago. Ultra-cool gas of extremely low excitation of <3\,K, i.e. comparable to temperature estimates of some species in CK\,Vul, have been measured in the Boomerang Nebula and indicate very efficient adiabatic cooling in some parts of the outflow \citep{sahai,adiabatic}. Similar processes could have taken place in CK\,Vul. 

\subsection{Polyatomic and complex organic species}\label{sect-originCOM}
Formaldehyde (H$_2$CO), which was the only polyatomic species identified in CK\,Vul in our earlier work, is accompanied in the current inventory by six other large compounds: isocyanic acid (HNCO), methaminine (CH$_3$NH), cyanoacetylene (HC$_3$N), methanol (CH$_3$OH), methyl cyanide (CH$_3$CN), and methylamine  (CH$_3$NH$_2$). Note that only three of them contain oxygen and all, except methanol and formaldehyde, contain nitrogen. Future observations at better sensitivity or at lower frequencies may reveal even more complex species. 

With the observed molecules, we trace the full hydrogenation sequence based on the cyanide radical: CN, HCN, H$_2$C=NH, H$_3$C=NH$_2$. A hydrogenation sequence is also apparent for CO-based species: CO, (HCO$^+$), H$_2$C=O, and H$_3$C=OH. The apparent hydrogenation sequences are likely to be the actual formation paths of the largest species observed in CK\,Vul. Long carbon chains are not detected in CK\,Vul at the current sensitivity but their building blocks, the simplest linear carbon chain CCH and the simplest cyanopolyyne HC$_3$N (and HCN), are observed. Our mm-wave spectra cover transitions of longer carbon chains but their $E_u$ values are too high for the emission to be detected if their excitation temperature is $\lesssim$12\,K. Their presence should be verified by observations at lower radio frequencies.  

\subsubsection{Origin of the polyatomic molecules in CK\,Vul}

In an attempt to understand the origin of polyatomic species in CK\,Vul, let us consider possible formation channels of CH$_3$OH only. Much theoretical work has been devoted to understand the formation of methanol in different interstellar and circumstellar environments. These environments have initial elemental compositions that differ from these of CK\,Vul. Nevertheless, it is instructive to briefly review known chemical pathways that lead to methanol and consider if they are adequate for CK\,Vul. 
\begin{itemize}
\item Efficient gas-phase production of methanol requires the presence of molecular ions. In the ISM, they are produced by cosmic rays (CRs) or ultraviolet (UV) radiation. Owing to low efficiency of these processes, the gas-phase production of CH$_3$OH is often considered insignificant in the local ISM \citep{coutens}. Note, however, that molecular ions are abundant in the remnant of CK\,Vul and ion chemistry may be a viable mechanism to produce the polyatomic species. 
\item  Methanol can be produced in icy mantles through hydrogenation of CO \citep{WC2002}. It is then released to the gas phase through desorption. Note that dust in CK\,Vul has a component of a temperature of 15\,K \citep{kamiNat} or even lower (7.5\,K; Tylenda et al., in prep.), at which the grains may contain icy mantles. Shocks in the fast molecular jets can naturally explain how methanol is liberated to the gas phase from ice \citep{hartquist}. Additionally, in the central parts of the nebula, thermal desorption may be taking place and release hydrogenated polyatomic species from grains to gas. This is a viable scenario for CK\,Vul as there is also dust of a temperature of about 50\,K at which many ices would be desorbed. If this is the case, methanol should be spatially collocated with the warmer dust. This can be verified by interferometric mapping. 
\item Methanol can also be produced in high-density neutral gas where three-body reactions are possible. It involves chemical reactions of a high order of endothermicity, i.e. of 6500 and 14\,700\,K \citep{hartquist}. This process requires the presence of OH and H$_2$O. \citet{coutens} consider such a formation channel for CH$_3$OH during grain explosions which have enough energy to break the reaction barrier. Grain explosions may occur in CK\,Vul, for instance after a passage of shocks. However, CK\,Vul does not appear  to be abundant in O-bearing molecules other than CO. In fact, the lack of OH features in the Effelsberg radio spectra \citep{kamiNat} may indicate that OH (and H$_2$O) is currently frozen onto grain mantles. If water was liberated from the mantles to allow formation of methanol, we should be able to observe it in the gas phase in quantities comparable to these of methanol. This can be verified by a search for thermal emission of water (given the low temperature of the gas, this is however impossible from the ground).   
\end{itemize}
%
Production of polyatomic species that involves icy mantles would require ({\it i}) quick formation of ice after the explosion of CK\,Vul or ({\it ii}) the presence of icy grains in older circumstellar or preexisting interstellar material. We cannot exclude the presence of dusty ISM surrounding the object before 1670 but icy grains are not expected there. In the direction of CK\,Vul, we observe only diffuse interstellar clouds (Sect.\,\ref{sec-obs}) where icy mantles are not supposed to form or survive. Very quick formation of ice, i.e. on time scales shorter than 350\,yr, is possible in CK\,Vul and would require substantial gas densities, i.e. >10$^6$\,cm$^{-3}$ at a kinetic temperature of 10\,K \citep{spitzer78}. This condition is not unrealistic.  


With the current knowledge of CK\,Vul, we cannot distinguish between the different formation scenarios of the complex molecules.  A detailed modeling of their origin is hampered by unknown physical parameters and the history of the remnant.  

\subsection{Origin of ionic species}\label{sect-ions}
Although HCO$^+$ and N$_2$H$^+$ are often observed in the ISM and star-forming regions, they are very rare in circumstellar envelopes. 
%
The HCO$^+$ ion has been observed in a number of O- and C-rich environments of evolved stars \citep{hcoplus}, mostly in pre-PNe \citep{bujarrabal88,contreras2015} and PNe \citep{ngc7027,zhang2016}. 
In O-rich media, the molecule is thought to be created in a reaction involving water, i.e. \citep{mamon87}
\begin{equation}\label{eq-hcop-water}
{\rm C}^+ + {\rm H}_2{\rm O} \to {\rm HCO}^+ +{\rm H}. 
\end{equation}
Carbon ions can be present in CK\,Vul as other atomic ions have been observed in its optical and IR spectra. 
%
However, gas-phase water does not appear to be abundant in CK\,Vul making the above formation channel questionable. 
An alternative channel occurs in the ISM and involves the protonated hydrogen molecule (H$_3^+$) \citep{mamon87},
\begin{equation}
{\rm H}_3^+ + {\rm CO} \to {\rm HCO}^+ + {\rm H}_2.
\end{equation}
The H$_3^+$ ion can be produced from H$_2$ (through H$_2^+$) by ionization by CRs or by energetic far-UV radiation. A chemical impact of CRs should be negligible in CK\,Vul owing to the short life time of the remnant compared to CR ionization time scales. HCO$^+$ could rather be produced by an internal UV source. \citet{contreras2015} consider yet another formation channel, where HCO$^+$ is formed through proton exchange with N$_2$H$^+$. As we discuss below, however, the formation of N$_2$H$^+$ requires the presence of H$_3^+$ and, again, implies a photoionizing source. 

Another ion seen in CK\,Vul, N$_2$H$^+$, has been observed in a PN, NGC\,7027, whose ion chemistry is strongly influenced by radiation of the hot (220\,000\,K) central star \citep{ngc7027}, and in two well known pre-PNe: O-rich OH231.8 \citep{contreras2015} and C-rich CRL\,618 \citep{Pardo2007}. Efficient formation of N$_2$H$^+$ requires the presence of H$_3^+$ \citep{mamon87}, i.e.
\begin{equation}
{\rm H}_3^+ + {\rm N}_2 \to {\rm N}_2{\rm H}^+ + {\rm H}_2.
\end{equation}
The relatively strong emission of N$_2$H$^+$ in CK\,Vul would suggest a high formation rate of H$_3^+$ (and H$_2^+$), and this -- in turn -- would imply the presence of a UV emitter in the system. \citet{contreras2015} propose another formation channel of N$_2$H$^+$ through a reaction of H$_2$ with N$_2^+$ and propose N$_2^+$ is produced in proton exchange reaction of N$_2$ with He$^+$.  The latter requires a direct ionization by UV photons. Therefore, both HCO$^+$ and N$_2$H$^+$, whatever their actual formation paths are, seem to imply the presence of a UV source in CK\,Vul.  



In order to explain high abundances of the two molecular ions in OH231.8, \citet{contreras2015} proposed a major role of shocks in ion formation. As noted in \citet{kamiNat}, there are many similarities between OH231.8 and CK\,Vul. However, no physio-chemical models exist to verify this formation channel. If shocks indeed can explain the presence of the two ions, also in environments where oxygen is not an abundant element, one would not need to postulate the presence of a central  UV source in CK\,Vul. Currently, we are not able to distinguish between the two alternatives.

\subsection{Chemistry and elemental abundances}
The elemental composition of the gas surrounding Nova\,1670 is of primary interest as it may provide crucial information necessary to decode the nature of the object. As mentioned in Sect.\,\ref{sec-intro}, some hints on the gas elemental composition have been found from observations of the atomic component of the nebula. It is unknown if the molecular gas has the same elemental composition as the recombining plasma. To independently derive the elemental composition of the molecular remnant, it is first necessary to establish accurate molecular abundances and constrain the chemical history of the object. Considering how little is known about the structure and history of the remnant, this undertaking is currently impossible. However, the composition of CK\,Vul can be analyzed on phenomenological grounds, e.g. from comparisons with well-known objects characterized by different types of chemistry. Below, we discuss our results for the most interesting elements.

\paragraph{Hydrogen} The high number of observed molecules containing hydrogen may suggest that the environment is not considerably depleted of hydrogen, against the claims in \citet{hajduk2007}. Unfortunately, we do not observe mm recombination lines and are unable to provide useful constraints on the amount of H in the recombining remnant. The detection of a MIR line of H$_2$ \citep{evans} in combination with observations of other H$_2$ lines might in future help determining the amount of molecular hydrogen in the remnant.

\paragraph{Carbon} From our inventory of molecules and the presence of the unidentified infrared emission bands \citep{evans}, one could conclude that the \emph{chemistry} of the remnant is dominated by carbon. Generally, this does not directly imply that the carbon \emph{abundance} is higher than that of oxygen and nitrogen. We attempted a quantitative analysis of the relative abundances by comparing the number of CNO atoms observed in the molecular gas. For all molecular carriers of the given element, we summed the products of the column density and source solid angle derived in our LTE analysis (Table\,\ref{tab-cassis}). The numbers were corrected for the isotopic composition (Sect.\,\ref{sec-isot}) and the number of atoms building the molecule. This yielded the C:N:O ratios of 43:1:33 which suggest that carbon dominates over the two other metals, in particular over nitrogen. These ratios are highly influenced by a very high abundance of CO. Because we do not observe all of the most abundant species---in particular  observations of C$_2$, O$_2$, and N$_2$ are missing---the ratios are only a suggestion of a dominance of carbon. In fact, the molecular inventory does not allow us to consider the environment as a typical `carbon-rich' one in the sense this term is used to classify AGB stars or pre-PNe. Additionally, there may be spatial variations of the relative elemental abundances across the remnant. 

We observe the $^3P_1$--$^3P_0$ fine-structure line of neutral carbon, [\ion{C}{I}], which is from an energy level at $E_u$=23.6\,K. The feature near 492.2\,GHz represents a tight blend of $^{12}$C and $^{13}$C transitions (Figs.\,\ref{atlas-apex} and \ref{fig-model-apex}). The spatial distribution of the atomic gas and its excitation temperature are currently unknown. In the ISM, neutral carbon is found to trace the same intermediate-density material as that seen in low to moderate opacity low-$J$ $^{13}$CO lines \citep[e.g.][]{Plume}. Assuming thus that the [\ion{C}{I}] emission originates from the same region as CO and the $^{12}$C to $^{13}$C ratio is of 3.8 (Sect.\,\ref{sec-isot}), the measured line intensity of 3.7\,K\,\kms\ ($T_A^*$) implies \ion{C}{I} column densities of the order of 10$^{17}$\,cm$^{-2}$, i.e. similar to that of CO, for excitation temperatures of 6--13\,K that are typical for most molecular species in Table\,\ref{tab-cassis}. At lower excitation temperatures, the \ion{$^{12}$C}{I} column densities would have to be considerably higher, e.g. of $5 \times 10^{19}$\,cm$^{-2}$ at 3.5\,K, and atomic gas would then be the dominant C-bearing component of the remnant. We therefore conclude that C is a major element in the remnant but there is currently no direct observational evidence that it is the dominant CNO element.

\paragraph{Oxygen} The lack of oxides typically omnipresent in spectra of O-rich envelopes of AGB stars and red supergiants \citep[e.g.][]{kami_surv,aro_surv} strongly suggests that the circumstellar gas of CK\,Vul is not dominated by oxygen. We observe strong lines of SiO, but this molecule is also observed in envelopes of carbon stars \citep{sio}. Some other O-bearing molecules we observe---HCO$^+$, H$_2$CO, and CO---are present in a broad range of chemically diverse objects and their presence does not constrain the dominant chemistry type. On the other hand, SO, SO$_2$, NO, HNCO, and CH$_3$OH are never observed in C-rich media. Their presence can be explained by non-equilibrium chemistry taking place in shocks, analogously to how O-bearing molecules are produced in C-rich stars \citep{Cherchneff2006}. These molecules form mainly after dissociation of CO which liberates oxygen for chemical reactions with other metals.  


\paragraph{Nitrogen} A striking feature of the mm and submm spectra of CK\,Vul is the rich variety of N-bearing species. In addition, radio observations revealed relatively strong emission of NH$_3$ \citep{kamiNat}. From all the N-bearing species predicted to be abundant in thermal-equilibrium (for gas dominated by N or O over C), i.e. N$_2$, NO, HCN, HNC, and NH$_3$ \citep{NOpaper}, only N$_2$ remains undetected in CK\,Vul. In addition to these abundant species, we also observe PN, NS, N$_2$H$^+$, SiN and several polyatomic organic species that contain N. Note that although we do not observe species like CP and SiC, we do observe their N-bearing analogs, PN and SiN. Does the enhanced abundance of N-bearing species indicate an enhanced abundance of elemental nitrogen in CK\,Vul? As discussed in \citet{contreras2015} for OH231.8, shock-induced chemistry can produce N-bearing species in environments where the abundance of N is not enhanced. Based only on observations of molecules, the evidence for an overabundance of elemental N  is currently circumstantial, at best. 
%



\paragraph{Fluorine} The presence of the bright emission of AlF in the spectrum of CK\,Vul is remarkable. Although we are not able to provide any quantitative evidence at this time, we believe this emission can be an indication of an overabundance of $^{19}$F relative to solar composition. Fluorine-bearing species are not easily traceable in circumstellar envelopes of cool stars at mm wavelengths. The only two other sources where rotational lines of AlF have been detected are the prototypical C-rich AGB star IRC+10216 \citep{cernicharoAlF,ziurysAlF} and a young PPN CRL\,2688 \citep{AlFinPPN}. The AlF emission in IRC+10216 indicates a substantial overabundance of fluorine which, most likely, was produced in He-burning thermal pulses \citep{jorissen,AlFinPPN}.

\paragraph{Sulfur} Sulfur-bearing species have a good representation in the molecular inventory of CK\,Vul: CS, NS, SO, SO$_2$, H$_2$S, H$_2$CS, and SiS.  Rich sulfur chemistry is thought to be triggered by shocks associated with outflows \citep[e.g.][]{shocksSchem,shockOutflow} but it is unclear if similar phenomena take place in CK\,Vul. Despite the clear presence of S-bearing molecules, there is no evidence for increased abundance of S in the remnant.   



\vspace{0.5cm}
The inventory of species in Table\,\ref{tab-mols} shows that CK\,Vul has molecules typical for all known chemical types of circumstellar envelopes and does not relate directly to any known object. The abundance of elemental carbon seems to exceed these of oxygen or nitrogen but complex chemistry stands in the way of us making definitive conclusions about the elemental composition of the molecular gas.

\subsection{Extraordinary isotopic composition}\label{sec-isot}
We are able to characterize the isotopic composition of the molecular gas much better than the elemental composition.
Table\,\ref{tab-iso} lists the isotopic ratios derived in the LTE analysis with CASSIS (Sect.\,\ref{sec-excit}). The values in the third column of the table specify the $3\sigma$ range of the $\chi^2$ distribution. The $\chi^2$ distribution is often asymmetric around the minimum. Also, in a few cases the minimum is found but the 3$\sigma$ range for the upper limit is indeterminate. For a majority of species, we also calculated line ratios which are given in the last column of the table. These are flux ratios corrected for the spectroscopic line strength ($S\mu^2$) and frequency but not for opacity. If several transitions are observed with a consistent ratio, a weighted mean is given in the table.     

For the species which we observed in multiple isotopic versions, the emission in the main isotopologue is typically very strong. It is often saturated, in particular in transitions to the ground state. Our CASSIS models apply a simple correction for opacity but the oversimplified treatment of the source structure makes this correction insufficient in some cases. Additionally, CASSIS simulations reproduce only very crudely the profile shapes, in particular for molecules that show complex kinematics (e.g. CO and HCN). Other shortcomings of our models mentioned in Sect.\,\ref{sec-excit} also influence the isotopic ratios. Below, we critically discuss the results for all relevant isotopic pairs. 

\subsubsection{$^{12}$C to $^{13}$C ratio} %
The values in Table\,\ref{tab-iso} typically yield $^{12}$C/$^{13}$C between 2 and 4 and have a weighted mean of 2.9$\pm$0.1 (3$\sigma$). However, the results are not free from the saturation effect which leads to underestimated values of the ratio. Strong emission features, e.g. of HCN, HNC, and HCO$^+$, imply values that are systematically lower ($\sim$2) than the much weaker emission of CS and HC$_3$N (with the ratio of $>$4). We believe that the weaker lines are optically thin, as they all have the same ratios, while for species with saturated emission, the line ratio is small (1.7--2.7) and systematically increases with $E_u$. However, we cannot entirely exclude that the differences in $^{12}$C/$^{13}$C for different species in Table\,\ref{tab-iso} arise due to isotopic fractionation (Sect.\,\ref{sect-isot-inter}). Ignoring these systematic uncertainties, we consider the median of all measurements of 3.8$\pm$1.0 as the best value on $^{12}$C/$^{13}$C (the uncertainty is one median absolute deviation). 

\subsubsection{$^{14}$N to $^{15}$N} %
The line ratios in Table\,\ref{tab-iso} yield a median $^{14}$N/$^{15}$N ratio of 16$\pm$6 (3$\sigma$). The corresponding CASSIS models lead to values with a large scatter but are generally consistent with the ratios derived from integrated line fluxes. The LTE analysis of CN and C$^{15}$N suggests a much higher ratio, i.e. of 24--190. Since CN is likely a photo-dissociation product of HCN, some fractionation enhancing CN over C$^{15}$N is possible. Also, the LTE models for CN species are among the worst and the results should be treated with more caution. The analysis of HNC also leads to a somewhat higher ratio, most likely owing to an imperfect fit constrained by only one firm detection of H$^{15}$NC.

Strong constraints on the $^{14}$N/$^{15}$N ratio are set the by line ratio and CASSIS models of HCN which combined give a value of $\approx$12. The models however overpredict the intensity of the cores of the HC$^{15}$N transitions indicating saturation in the lines of the main species. The actual isotopic ratios may be higher than derived for HCN isotopologues, possibly above the formal uncertainty calculated for the line ratios. 

In an attempt to reduce the saturation problem for CN, HCN, and HNC, we also compared the pairs of rare isotopologues substituted by $^{15}$N and $^{13}$C, for instance H$^{13}$CN and HC$^{15}$N. Ignoring the CASSIS simulations for CN and HNC, which unsatisfactorily reproduce the observations, we find a $^{13}$C/$^{15}$N ratio of 6.2 and the corresponding 3$\sigma$ range of 5.0--8.1. With the $^{12}$C/$^{13}$C ratio of 3.8 derived above, we obtain $^{14}$N/$^{15}$N of 23.6 (19.0--30.8). This value is close to ratios derived directly from line ratios of H$^{13}$CN and HC$^{15}$N, and of HN$^{13}$C and H$^{15}$NC. The weighted mean of these three measurements of 24$\pm$1 (3$\sigma$) is slightly higher than the value 12 and 16 deduced in the previous paragraphs but is affected by an extra uncertainty in the adopted $^{12}$C/$^{13}$C ratio. 


We conclude that the elemental $^{14}$N/$^{15}$N ratio is most likely in the range of about 12--24 and we arbitrarily adopt 20$\pm$10 as its most representative value. 

\subsubsection{$^{16}$O to $^{18}$O and $^{18}$O to $^{17}$O}
We use the median of all measurements and the median absolute deviation, 36$\pm$14, as the representative values for the $^{16}$O/$^{18}$O ratio and its uncertainty. Direct comparisons of emission fluxes of the main species with these where $^{16}$O is substituted by $^{18}$O give systematically lower values, but the saturation effect for these pairs must be most severe. We trust more the results obtained for isotopologue pairs involving the $^{13}$C isotope which give values of 32--40, close to the overall median. 

The C$^{17}$O and Si$^{17}$O isotopologues are only tentatively detected and yield upper limits on the elemental abundance of $^{17}$O. Because line intensities of $^{17}$O-bearing isotopologues are comparable in intensity to these of $^{18}$O, the $^{18}$O vs $^{17}$O pairs permit a more accurate determination of the isotopic ratio than $^{16}$O vs $^{17}$O. Our estimates yield the $^{18}$O/$^{17}$O ratio of $\gtrsim$5. Subsequently, $^{16}$O/$^{17}$O>110, using the lowest value of $^{16}$O/$^{18}$O of 22 within the uncertainties quoted above.

\subsubsection{Other isotopes}
We are able to constrain the relative abundance of the unstable isotope of carbon, $^{14}$C, with a half-life time of $t_{1/2}$=5730\,yr. The upper limit on the emission of $^{14}$CO(1--0) yields $^{13}$C/$^{14}$C>72. An analysis of the H$^{14}$CN(1--0) line provides an even more stringent 3$\sigma$ limit of $^{13}$C/$^{14}$C>141 or $^{12}$C/$^{14}$C>530. The line frequencies were taken from \citet{14C}.

Our LTE models and line ratios provide a very accurate ratio of $^{29}$SiO to $^{30}$SiO of 1.0$\pm$0.1 (3$\sigma$). The ratios of the two Si isotopes with respect to the main one are affected by saturation in the lines of the main isotopologue and are less certain. Nevertheless, our analysis of the $^{28}$Si/$^{29}$Si and $^{28}$Si/$^{30}$Si ratios implies a weighted mean of 6.7$\pm$0.4. Additionally, from an upper limit on the $J$=2--1 line of $^{32}$SiO---i.e. the isotopologue containing the radioactive nuclide of $^{32}$Si ($t_{1/2}$=153\,yr)---we derive a lower limit on the $^{30}$Si/$^{32}$Si ratio of 12, which converted to the $^{28}$Si/$^{32}$Si ratio gives a lower limit of about 80.  

The heaviest isotope seen in our data is $^{34}$S and it is $\sim$14 times less abundant than the main isotope of sulfur. We compared both CS and $^{13}$CS to C$^{34}$S and expect all lines to be optically thin. The isotope of $^{33}$S is even less abundant with a lower limit of $^{32}$S/$^{33}$S>34.


\begin{table*}[]\centering \small
\caption{Isotopic ratios resulting from our LTE analysis with CASSIS and directly from line ratios.}\label{tab-iso}
\begin{tabular}{cc cc}
\hline\hline
Isotopologues& Column-density& 3$\sigma$ ratio & Line ratio\\
             & ratio         & range           & and 3$\sigma$ error\\
\hline \hline
\multicolumn{1}{r}{\bf $^{12}$C/$^{13}$C}&\multicolumn{3}{l}{$=3.8\pm1.0$ [89.3]$_{\sun}$}\\
\hline
HC$_3$N, H$^{13}$CCCN, HC$^{13}$CCN+HCC$^{13}$CN & 4.3	& 3.3--5.6 & 4.1$\pm$1.1	\\ 
CCH, $^{13}$CCH, C$^{13}$CH 	                 & 3.0	& 2.2--3.4 & \\ 
CN, $^{13}$CN\tablefootmark{a}                   & 2.3	& 2.1--2.6 & \\			
CS, $^{13}$CS                                    & 4.6	& 3.1--7.0 & 3.4$\pm$0.2 \\	
HCO$^+$, H$^{13}$CO$^+$	                         & 2.5	& 1.8--4.2 & 2.1$\pm$0.1 \\	
HCN, H$^{13}$CN	                                 & 2.0	& 1.6--2.6 & 1.7--2.7\tablefootmark{b}\\
CO, $^{13}$CO                                    & 4.0	& 3.0--5.3 & \\ 
CH$_3$OH, $^{13}$CH$_3$OH\tablefootmark{c}       & 3.8	& 2.9--4.9 & 3.8$\pm$0.6\\	
CH$_3$CN, $^{13}$CH$_3$CN\tablefootmark{a,c}     & 1.5	& 1.2--2.8 & 4.3$\pm$0.1\\	
H$_2$CO, H$_2^{13}$CO\tablefootmark{d}           & 6.5	& 4.4--9.7 & 4.0$\pm$0.2\\ 
HNC, HN$^{13}$C	                                 & 6.2  & 4.3--8.5 & 2.1$\pm$0.1 \\[5pt]
C$^{18}$O, $^{13}$C$^{18}$O	                     & 5.0	 & 2.5--16.6 & \\	
\hline
\multicolumn{1}{r}{\bf $^{13}$C/$^{14}$C}&\multicolumn{3}{l}{>141}\\
\hline
H$^{13}$CN, H$^{14}$CN                           &       &           &>142 \\
$^{13}$CO, $^{14}$CO                             &       &           &>72  \\
\hline
\multicolumn{1}{r}{\bf $^{14}$N/$^{15}$N}&\multicolumn{3}{l}{$=20\pm10$ [441]$_{\sun}$\tablefootmark{e}}\\
\hline
CN, C$^{15}$N\tablefootmark{b}                  & 44.8	& 24.0--190.0 &              \\%
N$_2$H$^+$, $^{15}$NNH$^+$, N$^{15}$NH$^+$      & 14.6	& >5.3		  & 16.2$\pm$9.5 \\
HCN, HC$^{15}$N	                                & 12.8	& 10.1--17.0  &  11.1$\pm$0.3 \\
HNC, H$^{15}$NC                                 & 30.8	& 15.2--67.0   & 15.9$\pm$1.7 \\ 
NS, $^{15}$NS                                   &   	& >17.4       & \\ 
HC$_3$N, HC$_3\!^{15}$N	                        &    	& >7.6        & \\[5pt]	
$^{13}$CN, $^{13}$C$^{15}$N	                    &       & >37.0       & \\					
H$^{13}$CN, H$^{13}$C$^{15}$N	                & 18.4	& 13.8--27.5  &	16.3$\pm$1.4 \\	
HN$^{13}$C, H$^{15}$N$^{13}$C			        &       &             & 13.9$\pm$6.2 \\[5pt]	
HC$_3$N, HC$_3^{15}$N        			        &       & >7.6        &  \\[5pt]	
$^{13}$CN$\times$3.8\tablefootmark{f}, C$^{15}$N   & 69.1	    & 49.0--106.0 & \\ 			
H$^{13}$CN$\times$3.8\tablefootmark{f}, HC$^{15}$N &	23.6	& 19.0--30.8 &	24.4$\pm$1.1\\
HN$^{13}$C$\times$3.8\tablefootmark{f}, $^{15}$HNC &	55.1	& 30.4--84.4 &	26.2$\pm$2.3\\
\hline
\multicolumn{1}{r}{\bf $^{16}$O/$^{18}$O}&\multicolumn{3}{l}{$=36 \pm 14$ [498.8]$_{\sun}$}\\
\hline
CO, C$^{18}$O	                                  & 36.7	& 26.2--60.0 &	24.5$\pm$2.7\\
SiO, Si$^{18}$O	                                  & 12.1    &  9.3--16.7 &	11.5$\pm$2.3\\
HCO$^+$, HC$^{18}$O$^+$	                          & 53.1    &  >25.0     &  15.3$\pm$8.1\\
SO, S$^{18}$O	                                  &     	&  >16.5     &	\\
H$_2$CO, H$_2$C$^{18}$O\tablefootmark{d}          &     	&  >71.6     &	\\[5pt]
$^{13}$CO, $^{13}$C$^{18}$O	                      & 69.2	& >32.0      & 40.2$\pm$19.7\\[5pt]
$^{13}$CO$\times$3.8\tablefootmark{f}, C$^{18}$O  & 52.1	& 35.3--76.0 & 37.4$\pm$6.2\\
H$^{13}$CO$^+\times$3.8\tablefootmark{f}, HC$^{18}$O$^+$ 
                            	                  & 86.5	& 39.9--534.7 & 35.4$\pm$18.8\\
\hline
\multicolumn{1}{r}{\bf $^{18}$O/$^{17}$O}&\multicolumn{3}{l}{$\gtrsim 5$ [5.4]$_{\sun}$}\\
\hline
Si$^{18}$O, Si$^{17}$O	                          & >7.7	&            & 	2.8$\pm$2.3\\
C$^{18}$O, C$^{17}$O 	                          & 5.5	    & 4.9--8.3	 &  4.7$\pm$4.1\\
\hline
\multicolumn{1}{r}{\bf $^{28}$Si, $^{29}$Si, $^{30}$Si:}& \multicolumn{3}{l}{{\bf $^{28}$Si/$^{29}$Si}$=6.7 \pm 0.4$ [19.7]$_{\sun}$; {\bf $^{29}$Si/$^{30}$Si}$=1.0 \pm 0.1$ [0.7]$_{\sun}$} \\
\hline
SiO, $^{29}$SiO	                                  & 7.3	 & 5.7--9.4 &	6.6$\pm$0.5\\ 
SiS, $^{29}$SiS	                                  & 7.7  & >2.3 & \\[5pt]					
SiO, $^{30}$SiO	                                  & 8.9	 & 6.7--11.8 &	6.7$\pm$0.7\\[5pt]
$^{29}$SiO, $^{30}$SiO	                          & 1.2	 & 1.0--1.5 &	1.0$\pm$0.1\\
\hline
\multicolumn{1}{r}{\bf $^{32}$S, $^{33}$S, $^{34}$S:}&\multicolumn{3}{l}{{\bf $^{32}$S/$^{34}$S}$=14 \pm 3$ [22.5]$_{\sun}$; {\bf $^{32}$S/$^{33}$S}>34 [126.6]$_{\sun}$}\\
\hline
CS, C$^{33}$S                                     &	>34.0 & & \\[5pt]		
CS, C$^{34}$S	                                  & 23.5  &	14.4--41.5 & 13.9$\pm$3.3 \\
SO, $^{34}$SO                                     & 14.1  & >5.5 &  \\
NS, N$^{34}$S                                     & 35.3  & >15.0 & \\[5pt]
$^{13}$CS$\times$3.8\tablefootmark{f}, C$^{34}$S&	19.0  &	11.8--37.2 & 14.1$\pm$7.2\\
\hline\hline
\end{tabular}
\tablefoot{The headlines with isotopic ratios in boldface indicate the most likely values of the ratios in CK\,Vul, as discussed in the text. The values given in square brackets correspond to Solar System isotopic ratios from \citet{lodders}. 
\tablefoottext{a}{Very poor fit.}
\tablefoottext{b}{We obtain different ratios for different transition pairs and therefore the values were not averaged.}
\tablefoottext{c}{Analysis performed at the fixed $A$ to $E$ ratio of 1.}
\tablefoottext{d}{Only {\it ortho} transitions were analyzed.}
\tablefoottext{e}{\citet{Marty15N} measured a $^{14}$N/$^{15}$N ratio of 441 for the solar wind. The value of 272.3 in \citet{lodders} corresponds to terrestrial ratio.} 
\tablefoottext{f}{As explained in the text, the ratios were transformed with the average $^{12}$C/$^{13}$C ratio.}

}
\end{table*}

\subsection{Interpretation of the isotopic composition}\label{sect-isot-inter}
\paragraph{Fractionation}
In the above, we interpreted the molecular ratios as the elemental isotopic ratios and ignored isotopic fractionation processes that can alter the observational signatures of the isotopes and lead to biased isotopic ratios. We mentioned that photodissociation can cause fractionation that could enhance the abundance of the rare isotopologues relative to these of the main ones, but this effect seems to be only likely in the case of the CN isotopologues. However, fractionation can be also caused by a thermodynamic effect and isotope-exchange reactions. \citet{TH2000} and \citet{roueff2015} explored these fractionation effects on N- and $^{15}$N-bearing species in physical conditions typical for dense interstellar clouds. Temperatures in these clouds are similarly low ($\sim$10\,K) as in CK\,Vul's remnant. The studies included reactions with molecular ions but their chemical networks are not directly applicable to CK\,Vul. For most species, the expected fractionation is at a negligible level of a few \%. \citet{TH2000} found higher values of up to 39\% (typically 20\%) for CN, HCN and HNC but \citet{roueff2015} questioned some of these results and nitrogen fractionation is likely to be negligible. Roueff et al. showed, however, that carbon fractionation is important in nitriles and can considerably enhance (in the case of CN) or deplete (mainly in the case of HCN and HNC) $^{13}$C in molecules. The thermodynamic fractionation may be therefore non-negligible in CK\,Vul and may introduce extra uncertainties of a few tens of \% in our derived elemental isotopic ratios. A detailed reaction network explaining the formation of all species would be necessary to correct our results for this effect. This task is beyond our current possibilities.

The isotopic ratios derived here are more accurate than in our earlier study and the list of isotopic pairs is much longer. Below, we attempt to identify the origin of the isotopic ratios.


\paragraph{Hydrostatic CNO burning}
The enhanced abundance of the rare CNO isotopes in CK\,Vul suggests that the gas of the remnant was processed by hydrogen-burning in the CNO cycles. 
Let us consider first whether hydrostatic H burning via the CNO cycles can explain the observed isotopic ratios. Non-explosive CNO burning very quickly converts all the CNO nuclides available into $^{14}$N. The processed gas should be dominated by this isotope and nitrogen should be most abundant among the three CNO elements. Optical and our mm-wave data for CK\,Vul suggest that nitrogen may indeed be enhanced but quantitative estimates are missing. Predictions for the relative abundances of the other CNO elements depend on the burning temperature (mass of the star) and duration of the CNO burning \citep{arnould1999}. 
In most conditions, the $^{12}$C/$^{13}$C ratio quickly reaches its equilibrium value of about 4.0 \citep{Iliadis2007}. This ratio is in excellent agreement with what we found for CK\,Vul ($3.8\pm1.0$). Taken at face value, it could indicate that all the carbon we observe was processed by the CNO cycles. If that were the case, however, in a wide range of burning conditions $^{16}$O should be over $10^4$ times more abundant than $^{18}$O, while the CK\,Vul  ratio is much lower, i.e. of $\sim$36. 
The CNO cycles enhance the $^{17}$O abundance so that it can be even four orders of magnitude more abundant than $^{18}$O. The CK\,Vul's $^{18}$O/$^{17}$O ratio of $\gtrsim$5 is far from predictions for the CNO burning. In fact, $^{18}$O and similarly $^{19}$F are typically destroyed in hydrostatic CNO burning at temperatures less than 100 million K.  
Even more importantly, $^{15}$N should be $\sim10^4$ times less abundant than $^{14}$N as a result of hydrostatic burning while the observations indicate a low ratio of about 20 in CK\,Vul. 
We conclude that it is impossible to explain the observed isotopes with pure products of hydrostatic H burning in CNO cycles. 

\paragraph{Nova nucleosynthesis}
Hot (explosive) CNO cycles, taking place in classical novae, can proceed along many different scenarios (different chemical compositions of the accretor and donor stars, extent of the convective mixing, mass of the white dwarf, etc.) and it is impossible to define any standard isotopic yields. State of the art models of runaway nucleosynthesis show, however, that classical novae are the main Galactic producers of $^{13}$C, $^{15}$N, $^{17}$O and synthesize other important isotopes, e.g. $^7$Li, $^{19}$F, and $^{26}$Al \citep{JoseRev}. Most of these isotopes are enhanced in CK\,Vul. In particular, hot CNO nucleosynthesis at temperatures above 100 million K appears to be the only mechanism able to explain the low  $^{14}$N/$^{15}$N ratio measured in CK\,Vul. For instance, some CO nova models of \citet{jose2004} predict the right $^{14}$N/$^{15}$N ratio of about 20 but at the same time produce too much $^{13}$C and $^{17}$O (by factors of $\sim$13 and 34) and not enough $^{18}$O (by a factor of $\sim$5) compared to CK\,Vul. \citet{jose2004} consider also novae with a NeO white dwarf that produce $^{12}$C/$^{13}$C ratios close to the value measured in CK\,Vul but then other isotopic ratios are inconsistent with observations, most importantly the model yields too much $^{15}$N and $^{17}$O (by factors of $\sim$50, and >78, respectively). No nova model can explain consistently all isotopic ratios measured in CK\,Vul; in particular, most nova models do not produce enough $^{18}$O. It is worth noting that the detection of Li in the remnant \citep{hajduk2013} would be consistent with nova-type nucleosynthesis as $^7$Li is abundantly produced in nova runaway explosions \citep{7Be}. However, the measured CNO isotopic ratios strongly suggest that the molecular material does not originate from pure nova runaway nucleosynthesis. This adds to a large body of evidence that Nova\,1670 was not a classical nova \citep{kamiNat,evans}.

\paragraph{He burning}
We next discuss whether the gas of CK\,Vul was enriched in products of He burning. Partial He burning of material already processed in CNO cycles can add $^{12}$C and reduce $^{14}$N, possibly inverting the C/N ratio to values >1. It can also lower the $^{16}$O/$^{18}$O ratio as $^{14}$N is converted to $^{18}$O \citep{Iliadis2007} and increase the abundance of $^{19}$F \citep{jorissen}. Such partial He burning taking place in a merger event between a He- and a CO-white dwarf was introduced to explain very low $^{16}$O/$^{18}$O ratios, i.e. 1--20, observed in R Coronae Borealis (RCB) stars \citep{Clayton2007,jeffery,menon}. Although this ratio is also very low in CK\,Vul (34$\pm$14), other chemical signatures of CK\,Vul appear to be very different from RCB stars, e.g. RCB stars are H deficient and typically have relatively high $^{12}$C/$^{13}$C ratios of 40--100 \citep{hema} \citep[there are exceptions with $^{12}$C/$^{13}$C$\approx$3--4;][]{RaoLambert}. Partial He burning is an attractive scenario for CK\,Vul because it could explain the high abundance of elemental carbon, the presence of fluorine, and the low ratio of $^{16}$O/$^{18}$O. It would have to be activated, however, in astrophysical circumstances different than these proposed for progenitor systems of RCB stars.
 

\paragraph{CNO ashes as impurities}
We next consider if the observed composition can be explained by a mixture of processed and unprocessed material of solar composition. Given the high isotopic ratios of $^{12}$C/$^{13}$C, $^{14}$N/$^{15}$N, and $^{16}$O/$^{17,18}$O  in solar composition (Table\,\ref{tab-iso}), a dilution of nuclear ashes in unprocessed gas is equivalent to adding mainly $^{16}$O, $^{14}$N, and $^{12}$C to the mixture. For both hydrostatic and explosive CNO burning, such an addition would help explain the observed nitrogen and oxygen isotopic ratios but the added unprocessed mass would have to be orders of magnitude higher than the mass of the nuclear products. Also, $^{12}$C/$^{13}$C would then become much higher (hydrostatic burning) or much lower (explosive burning) than the observed ratio. However, we discuss below nova runaway models whose yields mixed into material of a solar composition give isotopic ratios that are close to these in CK\,Vul.  

Given that only hot CNO cycles can explain the measured $^{14}$N/$^{15}$N ratio and that partial He burning seems to be the only mechanism able to produce a $^{16}$O/$^{18}$O ratio close to the observed one, we consider a mixture of products of these different nucleosynthesis processes. Such a mixture would be in agreement with the high abundance of elemental carbon and fluorine -- originating mainly from He burning -- and nitrogen, originating from hot CNO burning (although, the elemental composition of CK\,Vul needs to be verified by observations). Most of $^{17}$O would come from the hot CNO cycles. The observed ratio of carbon isotopes ($^{12}$C/$^{13}$C$\approx$3.8) can be explained if carbon produced in CNO cycles with $^{12}$C/$^{13}$C of 0.53 \citep[Eq. 5.86 in][]{Iliadis2007}\footnote{The value of 0.53 should be treated here as an example only. Other values of the ratio are possible.} is mixed with $^{12}$C from He-burning products that contain no extra $^{13}$C (which is destroyed by $\alpha$-capture reactions). Helium burning would have to yield 1.7 times more $^{12}$C than H burning (if no extra admixture of solar-composition material is considered). We therefore conclude that hot H and partial He burning are very likely to explain the isotopic composition of CK\,Vul. Identifying the astrophysical scenario in which they were activated is beyond the scope of this paper.  




\paragraph{Presolar grains}
There is a rare type of SiC presolar dust grains known as "nova grains" but of unclear origin \citep[e.g.][]{Liu,NH2005}. Their measured isotopic ratios for C, N, and Si are remarkably close to these of CK\,Vul (oxygen is not present in these grains). The best explanation of nova grains seem to be nova runaway events involving hot nucleosynthesis on a massive (>1.2\,M$_{\sun}$) ONe white dwarf \citep[e.g.][]{Amari2001,JoseHernanz}. Pure nova ashes give isotopic ratios that are less extreme than those in nova grains but a very good match is achieved by diluting the theoretical nucleosynthesis products with unprocessed material of solar composition. The mixture has to be dominated by the unprocessed material so that only a few percent of the total mass originates from runaway products \citep[e.g.][]{Amari2001}. These models have several problems, though: ({\it i}) there is no optimal mixture that can simultaneously reproduce well all of the isotopes; ({\it ii}) it is unclear what would be the origin of the unprocessed material and how it would be mixed with the runaway products; and ({\it iii}) the mixture would be dominated by O-rich material that should not produce SiC dust. The very close match between isotopic ratios of nova grains with those of CK\,Vul is however striking and may provide clues on the true nature of Nova\,1670. It appears that the models attempting to explain nova grains in many respects match better the characteristics of CK\,Vul, which also does not have SiC dust \citep{evans}. The link of CK\,Vul to nova grains is very intriguing.


\paragraph{Nuclear burning during a merger} From the above discussion, it is clear that the remnant of CK\,Vul is enhanced in products of helium and CNO hydrogen burning that occurred under unusual, as yet unknown, conditions. Since the observational characteristics of the 1670--73 eruption and its remnant are consistent with a "red nova" \citep{kato,tyl-blg360,kamiNat}, we now consider how this thermonuclear burning can be accommodated with the stellar-merger scenario for CK\,Vul (cf. Sect.\,\ref{sec-intro}). The outbursts of red novae are thought to be powered mainly by the gravitational energy of the merging binary \citep{TS2006}. However, this does not exclude the possibility that during the collision some nuclear burning takes place. Because stellar mergers involve objects with active nuclear-burning regions, the nuclear reactions can be affected by the coalescence, particularly when the cores of the stars finally merge into one. Admittedly, it is not clear how products of such altered burning could be dispersed into circumstellar environment. What is the contribution of such nuclear burning to the overall energy of red novae eruptions and to the observational appearance of red novae transients are other intriguing and important problems that need to be addressed in future observational and theoretical studies. However, this would not change the general conclusion that red novae are powered by mergers.       

The chemical composition of CK\,Vul derived here is extraordinary, even among other red novae. Remnants of other red novae, as observed years to decades after their respective eruptions, are all oxygen-rich (i.e. O/C$>$1), have only minor chemical peculiarities (e.g. V838\,Mon has a lithium excess \citep{kamiV838} and V4332\,Sgr and V1309\,Sco have exceptionally strong bands of CrO \citep{kamiV4332,kamiV1309}), and their overall chemical compositions appear close to solar \citep[e.g.][]{kipper}. Their isotopic compositions remain largely unknown but based on the spectra available thus far, they are unlikely to be as extreme as in CK\,Vul \citep[e.g.][derived $^{12}$C/$^{13}$C$>$16 for V838\,Mon]{rushton}, if at all are non-standard. The distinctive chemical composition of the CK\,Vul's remnant does not necessarily indicate a very different, i.e. other than a merger, origin of this ancient transient. Stellar mergers are expected to occur in many different binary configurations and all across the Hertzsprung-Russell diagram. For instance, V838\,Mon was a young \citep[$<25$\,Myr;][]{afsar,kami_coecho} triple system \citep{tyl_v838} whereas V1309\,Sco was an evolved contact binary with both components probably having evolved off the main sequence \citep{stepien,v1309}. If CK\,Vul is a product of a merger, it could have involved stars in yet another configuration and evolutionary phases. Future studies exploring the merger scenario may be able to identify the progenitor system of CK\,Vul from the extraordinary chemical composition of the remnant.


\section{Summary and main conclusions}\label{sec-conclusion}
The extraordinary molecular spectrum of CK\,Vul acquired with single antennas reveals the cool molecular nebula which was most likely dispersed in the 1670--72 eruption. The observations lead to the following conclusions: 
\begin{itemize}
\item With 27 identified molecules and their associated isotopologues, the molecular remnant of CK\,Vul is one of the most chemically rich stellar-like objects known to mm-wave astronomy. Only a handful of envelopes of AGB and post-AGB stars have higher or comparable number of identified molecules. It is certainly the most chemically-rich environment associated with an eruptive object, although some other red novae are also surrounded by molecular gas of high complexity \citep{kamiV838,kamiV4332,kamiV1309}.   

\item The kinematics of the remnant traced in single-antenna line profiles is very complex and requires interferometric imaging for a full disclosure of its multiple components. Different molecules are associated with different kinematical components indicating that the chemical composition of the gas may vary across the remnant.

\item The excitation temperatures of the molecular gas are of 3--12\,K and may imply a kinetic temperature of about 12\,K for the remnant regions that give rise to the most intense molecular emission. Radiative losses and adiabatic expansion can be responsible for effective cooling of the molecular remnant since the 1670--72 eruption.   

\item The chemical and elemental compositions of the remnant are very peculiar. The molecular inventory includes a myriad of C-bearing species typical for carbon stars, a great abundance of N-bearing species, and several oxides which are never observed in carbon stars. CK\,Vul's chemistry cannot be categorized with the known classifications. It appears that carbon is the most abundant among the CNO elements but this finding requires further study. Nitrogen and fluorine appear enhanced, but the evidence is only circumstancial.

\item The remnant contains various polyatomic organic molecules whose emission lines are particularly apparent in the mm-wave spectra and were not seen in the submm-wave spectra analyzed in \citet{kamiNat}. The origin of these complex species in CK\,Vul is not understood but their formation requires ion chemistry or reactions on icy grain mantles.

\item The two observed ions, HCO$^+$ and N$_2$H$^+$, indicate the presence of a photo-ionization source in the remnant. With the current data, we are unable to distinguish whether the ionization is caused by the central stellar source or strong shocks. 

\item We derive isotopic ratios for five elements at a higher accuracy than in \citet{kamiNat}. No standard stellar-nucleosynthesis scenario can explain these ratios but the observed material was very likely processed in hot CNO cycles and partial and rapid He burning. Further studies are necessary to investigate the relation of these nucleosynthesis signatures to the outburst of CK\,Vul, including the stellar-merger scenario.

\item The isotopic ratios of CK\,Vul are very unusual and are close to these measured in presolar nova grains. Deeper investigation of the link may explain the nature of Nova\,1670 or origin of nova grains. 


\end{itemize}




\paragraph{Prospects} 
The current study determines the direction of future more detailed modeling and observational efforts. Decoding the remnant's chemistry and deriving better isotopic ratios is possible if detailed interferometric mapping is performed and a 3-dimensional radiative-transfer code is used to take into account the complex structure of CK\,Vul. Such efforts are pending and include observations from the NOEMA and ALMA arrays. They will allow us to look for spatial variations of the isotopic ratios and investigate the role of fractionation on the isotopic abundances. In parallel, the spatio-kinematical structure of the ionized and molecular regions should be investigated to understand better the dynamics of the remnant and establish dynamical ages of its components. This will shed light on the history of the remnant and its chemical composition, bringing us closer to solving the CK\,Vul puzzle.


\begin{acknowledgements}
We thank Alexander Breier for providing us with a line list of $^{32}$SiO and Jan-Martin Winters for reading and commenting on the manuscript.
Based on observations made with IRAM 30m and APEX telescopes. 
IRAM observations were carried out under project numbers 183-14, 161-15, and D07-14 with the IRAM 30m Telescope. IRAM is supported by INSU/CNRS (France), MPG (Germany) and IGN (Spain). The IRAM observations were supported by funding from the European Commission Seventh Framework Programme (FP/2007-2013) under grant agreement N\textsuperscript{\underline{o}} 283393 (RadioNet3).
Part of APEX data were collected under the programs 095.F-9543(A) and 296.D-5009(A). APEX is a collaboration between the Max-Planck-Institut f\"ur Radioastronomie, the European Southern Observatory, and the Onsala Space Observatory. 
Analysis was partly carried out with the CASSIS software. CASSIS has been developed by IRAP-UPS/CNRS.
\end{acknowledgements}

\appendix
\section{Observation logs}
\begin{table*}\centering \small
\caption{Details of APEX observations of CK\,Vul. Only observations longer than 5\,min are listed.}\label{tab-log-apex}
\begin{tabular}{lccccccc}
\hline\hline
\multicolumn{1}{c}{Central}&$T_{\rm sys}$&Time on&Bandwidth&rms\tablefootmark{a}&$\eta_{\rm mb}$&Beam     & Observation\\ 
\multicolumn{1}{c}{freq.  }&             &source &         &$T_A^*$             &               &FWHM     &dates\\
\multicolumn{1}{c}{(MHz)  }&          (K)&(min)  & (GHz)   &(mK)                &               &(\arcsec)& \\
\hline\hline
171311.9\tablefootmark{b}  &  141  &    23.8	&  4.0   &     1.2   &  0.78 &   36.4  & 6-Jun-2015 \\
172811.8                   &  142  &    20.8	&  4.0   &     1.2   &  0.78 &	 36.1  & 6-Jun-2015 \\
183310.1\tablefootmark{b}  &  915  &    23.8	&  4.0   &     7.7   &  0.77 &	 34.0  & 6-Jun-2015 \\
184810.0\tablefootmark{b}  &  540  &    20.8	&  4.0   &     6.9   &  0.77 &	 33.8  & 6-Jun-2015 \\
187400.3                   &  293  &   415.8	&  4.0   &     0.6   &  0.77 &	 33.3  & 4,5,6,7-Nov-2015, 9,10,11-Apr-2016 \\
199400.0                   &  162  &   415.8	&  4.0   &     0.3   &  0.76 &	 31.3  & 4,5,6,7-Nov-2015, 9,10,11-Apr-2016 \\
218800.0                   &  151  &    39.7	&  4.0   &     1.0   &  0.75 &	 28.5  & 9-May-2014 \\
226739.6\tablefootmark{b,c}&  183  &    15.1	&  4.0   &     3.7   &  0.75 &	 27.5  & 5-May-2014 \\
230538.0                   &  183  &    12.8	&  4.0   &     3.5   &  0.75 &	 27.1  & 4,5-May-2014 \\
243100.0                   &  177  &    41.6	&  4.0   &    19.8   &  0.74 &	 25.7  & 9-Jun-2015 \\
259011.8\tablefootmark{b}  &  258  &    93.5	&  4.0   &     1.0   &  0.73 &	 24.1  & 5,9,10-May-2014 \\
262900.0\tablefootmark{b}  &  293  &    58.0	&  4.0   &    27.4   &  0.73 &	 23.7  & 4-Jun-2015 \\
266721.9                   &  268  &		5.5	&  4.0   &     3.9   &  0.73 &	 23.4  & 5-May-2014 \\
271981.0                   &  225  &    97.1  &  4.0   &     0.9   &  0.73 &   22.9  & 18,19-Jul-2017\\
280500.0\tablefootmark{b}  &  156  &    53.4	&  4.0   &     0.7   &  0.72 &	 22.2  & 8-May-2014, 9-Jul-2014 \\
283979.7\tablefootmark{b}  &  217  &    97.1  &  4.0   &     0.8   &  0.72 &   22.0  & 18,19-Jul-2017\\
287364.0                   &  162  &    26.1	&  4.0   &     1.1   &  0.72 &	 21.7  & 19-May-2014 \\
292498.5\tablefootmark{b}  &  155  &    53.4	&  4.0   &     0.7   &  0.72 &	 21.3  & 8-May-2014, 9-Jul-2014 \\
293701.8\tablefootmark{b}  &  201  &    26.8	&  4.0   &    34.9   &  0.72 &	 21.2  & 5-Jun-2015 \\
293801.8\tablefootmark{b}  &  193  &    11.9	&  4.0   &    52.4   &  0.72 &	 21.2  & 5-Jun-2015 \\
296601.8\tablefootmark{b}  &  197  &    23.9	&  4.0   &    36.4   &  0.72 &	 21.0  & 5-Jun-2015 \\
299362.5\tablefootmark{b}  &  154  &    26.1	&  4.0   &     0.9   &  0.72 &	 20.8  & 19-May-2014 \\
305700.0                   &  202  &    26.8	&  4.0   &    38.0   &  0.71 &	 20.4  & 5-Jun-2015 \\
305800.0\tablefootmark{b}  &  193  &    11.9	&  4.0   &    52.9   &  0.71 &	 20.4  & 5-Jun-2015 \\
308600.0\tablefootmark{b}  &  204  &    23.9	&  4.0   &    40.1   &  0.71 &	 20.2  & 5-Jun-2015 \\
330588.0\tablefootmark{b}  &  393  &    21.9	&  4.0   &     2.6   &  0.70 &	 18.9  & 8-May-2014 \\
333797.5\tablefootmark{b}  &  240  &    14.0	&  4.0   &     2.3   &  0.70 &	 18.7  & 6-May-2014 \\
336301.5\tablefootmark{b}  &  186  &    40.0	&  4.0   &     0.9   &  0.70 &	 18.6  & 6,8,14-May-2014 \\
340247.8                   &  192  &    22.4	&  4.0   &     1.5   &  0.70 &	 18.3  & 6-May-2014, 9-Jul-2014 \\
342586.4\tablefootmark{b}  &  266  &    21.9	&  4.0   &     1.9   &  0.69 &	 18.2  & 8-May-2014 \\
343601.5\tablefootmark{b}  &  196  &    44.1	&  4.0   &     1.4   &  0.69 &	 18.2  & 6,14,15-May-2014 \\
345796.0\tablefootmark{b}  &  228  &    14.0	&  4.0   &     2.6   &  0.69 &	 18.0  & 6-May-2014 \\
348300.0\tablefootmark{b}  &  196  &    40.0	&  4.0   &     1.0   &  0.69 &	 17.9  & 6,8,14-May-2014 \\
352246.3\tablefootmark{b}  &  237  &    22.4	&  4.0   &     1.4   &  0.69 &	 17.7  & 6,9-Jul-2014 \\
355600.0                   &  228  &    44.1	&  4.0   &     1.3   &  0.69 &	 17.5  & 6,14,15-May-2014 \\
392500.0\tablefootmark{b}  &  870  &    64.1	&  4.0   &    74.6   &  0.67 &	 15.9  & 5-Jun-2015 \\
393749.9                   &  545  &    28.4	&  4.0   &     2.4   &  0.67 &	 15.8  & 9-Jul-2014 \\
398306.3                   &  484  &    56.4	&  4.0   &     1.4   &  0.67 &	 15.7  & 14,15-May-2014 \\
403086.0\tablefootmark{b}  &  452  &    14.2	&  4.0   &     3.1   &  0.67 &	 15.5  & 9-Jul-2014 \\
404498.2\tablefootmark{b}  &  575  &    64.1	&  4.0   &    42.6   &  0.66 &	 15.4  & 5-Jun-2015 \\
405748.8\tablefootmark{b}  &  416  &    28.4	&  4.0   &     1.9   &  0.66 &	 15.4  & 9-Jul-2014 \\
407963.5\tablefootmark{b}  &  442  &    26.1	&  4.0   &     2.6   &  0.66 &	 15.3  & 19-May-2014 \\
410304.8                   &  449  &    56.4	&  4.0   &     1.5   &  0.66 &	 15.2  & 14,15-May-2014 \\
415085.0                   &  449  &    14.2	&  4.0   &     2.7   &  0.66 &	 15.0  & 9-Jul-2014 \\
419962.0                   &  610  &    26.1	&  4.0   &     3.5   &  0.66 &	 14.9  & 19-May-2014 \\
428766.7                   & 1165  &    40.4	&  4.0   &     8.1   &  0.65 &	 14.6  & 6,8-May-2014 \\
440765.2                   & 5137  &    40.4	&  4.0   &    29.7   &  0.65 &	 14.2  & 6,8-May-2014 \\
461040.8                   & 1085  &    48.0	&  4.0   &     4.8   &  0.64 &	 13.5  & 6,8-May-2014 \\
473039.2                   & 4949  &    48.0	&  4.0   &    36.9   &  0.63 &	 13.2  & 6,8-May-2014 \\
479201.5                   & 1186  &		7.9	&  4.0   &    10.8   &  0.63 &	 13.0  & 6-May-2014 \\
491200.0\tablefootmark{b}  & 1659  &		7.9	&  4.0   &    17.3   &  0.62 &	 12.7  & 6-May-2014 \\
492160.0                   &  992  &     97.1 &  4.0   &     2.8   &  0.62 &   12.7  & 18,19-Jul-2017 \\
691473.1                   & 2728  &    88.3	&  2.9   &     4.9   &  0.53 &	  9.0  & 17-May-2014 \\
806651.8                   & 6236  &		8.2	&  2.9   &    36.2   &  0.47 &	  7.7  & 17-May-2014 \\
909158.8                   &17593  &    80.1	&  2.9   &    42.9   &  0.42 &	  6.9  & 17-May-2014 \\
\hline
\end{tabular}
\tablefoot{
\tablefoottext{a}{The rms noise level per 40\,\kms\ bin.}
\tablefoottext{b}{Spectral range overlaps with the next setup in the table.}
\tablefoottext{c}{The higher-frequency half of the spectrum was severely affected by bad performance of the receiver.}
}
\end{table*}

\begin{table*}\centering \small
\caption{Same as Table\,\ref{tab-log-apex} but for IRAM 30\,m observations.}\label{tab-log-iram}
\begin{tabular}{rccccccc}
\hline\hline
\multicolumn{1}{c}{Central}&$T_{\rm sys}$&Time on&Bandwidth&rms\tablefootmark{a}&$\eta_{\rm mb}$&Beam     & Observation\\ 
\multicolumn{1}{c}{freq.  }&             &source &         &$T_A^*$             &               &FWHM     &dates\\
\multicolumn{1}{c}{(MHz)  }&          (K)&(min)& (GHz)   &(mK)                &               &(\arcsec)& \\
\hline\hline
 79587.0\tablefootmark{b}   &   99  &   184.6 &   7.8  &   0.5  & 0.88   & 30.8  & 18-May-2016 \\
 85888.0\tablefootmark{b}   &  112  &    26.6 &	  7.8  &   1.3  & 0.87   & 28.5  & 11-Aug-2015  \\   
 86036.7\tablefootmark{b}   &  143  &    86.1 &	  7.8  &   0.9  & 0.87   & 28.4  & 11-Aug-2015  \\   
 86886.9\tablefootmark{b}   &   75  &    98.9 &	  7.8  &   0.4  & 0.87   & 28.1  & 8-May-2015  \\   
 89288.8\tablefootmark{b}   &   72  &   134.0 &	  7.8  &   0.4  & 0.86   & 27.4  & 22-Feb-2015  \\   
 95267.0\tablefootmark{b}   &  108  &   184.6 &   7.8  &   0.5  & 0.85   & 24.8  & 18-May-2016 \\
 95911.8\tablefootmark{b}   &   76  &   119.3 &	  7.8  &   0.4  & 0.85   & 24.6  & 22-Feb-2015  \\   
101563.3\tablefootmark{b}   &  130  &    26.6 &	  7.8  &   1.4  & 0.84   & 23.3  & 11-Aug-2015  \\   
101714.5\tablefootmark{b}   &  172  &    86.1 &	  7.8  &   1.2  & 0.84   & 23.2  & 11-Aug-2015  \\   
102565.6\tablefootmark{b}   &   82  &    98.9 &	  7.8  &   0.4  & 0.84   & 23.0  & 8-May-2015  \\   
104961.8\tablefootmark{b}   &   83  &   134.0 &	  7.8  &   0.3  & 0.83   & 22.5  & 22-Feb-2015  \\   
111582.8\tablefootmark{~}  &  112  &   119.3 &	  7.8  &   0.5  & 0.83   & 21.8  & 22-Feb-2015  \\[4pt]   
130589.4\tablefootmark{b}   &  100  &    62.6 &	  7.8  &   0.6  & 0.80   & 18.6  & 7-May-2015  \\   
138419.3\tablefootmark{b}   &   78  &    70.3 &	  7.8  &   0.4  & 0.79   & 17.5  & 8-May-2015  \\   
146260.5\tablefootmark{b}   &  107  &    62.6 &	  7.8  &   0.6  & 0.77   & 16.2  & 7-May-2015  \\   
151180.2\tablefootmark{b}   &  101  &    65.8 &	  7.8  &   0.5  & 0.77   & 16.0  & 7-May-2015  \\   
154099.7\tablefootmark{b}   &   88  &    70.3 &	  7.8  &   0.5  & 0.76   & 15.4  & 8-May-2015  \\   
158405.7\tablefootmark{b}   &   98  &    16.7 &	  7.8  &   1.2  & 0.75   & 15.0  & 6-May-2015  \\   
158904.3\tablefootmark{~}  &  104  &    85.9 &	  7.8  &   0.5  & 0.75   & 15.0  & 6-May-2015  \\   
166850.3\tablefootmark{b}   &  133  &    65.8 &	  7.8  &   0.6  & 0.74   & 14.2  & 7-May-2015  \\   
174089.6\tablefootmark{b}   &  158  &    16.7 &	  7.8  &   2.0  & 0.73   & 13.9  & 6-May-2015  \\   
174578.3\tablefootmark{~}  &  168  &    85.9 &	  7.8  &   0.8  & 0.73   & 13.9  & 6-May-2015  \\[4pt]  
234405.8\tablefootmark{b}   &  179  &   158.9 &	  7.8  &   0.5  & 0.64   & 10.3  & 12-Jun-2015, 1-Apr-2016  \\   
234504.1\tablefootmark{b}   &  192  &   132.3 &	  7.8  &   0.6  & 0.64   & 10.3  & 12-Jun-2015, 1-Apr-2016  \\   
236910.9\tablefootmark{b}   &  252  &    13.2 &	  7.8  &   2.1  & 0.63   & 10.2  & 2-Apr-2016  \\   
237009.3\tablefootmark{b}   &  275  &  	6.6 &	  7.8  &   3.1  & 0.63   & 10.2  & 2-Apr-2016  \\
237402.7\tablefootmark{b}   &  293  &   147.9 &	  7.8  &   0.6  & 0.63   & 10.2  & 2-Apr-2016, 19-May-2016 \\  
237501.0\tablefootmark{~}   &  212  &   102.9 &   7.8  &   0.8  & 0.63   & 10.2  & 19-May-2016 \\[4pt] 
250086.8\tablefootmark{b}  &  196  &   158.9 &	  7.8  &   0.7  & 0.61   &  9.5  & 1-Apr-2016  \\   
250185.3\tablefootmark{b}   &  210  &   132.3 &	  7.8  &   0.6  & 0.61   &  9.5  & 1-Apr-2016  \\   
252581.7\tablefootmark{b}   &  280  &    13.2 &	  7.8  &   2.7  & 0.60   &  9.4  & 2-Apr-2016  \\   
252680.2\tablefootmark{b}   &  303  &	6.6 &	  7.8  &   4.0  & 0.60   &  9.4  & 2-Apr-2016  \\
253091.0\tablefootmark{b}  &  328  &   147.9 &	  7.8  &   0.7  & 0.60   &  9.4  & 2-Apr-2016, 19-May-2016 \\	
253189.5\tablefootmark{~}  &  229  &   102.9 &    7.8  &   0.6  & 0.60   &  9.4  & 19-May-2016 \\
\hline
\end{tabular}
\tablefoot{
\tablefoottext{a}{The rms noise level per 40\,\kms\ bin.}
\tablefoottext{b}{Spectral range overlaps with the next setup in the table.}
}
\end{table*}

\section{An atlas of IRAM and APEX spectra}
\clearpage
\begin{figure*} [tbh]
\centering
\includegraphics[angle=270,width=0.85\textwidth]{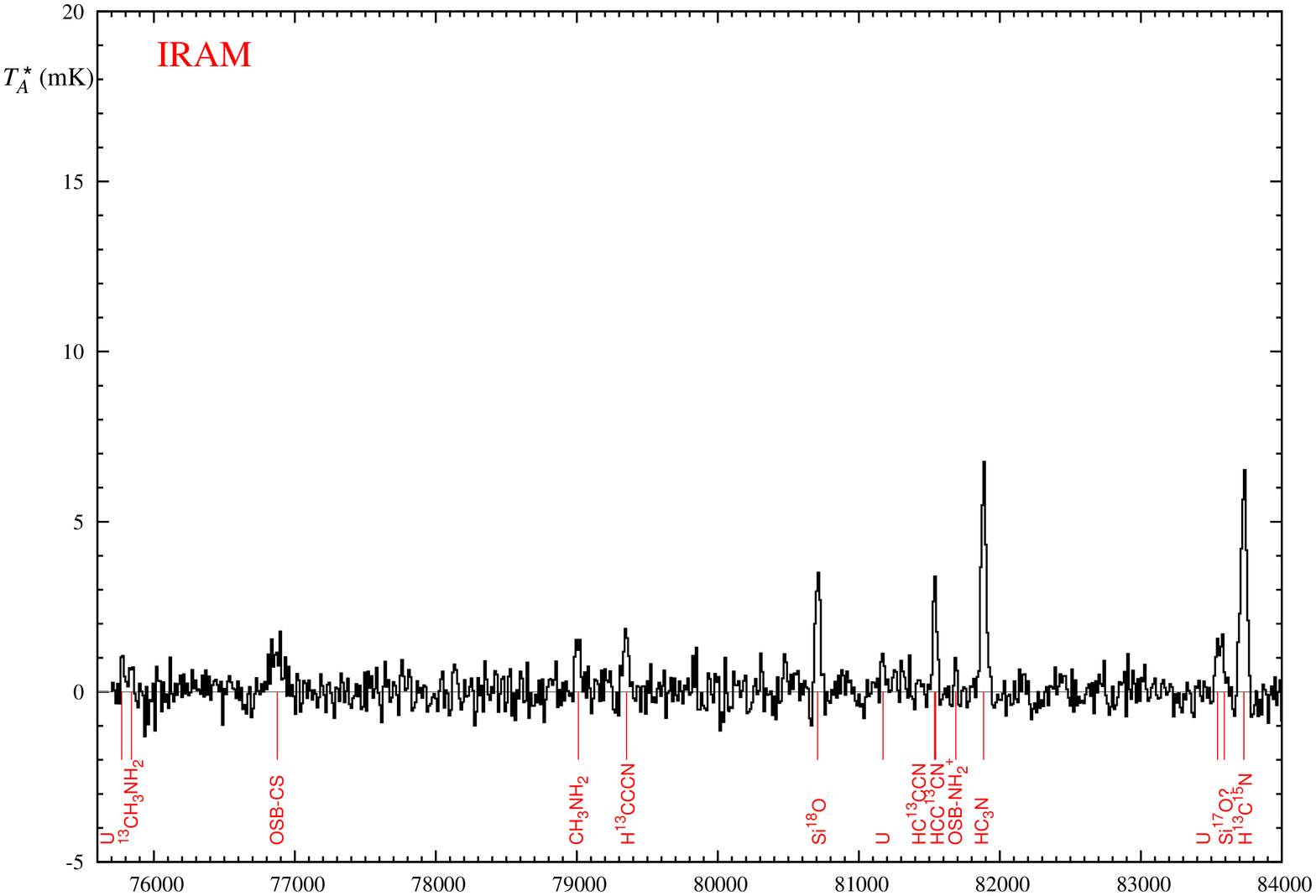}
\includegraphics[angle=270,width=0.85\textwidth]{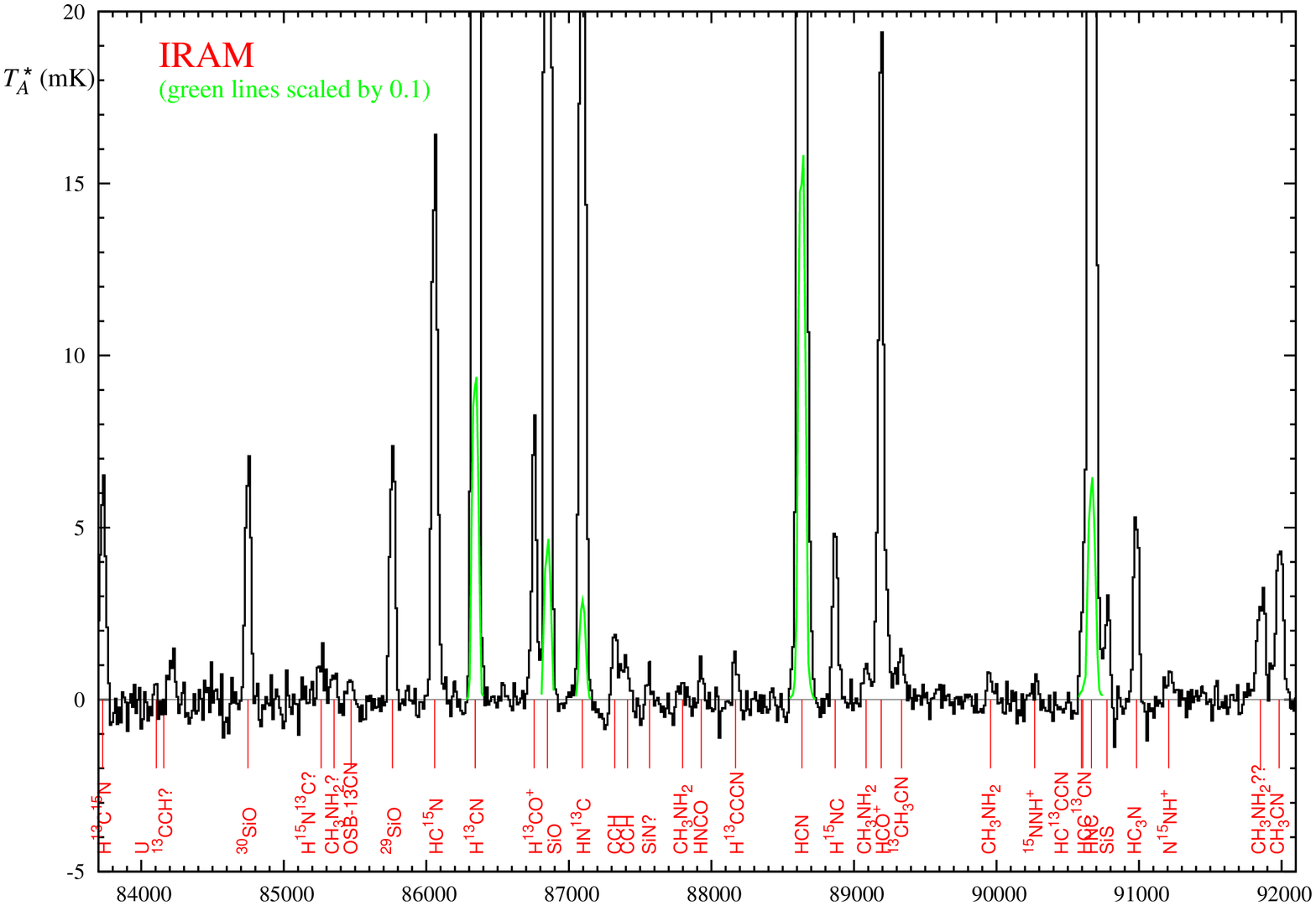}
\caption{IRAM spectra (black) and assigned spectral features. Shown with green line are profiles of bright lines scaled by 0.1. The spectra are displayed in rest frequency (i.e. are corrected for $V_{LSR}$=--10\,\kms) in MHz and in the the antenna-temperature scale in mK.}\label{atlas-iram}
\end{figure*}

  \setcounter{figure}{0}%

\begin{figure*} [tbh]
\centering
\includegraphics[angle=270,width=0.85\textwidth]{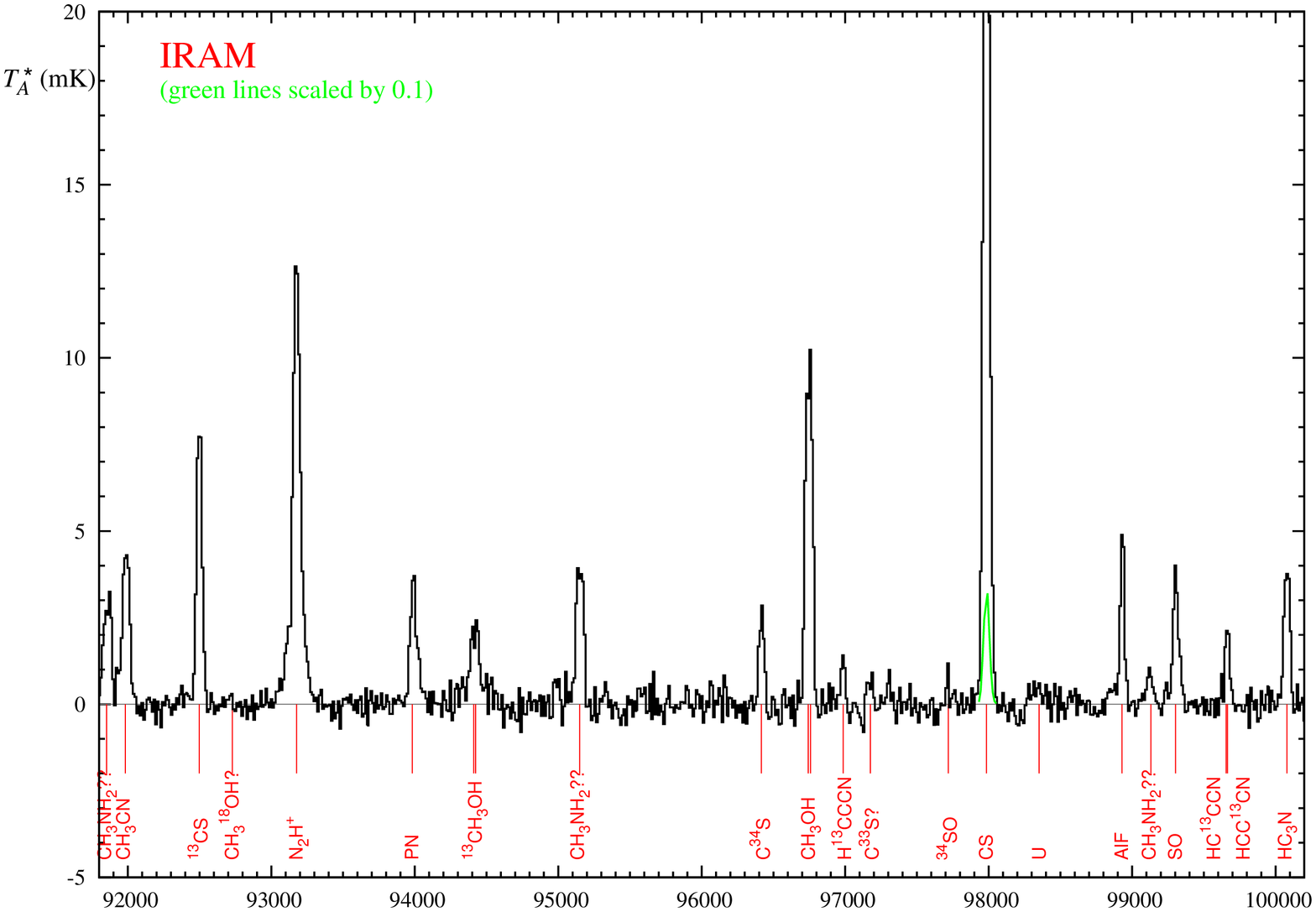}
\includegraphics[angle=270,width=0.85\textwidth]{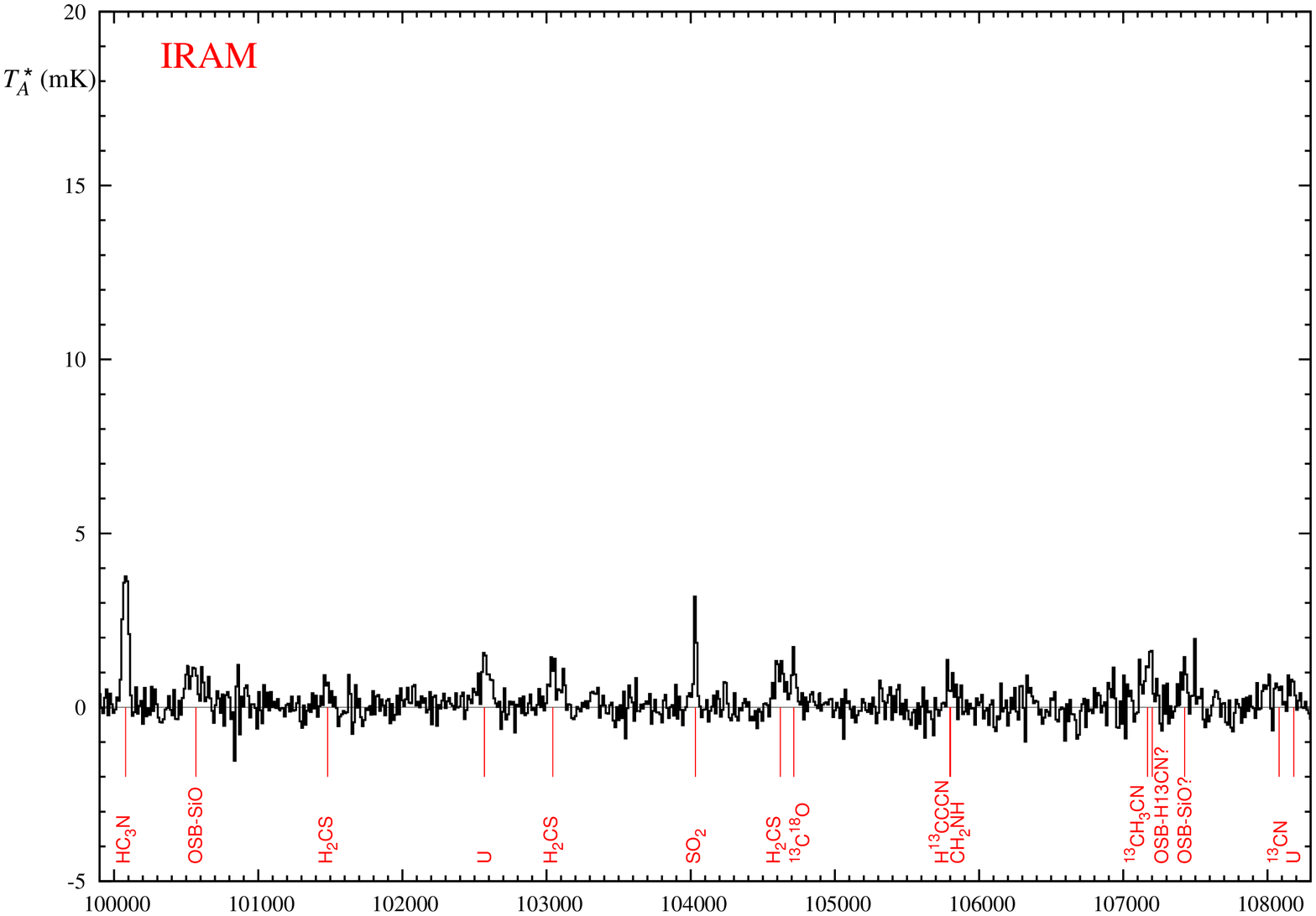}
\caption{Continued.}
\end{figure*}

  \setcounter{figure}{0}%

\begin{figure*} [tbh]
\centering
\includegraphics[angle=270,width=0.85\textwidth]{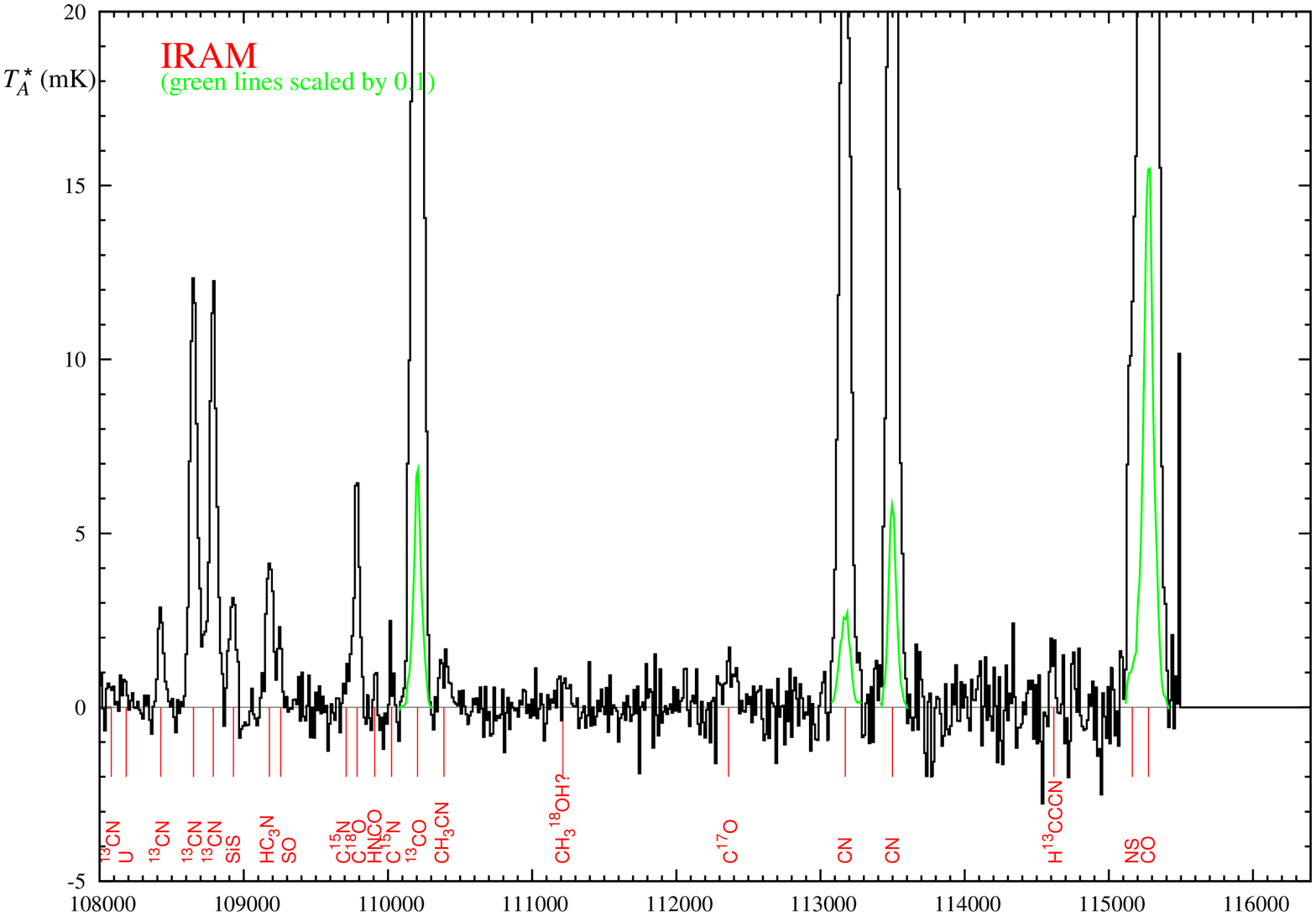}
\includegraphics[angle=270,width=0.85\textwidth]{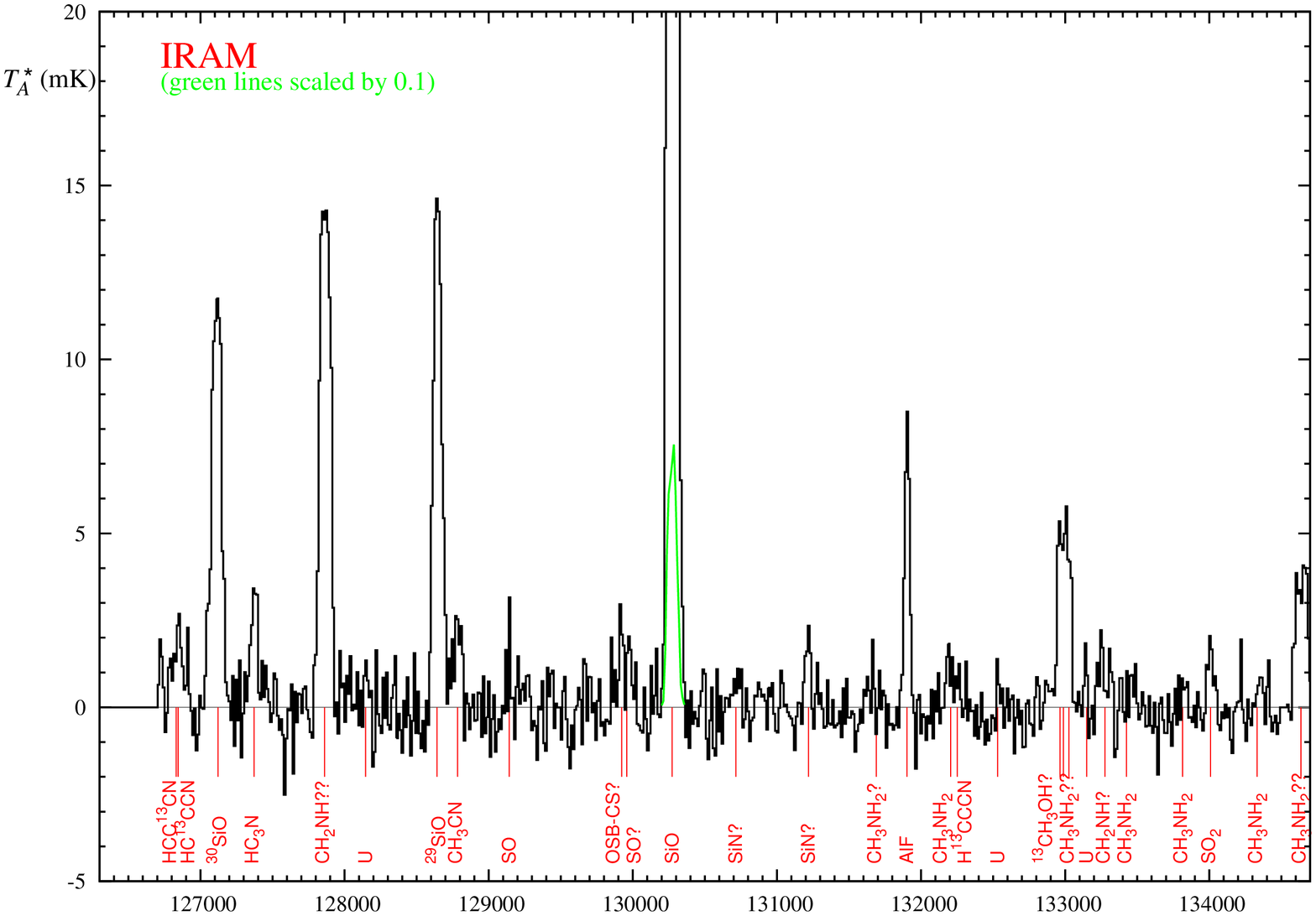}
\caption{Continued.}
\end{figure*}

  \setcounter{figure}{0}%

\begin{figure*} [tbh]
\centering
\includegraphics[angle=270,width=0.85\textwidth]{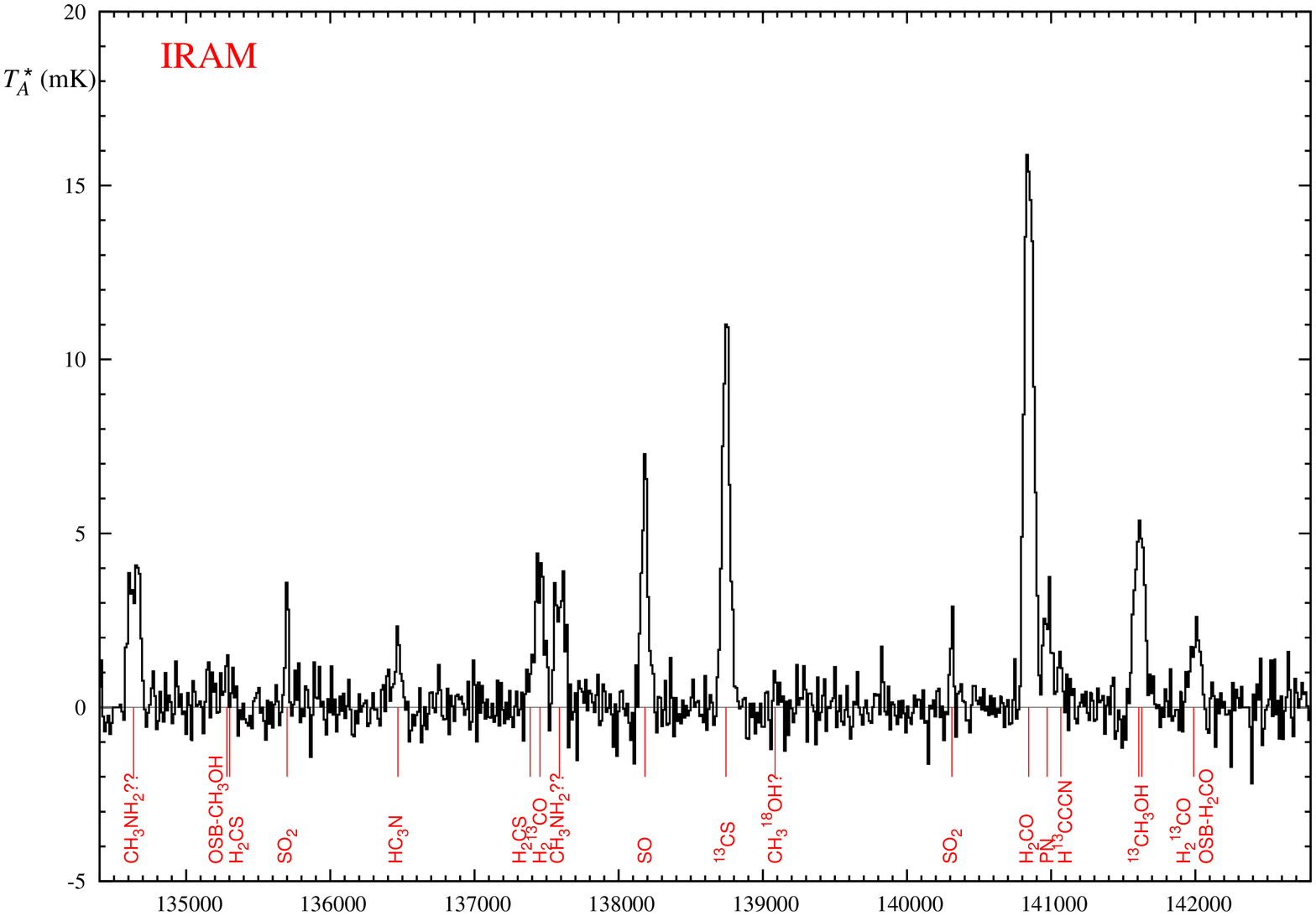}
\includegraphics[angle=270,width=0.85\textwidth]{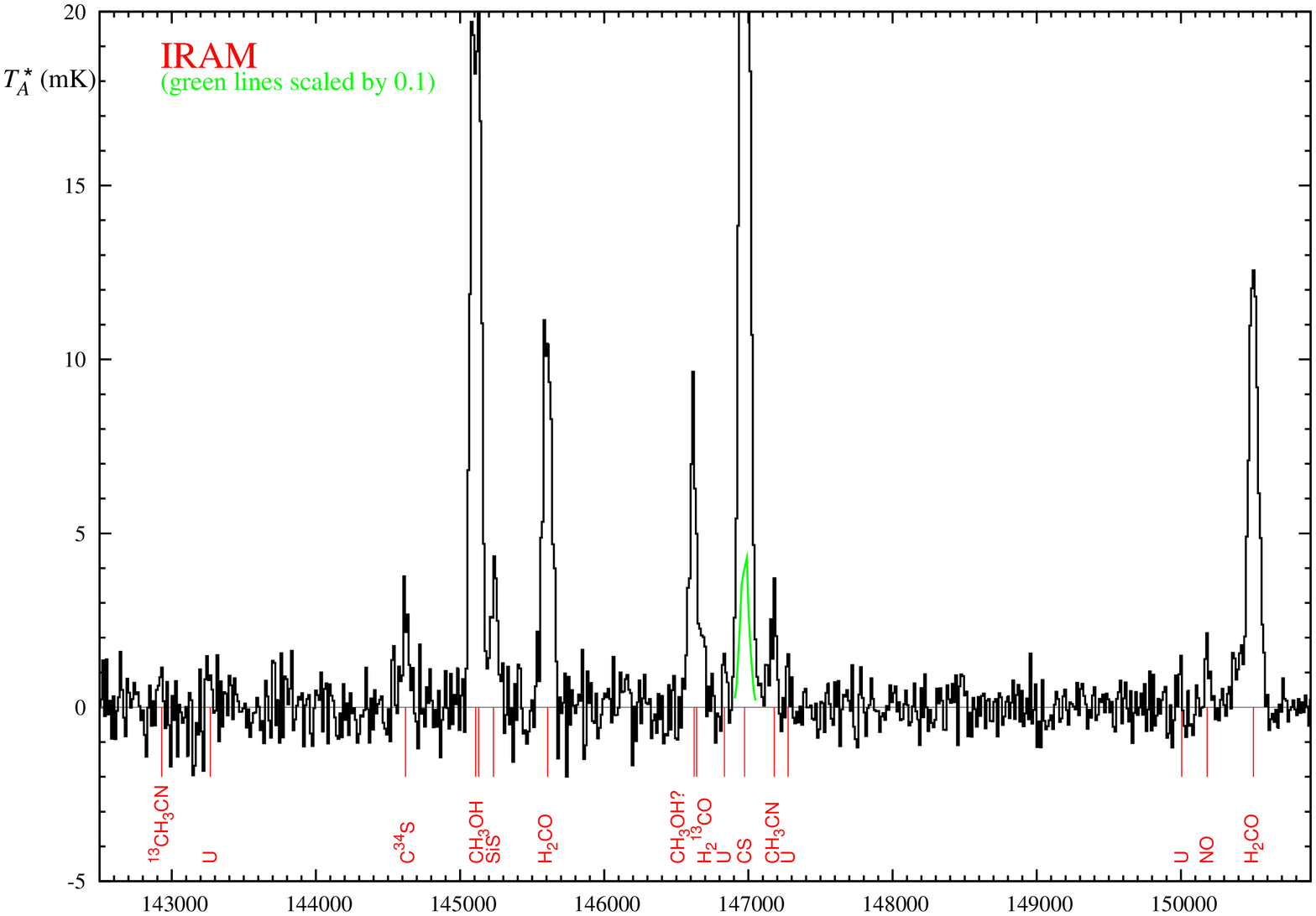}
\caption{Continued.}
\end{figure*}

  \setcounter{figure}{0}%

\begin{figure*} [tbh]
\centering
\includegraphics[angle=270,width=0.85\textwidth]{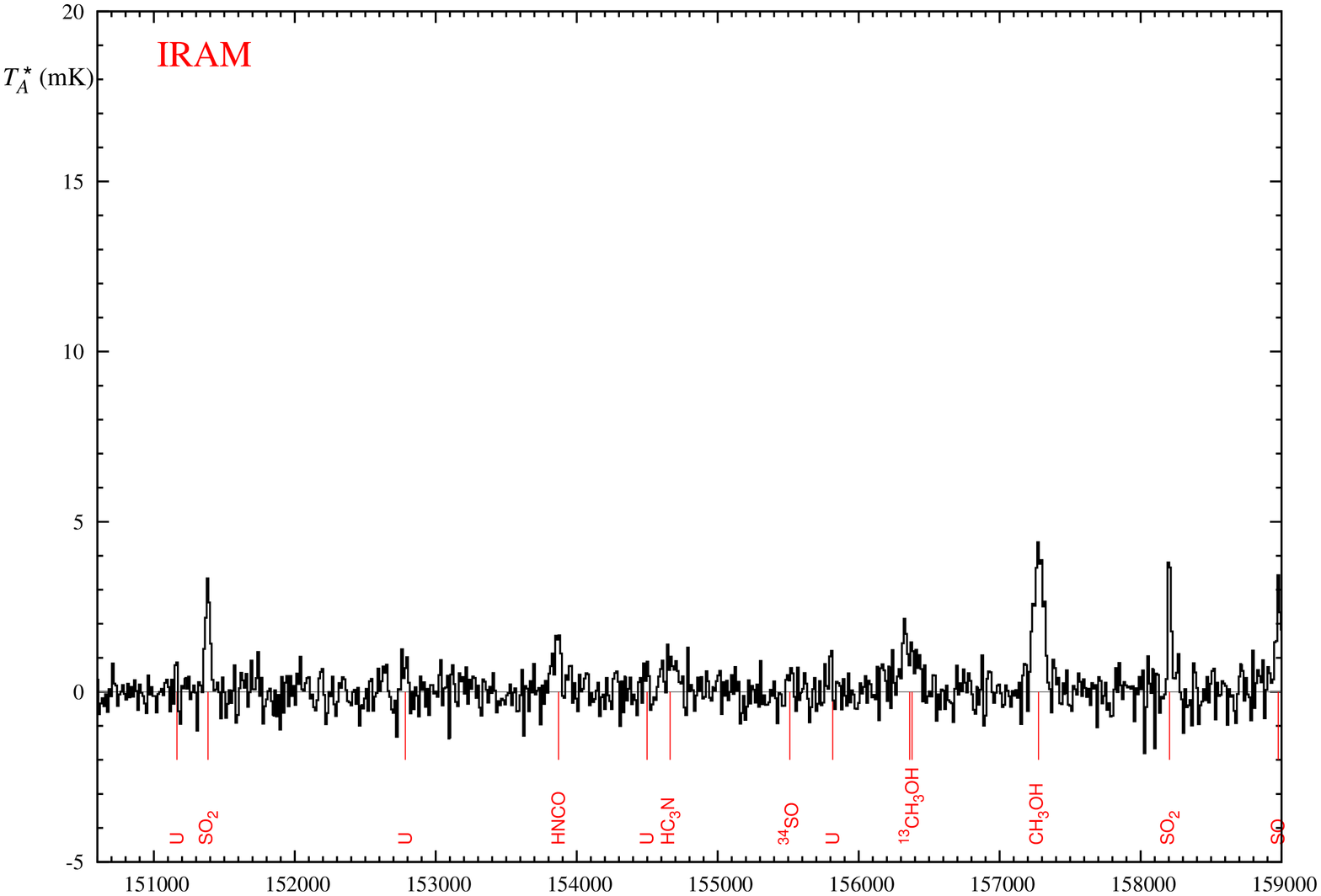}
\includegraphics[angle=270,width=0.85\textwidth]{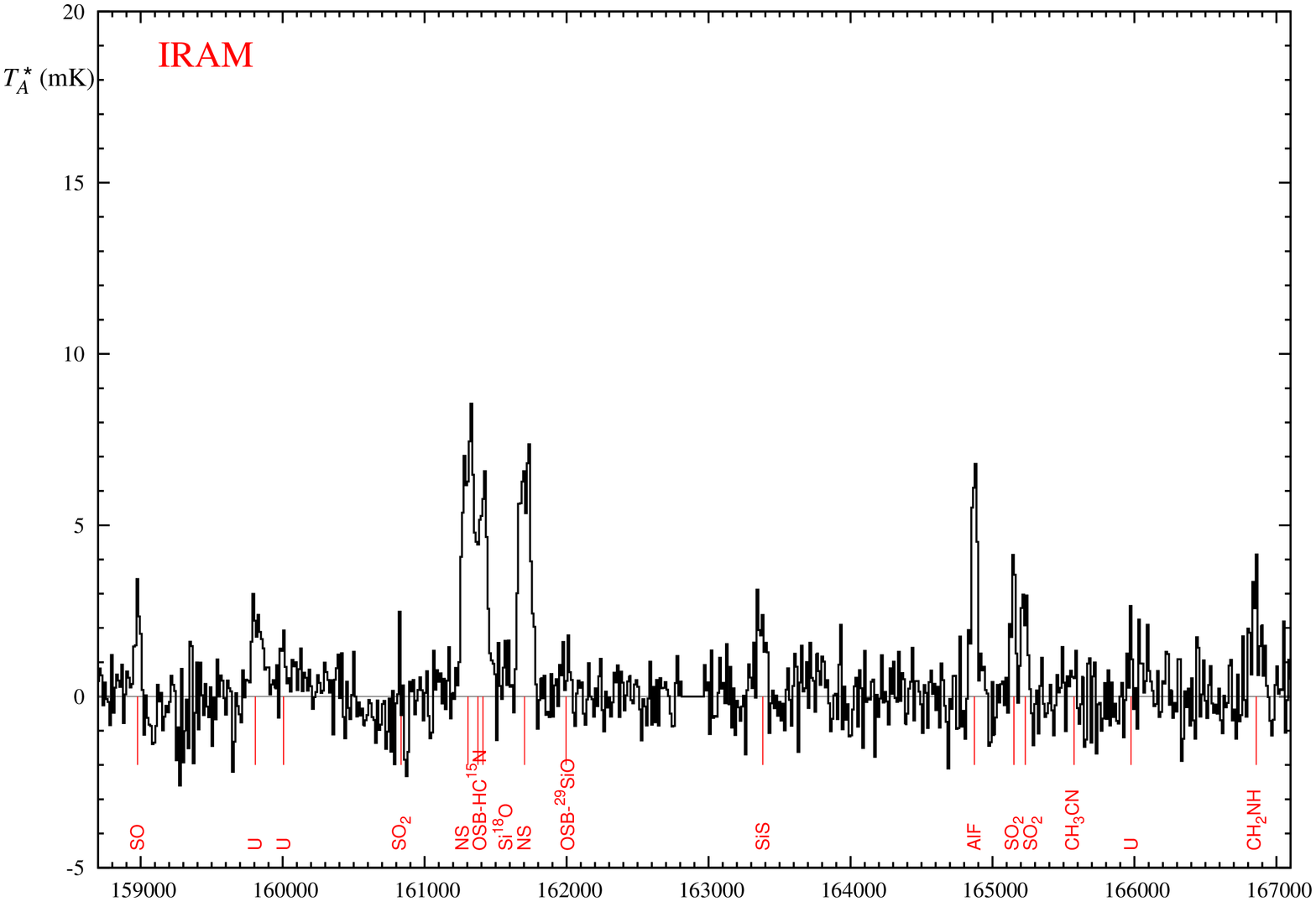}
\caption{Continued.}
\end{figure*}

  \setcounter{figure}{0}%

\begin{figure*} [tbh]
\centering
\includegraphics[angle=270,width=0.85\textwidth]{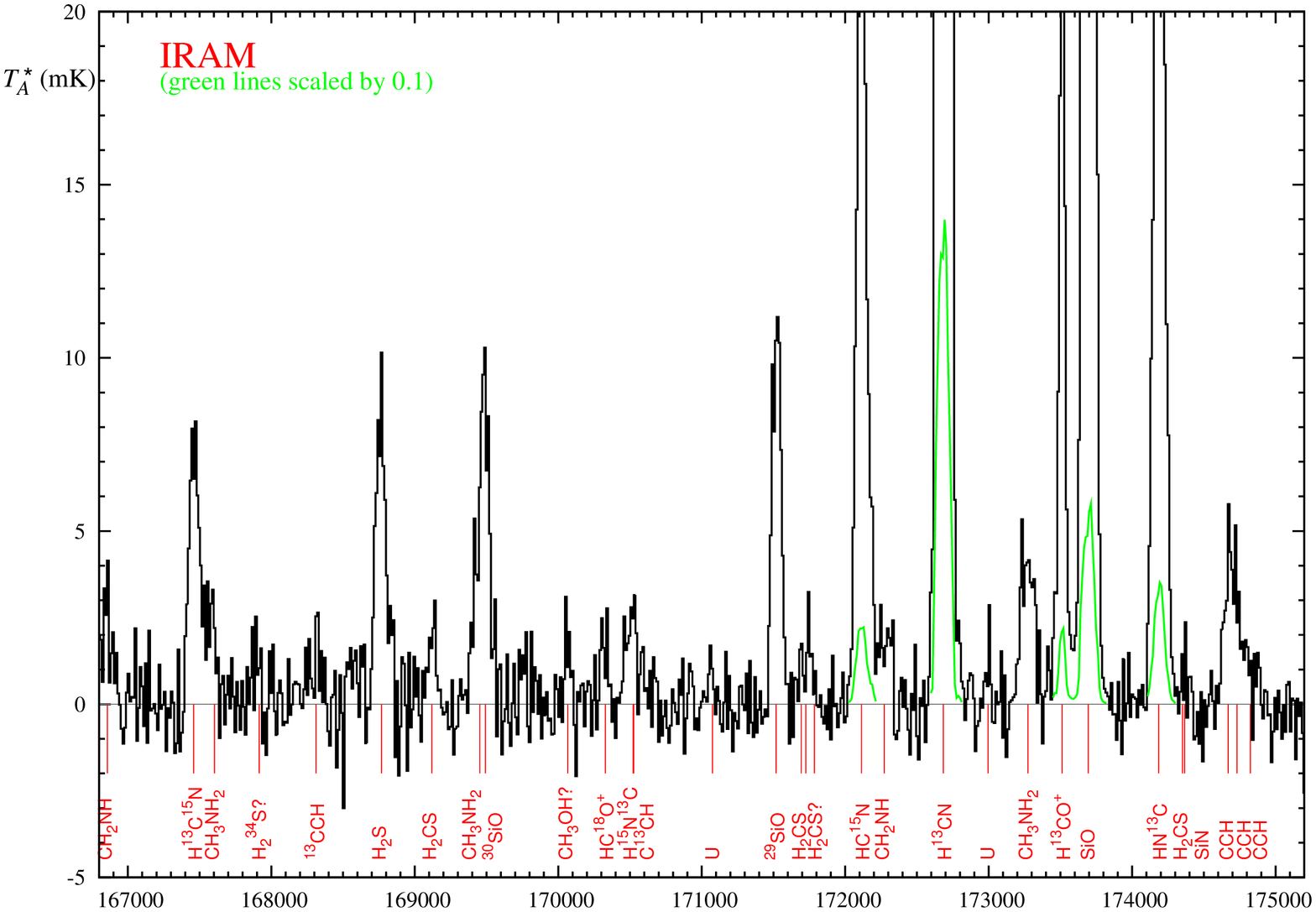}
\includegraphics[angle=270,width=0.85\textwidth]{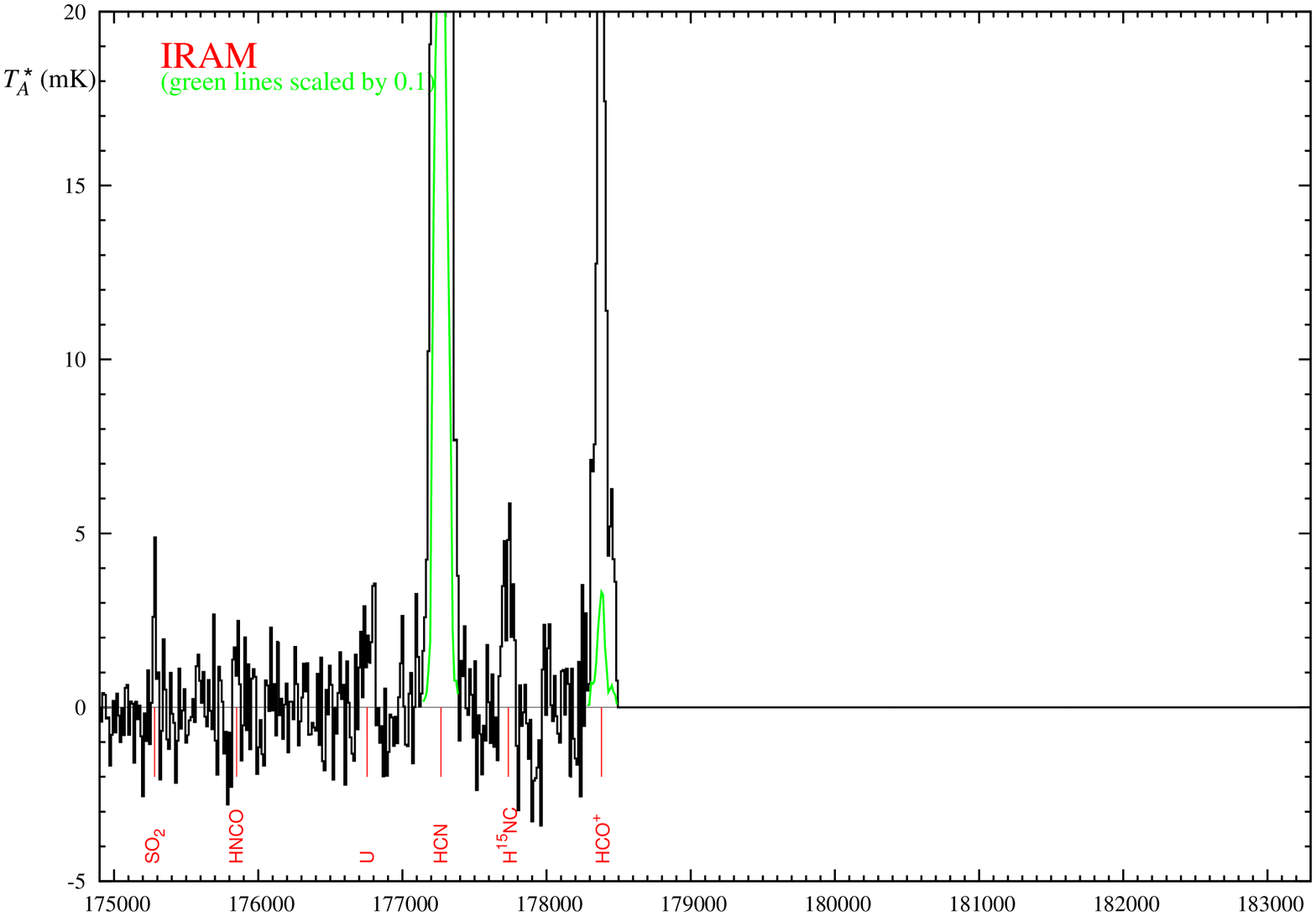}
\caption{Continued.}
\end{figure*}

  \setcounter{figure}{0}%

\begin{figure*} [tbh]
\centering
\includegraphics[angle=270,width=0.85\textwidth]{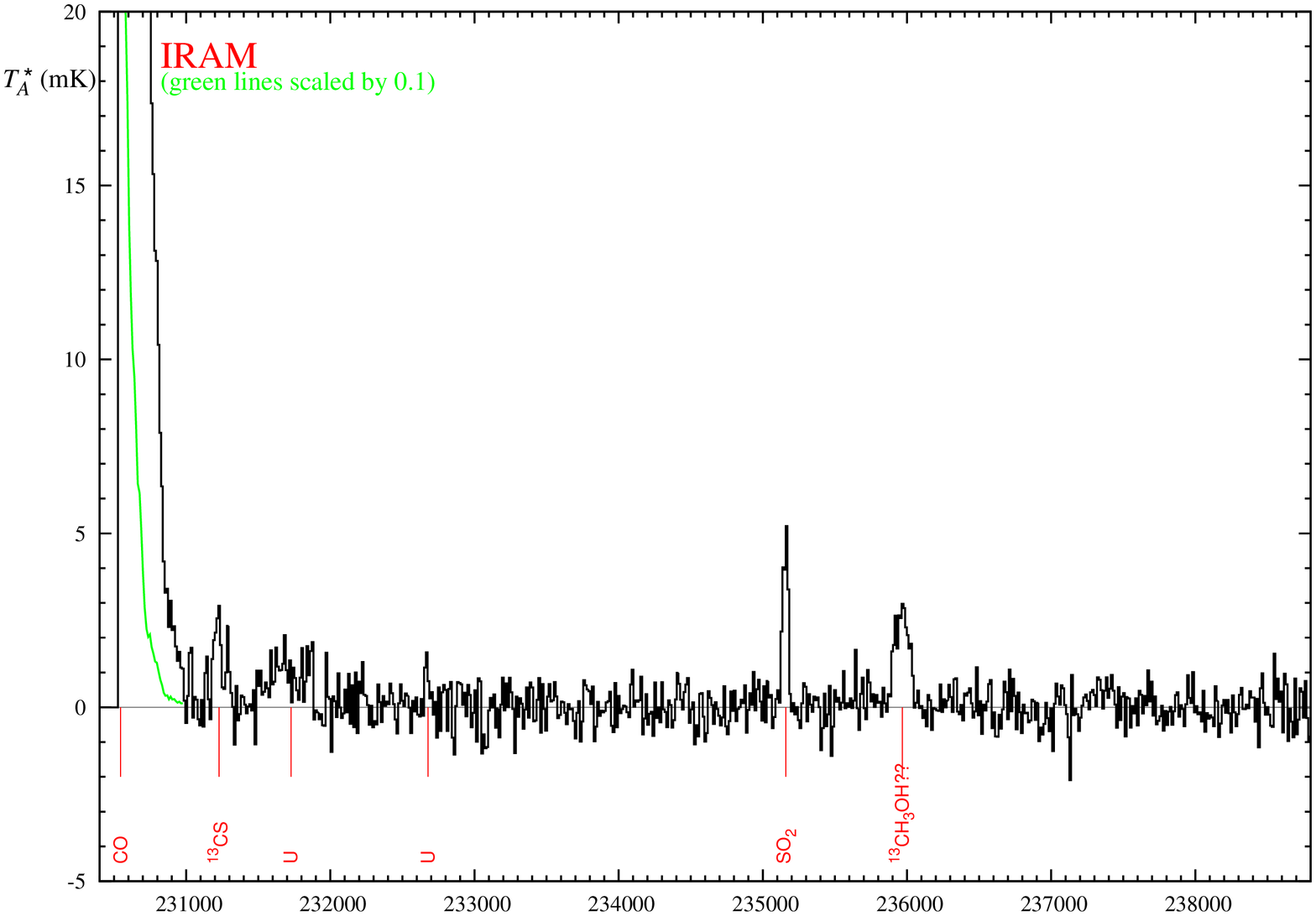}
\includegraphics[angle=270,width=0.85\textwidth]{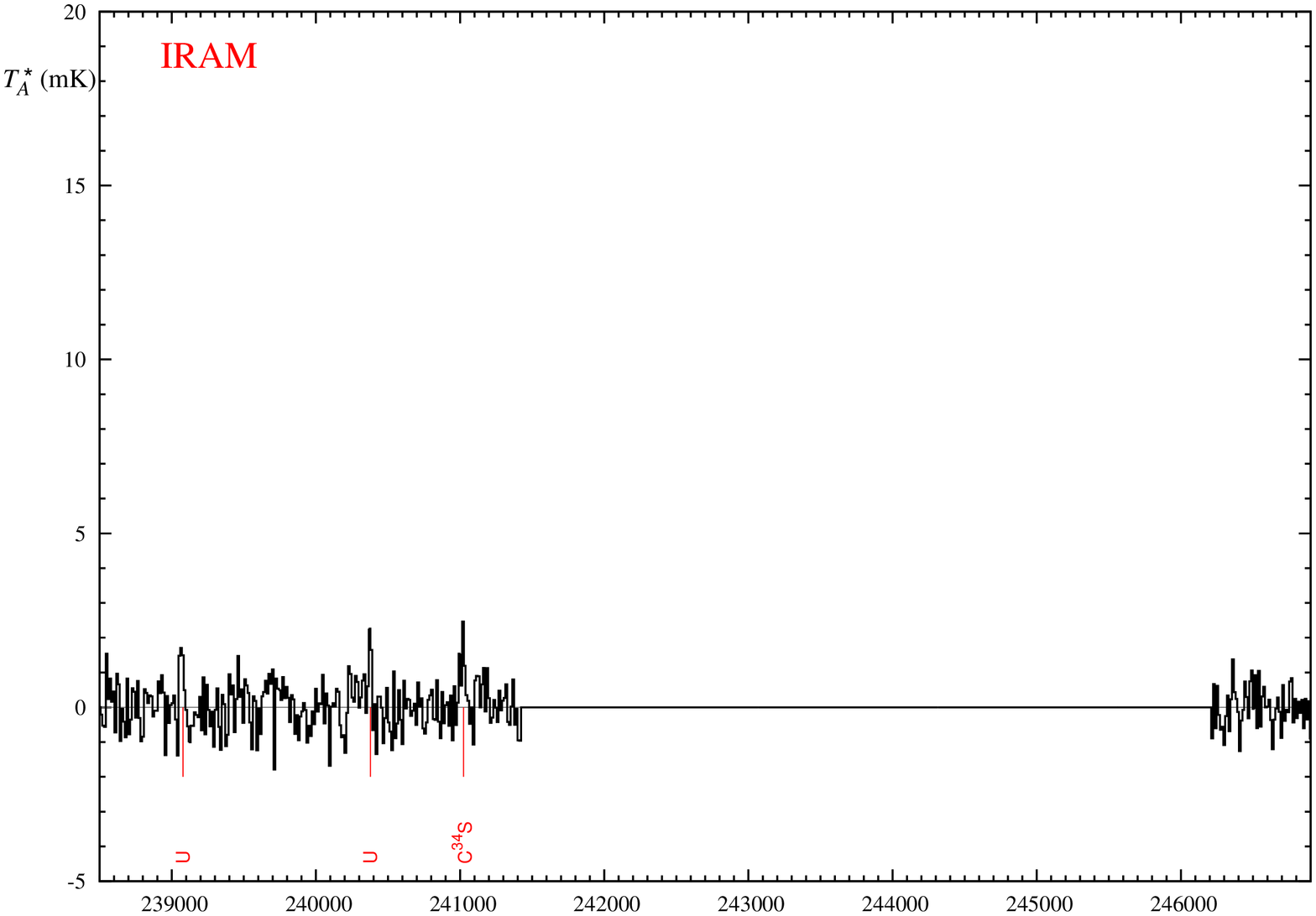}
\caption{Continued.}
\end{figure*}

  \setcounter{figure}{0}%

\begin{figure*} [tbh]
\centering
\includegraphics[angle=270,width=0.85\textwidth]{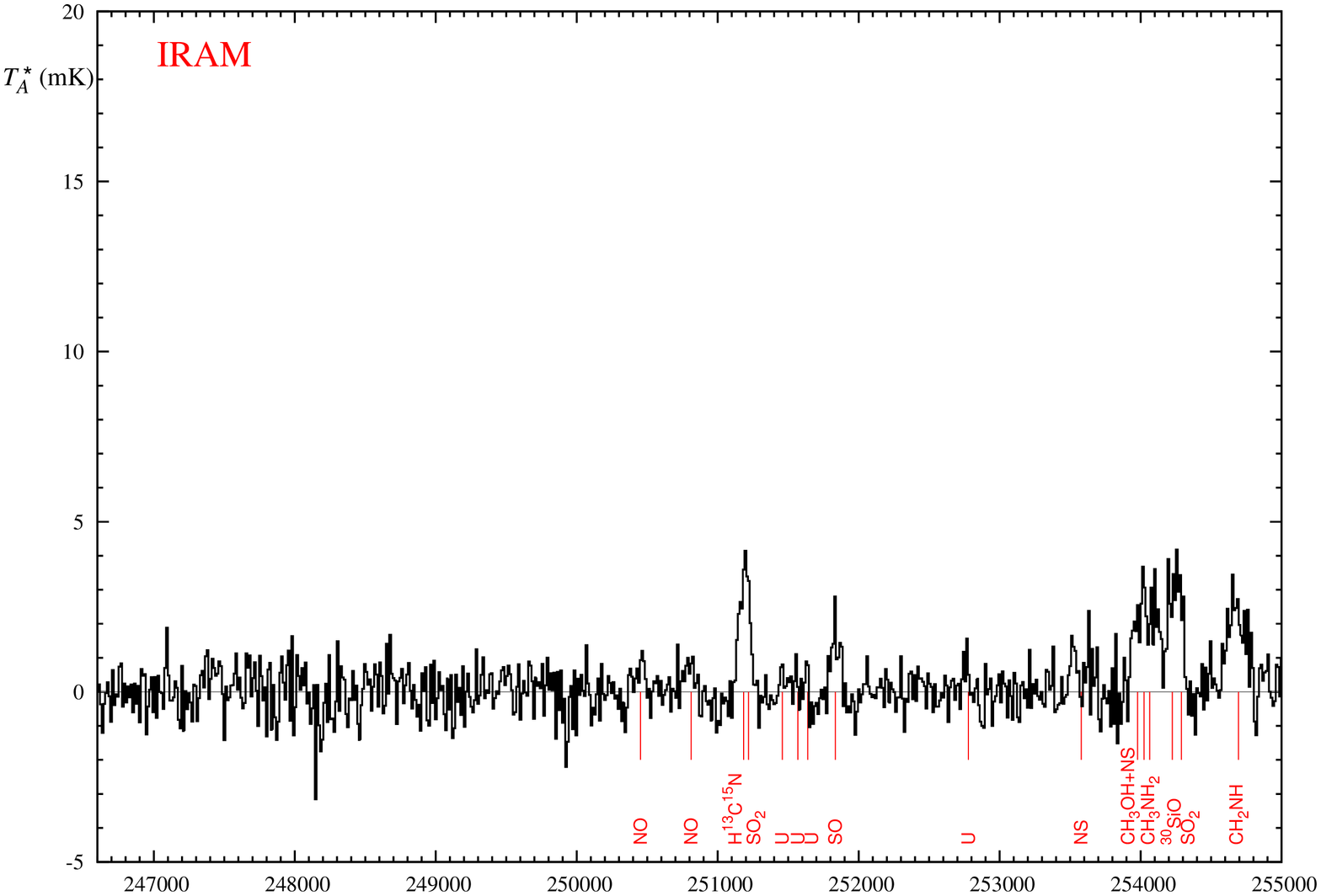}
\includegraphics[angle=270,width=0.85\textwidth]{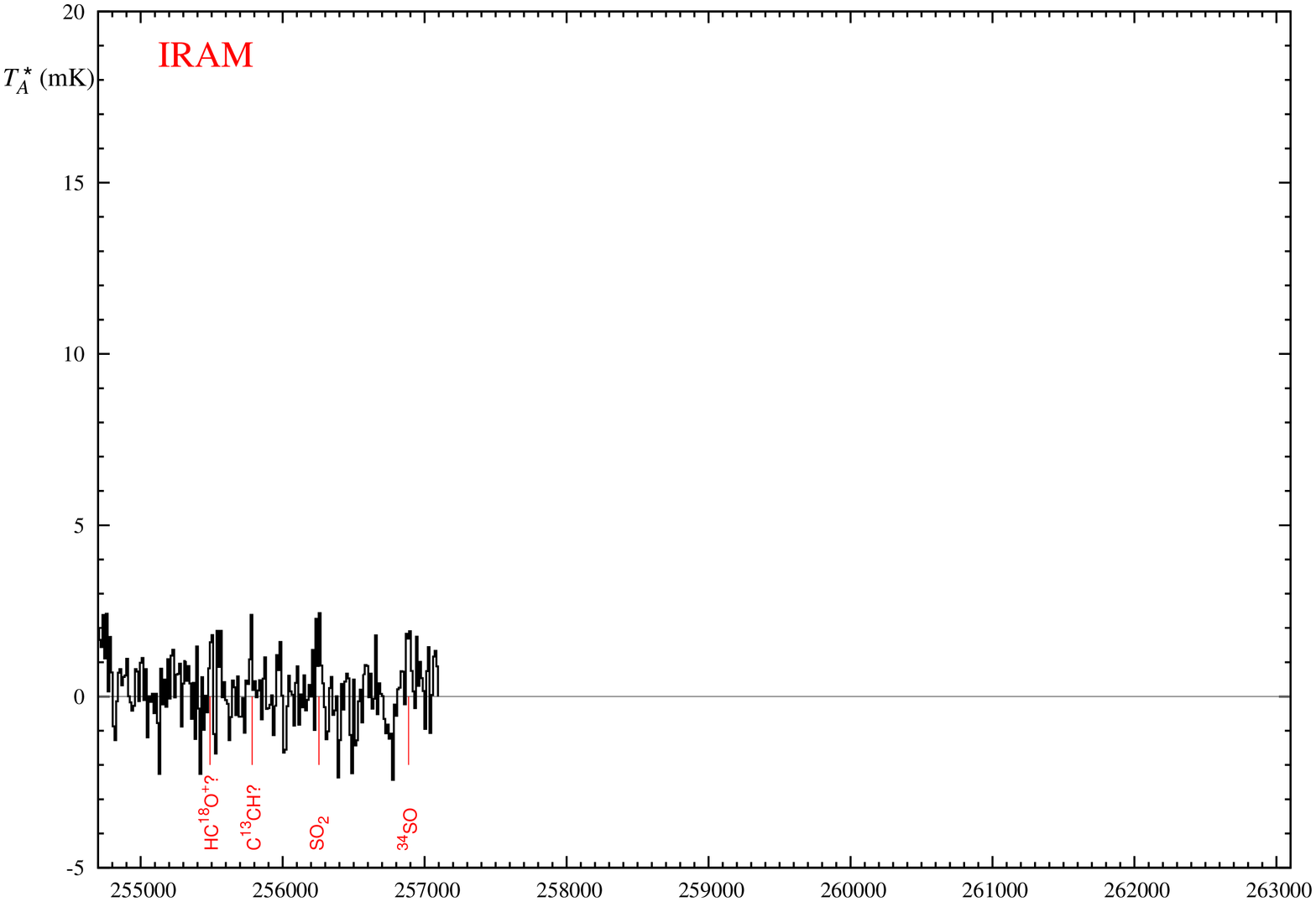}
\caption{Continued.}
\end{figure*}

  \setcounter{figure}{1}%

\clearpage
\begin{figure*} [tbh]
\centering
\includegraphics[angle=270,width=0.85\textwidth]{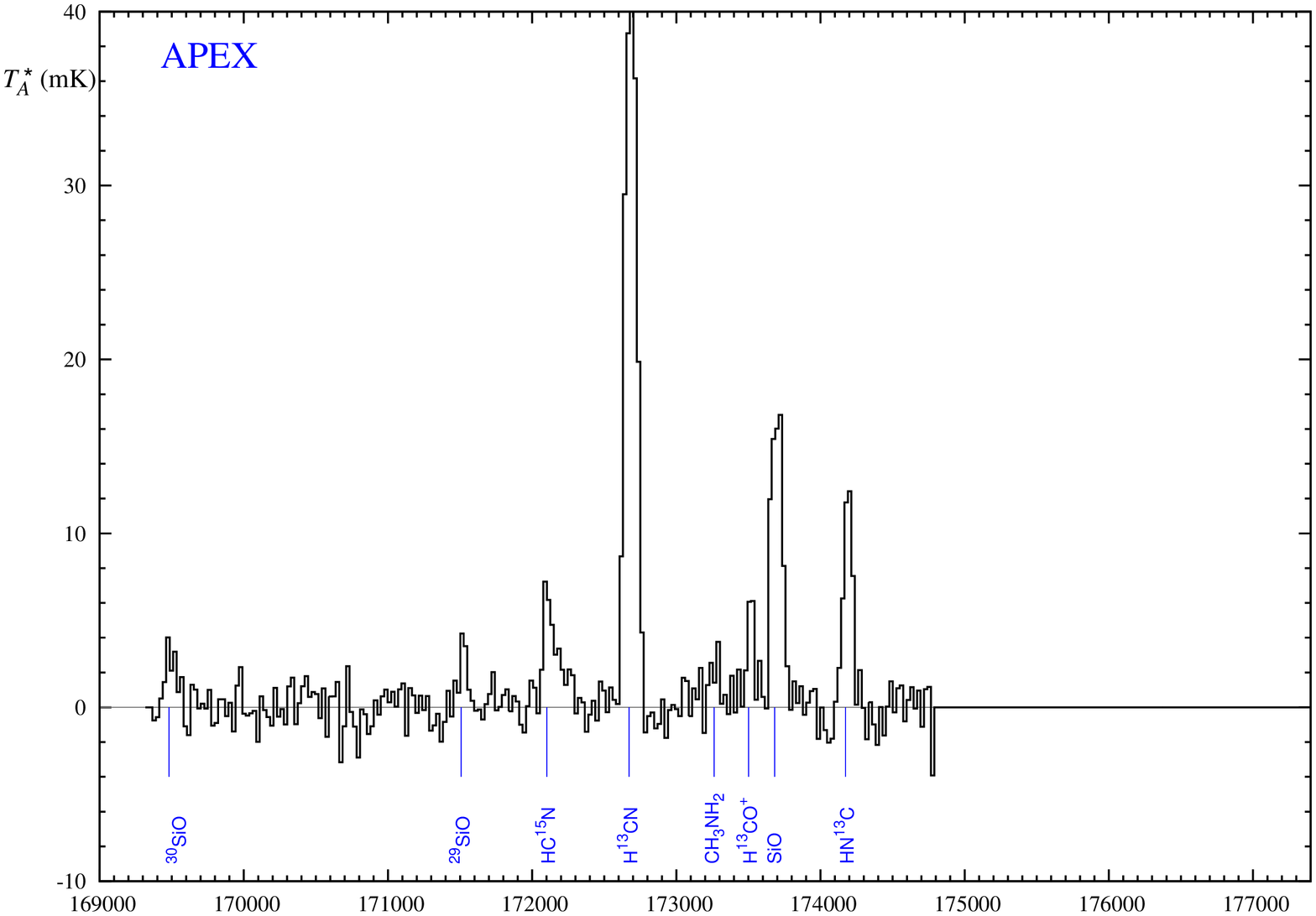}
\includegraphics[angle=270,width=0.85\textwidth]{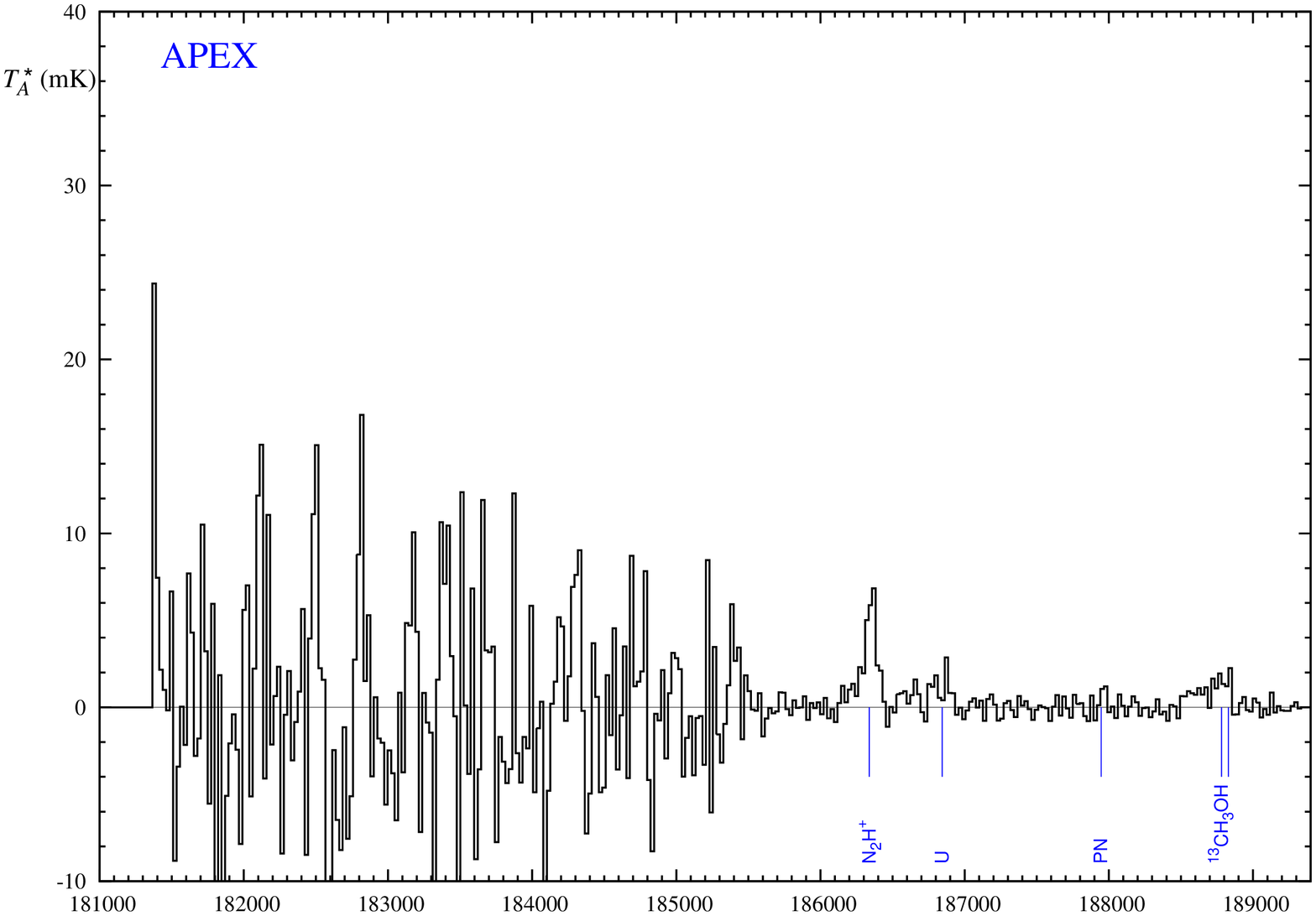}
\caption{Same as Fig.\,\ref{atlas-iram} but for APEX spectra. Magenta lines mark regions affected by telluric features or instrumental artifacts.}\label{atlas-apex}

\end{figure*}

  \setcounter{figure}{1}%

\begin{figure*} [tbh]
\centering
\includegraphics[angle=270,width=0.85\textwidth]{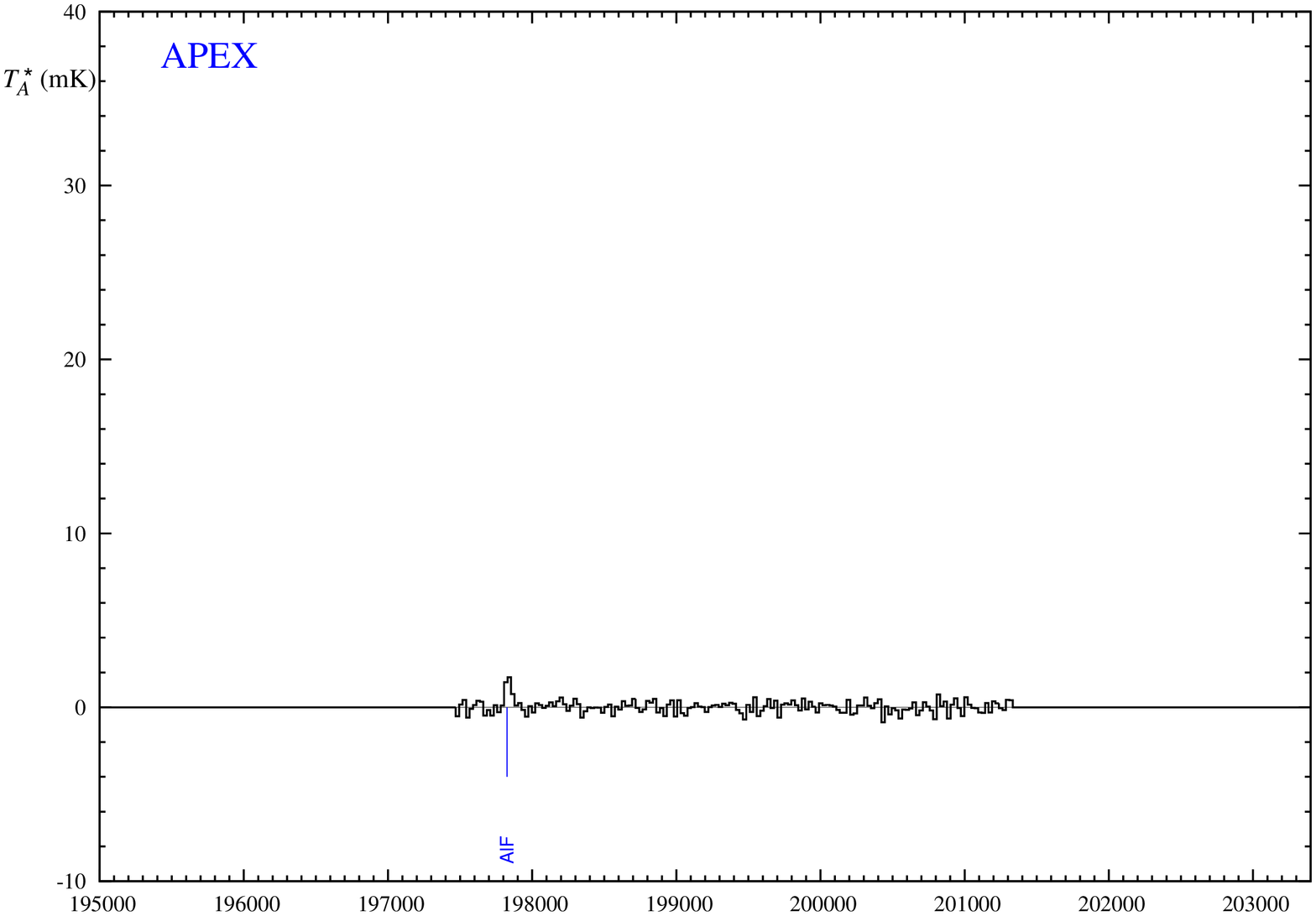}
\includegraphics[angle=270,width=0.85\textwidth]{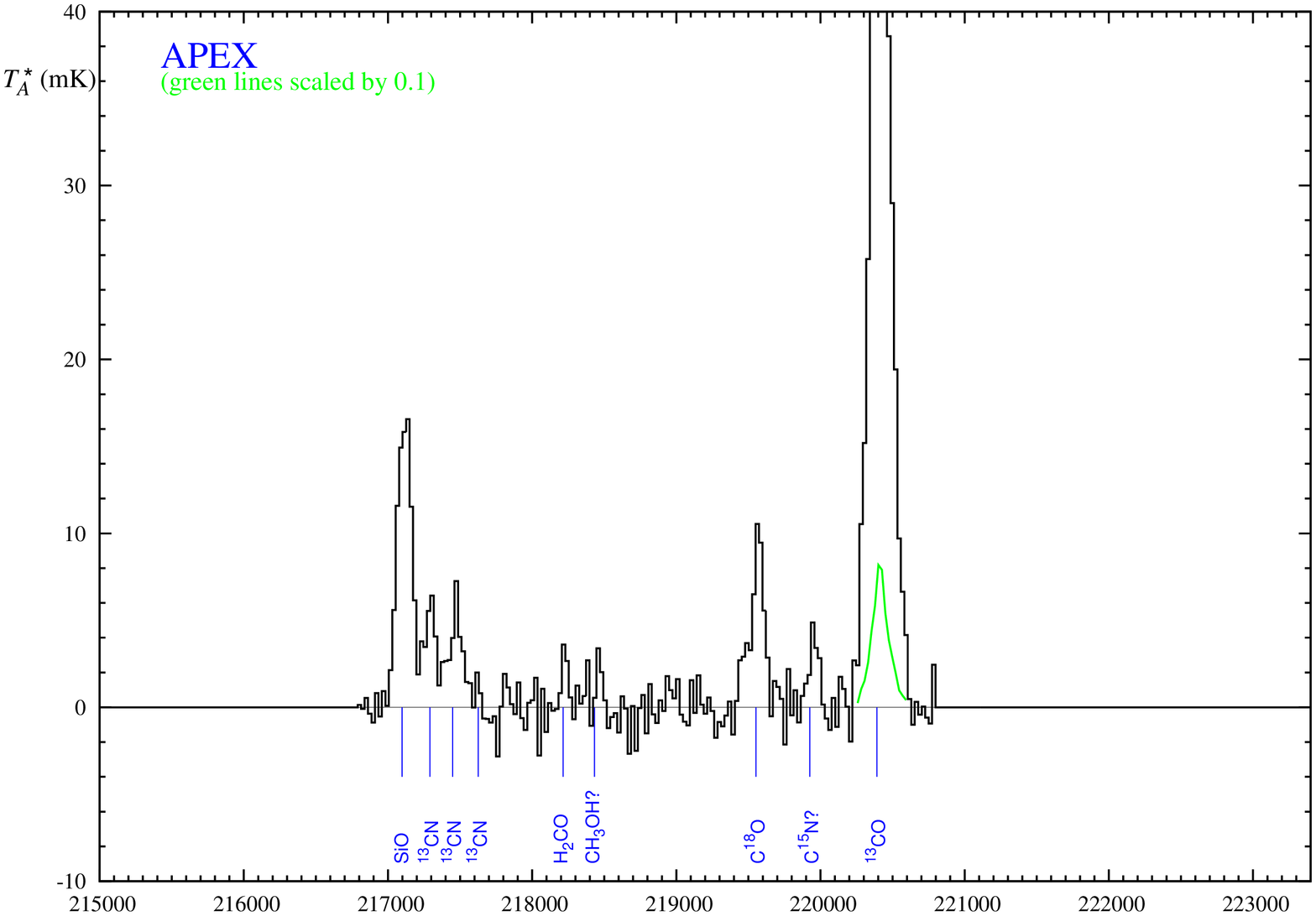}
\caption{Continued.}
\end{figure*}

  \setcounter{figure}{1}%

\begin{figure*} [tbh]
\centering
\includegraphics[angle=270,width=0.85\textwidth]{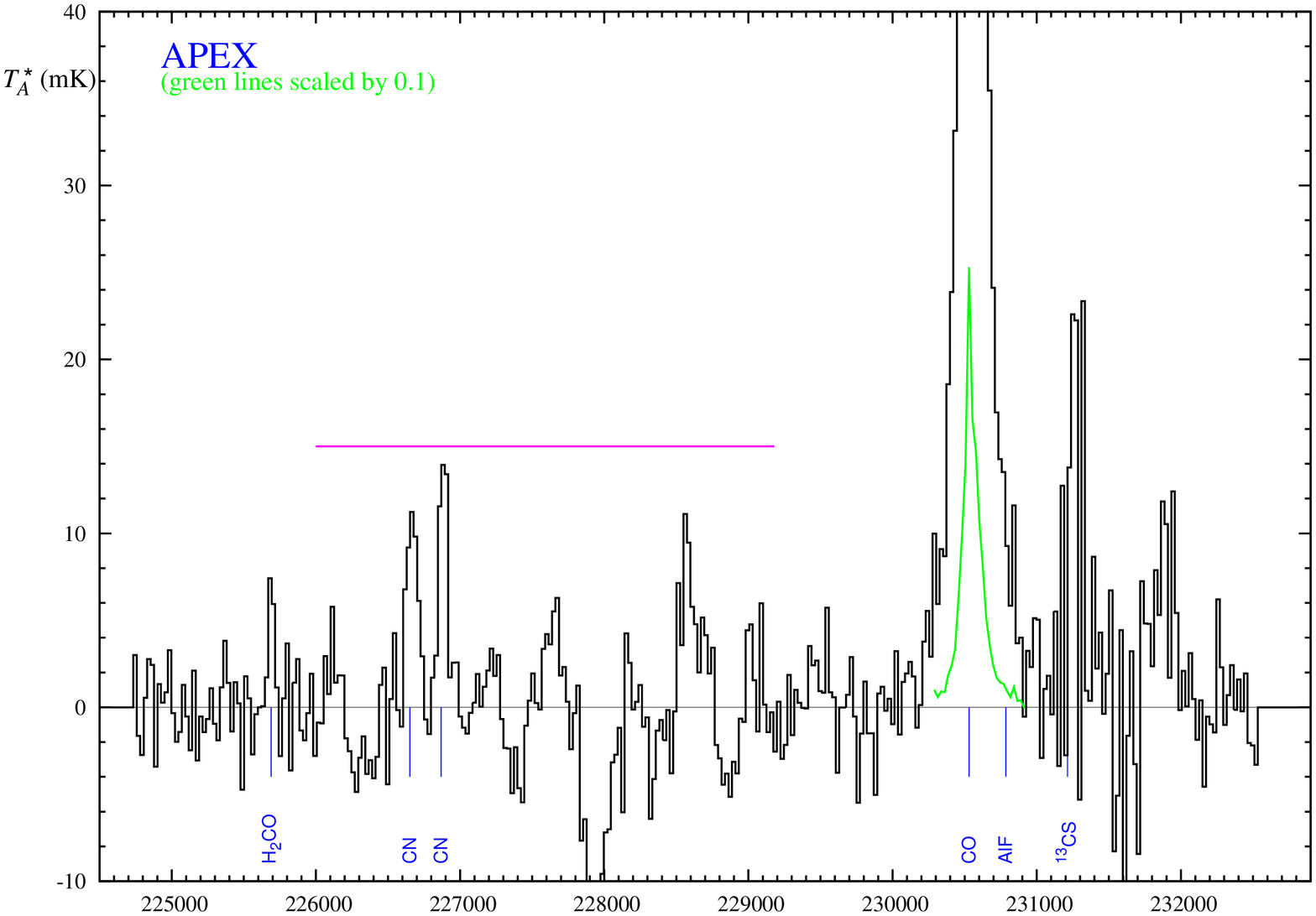}
\includegraphics[angle=270,width=0.85\textwidth]{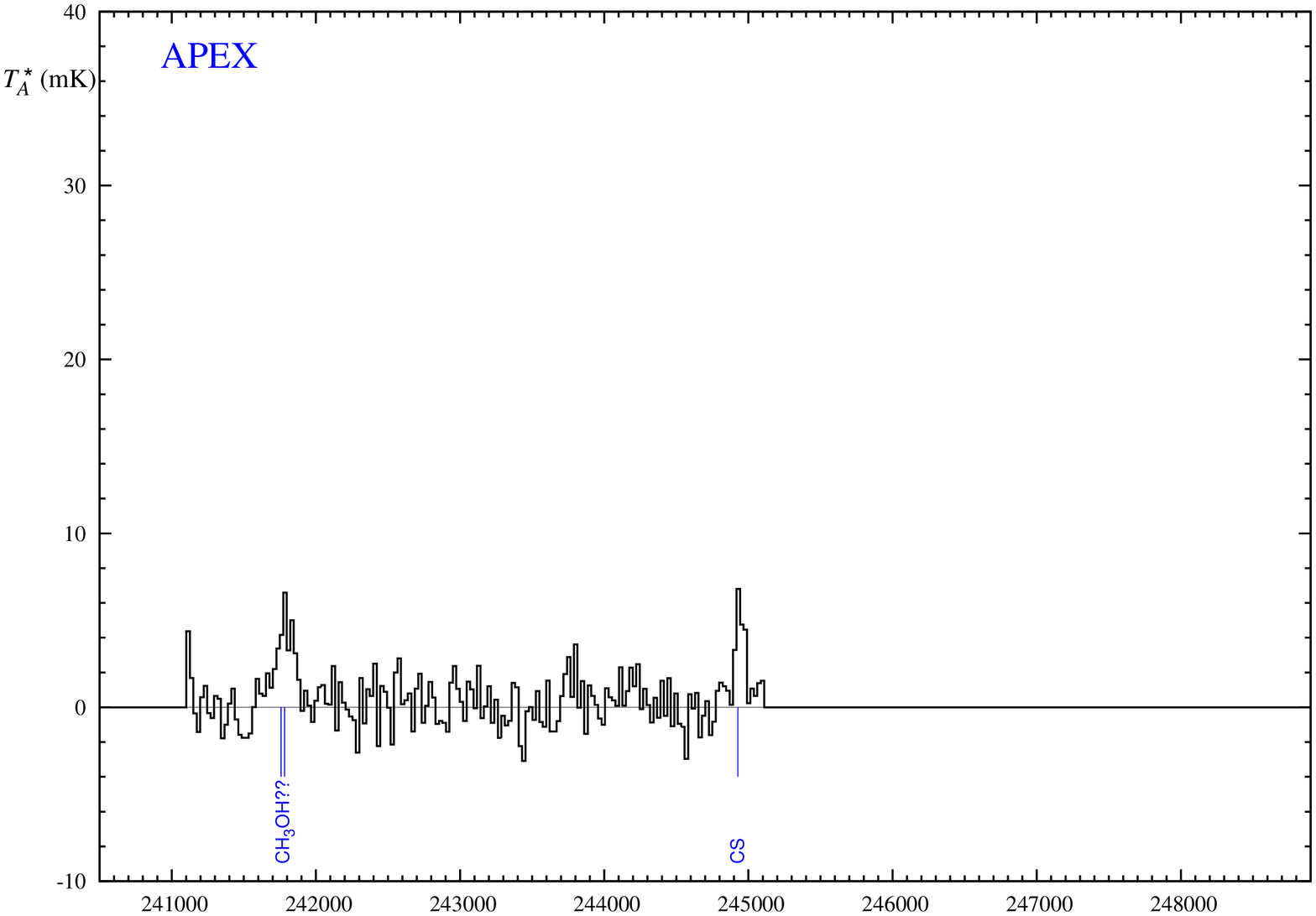}
\caption{Continued.}
\end{figure*}

  \setcounter{figure}{1}%

\begin{figure*} [tbh]
\centering
\includegraphics[angle=270,width=0.85\textwidth]{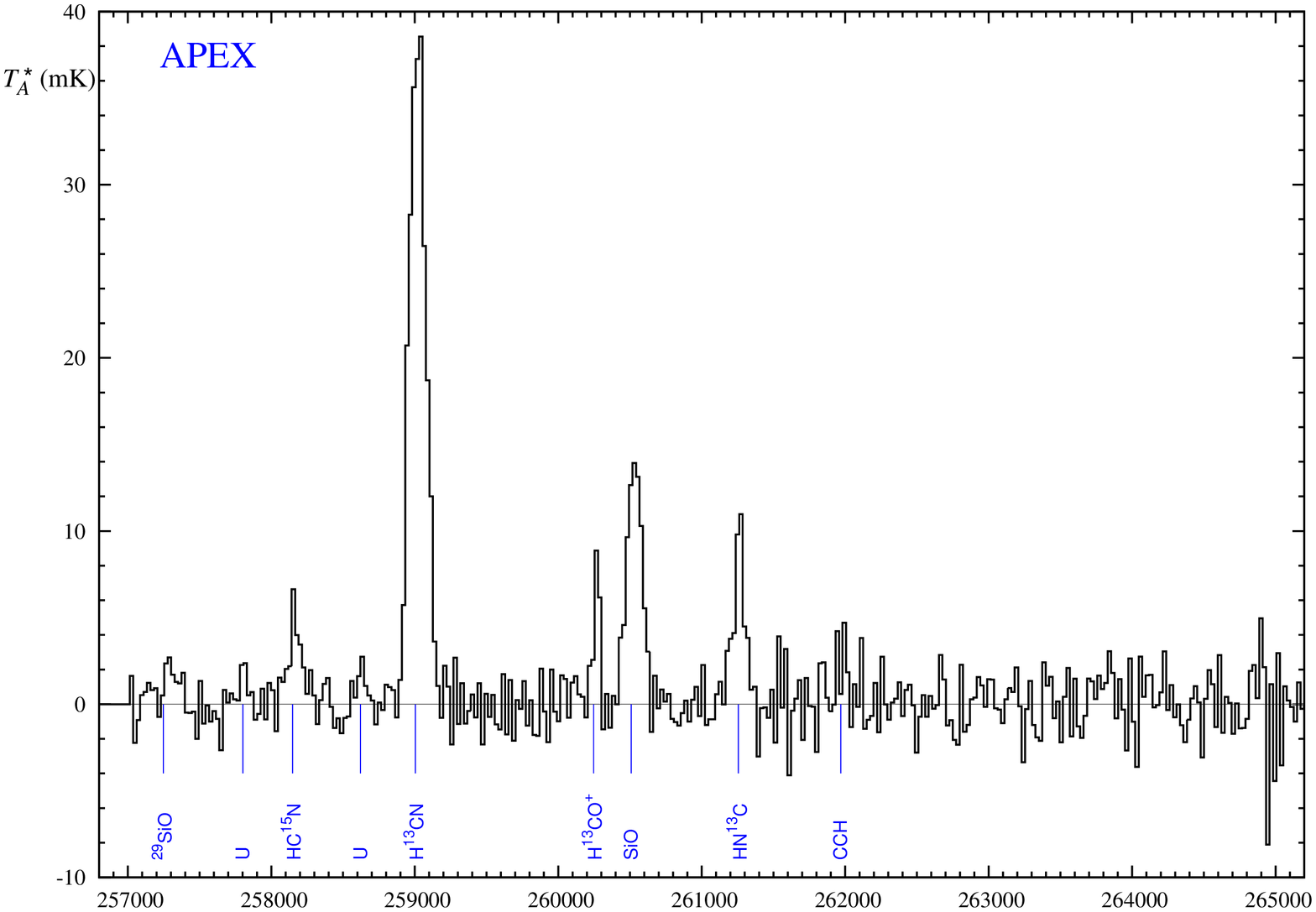}
\includegraphics[angle=270,width=0.85\textwidth]{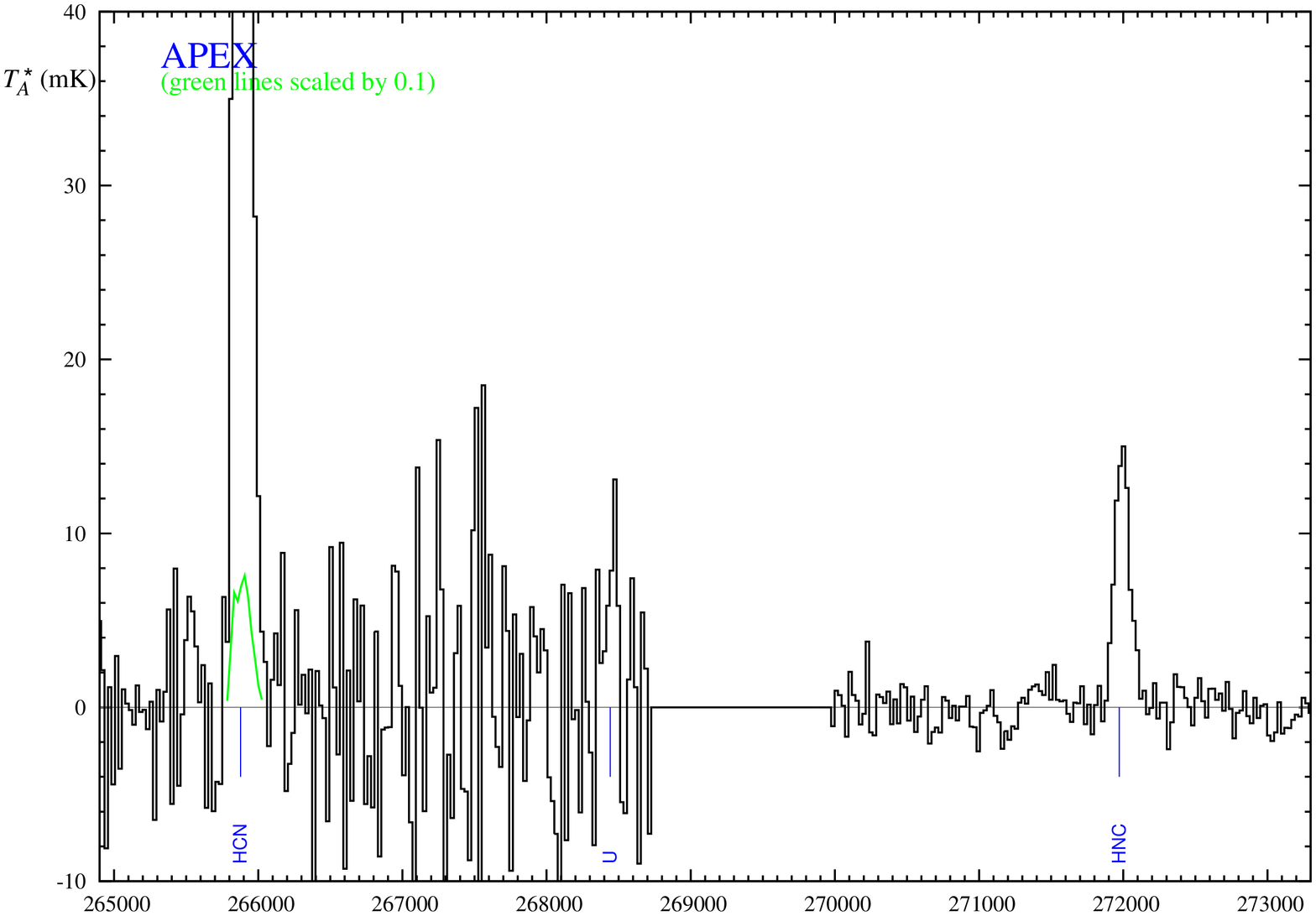}
\caption{Continued.}
\end{figure*}

  \setcounter{figure}{1}%

\begin{figure*} [tbh]
\centering
\includegraphics[angle=270,width=0.85\textwidth]{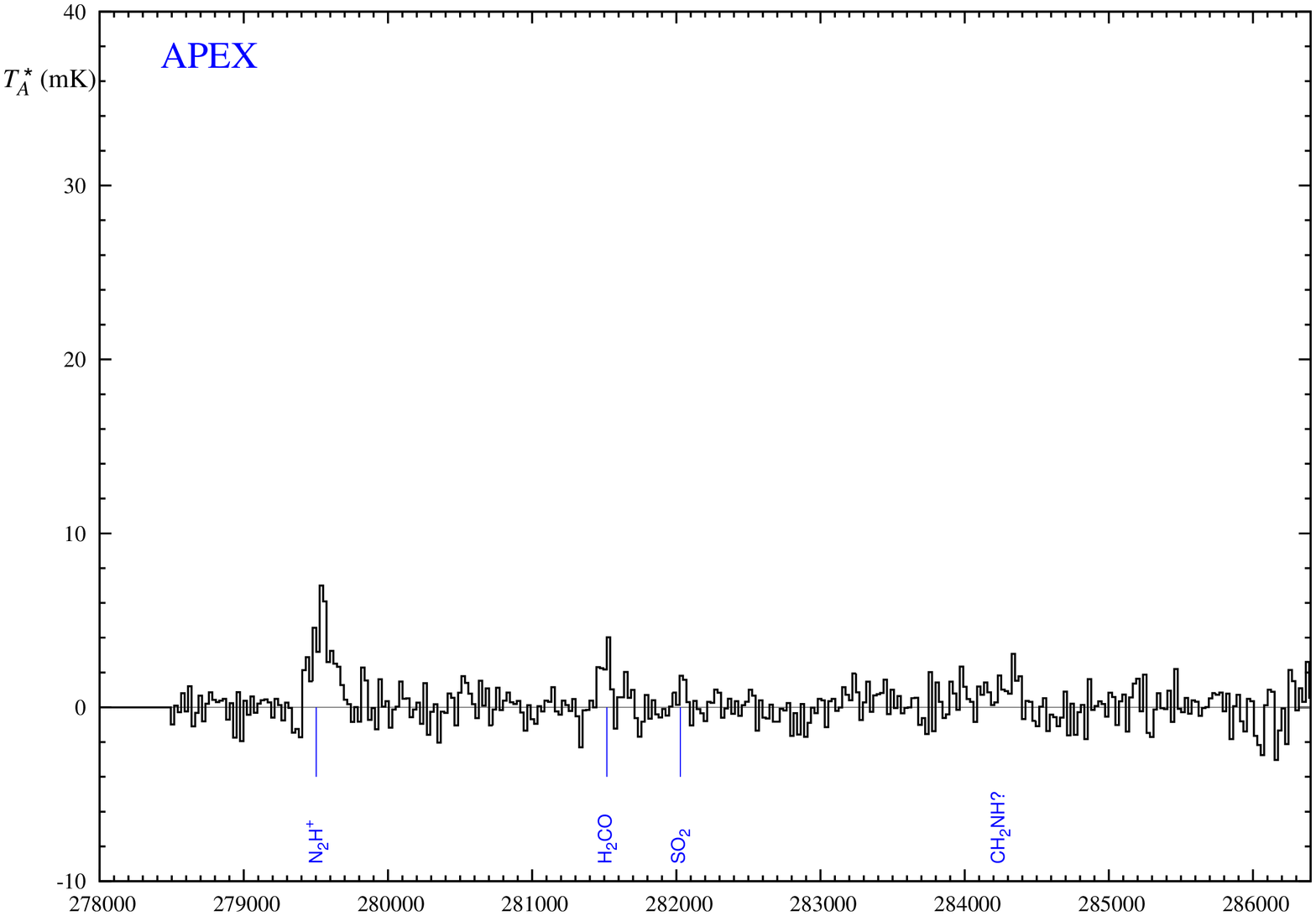}
\includegraphics[angle=270,width=0.85\textwidth]{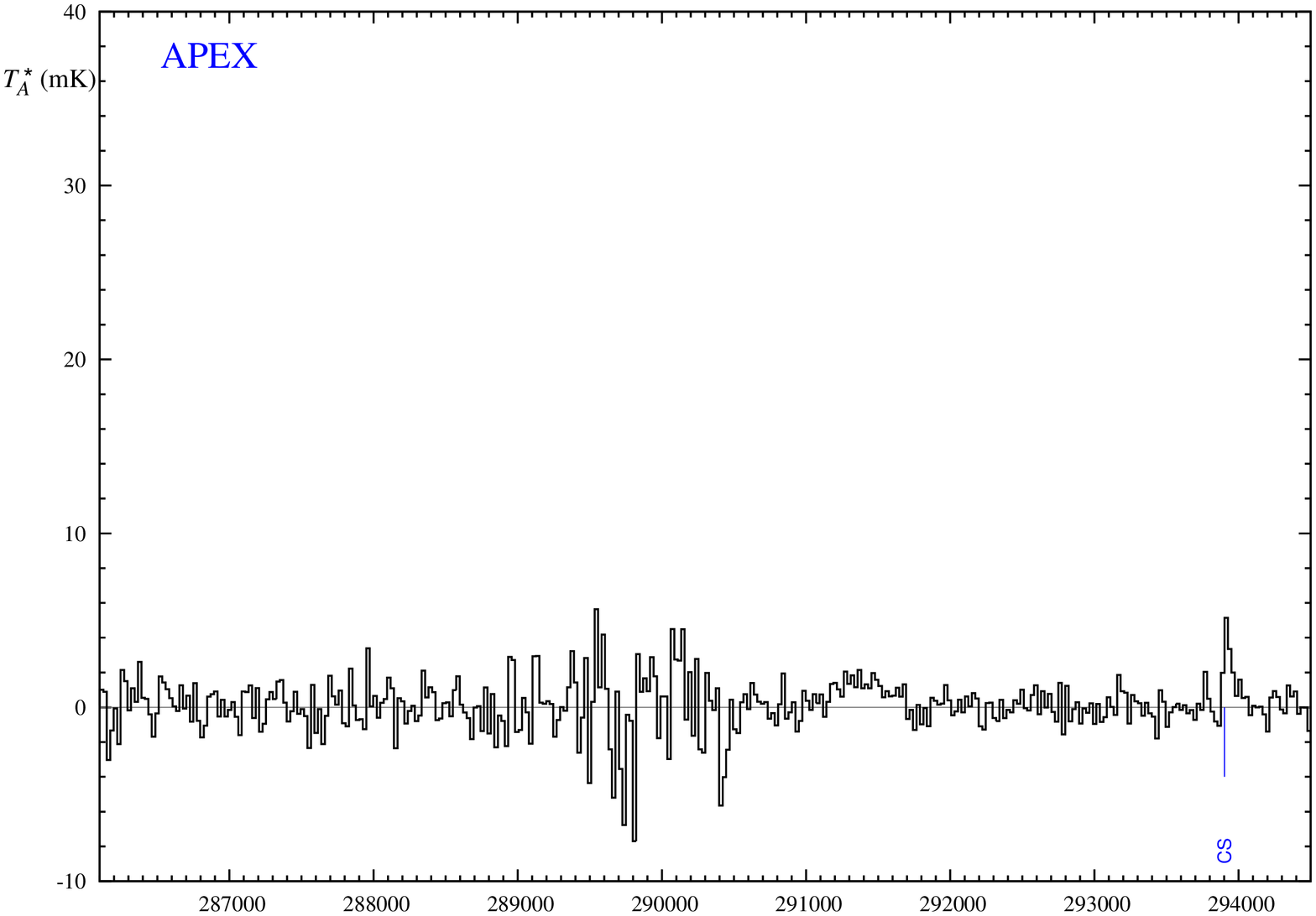}
\caption{Continued.}
\end{figure*}

  \setcounter{figure}{1}%

\begin{figure*} [tbh]
\centering
\includegraphics[angle=270,width=0.85\textwidth]{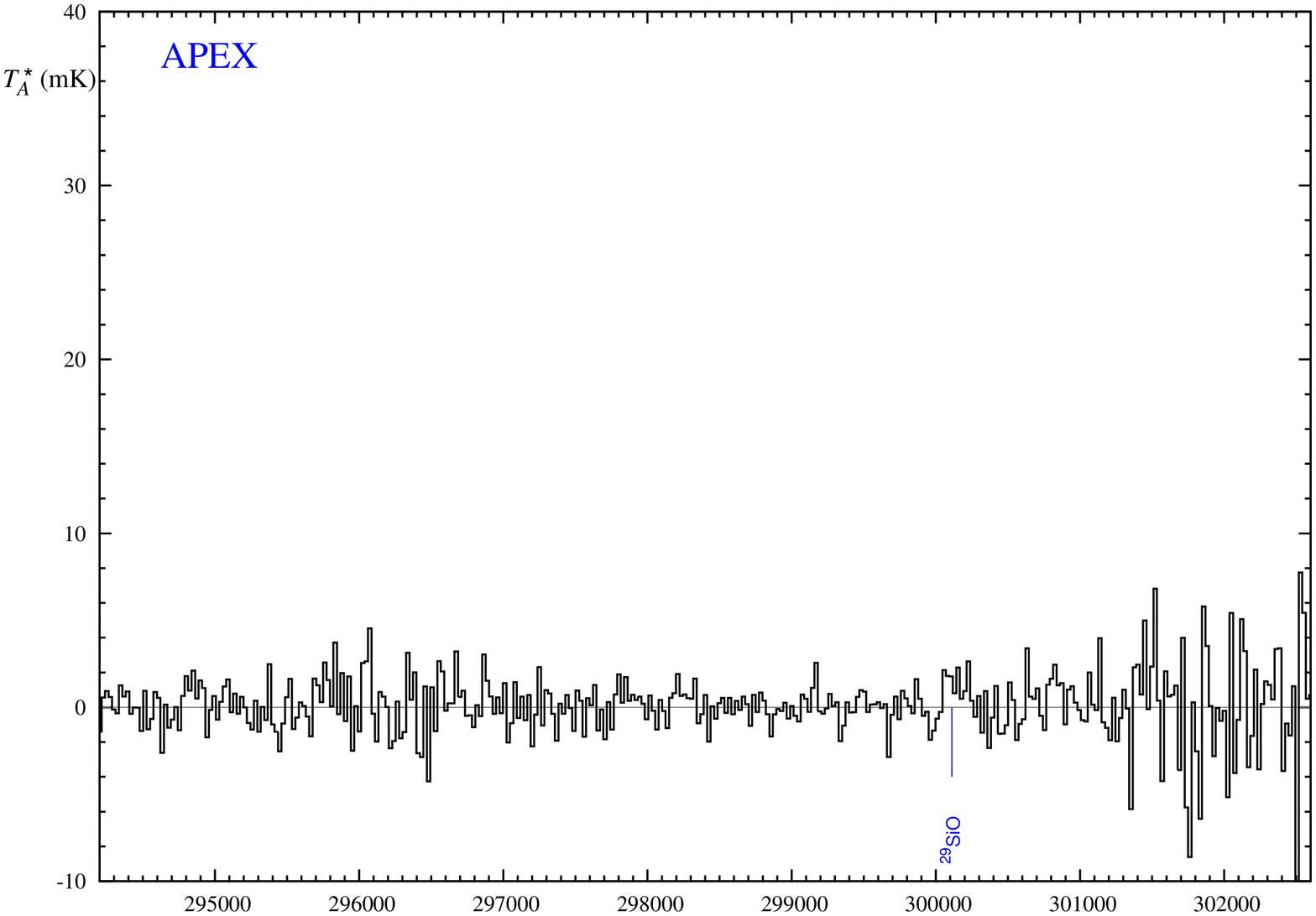}
\includegraphics[angle=270,width=0.85\textwidth]{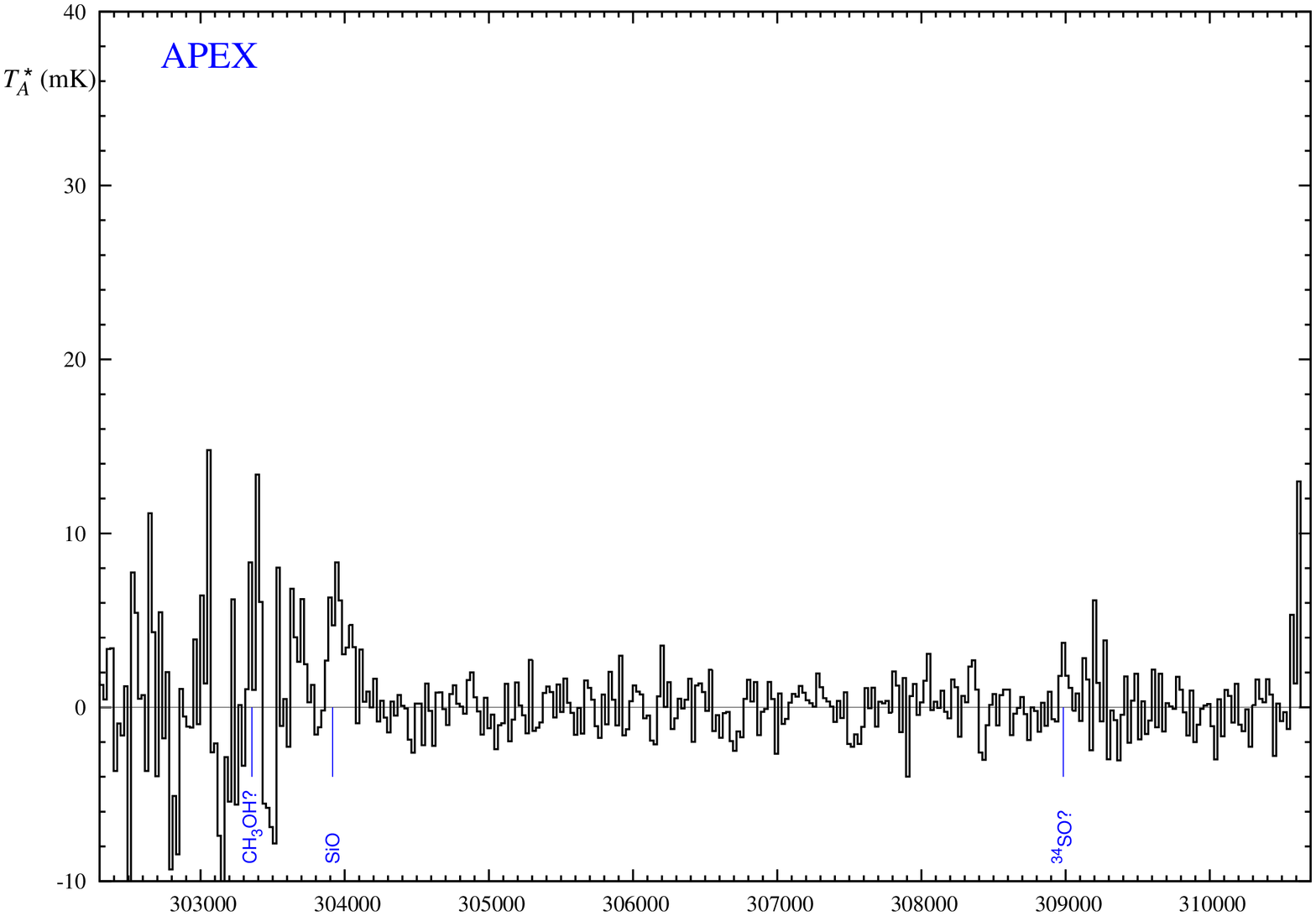}
\caption{Continued.}
\end{figure*}

  \setcounter{figure}{1}%

\begin{figure*} [tbh]
\centering
\includegraphics[angle=270,width=0.85\textwidth]{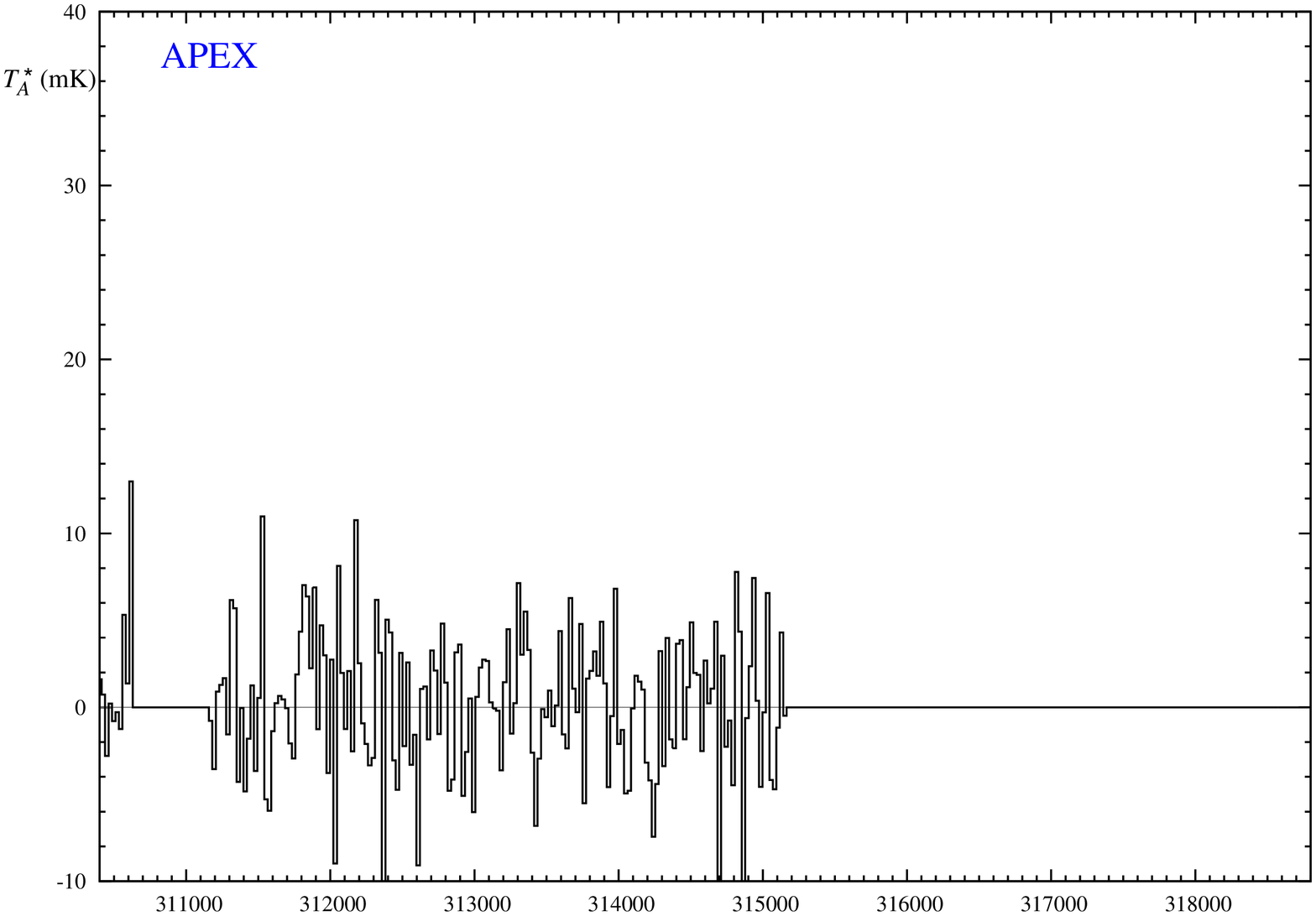}
\includegraphics[angle=270,width=0.85\textwidth]{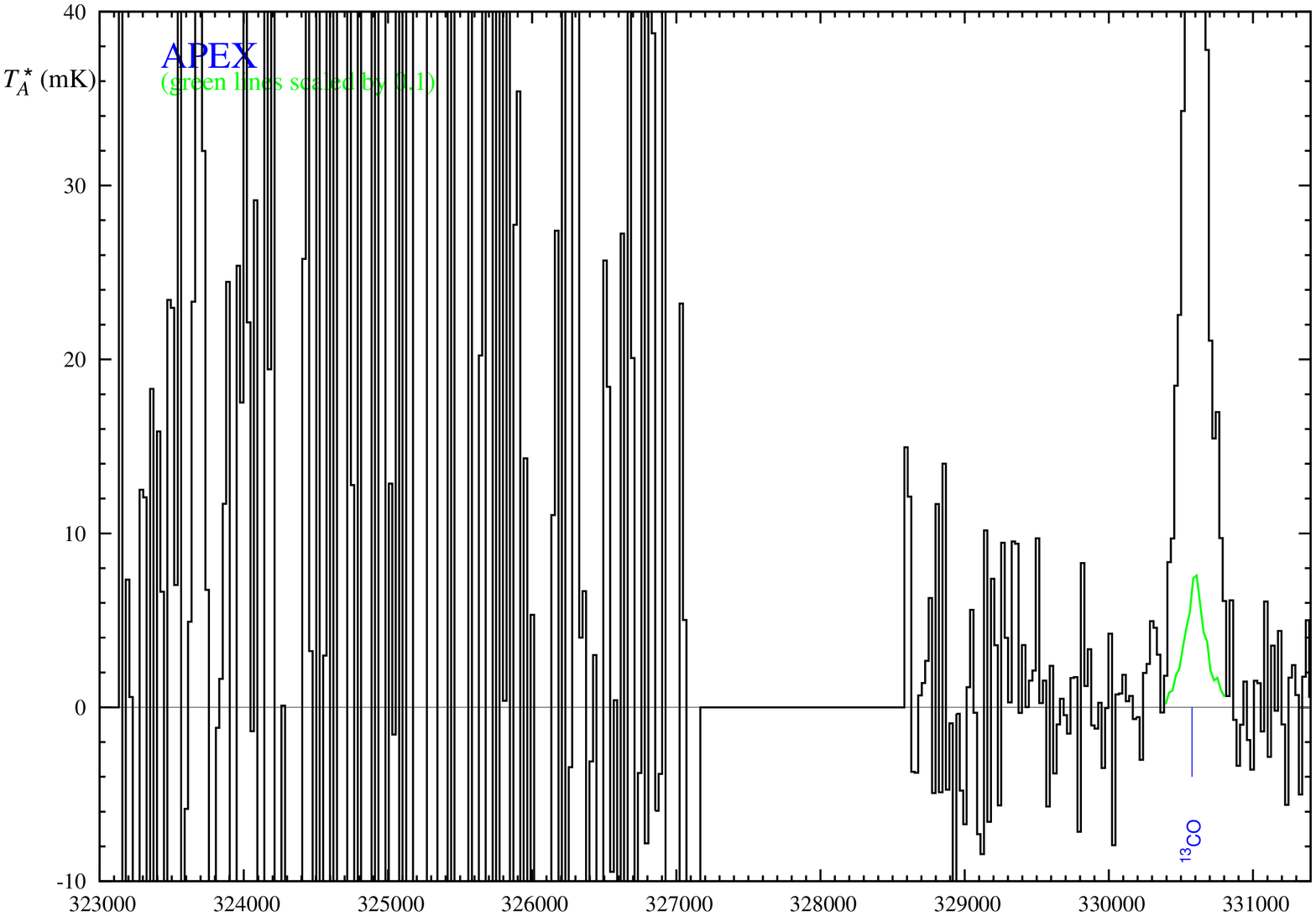}
\caption{Continued.}
\end{figure*}

  \setcounter{figure}{1}%

\begin{figure*} [tbh]
\centering
\includegraphics[angle=270,width=0.85\textwidth]{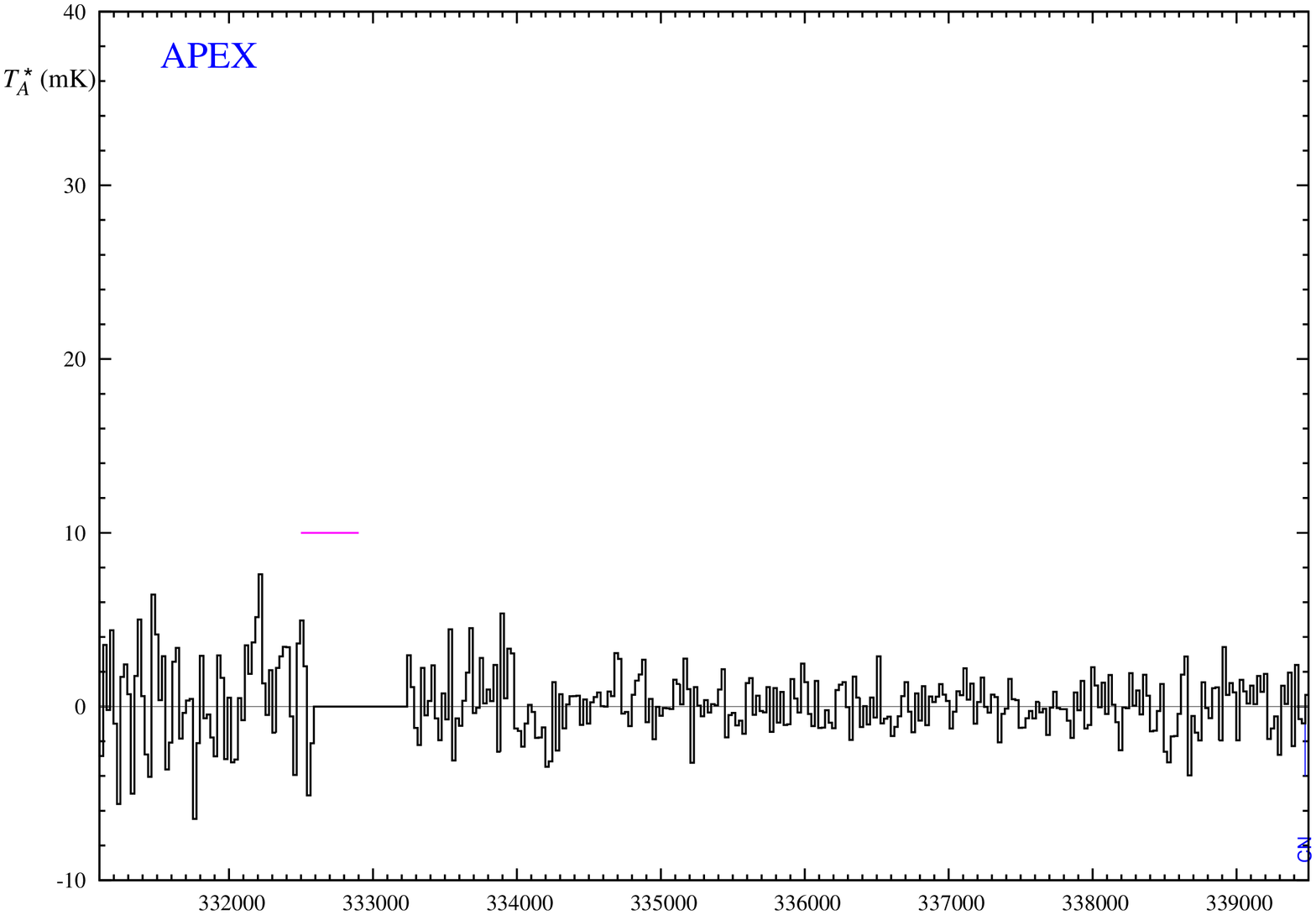}
\includegraphics[angle=270,width=0.85\textwidth]{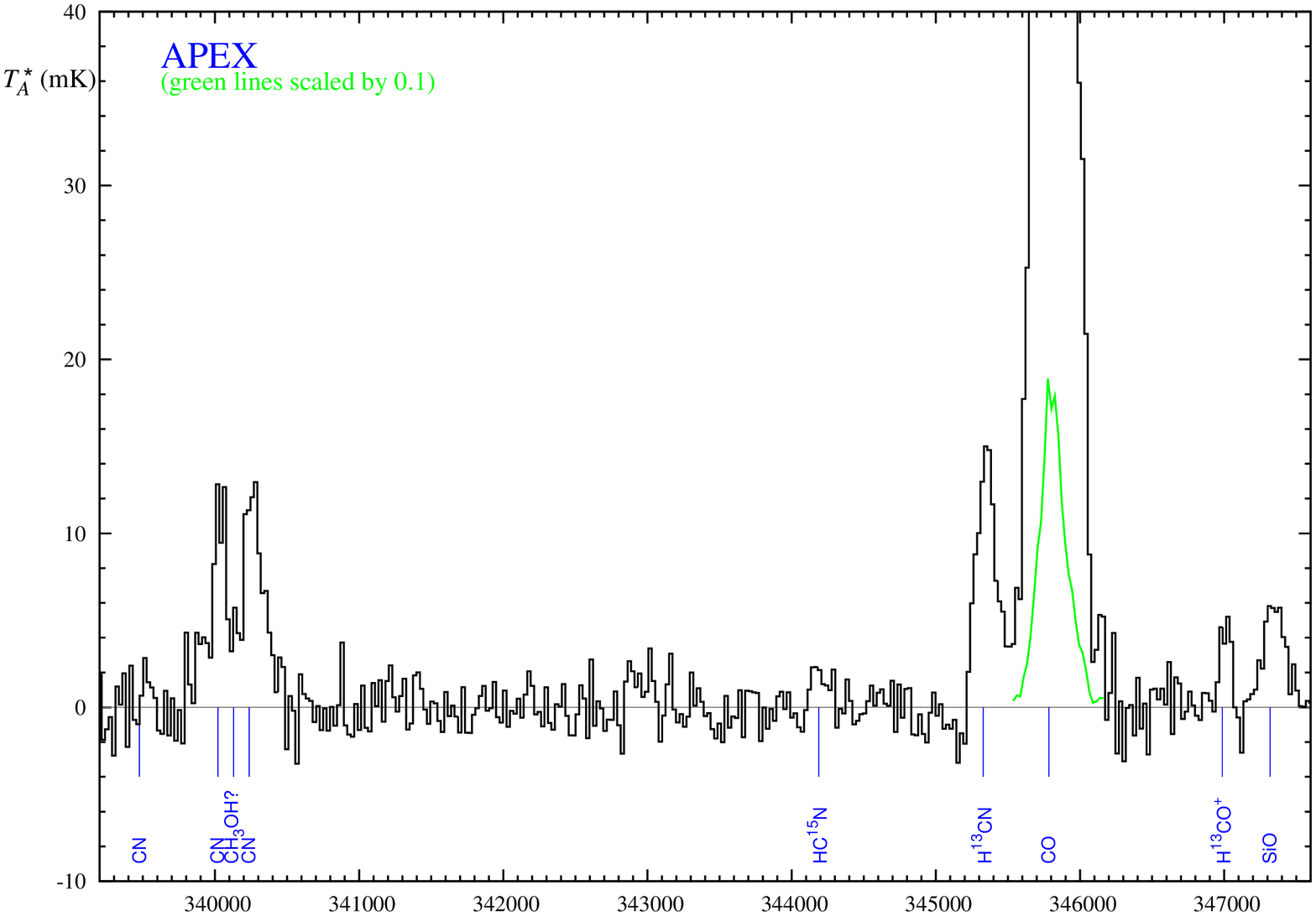}
\caption{Continued.}
\end{figure*}

  \setcounter{figure}{1}%

\begin{figure*} [tbh]
\centering
\includegraphics[angle=270,width=0.85\textwidth]{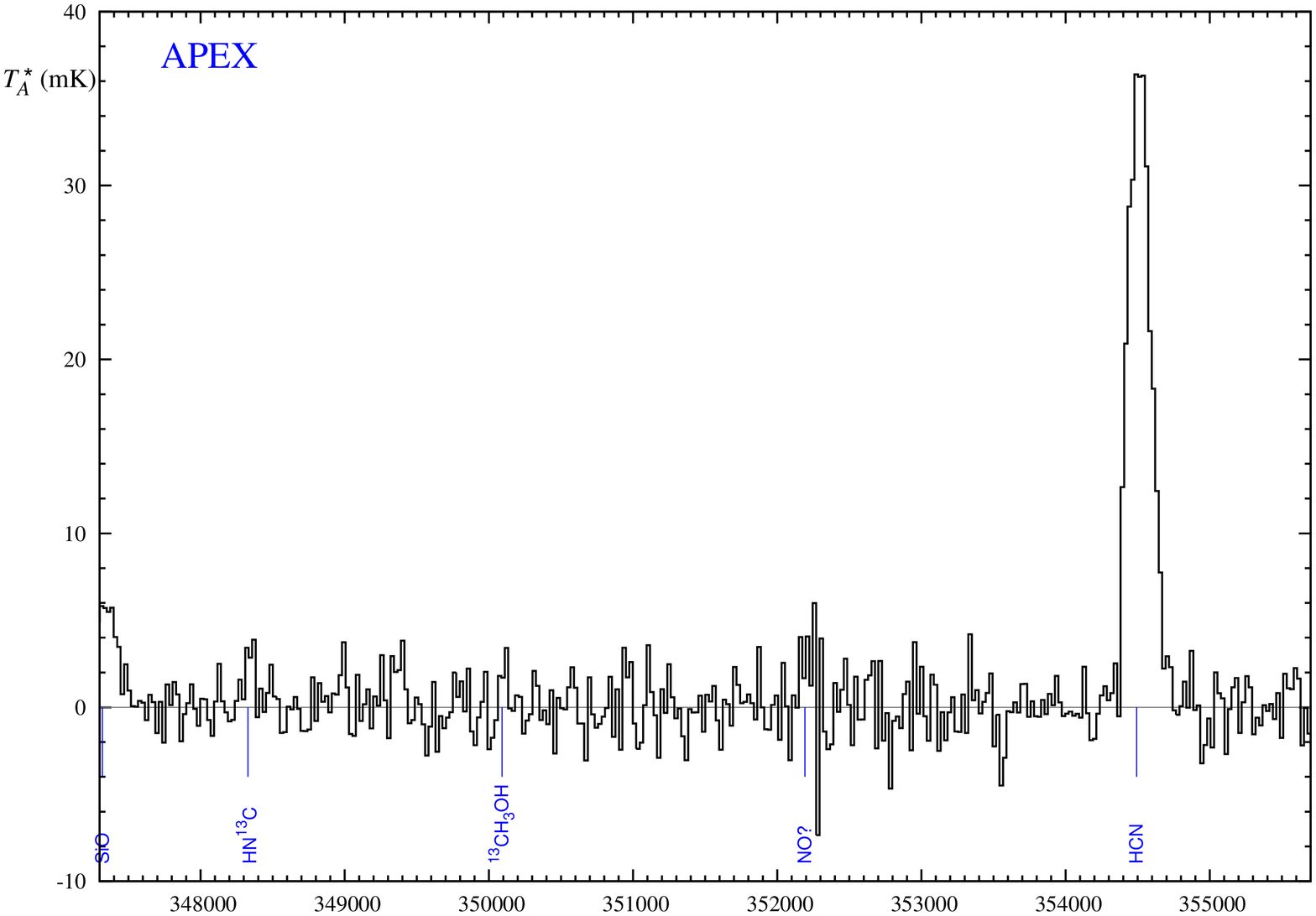}
\includegraphics[angle=270,width=0.85\textwidth]{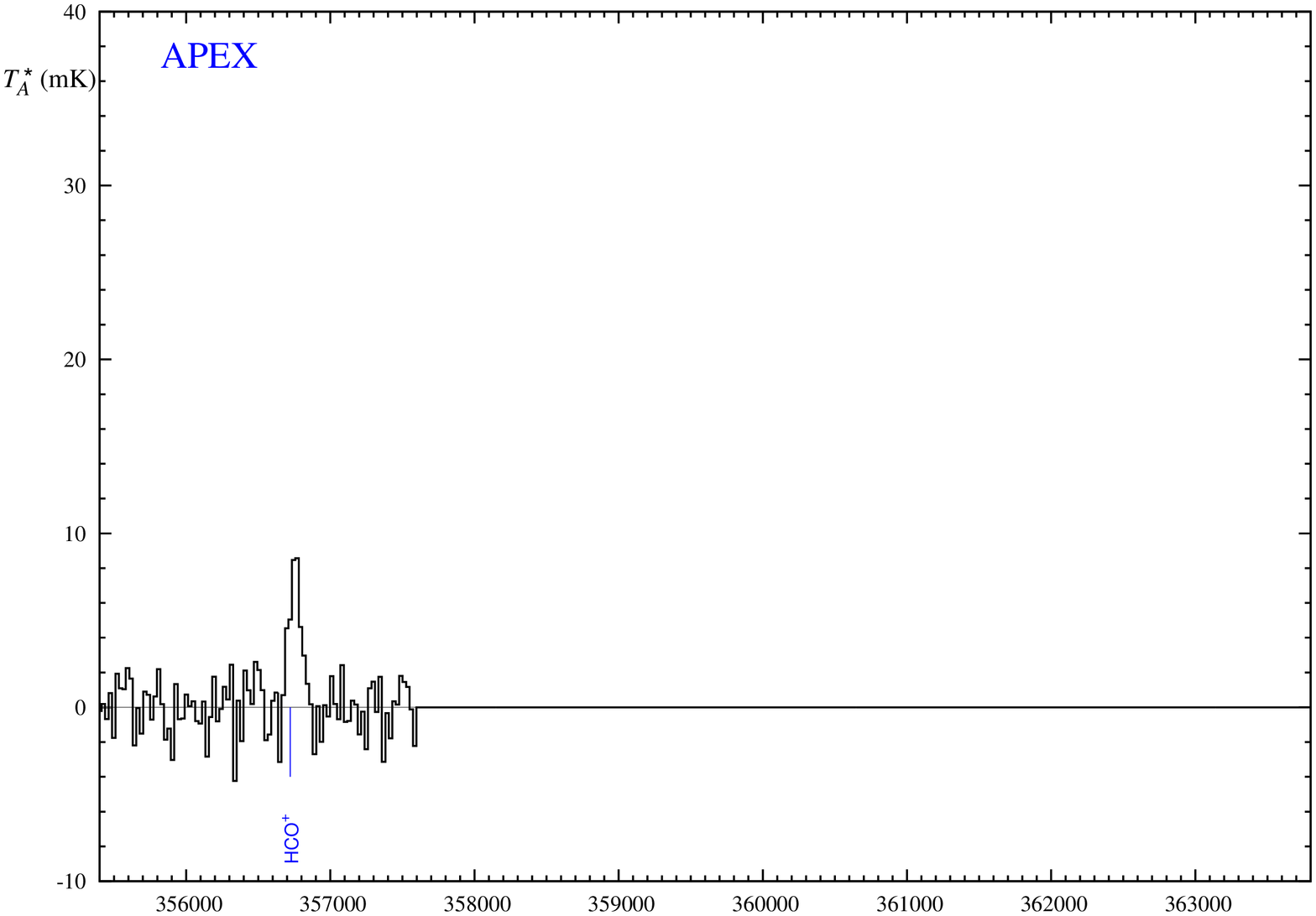}
\caption{Continued.}
\end{figure*}

  \setcounter{figure}{1}%

\begin{figure*} [tbh]
\centering
\includegraphics[angle=270,width=0.85\textwidth]{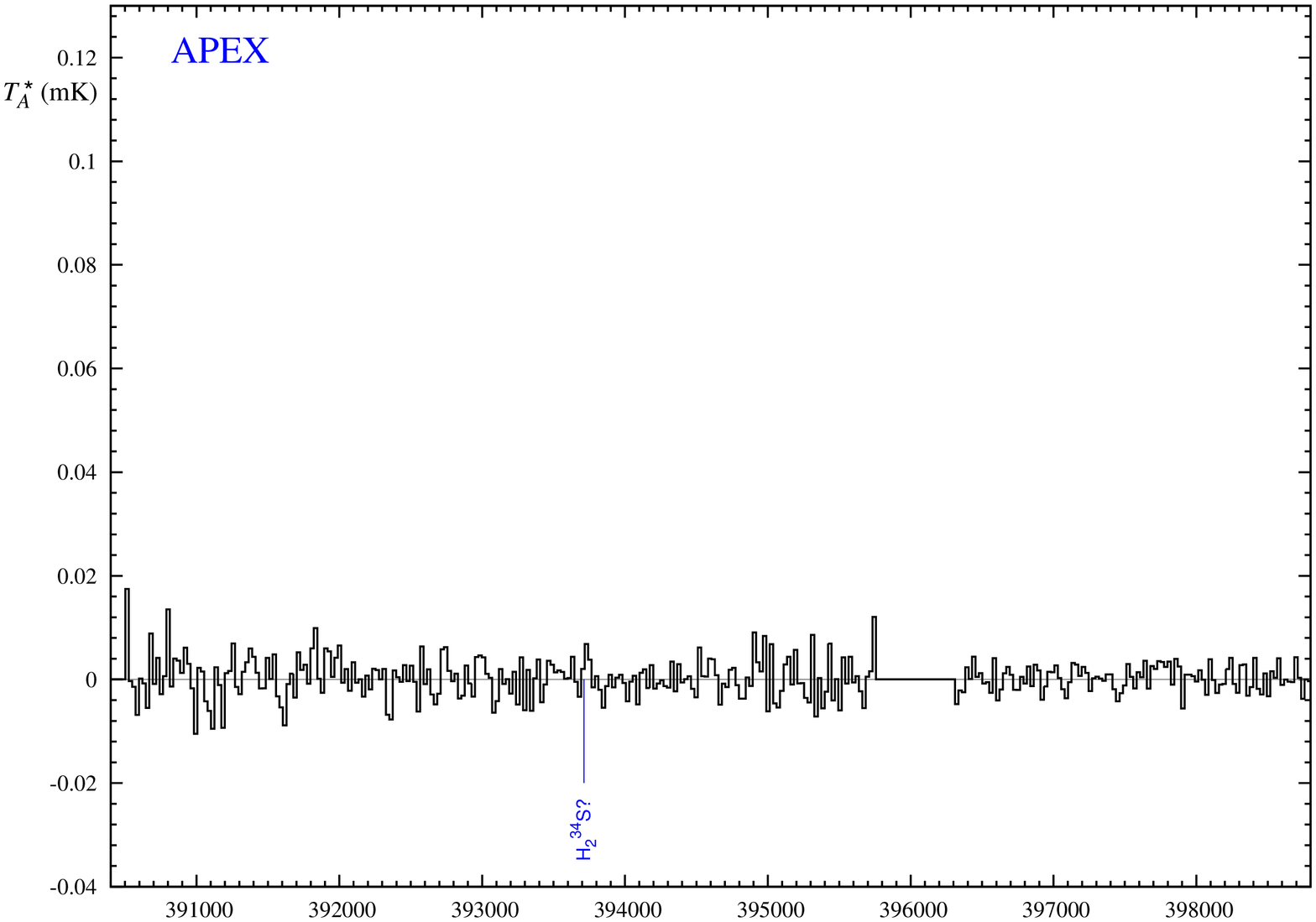}
\includegraphics[angle=270,width=0.85\textwidth]{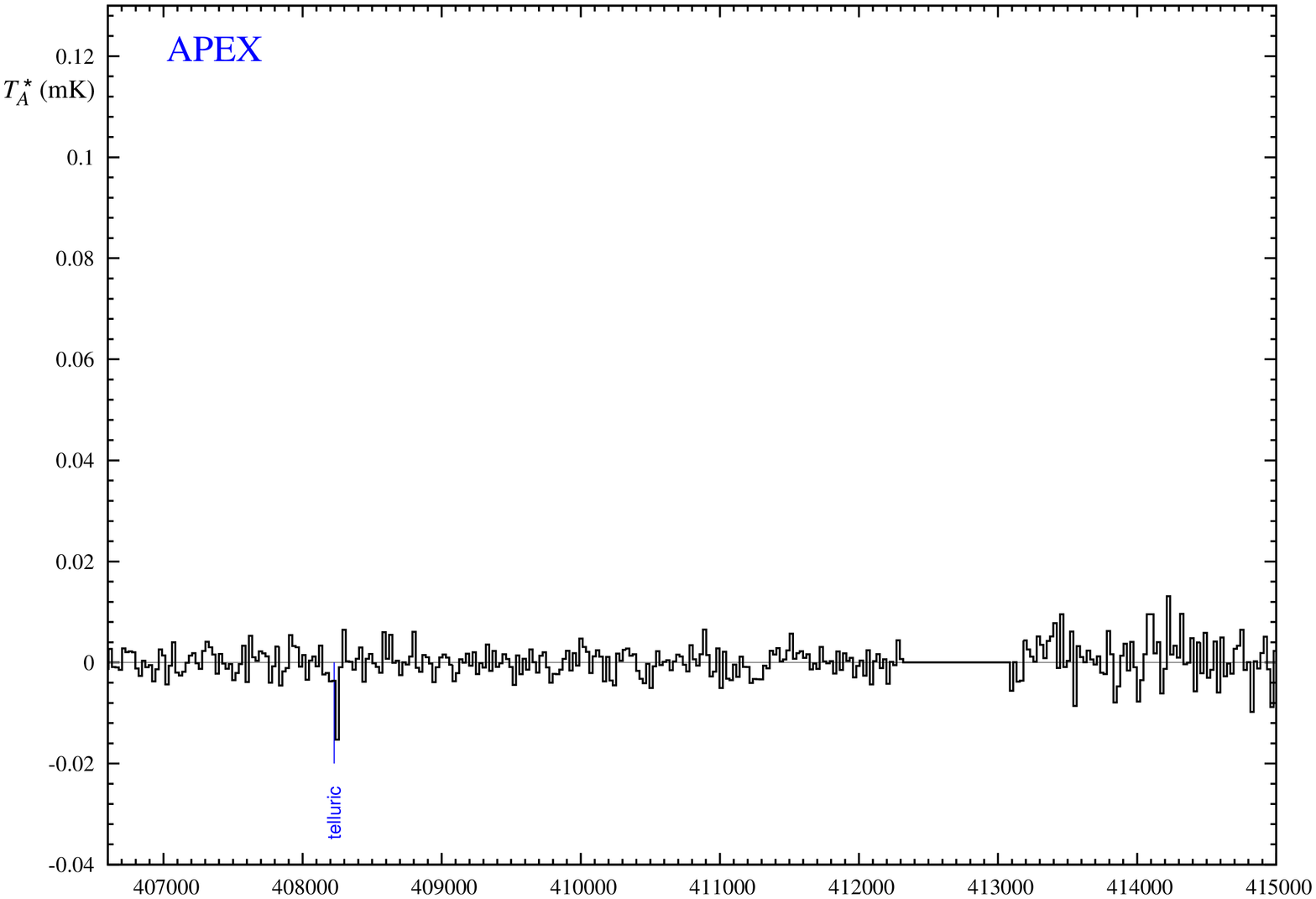}
\caption{Continued.}
\end{figure*}

  \setcounter{figure}{1}%

\begin{figure*} [tbh]
\centering
\includegraphics[angle=270,width=0.85\textwidth]{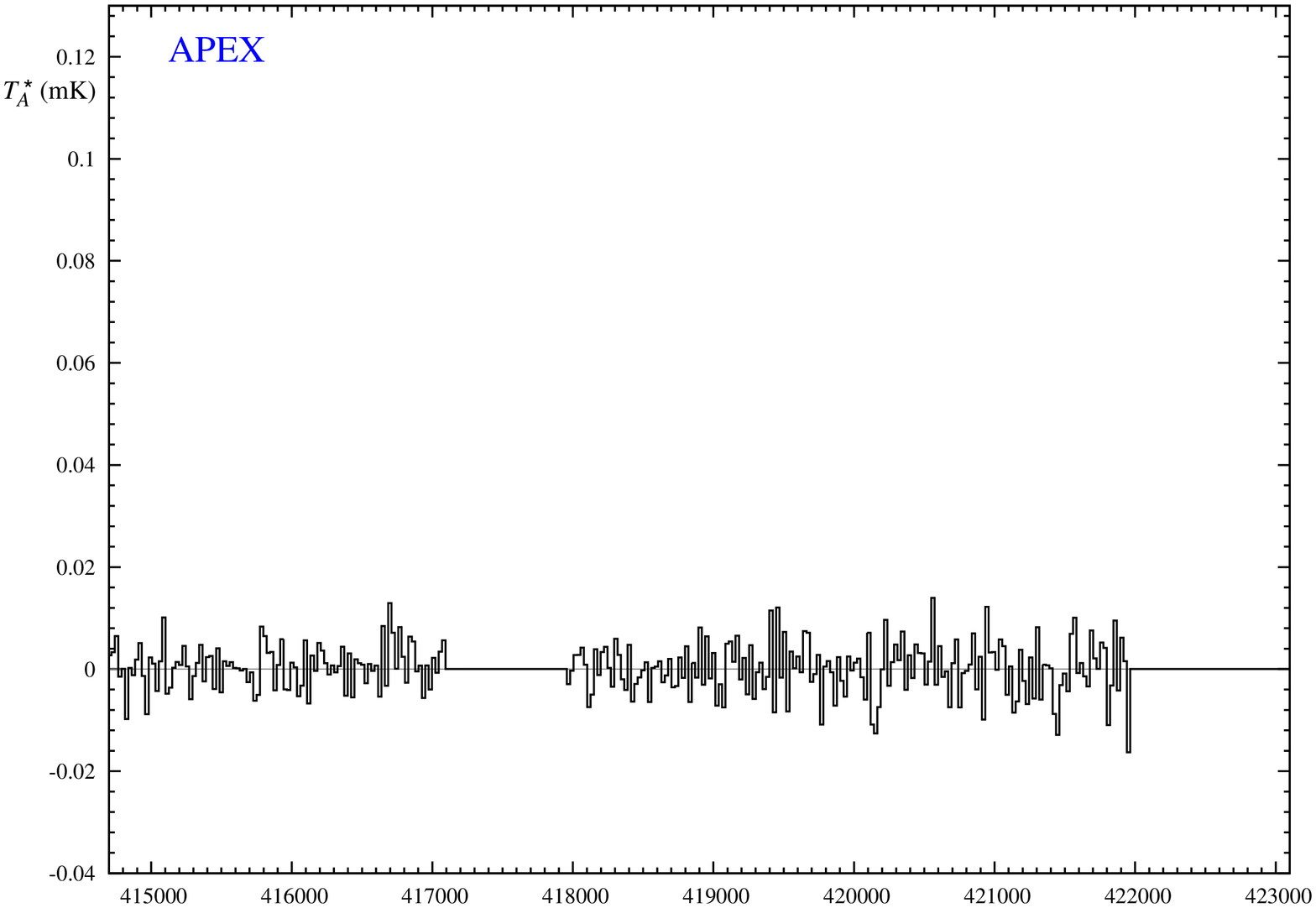}
\includegraphics[angle=270,width=0.85\textwidth]{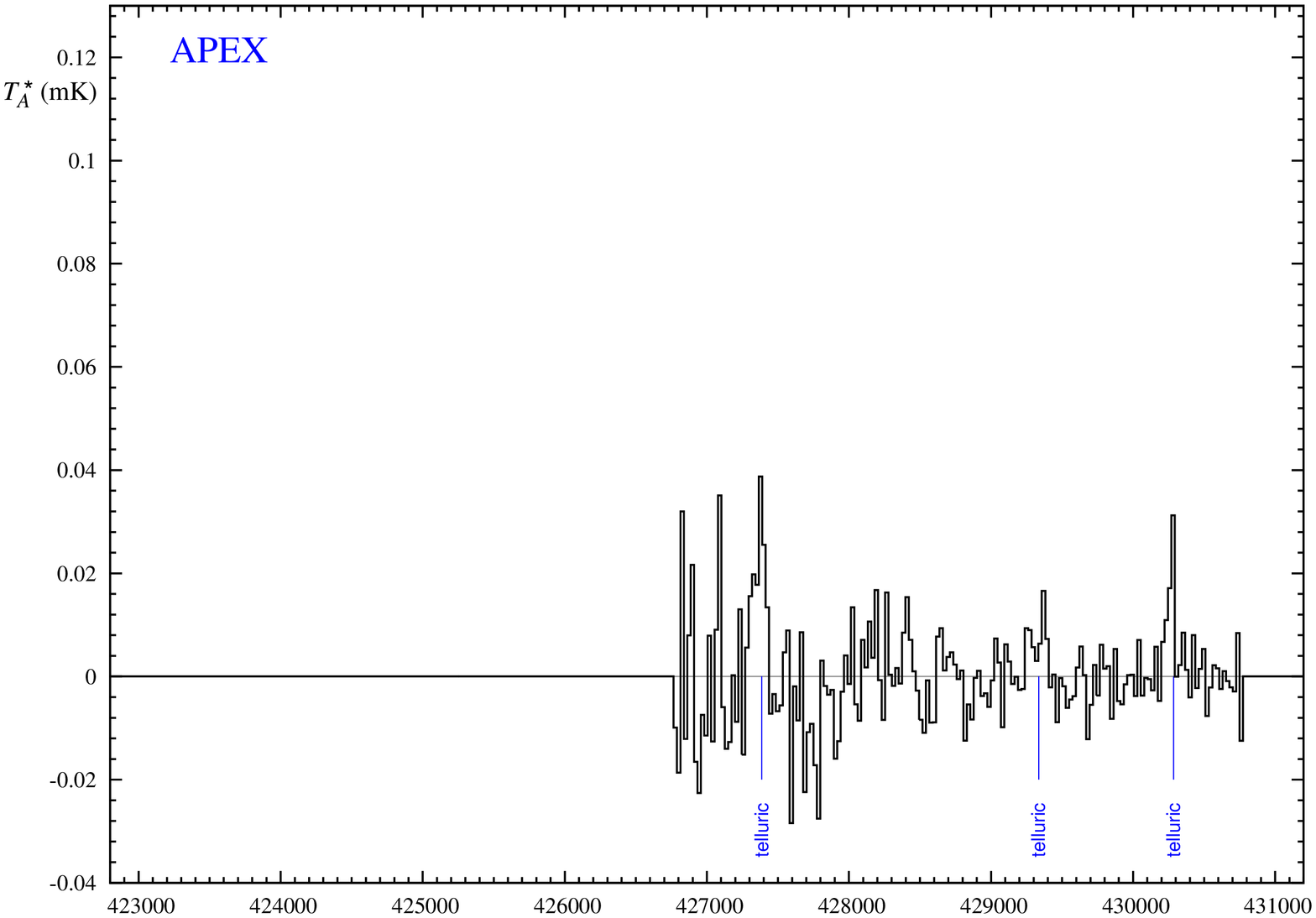}
\caption{Continued.}
\end{figure*}

  \setcounter{figure}{1}%

\begin{figure*} [tbh]
\centering
\includegraphics[angle=270,width=0.85\textwidth]{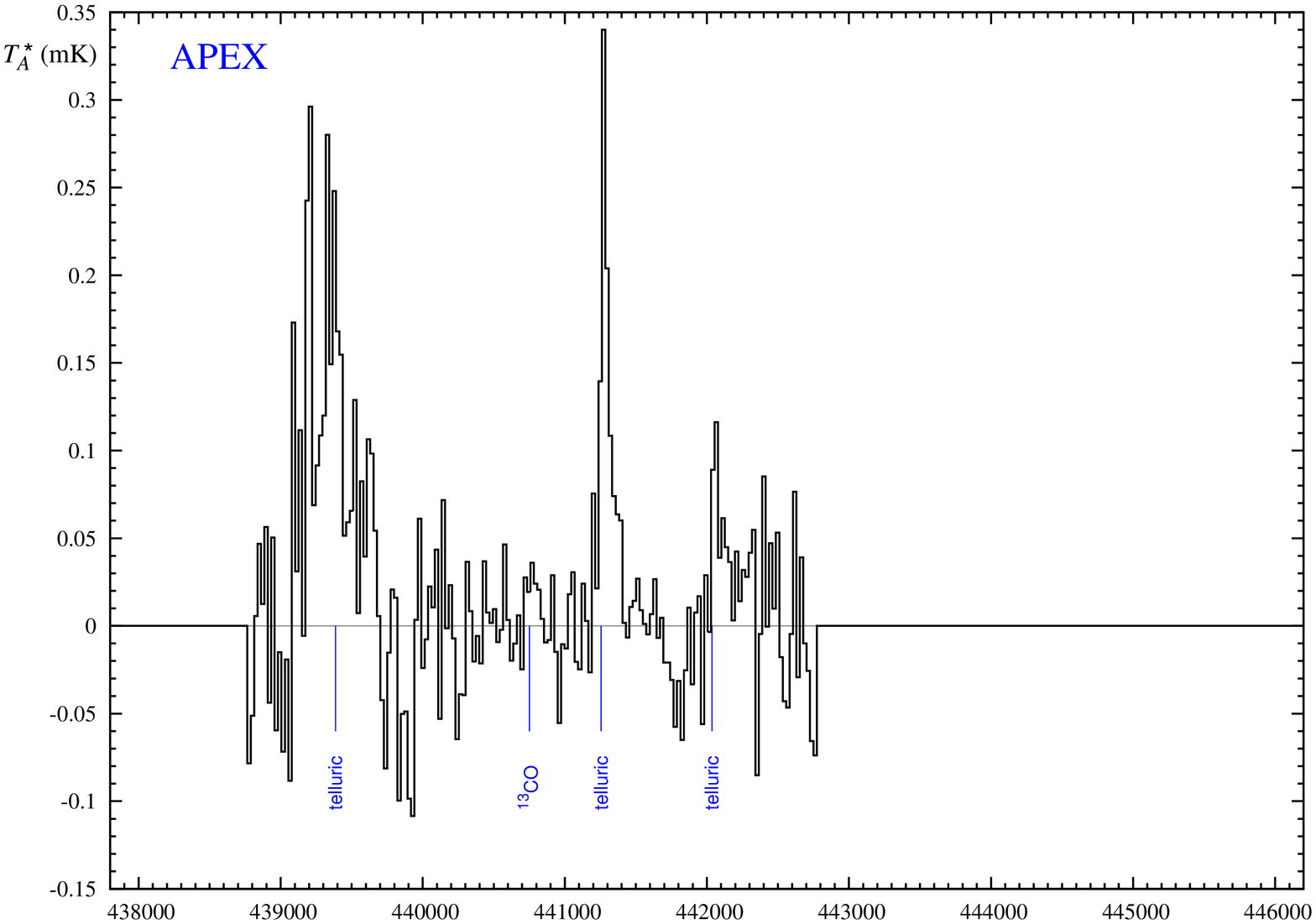}
\includegraphics[angle=270,width=0.85\textwidth]{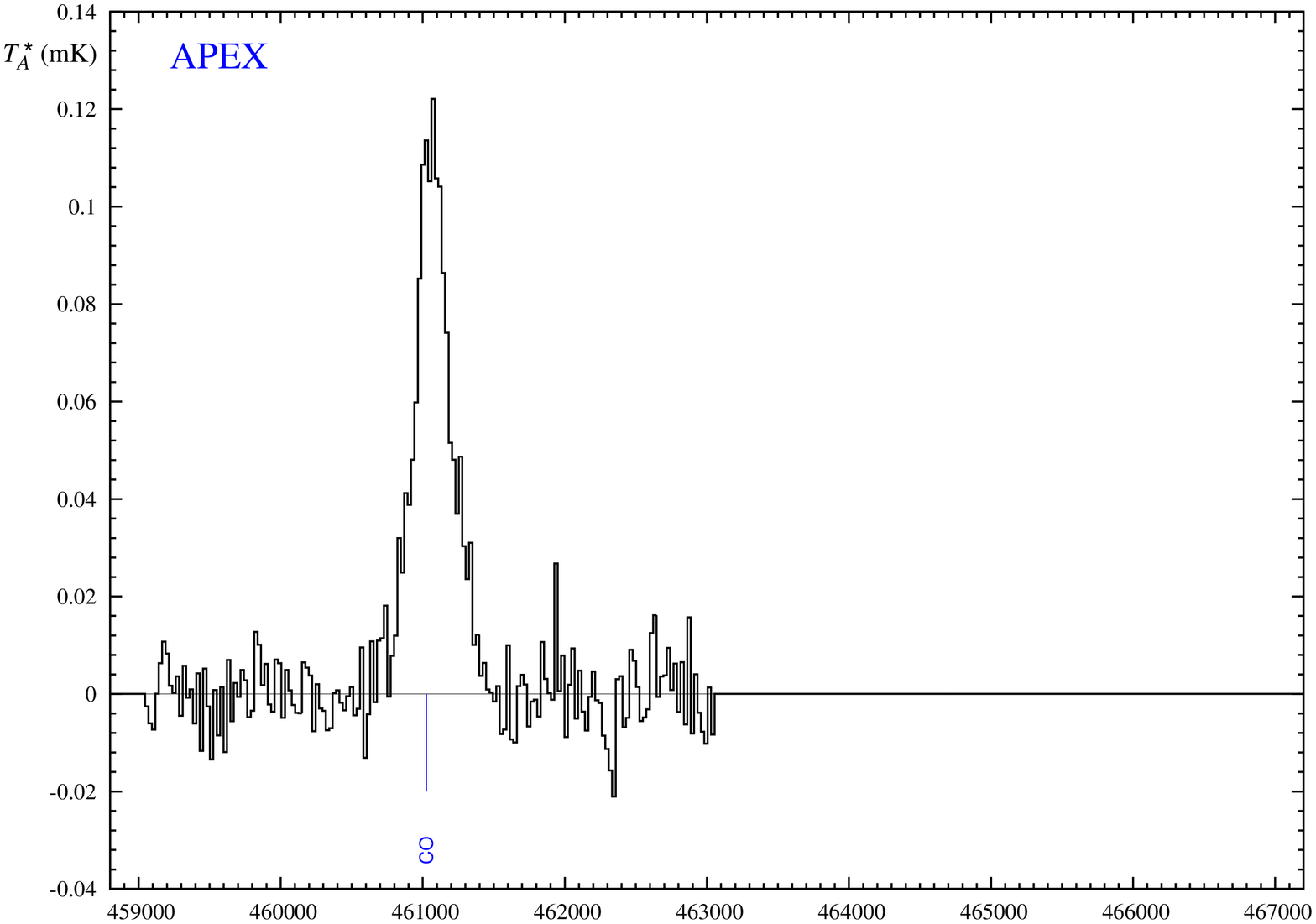}
\caption{Continued.}
\end{figure*}

  \setcounter{figure}{1}%

\begin{figure*} [tbh]
\centering
\includegraphics[angle=270,width=0.85\textwidth]{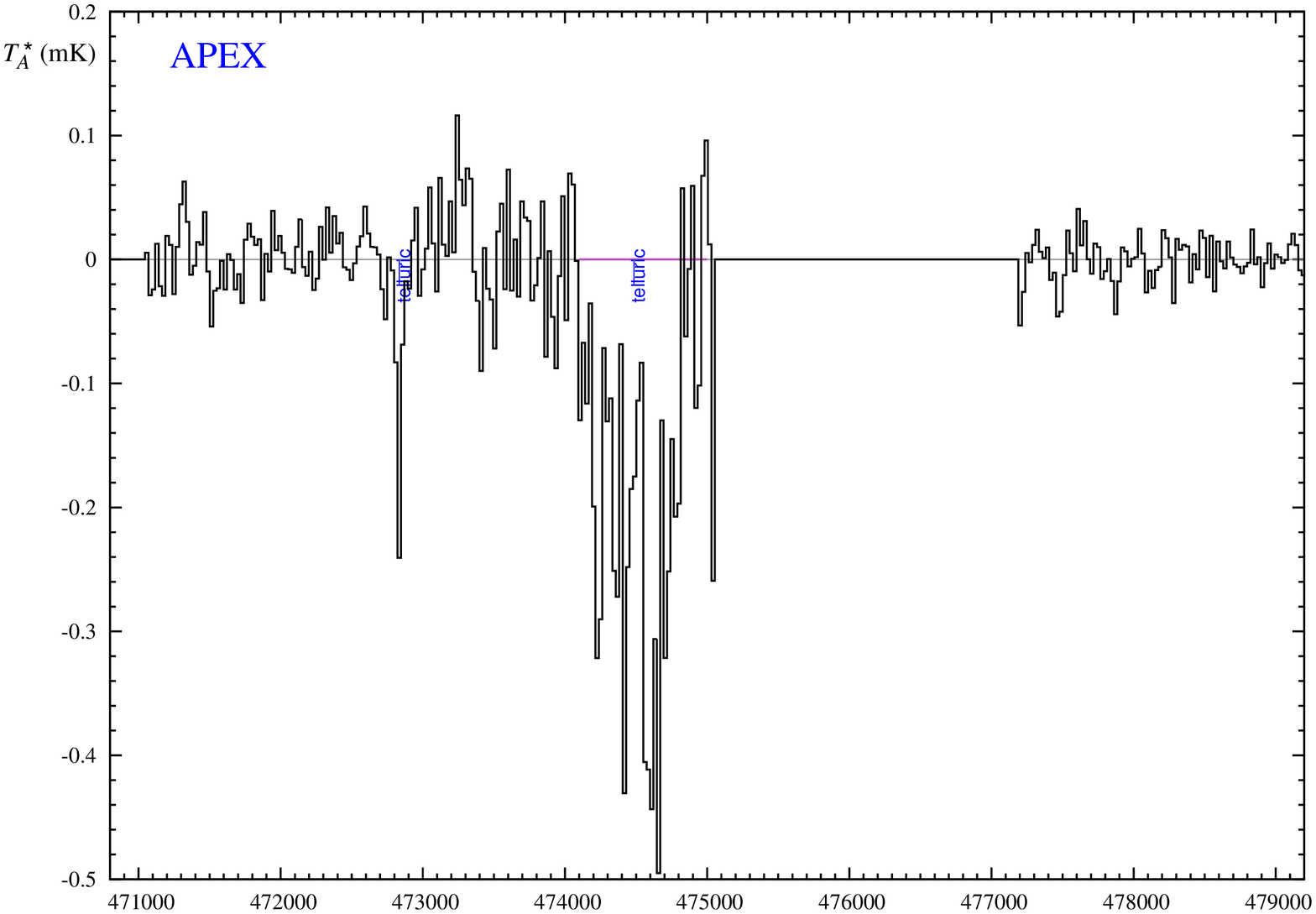}
\includegraphics[angle=270,width=0.85\textwidth]{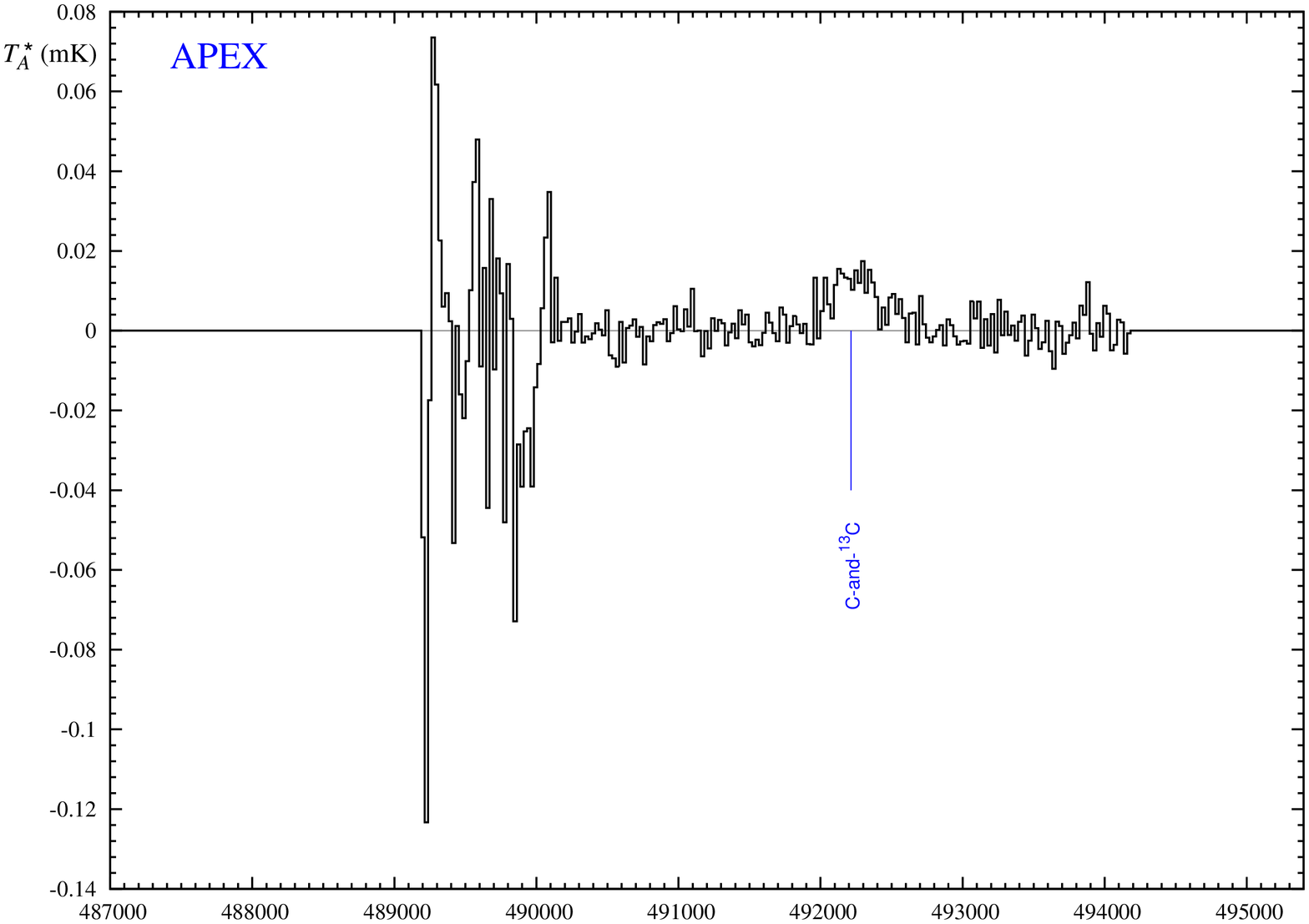}
\caption{Continued.}
\end{figure*}

  \setcounter{figure}{1}%

\begin{figure*} [tbh]
\centering
\includegraphics[angle=270,width=0.85\textwidth]{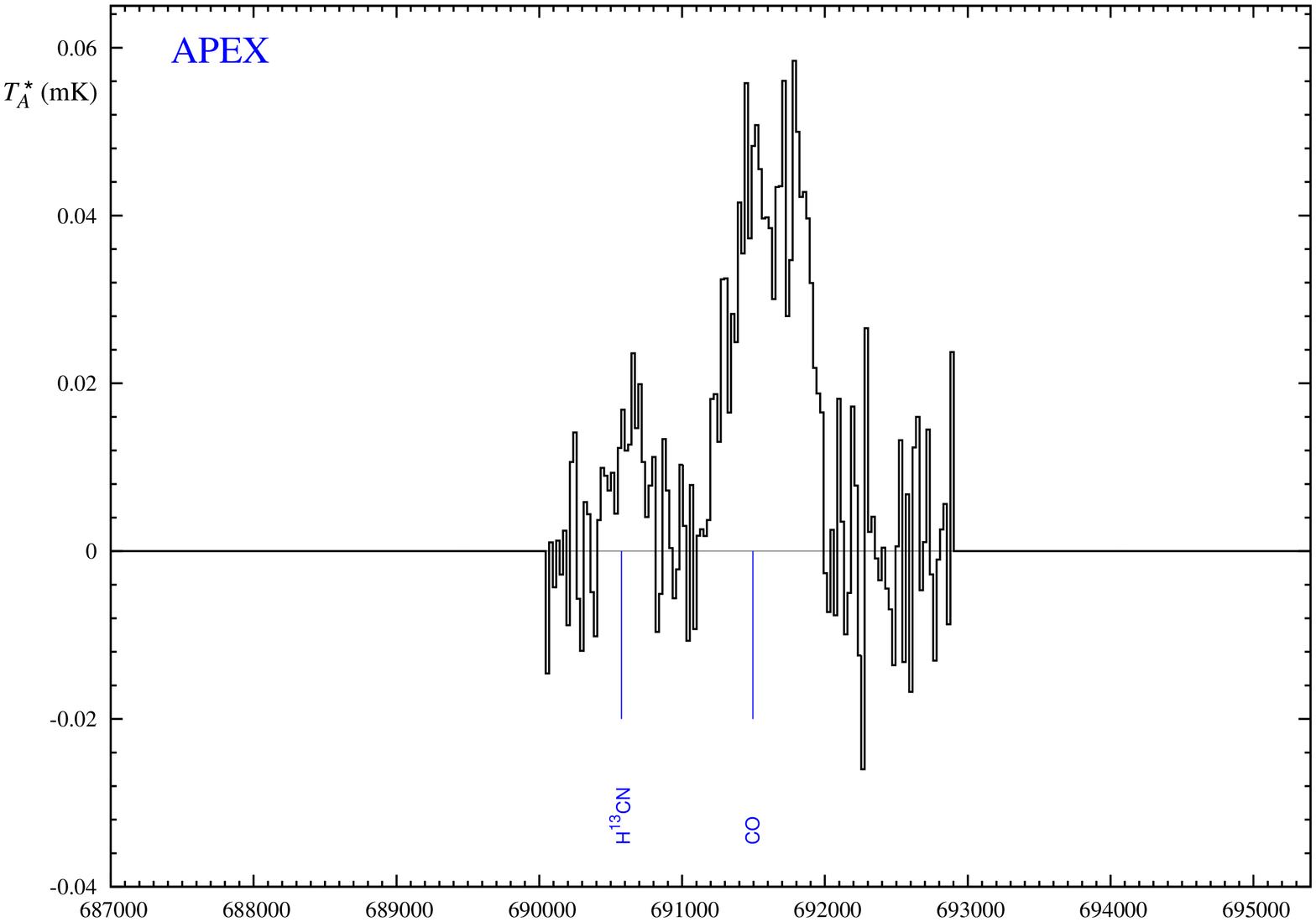}
\includegraphics[angle=270,width=0.85\textwidth]{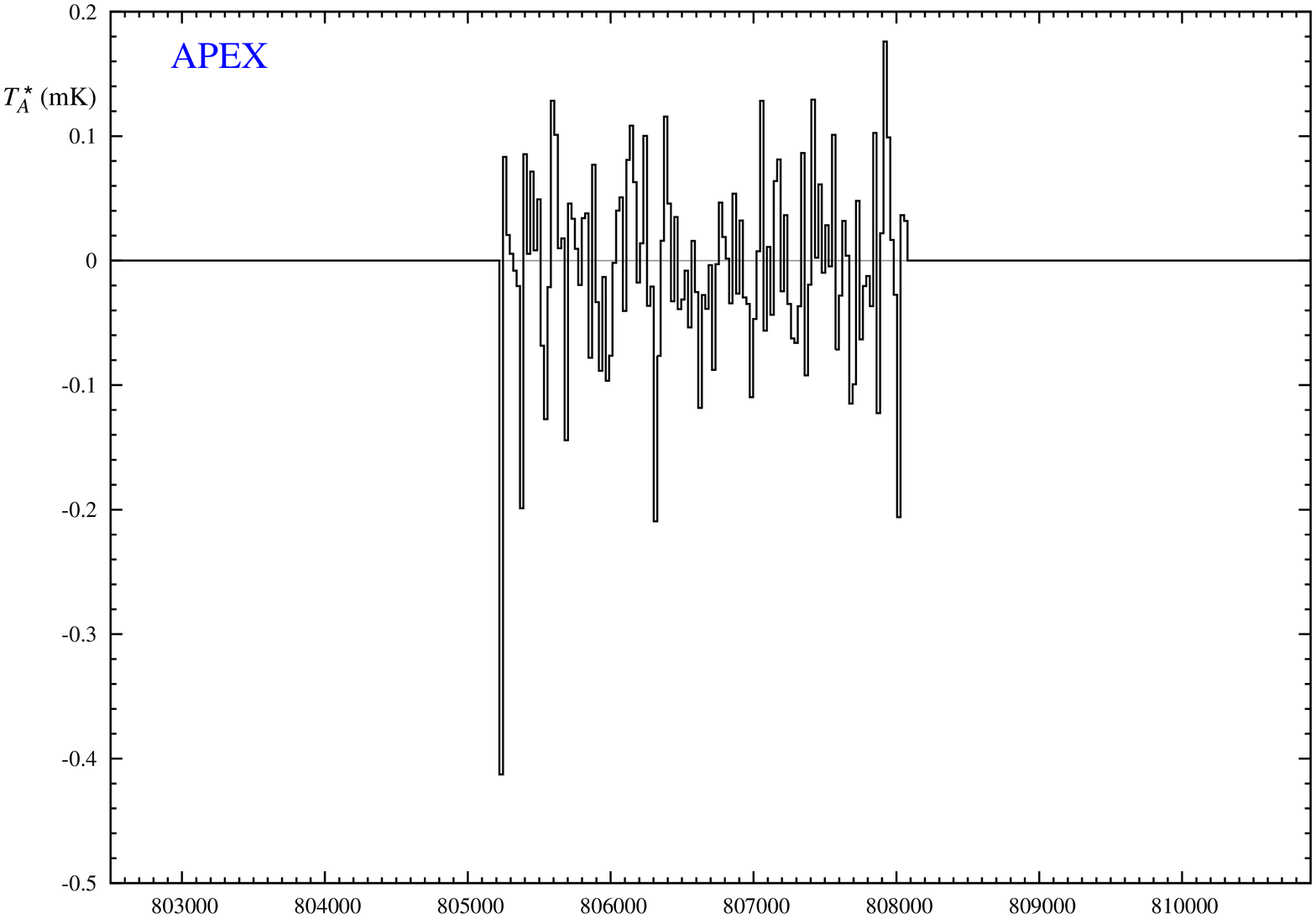}
\caption{Continued.}
\end{figure*}

  \setcounter{figure}{1}%

\begin{figure*} [tbh]
\centering
\includegraphics[angle=270,width=0.85\textwidth]{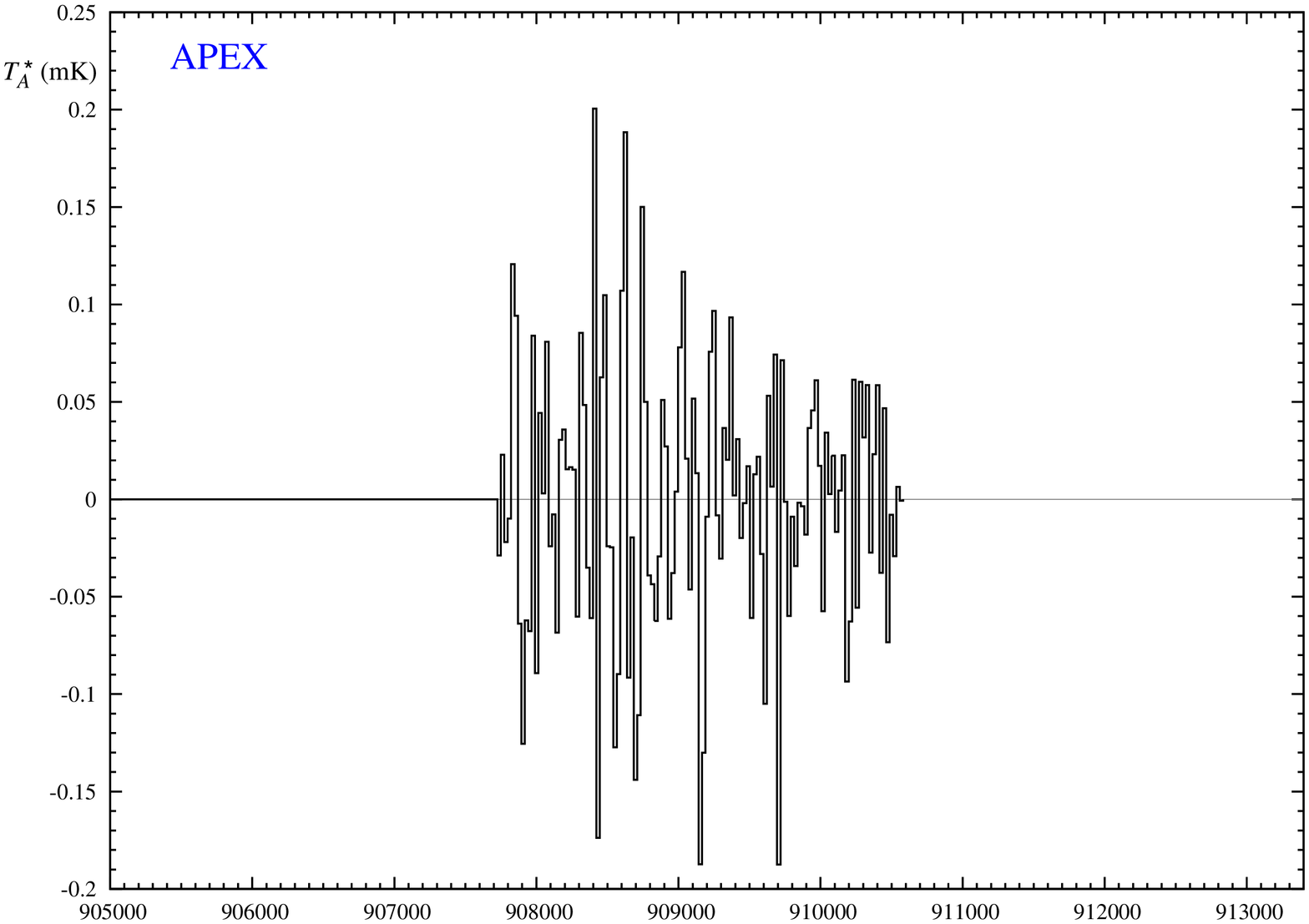}
\caption{Continued.}
\end{figure*}

\section{A list of identified transitions}

\longtab[1]{                       
\begin{longtable}{lcr}
\caption{\label{tab-ID-iram}Identified transitions in IRAM spectra.}\\
\hline\hline
Molecule& Transition & Rest frequency (MHz)\\
\hline
\endfirsthead
\caption{continued.}\\
\hline\hline
Molecule& Transition & Rest frequency (MHz)\\
\hline
\endhead
\hline
\endfoot
%
U               & U               &   75770.00\\ 
$^{13}$CH$_3$NH$_2$ & U               &   75838.45 \\ 
OSB-CS          & U               &   76872.43 \\ 
CH$_3$NH$_2$    & (1, 1, 2--1, 0, 3)  &   79008.42 \\ 
H$^{13}$CCCN    & (9--8)          &   79350.46 \\ 
Si$^{18}$O      & (2--1)          &   80704.94 \\ 
U               & U               &   81168.86 \\ 
HC$^{13}$CCN    & (9--8)          &   81534.11 \\ 
HCC$^{13}$CN    & (9--8)          &   81541.98 \\ 
OSB-N$_2$H$^+$  & U               &   81684.39 \\ 
HC$_3$N         & (9--8)          &   81881.47 \\ 
U               & U               &   83541.41 \\ 
Si$^{17}$O?     & (2--1)          &   83588.65 \\ 
H$^{13}$C$^{15}$N & (1--0)          &   83727.58 \\ 
U               & U               &   84102.78 \\ 
CH$_3$NH$_2$?   & (4, 0, 0--3, -1, 1) &   84215.03 \\ 
$^{13}$CCH?     & (1--0)          &   84154.51 \\ 
$^{30}$SiO      & (2--1)          &   84746.17 \\ 
H$^{15}$N$^{13}$C? & (1--0)          &   85258.92 \\ 
CH$_3$NH$_2$?   & (5, 1, 6--5, 0, 7)  &   85350.10 \\ 
OSB-13CN        & U               &   85469.39 \\ 
$^{29}$SiO      & (2--1)          &   85759.19 \\ 
HC$^{15}$N      & (1--0)          &   86054.97 \\ 
H$^{13}$CN      & (1--0)          &   86339.92 \\ 
H$^{13}$CO$^+$  & (1--0)          &   86754.29 \\ 
SiO             & (2--1)          &   86846.98 \\ 
HN$^{13}$C      & (1--0)          &   87090.83 \\ 
CCH             & (1, 1.5, $F$--0, 0.5, $F$') &   87318.45 \\ 
CCH             & (1, 0.5, $F$--0, 0.5, $F$') &   87408.95 \\ 
SiN?            & (2, 2.5, $F$--1, 1.5, $F$') &   87562.12 \\ 
CH$_3$NH$_2$    & (2, -1, 1--1, 1, 0) &   87795.01 \\ 
HNCO            & (4, 0, 4--3, 0, 3)  &   87925.24 \\ 
H$^{13}$CCCN    & (10--9)         &   88166.83 \\ 
HCN             & (1--0)          &   88631.60 \\ 
H$^{15}$NC      & (1--0)          &   88865.72 \\ 
CH$_3$NH$_2$    & (2, -1, 1--2, 0, 0) &   89081.46 \\ 
HCO$^+$         & (1--0)          &   89188.52 \\ 
$^{13}$CH$_3$CN & (5--4)          &   89331.28 \\ 
CH$_3$NH$_2$    & (1, 1, 0--1, 0, 1)  &   89956.07 \\ 
$^{15}$NNH$^+$  & (1--0)          &   90263.84 \\ 
HC$^{13}$CCN    & (10--9)         &   90593.06 \\ 
HCC$^{13}$CN    & (10--9)         &   90601.78 \\ 
HNC             & (1, 0, 0--0, 0, 0)  &   90663.57 \\ 
SiS             & (5--4)          &   90771.56 \\ 
HC$_3$N         & (10--9)         &   90979.02 \\ 
N$^{15}$NH$^+$  & (1--0)          &   91205.70 \\ 
CH$_3$NH$_2$??  & (4, 0, 5--3, 1, 5)  &   91848.43 \\ 
CH$_3$CN        & (5--4)          &   91979.99 \\ 
$^{13}$CS       & (2--1)          &   92494.31 \\ 
CH$_3^{18}$OH? & (2, $K_a$, 2--1, $K_a$', 1) &   92725.47 \\ 
N$_2$H$^+$      & (1--0)          &   93173.40 \\ 
PN              & (2--1)          &   93979.77 \\ 
$^{13}$CH$_3$OH & (2, 0, 2--1, 0, 1)  &   94407.13 \\ 
$^{13}$CH$_3$OH & (2, 1, 1--1, 1, 0)  &   94420.45 \\ 
CH$_3$NH$_2$??  & (4, 0, 2--3, -1, 3) &   95145.81 \\ 
C$^{34}$S       & (2--1)          &   96412.95 \\ 
CH$_3$OH        & (2, $K_a$, $K_c$--1, $K_a$', $K_c$') &   96739.36 \\ 
CH$_3$OH        & (2, 1, 1--1, 1, 0)  &   96755.51 \\ 
H$^{13}$CCCN    & (11--10)        &   96983.00 \\ 
C$^{33}$S?      & (2--1)          &   97172.06 \\ 
$^{34}$SO       & (2, 3--1, 2)      &   97715.32 \\ 
CS              & (2--1)          &   97980.95 \\ 
U               & U               &   98347.65 \\ 
AlF             & (3--2)          &   98926.73 \\ 
CH$_3$NH$_2$??  & (4, 0, 5--3, 1, 4)  &   99127.25 \\ 
SO              & (2, 3--1, 2)      &   99299.87 \\ 
HC$^{13}$CCN    & (11--10)        &   99651.85 \\ 
HCC$^{13}$CN    & (11--10)        &   99661.47 \\ 
HC$_3$N         & (11--10)        &  100076.39 \\ 
OSB-SiO         & U               &  100564.77 \\ 
H$_2$CS         & (3, 1, 3--2, 1, 2)  &  101477.81 \\ 
U               & U               &  102565.87 \\ 
H$_2$CS         & (3, 0, 3--2, 0, 2)  &  103040.45 \\ 
SO$_2$          & (3, 1, 3--2, 0, 2)  &  104029.42 \\ 
H$_2$CS         & (3, 1, 2--2, 1, 1)  &  104617.04 \\ 
$^{13}$C$^{18}$O & (1--0)          &  104711.40 \\ 
H$^{13}$CCCN    & (12--11)        &  105799.11 \\ 
CH$_2$NH        & (4, 0, 4, $F$--3, 1, 3, $F$') &  105793.92 \\ 
$^{14}$CO?      & (1--0)          &  105871.10 \\ 
$^{13}$CH$_3$CN & (6--5)          &  107196.57 \\ 
OSB-H13CN?      & U               &  107164.67 \\ 
OSB-SiO?        & U               &  107422.62 \\ 
U               & U               &  108180.00 \\ 
$^{13}$CN       & (1, 0.5, 1, $F$--0, 0.5, 1, $F$') &  108076.87 \\ 
$^{13}$CN       & (1, 0.5, 0, $F$--0, 0.5, 1, $F$') &  108420.73 \\ 
$^{13}$CN       & (1, 0.5 or 1.5, 1, $F$--0, 0.5, 0 or 1, $F$') &  108647.52 \\ 
$^{13}$CN       & (1, 1.5, 2, $F$--0, 0.5, 1, $F$') &  108784.21 \\ 
SiS             & (6--5)          &  108924.30 \\ 
HC$_3$N         & (12--11)        &  109173.63 \\ 
SO              & (3, 2--2, 1)      &  109252.22 \\ 
C$^{15}$N       & (1, 0.5, $F$--0, 0.5, $F$') &  109705.57 \\ 
C$^{18}$O       & (1--0)          &  109782.17 \\ 
HNCO            & (5, 0, 5--4, 0, 4)  &  109904.92 \\ 
C$^{15}$N       & (1, 1.5, $F$--0, 0.5, $F$') &  110021.71 \\ 
$^{13}$CO       & (1--0)          &  110201.35 \\ 
CH$_3$CN        & (6--5)          &  110383.50 \\ 
CH$_3^{18}$OH? & (3, 1, 3--4, 0, 4)  &  111209.73 \\ 
C$^{17}$O       & (1--0)          &  112358.78 \\ 
CN              & (1, 0.5, $F$--0, 0.5, $F$') &  113168.78 \\ 
CN              & (1, 1.5, $F$--0, 0.5, $F$') &  113494.86 \\ 
H$^{13}$CCCN    & (13--12)        &  114615.00 \\ 
NS              & $\Omega=$\,0.5 (2.5, $F$--1.5, $F$') $l=e$ &  115160.07 \\ 
CO              & (1--0)          &  115271.20 \\ 
gap             && 115490--126676     \\ 
HCC$^{13}$CN    & (14--13)        &  126839.59 \\ 
HC$^{13}$CCN    & (14--13)        &  126827.35 \\ 
$^{30}$SiO      & (3--2)          &  127117.54 \\ 
HC$_3$N         & (14--13)        &  127367.67 \\ 
CH$_2$NH        & (2, 0, 2, $F$--1, 0, 1, $F$') &  127856.55 \\ 
U?              & U               &  127856.00 \\ 
U               & U               &  128141.98 \\ 
$^{29}$SiO      & (3--2)          &  128637.04 \\
CH$_3$CN??      & (7--6)          &  128779.36 \\ 
SO              & (3, 3--2, 2)      &  129138.92 \\ 
OSB-CS?         & U               &  129918.62 \\ 
SO?             & (8, 8--8, 7)      &  129953.66 \\ 
SiO             & (3--2)          &  130268.68 \\ 
SiN?            & (3, 2.5, $F$--2, 1.5, $F$') &  130710.83 \\ 
SiN?            & (3, 3.5, $F$--2, 2.5, $F$') &  131215.24 \\ 
CH$_3$NH$_2$?   & (3, 1, 0--2, -1, 1) &  131685.75 \\ 
AlF             & (4--3)          &  131898.76 \\ 
CH$_3$NH$_2$    & (3, 1, 4--2, 1, 4)  &  132201.25 \\ 
H$^{13}$CCCN    & (15--14)        &  132246.36 \\ 
U               & U               &  132526.88 \\ 
$^{13}$CH$_3$OH? & (5, 1, 4--4, -2, 3) &  132960.07 \\ 
CH$_3$NH$_2$    & (3, 0, $\Gamma$--2, 0, $\Gamma$') &  132983.06 \\ 
CH$_3$NH$_2$    & (3, 2, $\Gamma$--2, 2, $\Gamma$') &  133022.24 \\ 
U               & U               &  133144.76 \\ 
CH$_2$NH?       & (2, 1, 1, 3--1, 1, 0, 2) &  133272.13 \\ 
CH$_3$NH$_2$    & (3, 1, 6--2, 1, 6)  &  133421.06 \\ 
CH$_3$NH$_2$    & (3, 1, 5--2, 1, 5)  &  133810.68 \\ 
SO$_2$          & (8, 2, 6--8, 1, 7)  &  134004.86 \\ 
CH$_3$NH$_2$    & (3, -1, 1 or 3--2, 1, 0 or 2) &  134327.05 \\ 
CH$_3$NH$_2$??  & (5, 0, 5--4, 1, 5)  &  134631.73 \\ 
OSB-CH$_3$OH    & U               &  135277.80 \\ 
H$_2$CS         & (4, 1, 4--3, 1, 3)  &  135298.26 \\ 
SO$_2$          & (5, 1, 5--4, 0, 4)  &  135696.02 \\ 
HC$_3$N         & (15--14)        &  136464.41 \\ 
H$_2$CS         & (4, 2, 3--3, 2, 2)  &  137382.12 \\ 
H$_2^{13}$CO  & (2, 1, 2--1, 1, 1)  &  137449.95 \\ 
CH$_3$NH$_2$??  & (5, 0, 3--4, 1, 2)  &  137583.91 \\ 
SO              & (3, 4--2, 3)      &  138178.60 \\ 
$^{13}$CS       & (3--2)          &  138739.34 \\ 
CH$_3^{18}$OH? & (3, $K_a$, 3--2, $K_a$', 2) &  139080.31 \\ 
SO$_2$          & (6, 2, 4--6, 1, 5)  &  140306.17 \\ 
H$_2$CO         & (2, 1, 2--1, 1, 1)  &  140839.50 \\ 
PN              & (3--2)          &  140967.69 \\ 
H$^{13}$CCCN    & (16--15)        &  141061.78 \\ 
$^{13}$CH$_3$OH & (3, $K_a$, 3--2, $K_a$, 2) &  141602.53 \\ 
$^{13}$CH$_3$OH & (3, $K_a$, $K_c$--2, $K_a$, $K_c$) &  141623.55 \\ 
H$_2^{13}$CO  & (2, 0, 2--1, 0, 1)  &  141983.74 \\ 
OSB-H$_2$CO     & U               &  141983.74 \\ 
$^{13}$CH$_3$CN & (8--7)          &  142925.60 \\ 
U               & U               &  143262.39 \\ 
C$^{34}$S       & (3--2)          &  144617.10 \\ 
CH$_3$OH        & (3, $K_a$, 3--2, $K_a$, 2) &  145103.15 \\ 
CH$_3$OH        & (3, $K_a$, $K_c$--2, $K_a$, $K_c$) &  145124.41 \\ 
SiS             & (8--7)          &  145227.05 \\ 
H$_2$CO         & (2, 0, 2--1, 0, 1)  &  145602.95 \\ 
CH$_3$OH?       & (9, 0, 9--8, 1, 1)  &  146618.70 \\ 
H$_2^{13}$CO  & (2, 1, 1--1, 1, 0)  &  146635.67 \\ 
U               & U               &  146828.37 \\ 
CS              & (3--2)          &  146969.03 \\ 
CH$_3$CN        & (8--7)          &  147174.59 \\ 
U               & U               &  147270.00 \\ 
U               & U               &  150000.00 \\ 
NO              & $\Omega=$\,0.5 (2, $F$--1, $F$') &  150176.48 \\ 
H$_2$CO         & (2, 1, 1--1, 1, 0)  &  150498.33 \\ 
U               & U               &  151158.24 \\ 
SO$_2$          & (2, 2, 0--2, 1, 1)  &  151378.63 \\ 
U               & U               &  152779.52 \\ 
HNCO            & (7, 0, 7--6, 0, 6)  &  153865.09 \\ 
U               & U               &  154493.70 \\ 
HC$_3$N         & (17--16)        &  154657.28 \\ 
$^{34}$SO       & (4, 3--3, 2)      &  155506.80 \\ 
U               & U               &  155810.60 \\ 
$^{13}$CH$_3$OH & (4, 0, 4--4, -1, 4) &  156356.39 \\ 
$^{13}$CH$_3$OH & ($K$, 0, $K_c$--$K$', -1, $K_c$')$K_c=K$ &  156373.66 \\ 
CH$_3$OH        & ($K$, 0, $K_c$--$K$', -1, $K_c$)$K_c=K$ &  157270.85 \\ 
SO$_2$          & (3, 2, 2--3, 1, 3)  &  158199.74 \\ 
SO              & (4, 3--3, 2)      &  158971.81 \\ 
U               & U               &  159802.16 \\ 
U               & U               &  160001.87 \\ 
SO$_2$          & (10, 0, 10--9, 1, 9) &  160827.88 \\ 
NS              &$\Omega=$\,0.5 (3.5, $F$--2.5, $F$') $l=e$ &  161300.59 \\ 
OSB-HC$^{15}$N  & U               &  161370.00 \\ 
Si$^{18}$O      & (4--3)          &  161404.88 \\ 
NS              & $\Omega=$\,0.5 (3.5, $F$--2.5, $F$') $l=f$ &  161697.91 \\ 
OSB-$^{29}$SiO  & U               &  161990.65 \\ 
gap             && 162810--162980 \\ 
SiS             & (9--8)          &  163376.78 \\ 
AlF             & (5--4)          &  164867.70 \\ 
SO$_2$          & (5, 2, 4--5, 1, 5)  &  165144.65 \\ 
SO$_2$          & (7, 1, 7--6, 0, 6)  &  165225.45 \\ 
U               & U               &  165970.00 \\ 
CH$_3$CN        & (9, 0--8, 0)      &  165569.08 \\ 
CH$_2$NH        & (1, 1, 0, $F$--1, 0, 1, $F$') &  166851.87 \\ 
H$^{13}$C$^{15}$N & (2--1)          &  167453.28 \\ 
CH$_3$NH$_2$    & (2, 1, 4--1, 0, 5)  &  167598.40 \\ 
H$_2^{34}$S?  & (1, 1, 0--1, 0, 1)  &  167910.52 \\ 
$^{13}$CCH      & (2--1)          &  168305.82 \\ 
H$_2$S          & (1, 1, 0--1, 0, 1)  &  168762.76 \\ 
H$_2$CS         & (5, 1, 5--4, 1, 4)  &  169114.08 \\ 
CH$_3$NH$_2$    & (2,1,2--1,0,3)      &  169447.51 \\  
$^{30}$SiO      & (4--3)          &  169486.87 \\ 
CH$_3$OH?       & (3, 2, 1--2, 1, 1)  &  170060.58 \\ 
HC$^{18}$O$^+$  & (2--1)          &  170322.63 \\ 
H$^{15}$N$^{13}$C  & (2--1)          &  170515.68 \\ 
C$^{13}$CH      & (2--1)          &  170520.94 \\ 
U               & U               &  171070.00 \\ 
$^{29}$SiO      & (4--3)          &  171512.80 \\ 
H$_2$CS         & (5, 0, 5--4, 0, 4)  &  171688.12 \\ 
H$_2$CS?        & (5, 2, 4--4, 2, 3)  &  171720.51 \\ 
H$_2$CS?        & (5, 2, 3--4, 2, 2)  &  171780.17 \\ 
HC$^{15}$N      & (2--1)          &  172107.96 \\ 
CH$_2$NH        & (2, 1, 1, $F$--2, 0, 2, $F$') &  172267.18 \\ 
H$^{13}$CN      & (2--1)          &  172677.85 \\ 
U               & U               &  172990.48 \\ 
CH$_3$NH$_2$    & (2, 1, 5--1, 0, 5)  &  173267.80 \\ 
H$^{13}$CO$^+$  & (2--1)          &  173506.70 \\ 
SiO             & (4--3)          &  173688.24 \\ 
HN$^{13}$C      & (2--1)          &  174179.41 \\ 
H$_2$CS         & (5, 1, 4--4, 1, 3)  &  174345.22 \\ 
SiN             & (4, 4.5, $F$--3, 3.5, $F$') &  174360.89 \\ 
CCH             & (2, 2.5, $F$--1, 1.5, $F$') &  174664.11 \\ 
CCH             & (2, 1.5, $F$--1, 0.5, $F$') &  174725.12 \\ 
CCH             & (2, 1.5, $F$--1, 1.5, $F$') &  174818.11 \\ 
SO$_2$          & (7, 2, 6--7, 1, 7)  &  175275.72 \\ 
HNCO            & (8, 0, 8--7, 0, 7)  &  175843.70 \\ 
U               & U               &  176749.10 \\ 
HCN             & (2--1)          &  177261.11 \\ 
H$^{15}$NC      & (2--1)          &  177729.09 \\ 
HCO$^+$         & (2--1)          &  178375.06 \\ 
gap             && 178480--230660 \\ 
CO              & (2--1)          &  230538.00 \\ 
$^{13}$CS       & (5--4)          &  231220.69 \\ 
U               & U               &  231720.00 \\ 
U               & U               &  232670.00 \\ 
SO$_2$          & (4, 2, 2--3, 1, 3)  &  235151.72 \\ 
$^{13}$CH$_3$OH??& (5, 0, 5--4, 0, 4)  &  235960.37 \\ 
U               & U               &  239070.00 \\ 
U               & U               &  240370.00 \\ 
C$^{34}$S       & (5--4)          &  241016.09 \\ 
gap             && 241410--246210   \\ 
NO              & $\Omega=$\,0.5 (3, $F$--2, $F$') &  250444.39 \\ 
NO              & $\Omega=$\,0.5 (3, $F$--2, $F$') &  250803.29 \\ 
H$^{13}$C$^{15}$N & (3--2)          &  251175.23 \\ 
SO$_2$          & (8, 3, 5--8, 2, 6)  &  251210.59 \\ 
U               & U               &  251450.00 \\ 
U               & U               &  251560.00 \\ 
U               & U               &  251630.00 \\ 
SO              & (6, 5--5, 4)      &  251825.77 \\ 
U               & U               &  252770.00 \\ 
NS              & $\Omega=$\,0.5 (5.5, $F$--4.5, $F$') $l=e$ &  253570.94 \\ 
NS              & $\Omega=$\,0.5 (5.5, $F$--4.5, $F$') $l=f$ &  253969.71 \\ 
CH$_3$OH        & (2, 0, 2--1, -1, 1) &  254015.34 \\ 
CH$_3$NH$_2$    & (4, 1, 4--3, 0, 5)  &  254055.85 \\ 
$^{30}$SiO      & (6--5)          &  254216.66 \\ 
SO$_2$          & (6, 3, 3--6, 2, 4)  &  254280.54 \\ 
CH$_2$NH        & (4, 0, 4, $F$--3, 0, 3, $F$') &  254685.26 \\ 
HC$^{18}$O$^+$? & (3--2)          &  255479.39 \\ 
C$^{13}$CH?     & (3--2)          &  255776.41 \\ 
SO$_2$          & (5, 3, 3--5, 2, 4)  &  256246.95 \\ 
$^{34}$SO       & (6, 7--5, 6)      &  256877.81 \\ 
\hline
\multicolumn{3}{l}{{\bf Note.} Quantum assignments to transitions follow the notation in JPL or CDMS.}
\end{longtable}
}

\longtab[2]{                       
\begin{longtable}{lcr}
\caption{\label{tab-ID-apex}The same as Table\, \ref{tab-ID-iram} but for APEX spectra.}\\
\hline\hline
Molecule& Transition & Rest frequency (MHz)\\
\hline
\endfirsthead
\caption{continued.}\\
\hline\hline
Molecule& Transition & Rest frequency (MHz)\\
\hline
\endhead
\hline
\endfoot
%
$^{30}$SiO      & (4--3)          &  169486.87 \\ 
$^{29}$SiO      & (4--3)          &  171512.80 \\ 
HC$^{15}$N      & (2--1)          &  172107.96 \\ 
H$^{13}$CN      & (2--1)          &  172677.85 \\ 
CH$_3$NH$_2$    & (2, 1, 5--1, 0, 5)  &  173267.80 \\ 
H$^{13}$CO$^+$  & (2--1)          &  173506.70 \\ 
SiO             & (4--3)          &  173688.24 \\ 
HN$^{13}$C      & (2--1)          &  174179.41 \\ 
gap             && 174770--181380  \\ 
N$_2$H$^+$      & (2--1)          &  186344.68 \\ 
U               & U               &  186850.00 \\ 
PN              & (4--3)          &  187953.26 \\ 
$^{13}$CH$_3$OH & (4, $K_a$, 4--3, $K_a$, 3) &  188788.03 \\ 
$^{13}$CH$_3$OH & (4, $K_a$, 3--3, $K_a$, 2) &  188836.43 \\ 
gap             && 189320--19.480   \\ 
AlF             & (6--5)          &  197833.15 \\ 
gap             && 201330--216820 \\ 
SiO             & (5--4)          &  217104.92 \\ 
$^{13}$CN       & (2, 1.5 or 2.5, 1 or 2, $F$--1, 0.5 or 1.5, 1, $F$') &  217297.72 \\ 
$^{13}$CN       & (2, 2.5, 2 or 3, $F$--1, 1.5, 1 or 2, $F$') &  217456.59 \\ 
$^{13}$CN       & (2, 1.5, 1, $F$--1, 0.5, 1, $F$') &  217633.04 \\ 
H$_2$CO         & (3, 0, 3--2, 0, 2)  &  218222.19 \\ 
CH$_3$OH?       & (4, 2, 2--3, 1, 2)  &  218440.05 \\ 
C$^{18}$O       & (2--1)          &  219560.35 \\ 
C$^{15}$N?      & (2, 2.5, $F$--1, 1.5, $F$') &  219933.63 \\ 
$^{13}$CO       & (2--1)          &  220398.68 \\ 
gap             & & 220790--224750   \\ 
H$_2$CO         & (3, 1, 2--2, 1, 1)  &  225697.78 \\ 
CN              & (2, 1.5, $F$--1, 0.5, $F$') &  226658.92 \\ 
CN              & (2, 2.5, $F$--1, 1.5, $F$') &  226876.39 \\ 
instr.          && 227260--229180  \\ 
CO              & (2--1)          &  230538.00 \\ 
AlF             & (7--6)          &  230793.89 \\ 
$^{13}$CS       & (5--4)          &  231220.69 \\ 
gap             && 232540--241110  \\ 
CH$_3$OH??      & (5, -1, 5--4, -1, 4) &  241767.22 \\ 
CH$_3$OH??      & (5, 0, 5--4, 0, 4)  &  241791.43 \\ 
CS              & (5--4)          &  244935.56 \\ 
gap             && 245100--257000 \\ 
$^{29}$SiO      & (6--5)          &  257255.21 \\ 
HC$^{15}$N      & (3--2)          &  258157.00 \\ 
U               & U               &  257810.00 \\ 
U               & U               &  258630.00 \\ 
H$^{13}$CN      & (3--2)          &  259011.80 \\ 
H$^{13}$CO$^+$  & (3--2)          &  260255.34 \\ 
SiO             & (6--5)          &  260518.01 \\ 
HN$^{13}$C      & (3--2)          &  261263.51 \\ 
CCH             & (3, 3.5, 3--2, 2.5, 3) &  261978.12 \\ 
HCN             & (3--2)          &  265886.43 \\ 
U               & U               &  268450.00 \\ 
gap             && 268730--270000 \\
HNC             & (3--2)          &  271981.142\\
gap             && 273970--278489 \\
N$_2$H$^+$      & (3--2)          &  279511.75 \\ 
H$_2$CO         & (4, 1, 4--3, 1, 3)  &  281526.93 \\ 
SO$_2$          & (6, 2, 4--5, 1, 5)  &  282036.57 \\
CH$_2$NH?       &(2,1,2,$F$--1,0,1,$F$')  &  284254.50 \\ 
CS              & (6--5)          &  293912.09 \\ 
$^{29}$SiO      & (7--6)          &  300120.48 \\ 
CH$_3$OH?       & (1, 1, 0--1, 0, 1)  &  303366.89 \\ 
SiO             & (7--6)          &  303926.81 \\ 
$^{34}$SO?      & (2, 2--1, 2)      &  308993.58 \\ 
gap             && 310610--311160  \\ 
$^{13}$CO       & (3--2)          &  330587.97 \\ 
telluric        && 332500--332900     \\ 
CN              & (3, 2.5, $F$--2, 2.5, $F$') &  339487.80 \\ 
CN              & (3, 2.5, $F$--2, 1.5, $F$') &  340031.29 \\ 
CH$_3$OH?       & (2, 2, 0, $+$0--3, 1, 3, $+$0) &  340141.22 \\ 
CN              & (3, 3.5, $F$--2, 2.5, $F$') &  340248.80 \\ 
HC$^{15}$N      & (4--3)          &  344200.11 \\ 
H$^{13}$CN      & (4--3)          &  345339.77 \\ 
CO              & (3--2)          &  345795.99 \\ 
H$^{13}$CO$^+$  & (4--3)          &  346998.34 \\ 
SiO             & (8--7)          &  347330.58 \\ 
HN$^{13}$C      & (4--3)          &  348340.90 \\ 
$^{13}$CH$_3$OH & (1, 1, 1--0, 0, 0)  &  350103.12 \\ 
NO?             & (4, 1, 3.5, 4.5--3, 1, 2.5, 3.5) &  352204.30 \\ 
HCN             & (4--3)          &  354505.48 \\ 
HCO$^+$         & (4--3)          &  356734.22 \\ 
gap             && 356600--390520\\ 
H$_2^{34}$S?  & (2, 1, 1--2, 0, 2)  &  393725.70 \\ 
gap             && 395720--396300  \\ 
gap             && 400300--401100  \\ 
CH$_3$NH$_2$?   & (5, -3, 1--5, 2, 0) &  402354.89 \\ 
U               & U               &  402500.00 \\ 
telluric        &                &  408240.00 \\ 
gap             && 412300--413100  \\ 
gap             && 417080--417970  \\ 
gap             && 421960--426760  \\ 
telluric        &                 &  427400.00 \\ 
telluric        &                &  429350.00 \\ 
telluric        &                &  430300.00 \\ 
gap             && 430750--438780  \\ 
telluric        &                &  439400.00 \\ 
telluric        &                &  441270.00 \\ 
$^{13}$CO       & (4--3)          &  440765.17 \\ 
telluric        &                &  442050.00 \\ 
gap             && 442750--459050   \\ 
CO              & (4--3)          &  461040.77 \\ 
gap             && 463050--471070  \\ 
telluric        &                &  472850.00 \\ 
telluric        &                &  474500.00 \\ 
telluric        && 474100--475000   \\ 
gap             && 475040--477200   \\ 
gap             && 481200--489200   \\ 
C and $^{13}$C  & (1--0)           & 492160.651 \\ 
gap             && 494180--690060   \\ 
H$^{13}$CN      & (8--7)          &  690552.08 \\ 
CO              & (6--5)          &  691473.08 \\ 
gap             && 692880--805230   \\ 
gap             && 808060--907730   \\ 
end             && 906600            \\ 
\end{longtable}
}

\section{Results of LTE excitation analysis}

\begin{table*}[]\centering \small
\caption{Results of CASSIS analysis of the main isotopologues.}\label{tab-cassis}
\begin{tabular}{cc rcc rr}
\hline\hline
Species & $N$&\multicolumn{1}{c}{$T_{\rm ex}$}&3$\sigma$ range in& \multicolumn{1}{c}{FWHM  }& \multicolumn{1}{c}{Size     }& \multicolumn{1}{c}{$V_{\rm LSR}$}\\ 
& (cm$^{-2}$)&\multicolumn{1}{c}{ (K)        }&$T_{\rm ex}$ (K)  & \multicolumn{1}{c}{(\kms)}& \multicolumn{1}{c}{(\arcsec)}& \multicolumn{1}{c}{(\kms)}\\
\hline\hline
AlF\tablefootmark{a}      
        & 6.9e15 &  7.1 &6.6--7.6  &  76 &  1.1 & --10.5\\
CN      & 3.8e15 &  5.7 &5.3--6.1  & 120 & 11.1 & --10.0\\
CO      & 3.0e17 & 11.3 &10.2--12.3& 206 &  9.6 & --18.0\\
CS\tablefootmark{a}        
        & 2.7e15 &  5.1 &4.4--5.7  & 138 &  4.5 & --11.5\\
NO      & 4.0e15 &  4.9 &4.0--5.7  & 105 &  4.1 &    5.0\\
NS      & 4.7e14 &  6.0 &5.5--6.4  & 185 &  5.0 & --11.0\\
PN      & 1.7e15 &  4.5 &3.9--5.4  & 135 &  1.1 & --17.0\\
SO      & 2.3e15 &  9.8 &3.2--12.5 &  90 &  1.4 &  --7.0\\
SiO     & 8.1e13 & 11.8 &9.5--13.7 & 163 & 11.6 & --12.5\\
SiN     & 2.5e14 &  5.1 &3.9--6.8  &  65 &  2.2 &  --8.0\\
SiS     & 7.9e15 &  9.2 &6.6--12.0 & 133 &  0.9 &  --9.0\\
CCH          & 6.4e14 &  3.5 &3.5--3.9& 154 &  7.0 &  --9.5\\
H$_2$S\tablefootmark{b}            
             & 8.9e15 &  5.5 & $>$5.1 & 145 &  2.0 &    8.4\\
HCN          & 2.8e14 &  8.4 &7.5--9.3& 145 &  9.6 & --14.3\\
HCO$^+$      & 5.3e14 &  3.7 &3.6--3.9&  77 &  4.5 & --10.5\\
HNC\tablefootmark{a}            
             & 5.0e14 &  3.9 &3.3--5.0& 140 &  9.5 & --15.0\\
N$_2$H$^+$   & 1.5e14 &  6.3 &5.6--6.8& 173 &  4.5 & --15.5\\
SO$_2$       & 9.4e14 &  9.4 &6.2--15.7& 70 &  1.9 & --11.1\\
H$_2$CO\tablefootmark{c}        
             & 1.8e15 &  3.0 &2.9--3.0 & 130 &  7.2 & --8.5\\
HNCO\tablefootmark{a}            
             & 3.1e13 & 12.5 &9.7--19.5&  75 &  4.5 & --6.0\\
H$_2$CS\tablefootmark{b}            
             & 1.8e15 &  9.0 &6.6--12.9& 150 &  1.0 & --10.0\\
CH$_2$NH\tablefootmark{e}      
             & 9.5e13 & 10.4 &7.0--21.0& 160 &  4.5 & --8.0\\
HC$_3$N      & 1.8e14 &  8.6 &6.5--10.8& 130 &  6.8 & --9.6\\
CH$_3$CN\tablefootmark{a,d,e}       
             & 2.4e14 &  6.7 &6.4--6.9 & 200 &  2.9 & --2.7\\
CH$_3$OH\tablefootmark{d,e}      
             & 1.0e16 &  3.3 &3.1--3.7& 130 &  4.5 & --15.6\\
CH$_3$NH$_2$\tablefootmark{a,e} 
             & 1.6e16 &  3.9 &3.2--5.0& 183 &  3.4 &  --7.0\\
\hline
\end{tabular}
\tablefoot{The species listed here were simulated simultaneously with their rare istotopic versions. Isotopic ratios are given separately in Table\,\ref{tab-iso}.
\tablefoottext{a}{Species analyzed using spectroscopic data from JPL. CDMS was used for all other species.}
\tablefoottext{b}{{\it Ortho} to {\it para} ratio assumed to be 1.}
\tablefoottext{c}{{\it Ortho} to {\it para} ratio is $2.5 \pm 0.8$ (3$\sigma$). VASTEL database was used in the simulations.}
\tablefoottext{d}{The ratio of $A$ to $E$ type species was assumed to be 1.}
\tablefoottext{e}{Non-LTE excitation possible.}
}
\end{table*}

\section{CASSIS LTE simulation}
\clearpage
\begin{figure*} [tbh]
\centering
\includegraphics[angle=270,width=0.85\textwidth]{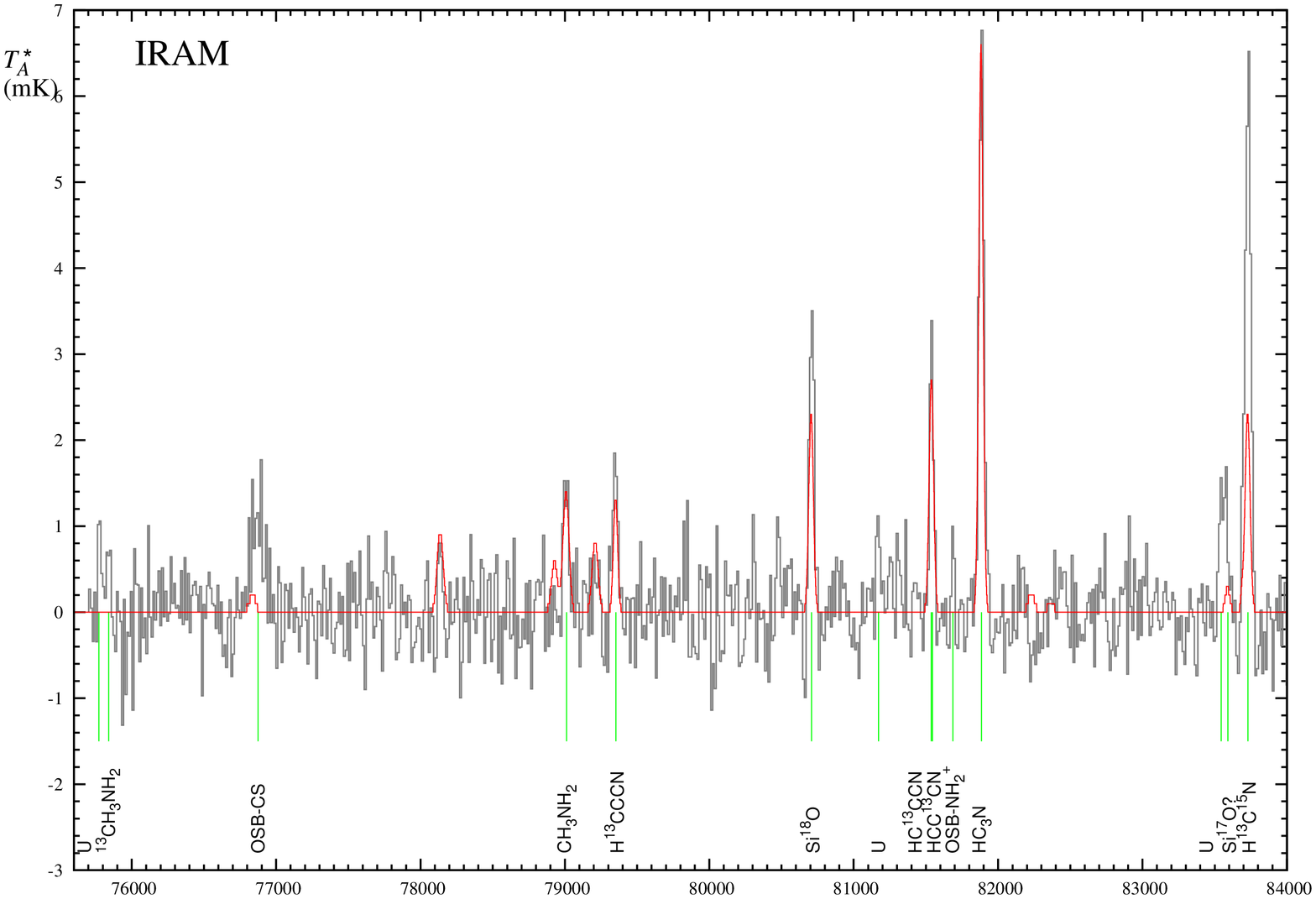}
\includegraphics[angle=270,width=0.85\textwidth]{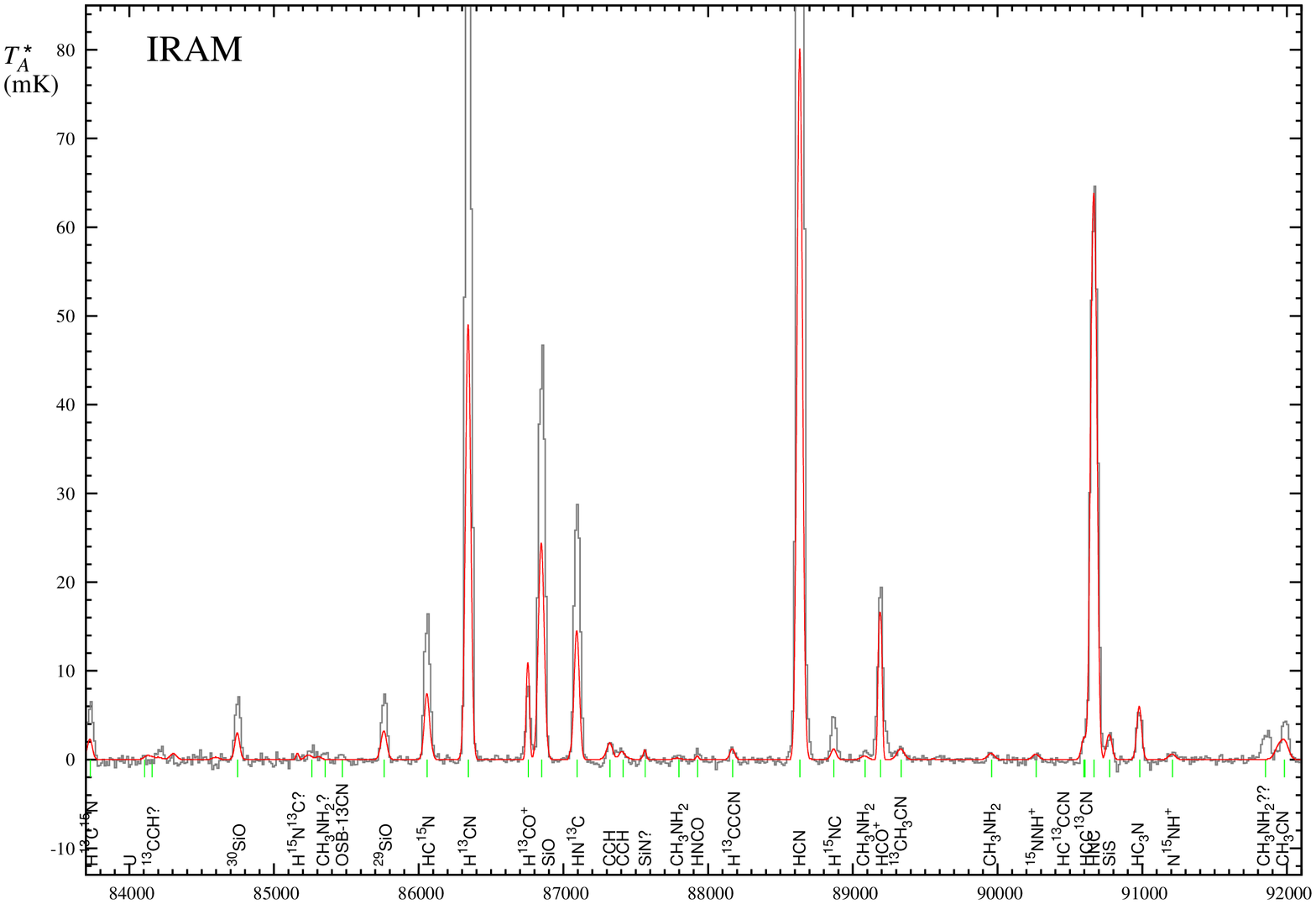}
\caption{IRAM spectra (grey) and a CASSIS LTE simulation of the spectrum (red). Some parts of the specta with no detected lines were skipped.}\label{fig-model-iram}
\end{figure*}

  \setcounter{figure}{0}%

\begin{figure*} [tbh]
\centering
\includegraphics[angle=270,width=0.85\textwidth]{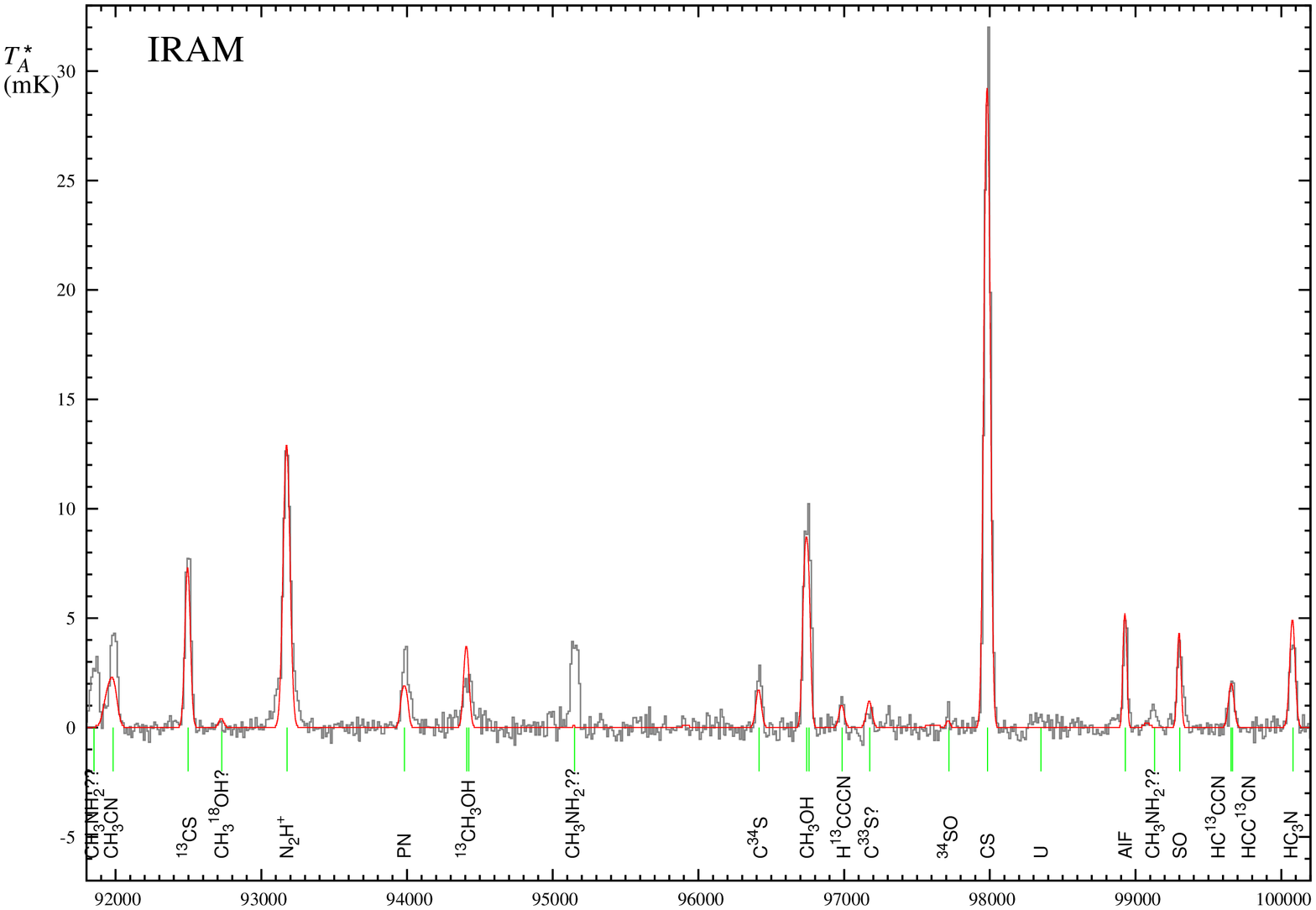}
\includegraphics[angle=270,width=0.85\textwidth]{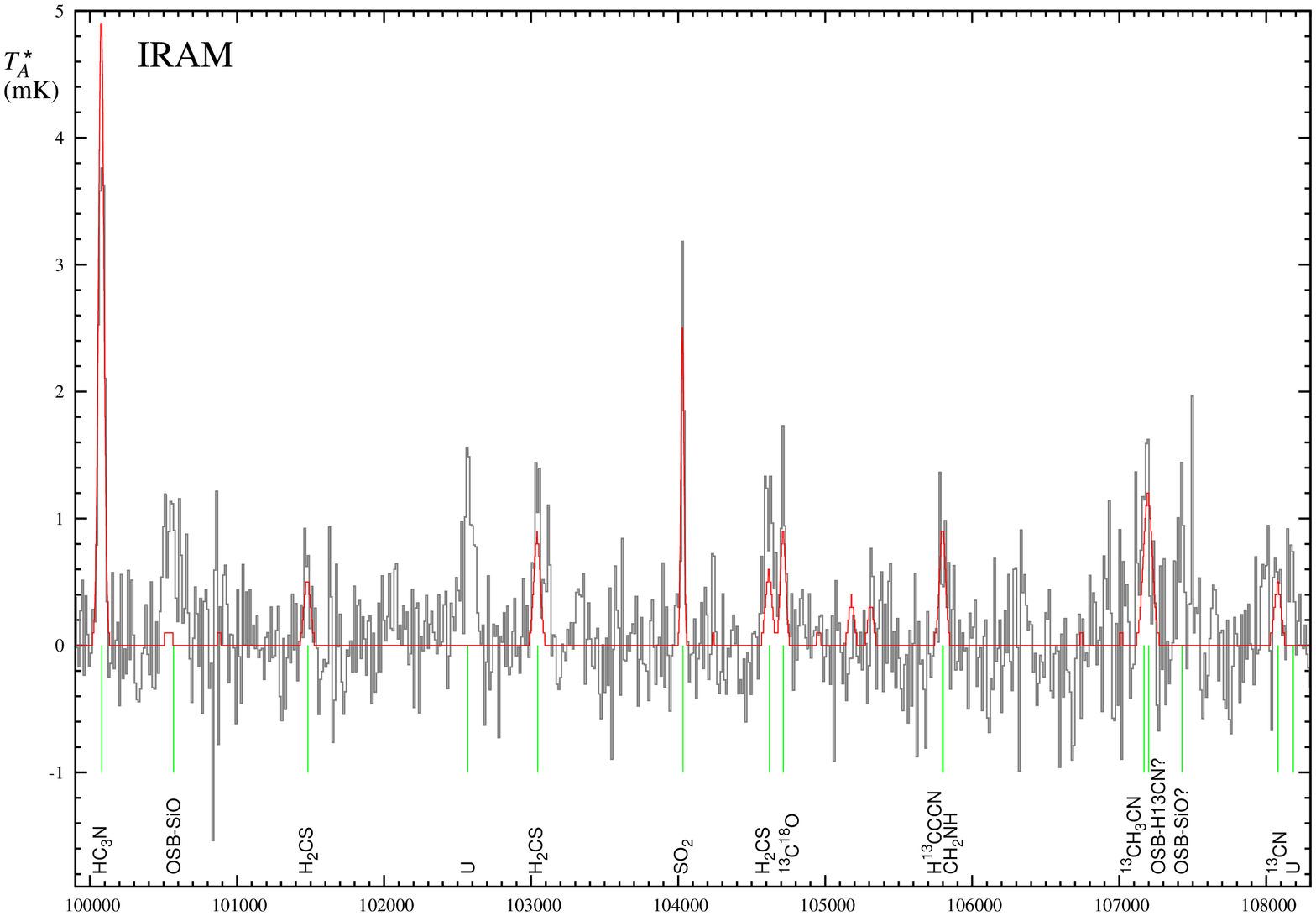}
\caption{Continued.}
\end{figure*}

  \setcounter{figure}{0}%

\begin{figure*} [tbh]
\centering
\includegraphics[angle=270,width=0.85\textwidth]{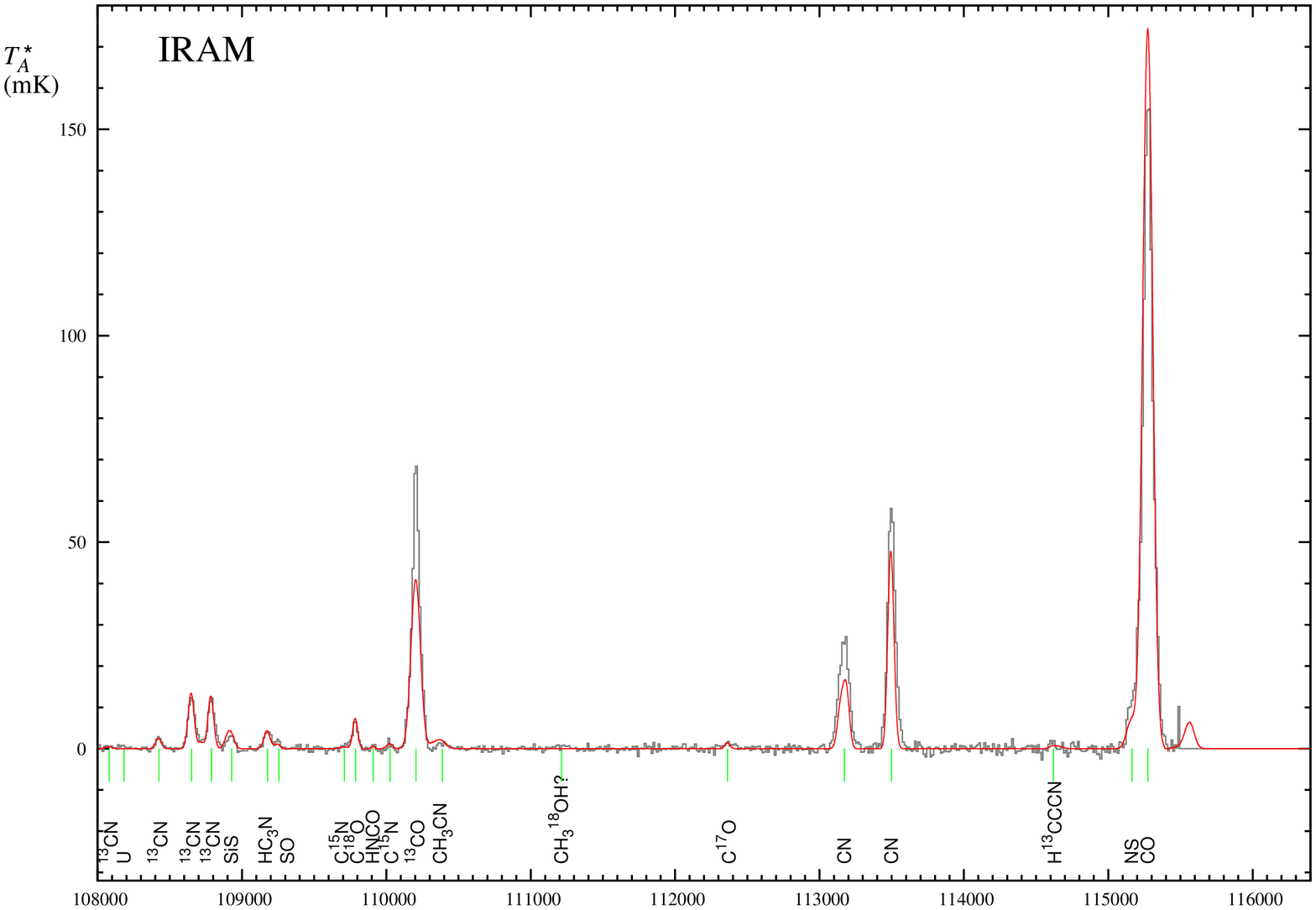}
\includegraphics[angle=270,width=0.85\textwidth]{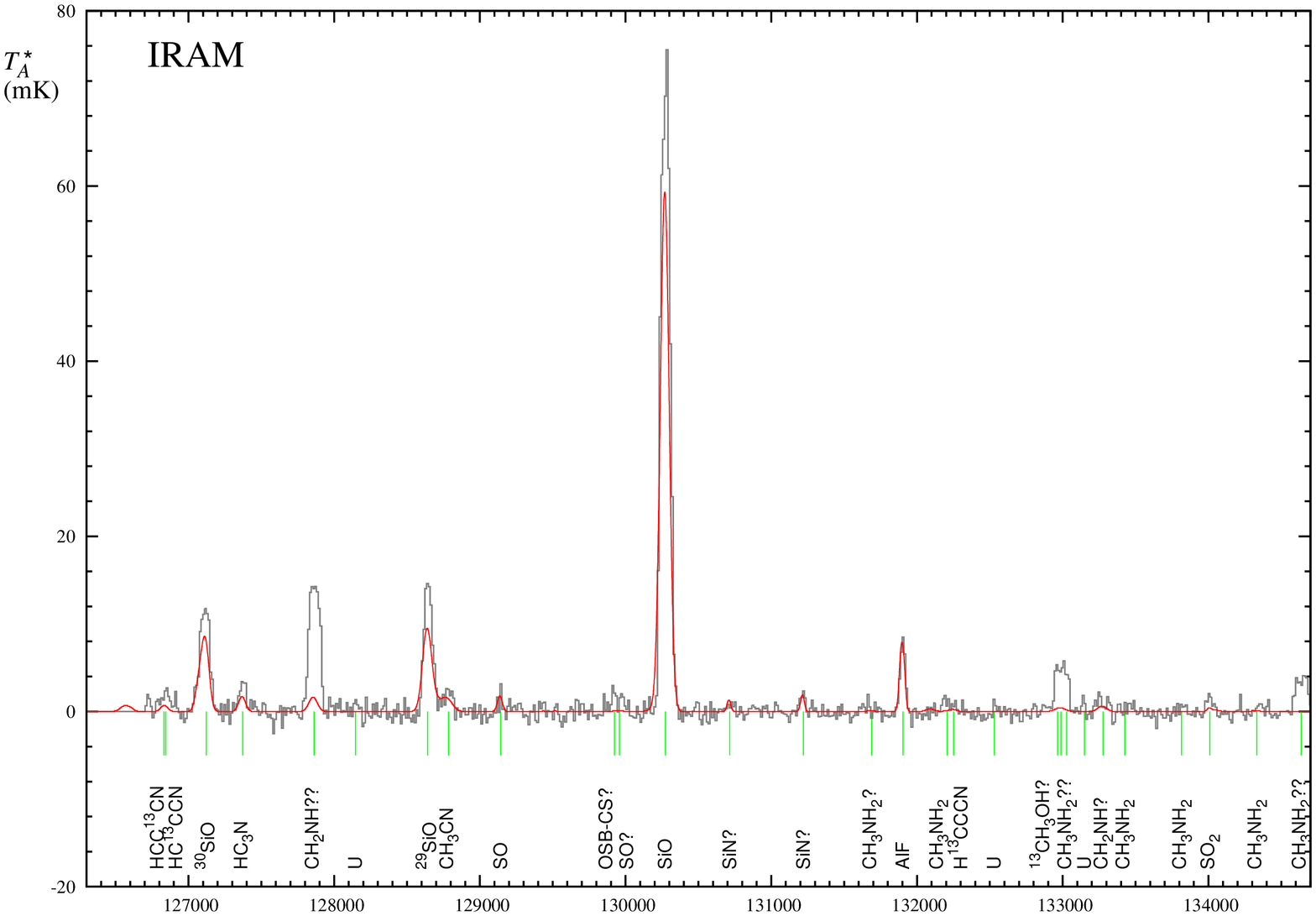}
\caption{Continued.}
\end{figure*}

  \setcounter{figure}{0}%

\begin{figure*} [tbh]
\centering
\includegraphics[angle=270,width=0.85\textwidth]{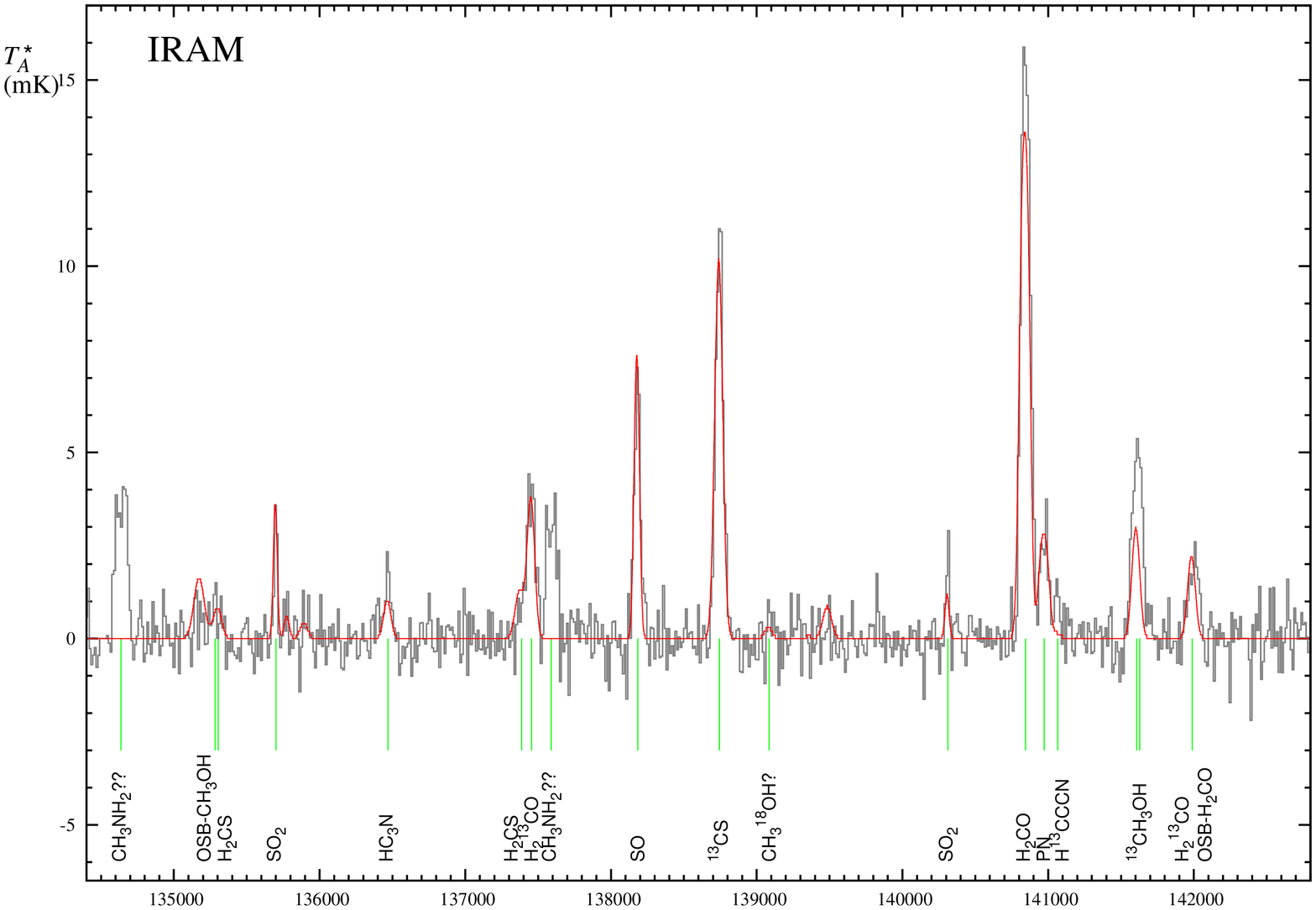}
\includegraphics[angle=270,width=0.85\textwidth]{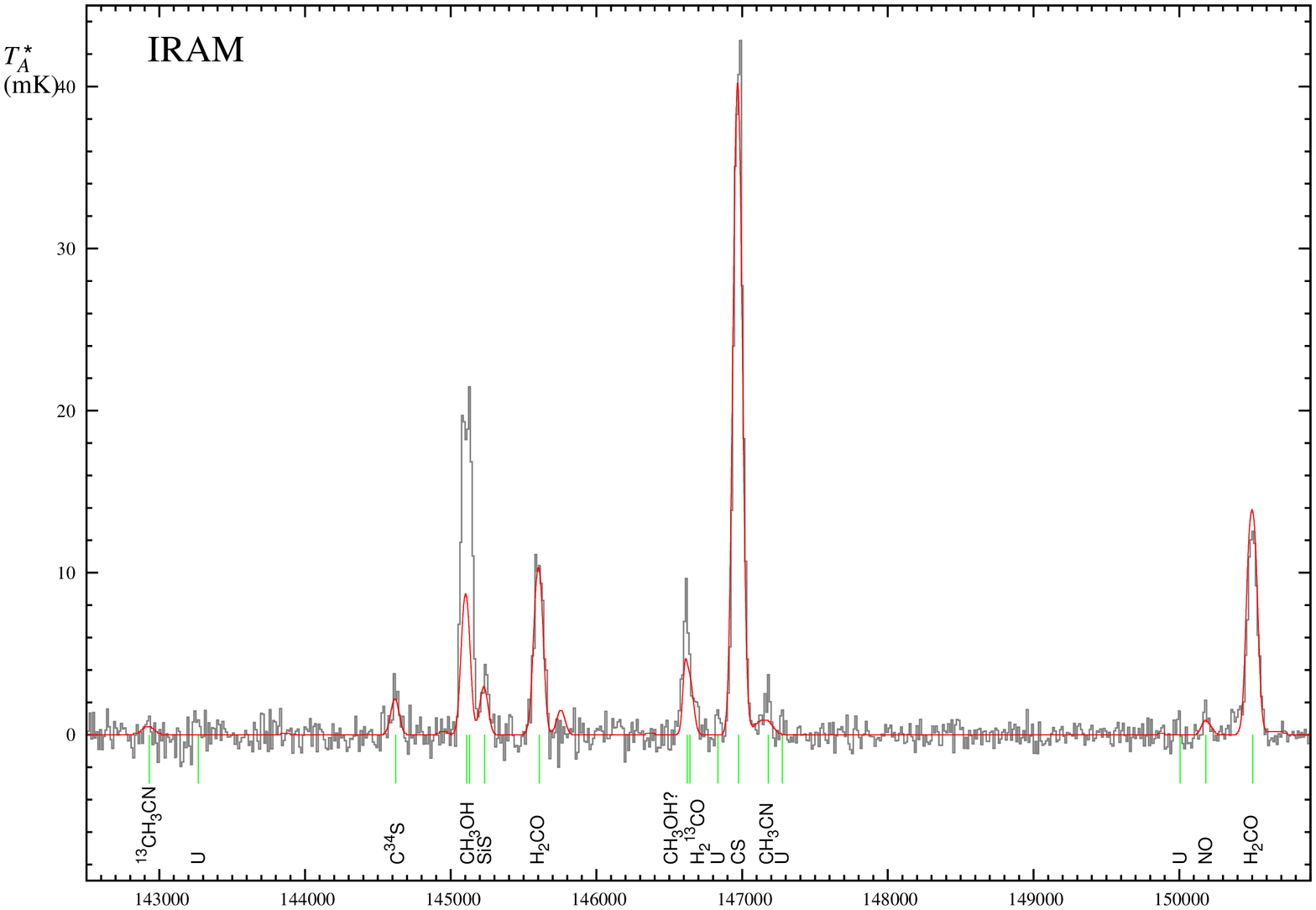}
\caption{Continued.}
\end{figure*}

  \setcounter{figure}{0}%

\begin{figure*} [tbh]
\centering
\includegraphics[angle=270,width=0.85\textwidth]{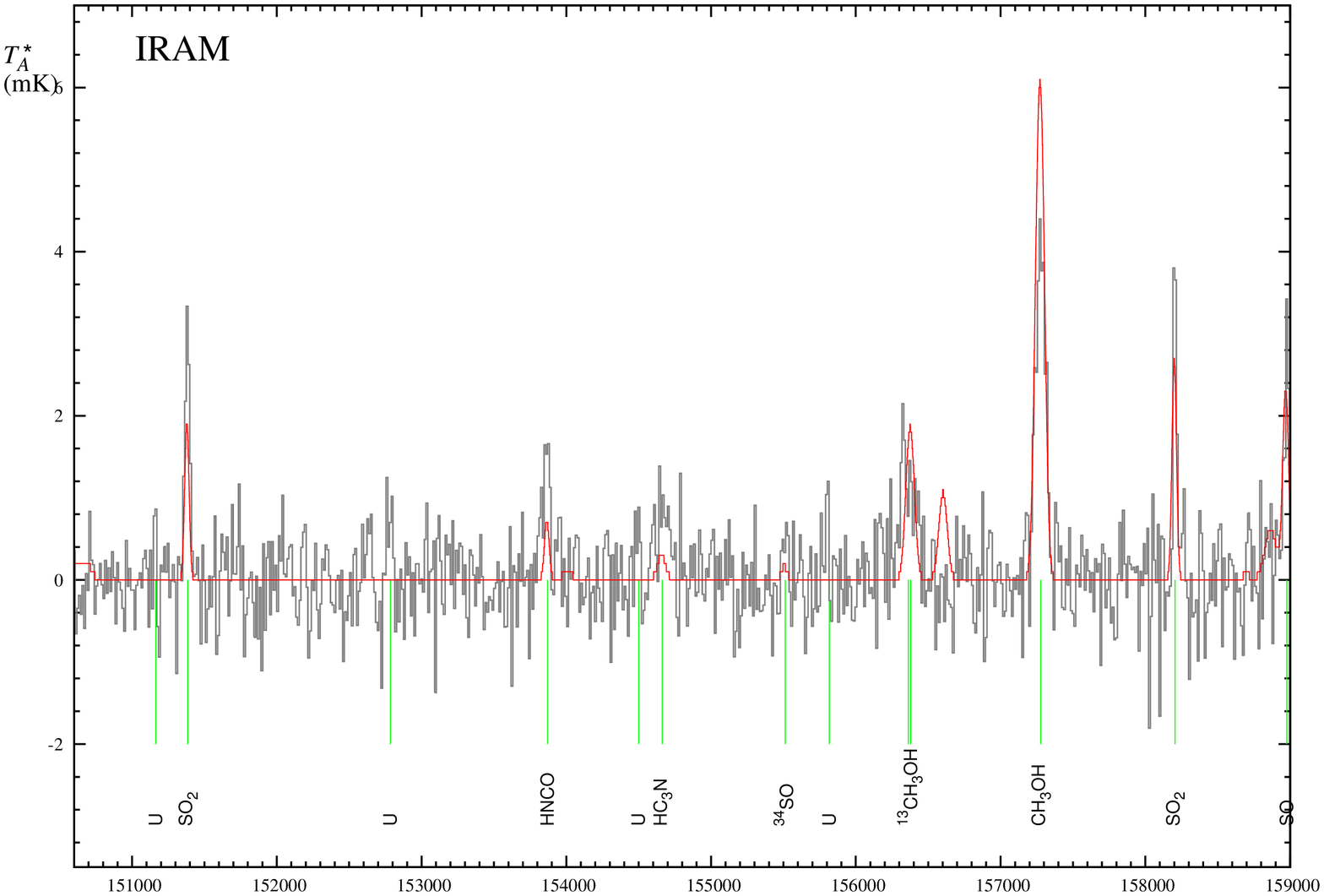}
\includegraphics[angle=270,width=0.85\textwidth]{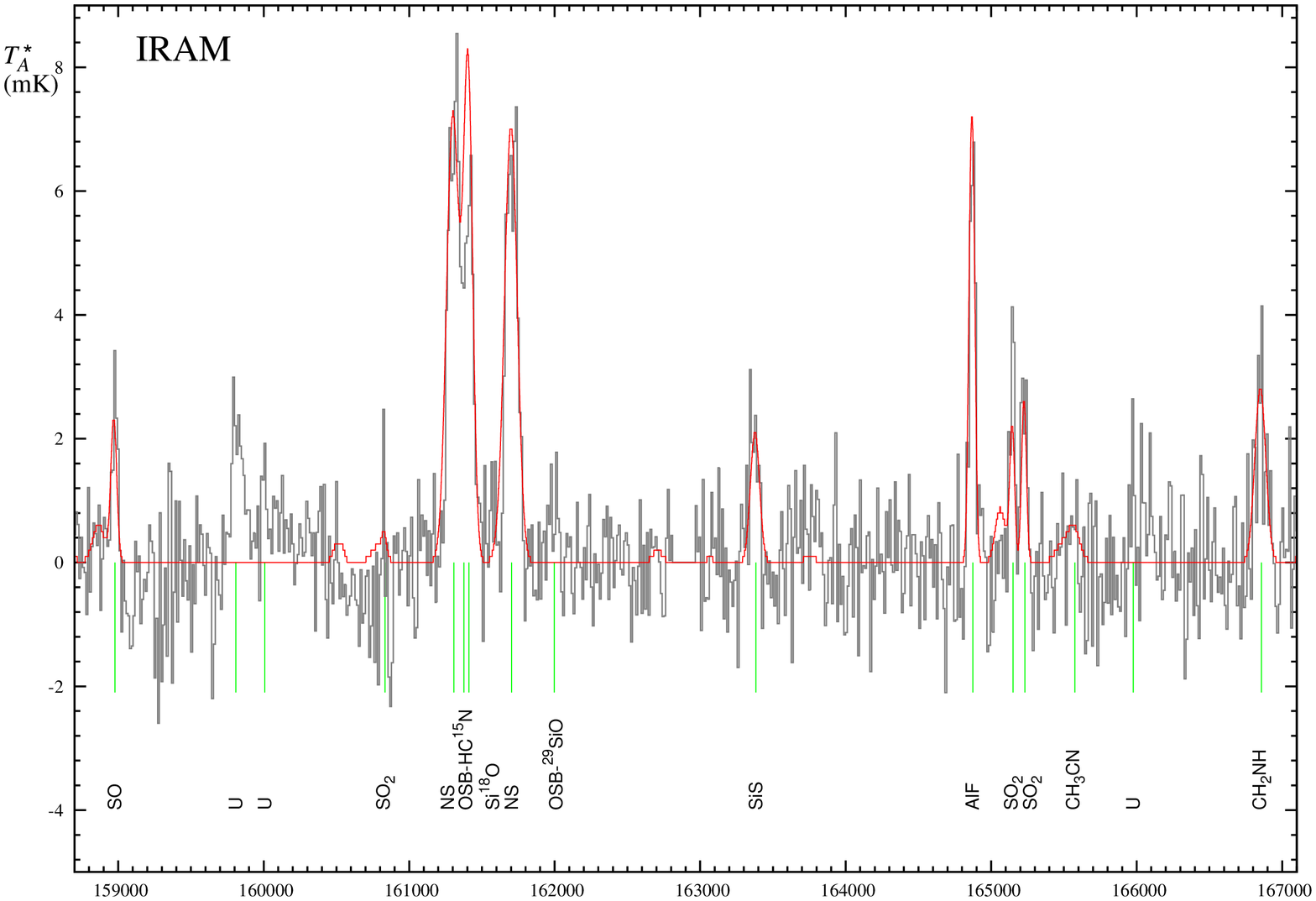}
\caption{Continued.}
\end{figure*}

  \setcounter{figure}{0}%

\begin{figure*} [tbh]
\centering
\includegraphics[angle=270,width=0.85\textwidth]{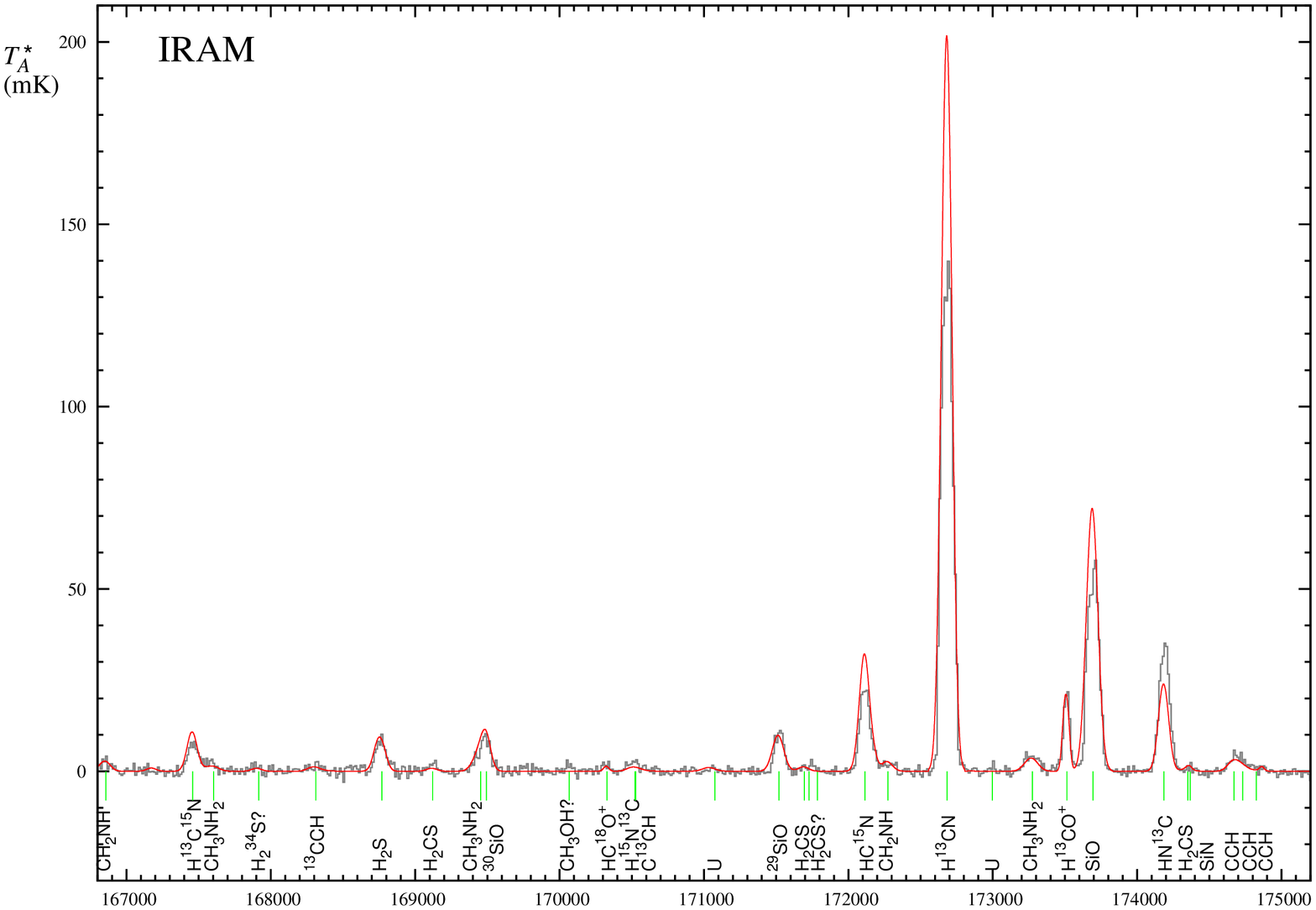}
\includegraphics[angle=270,width=0.85\textwidth]{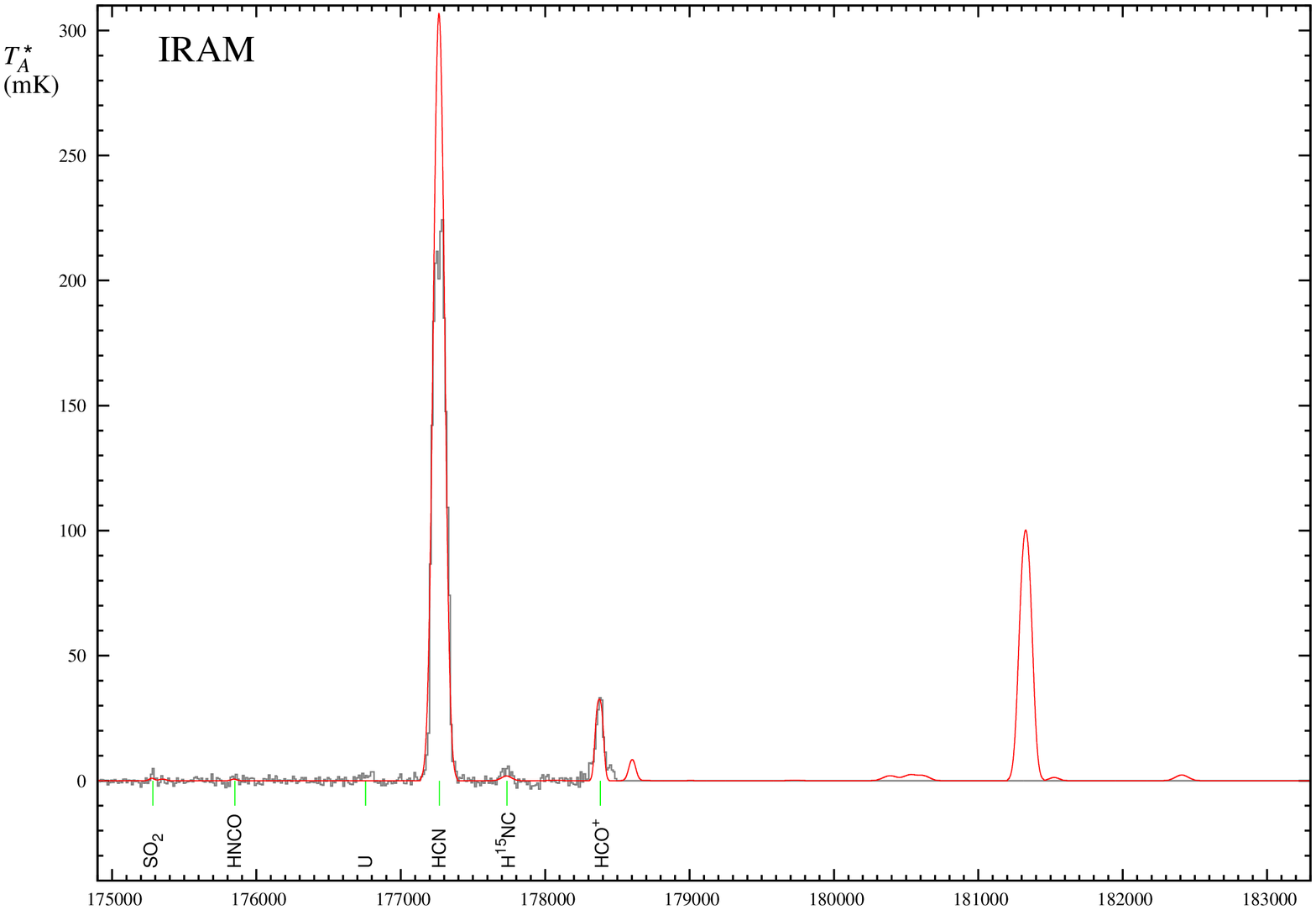}
\caption{Continued.}
\end{figure*}

  \setcounter{figure}{0}%

\begin{figure*} [tbh]
\centering
\includegraphics[angle=270,width=0.85\textwidth]{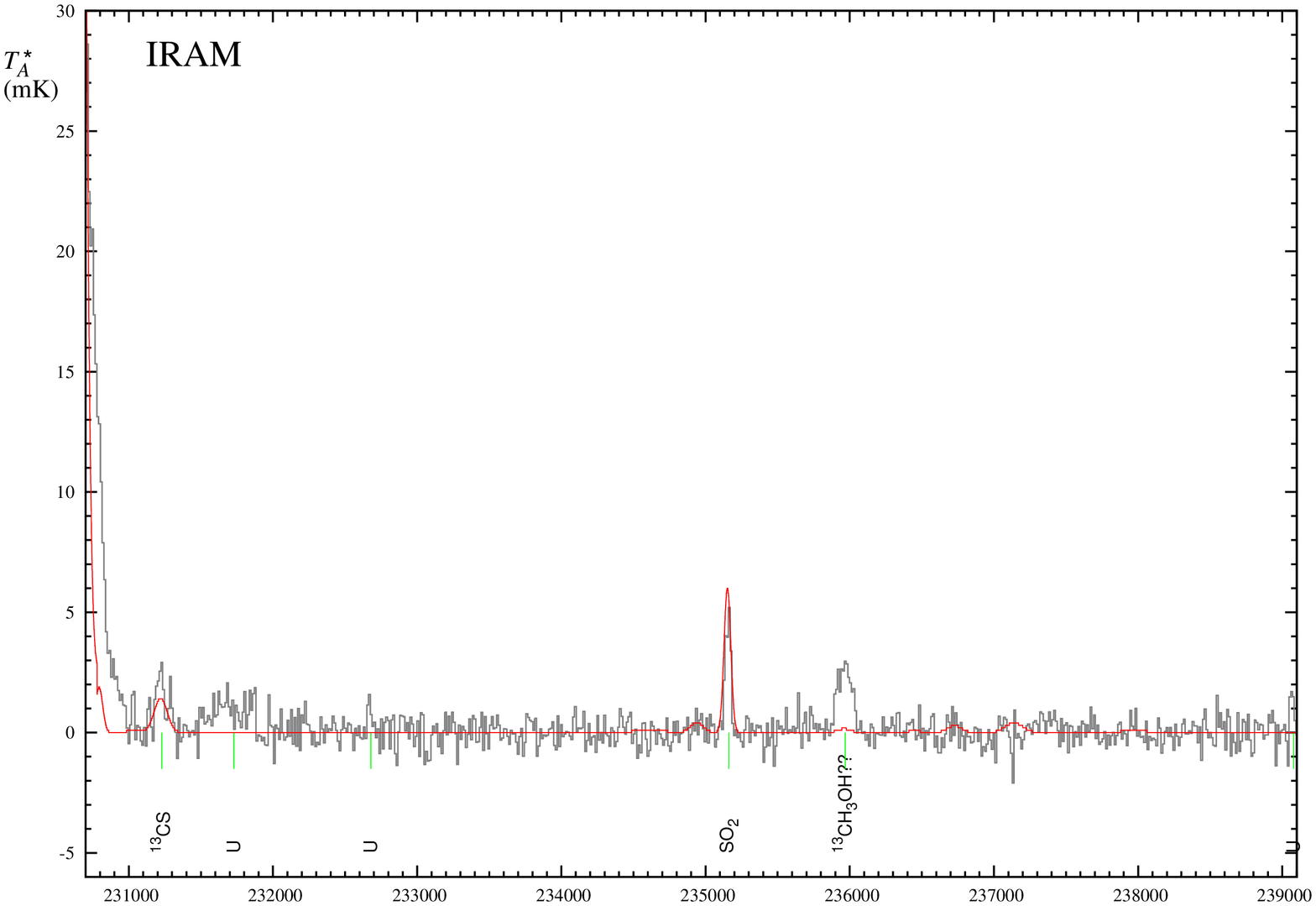}
\includegraphics[angle=270,width=0.85\textwidth]{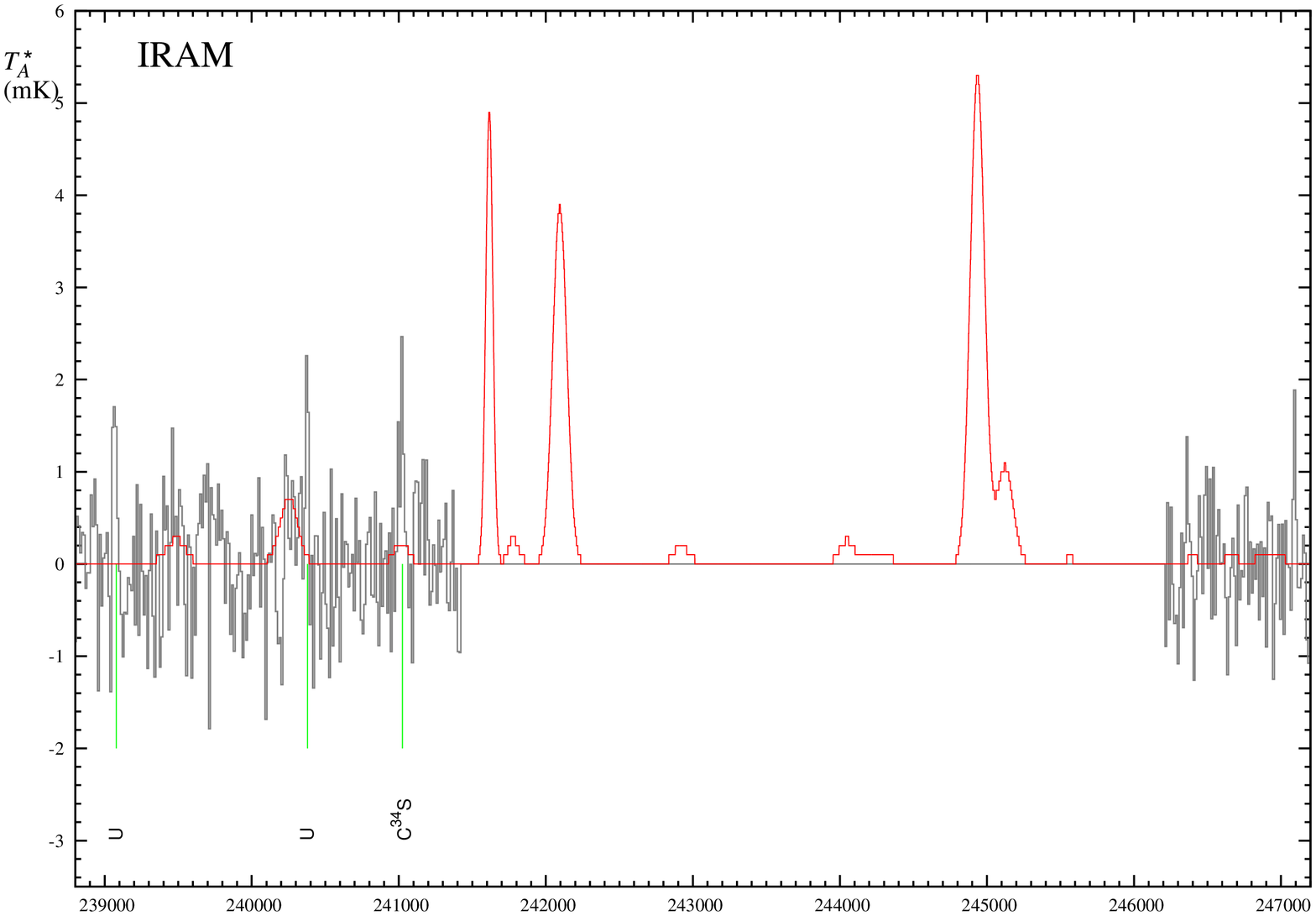}
\caption{Continued.}
\end{figure*}

  \setcounter{figure}{0}%

\begin{figure*} [tbh]
\centering
\includegraphics[angle=270,width=0.85\textwidth]{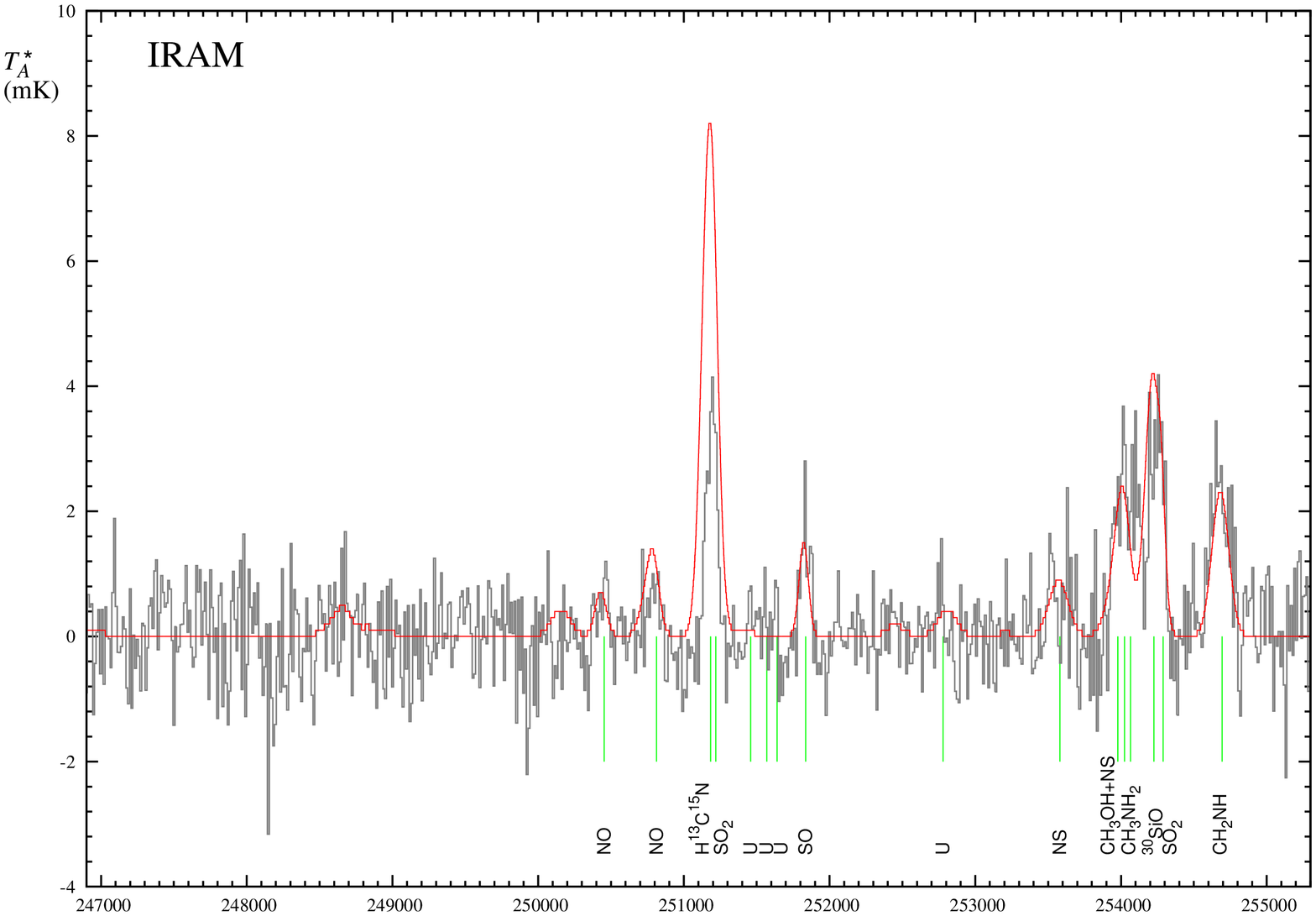}
\includegraphics[angle=270,width=0.85\textwidth]{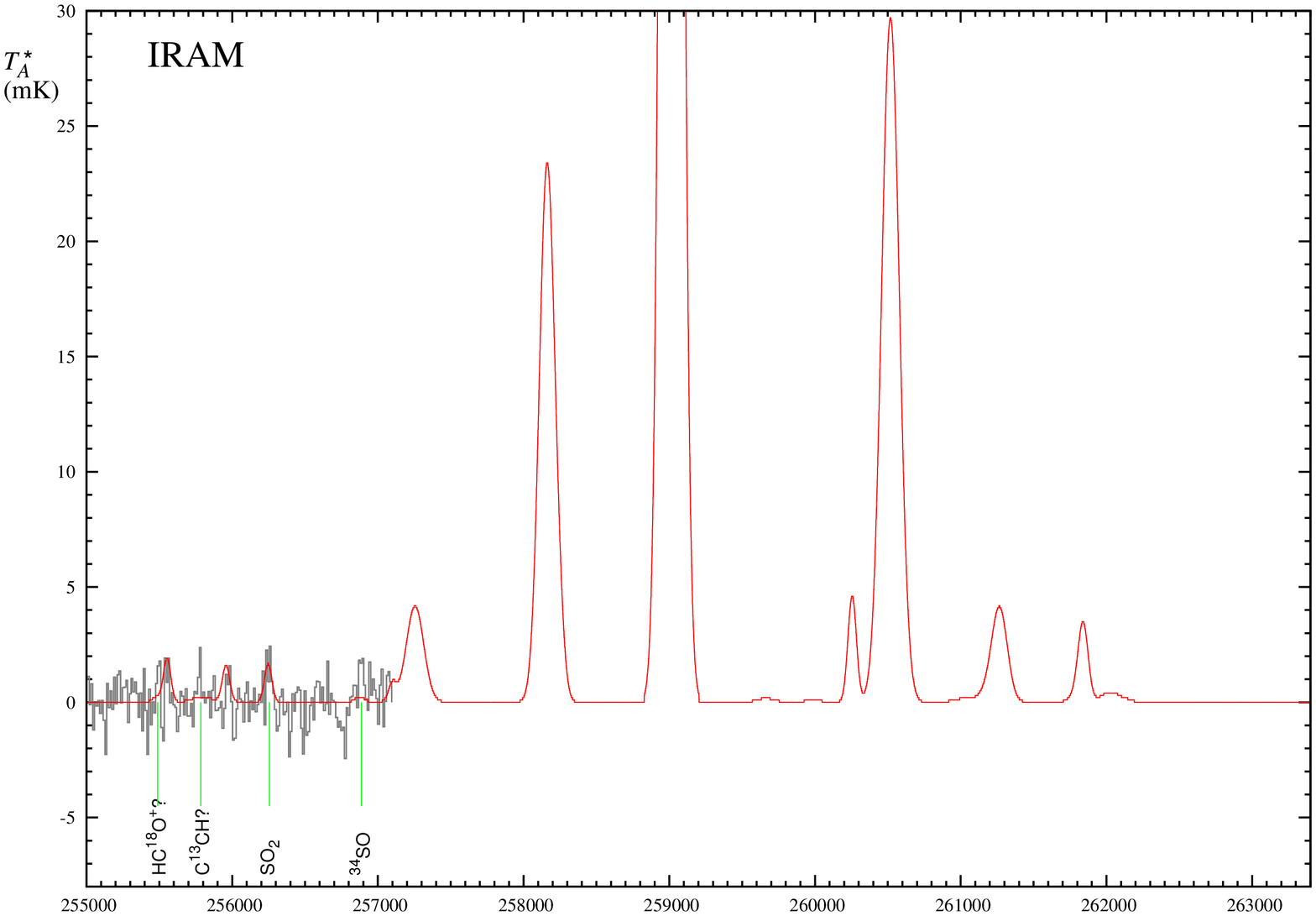}
\caption{Continued.}
\end{figure*}

  \setcounter{figure}{1}%

\clearpage
\begin{figure*} [tbh]
\centering
\includegraphics[angle=270,width=0.85\textwidth]{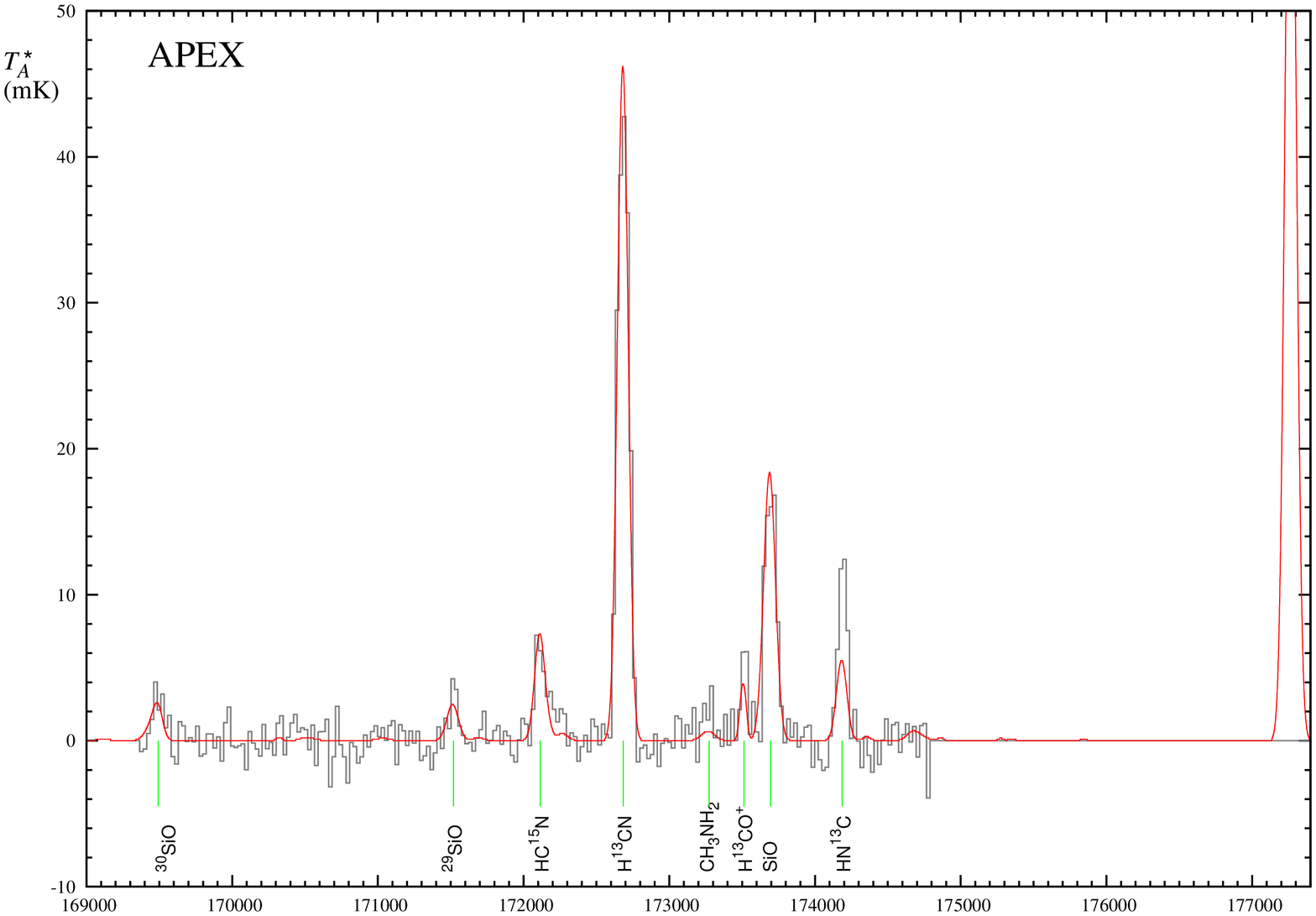}
\includegraphics[angle=270,width=0.85\textwidth]{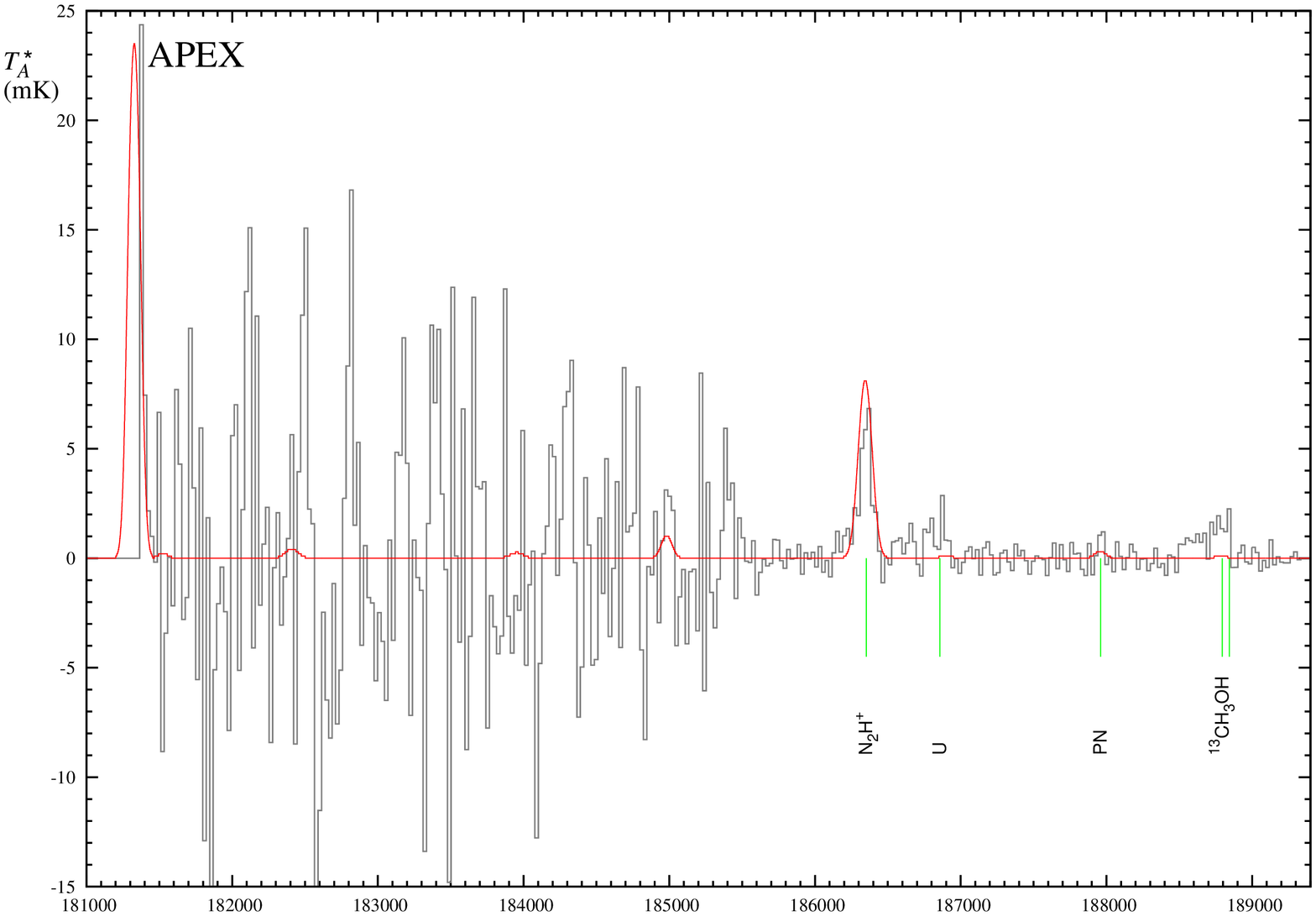}
\caption{Same as Fig.\,\ref{fig-model-iram} but for APEX spectra.}\label{fig-model-apex}

\end{figure*}

  \setcounter{figure}{1}%

\begin{figure*} [tbh]
\centering
\includegraphics[angle=270,width=0.85\textwidth]{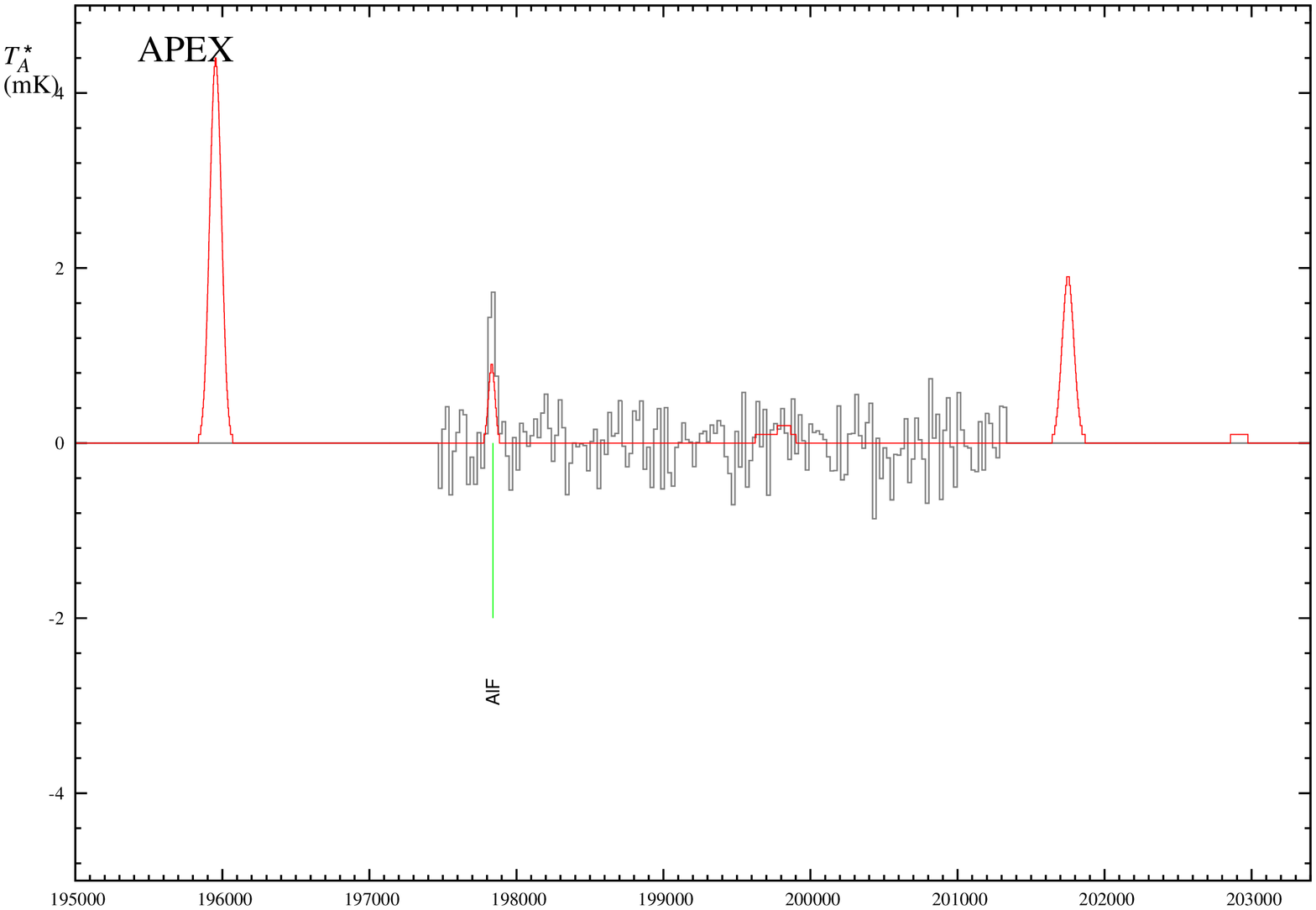}
\includegraphics[angle=270,width=0.85\textwidth]{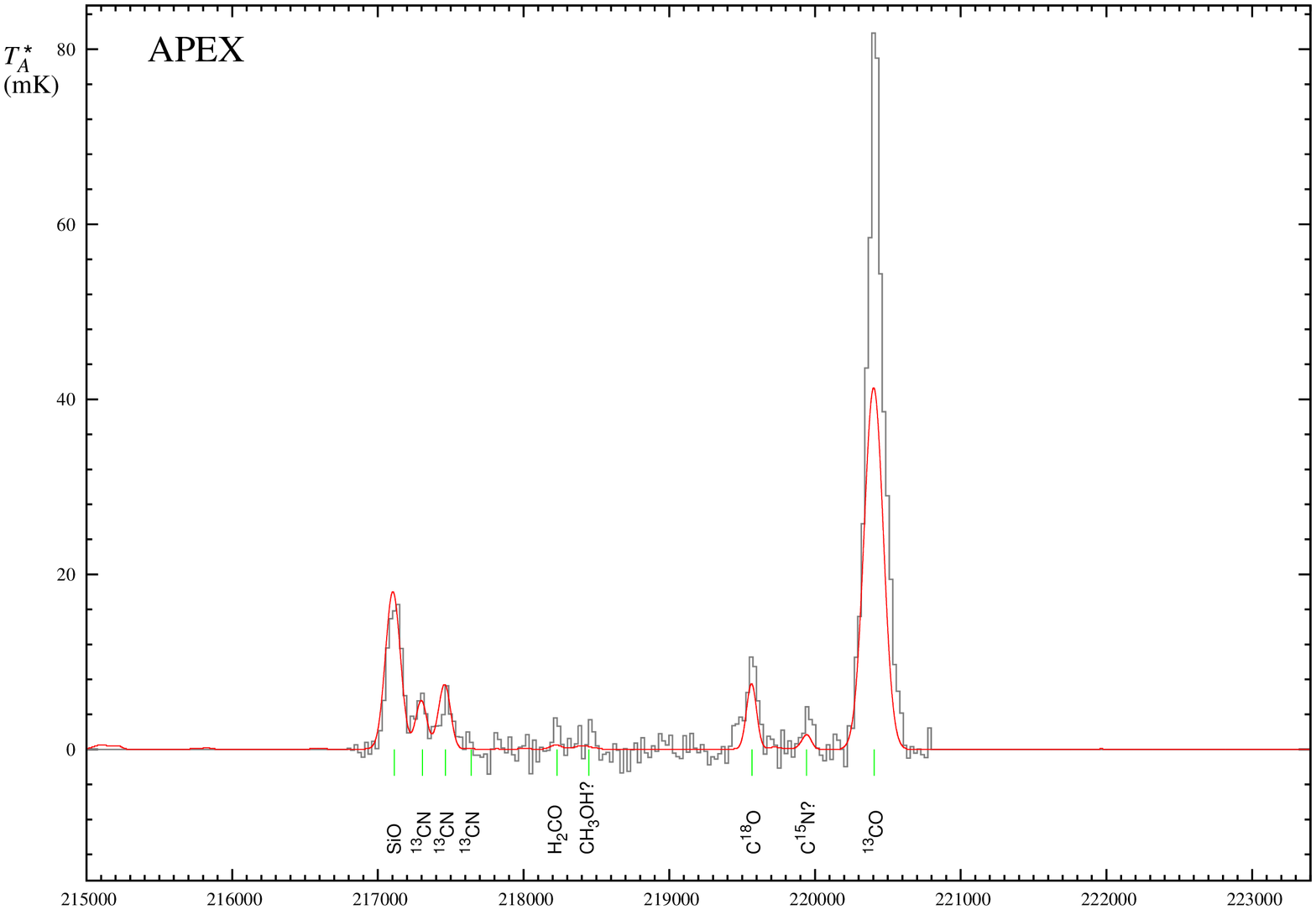}
\caption{Continued.}
\end{figure*}

  \setcounter{figure}{1}%

\begin{figure*} [tbh]
\centering
\includegraphics[angle=270,width=0.85\textwidth]{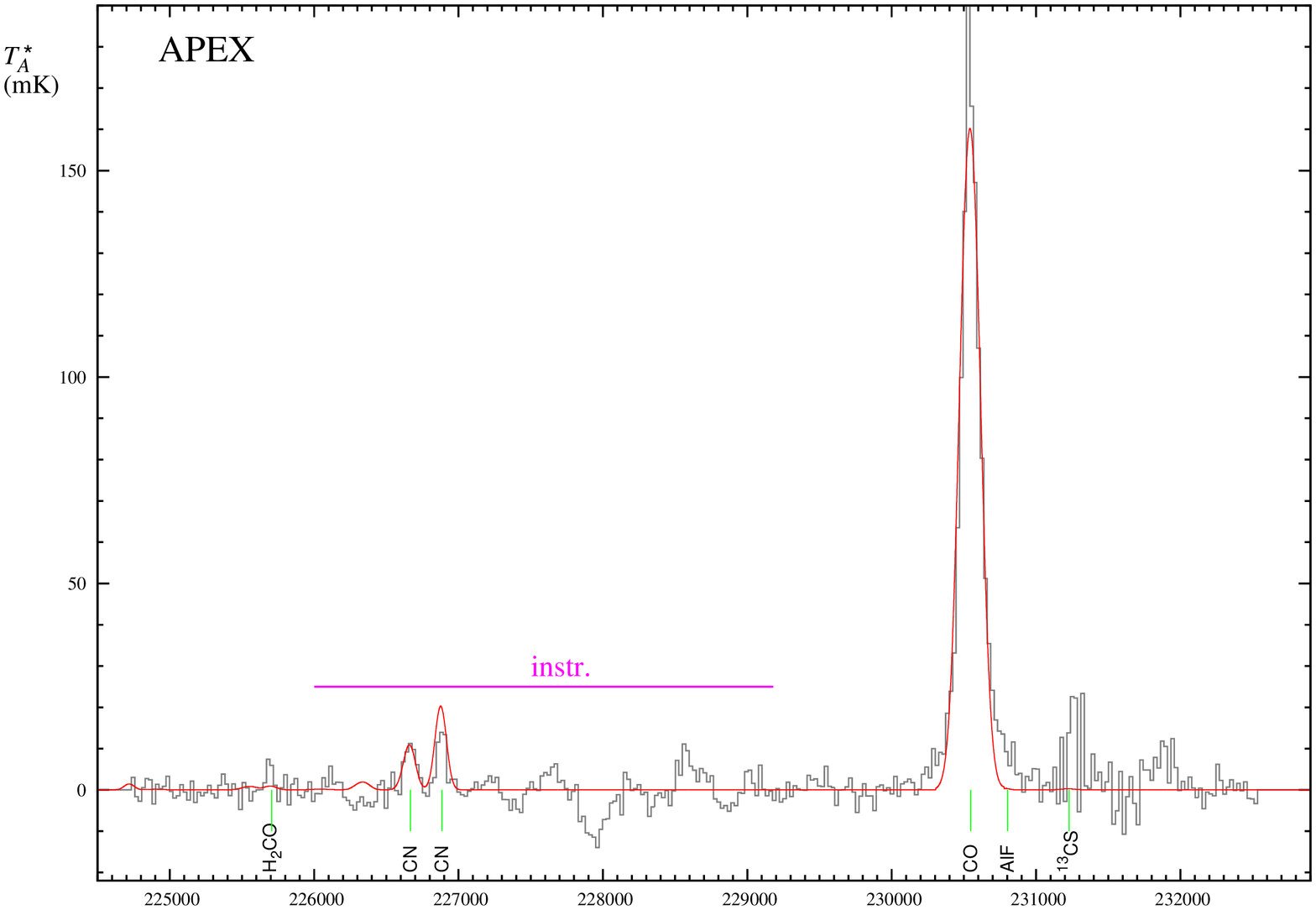}
\includegraphics[angle=270,width=0.85\textwidth]{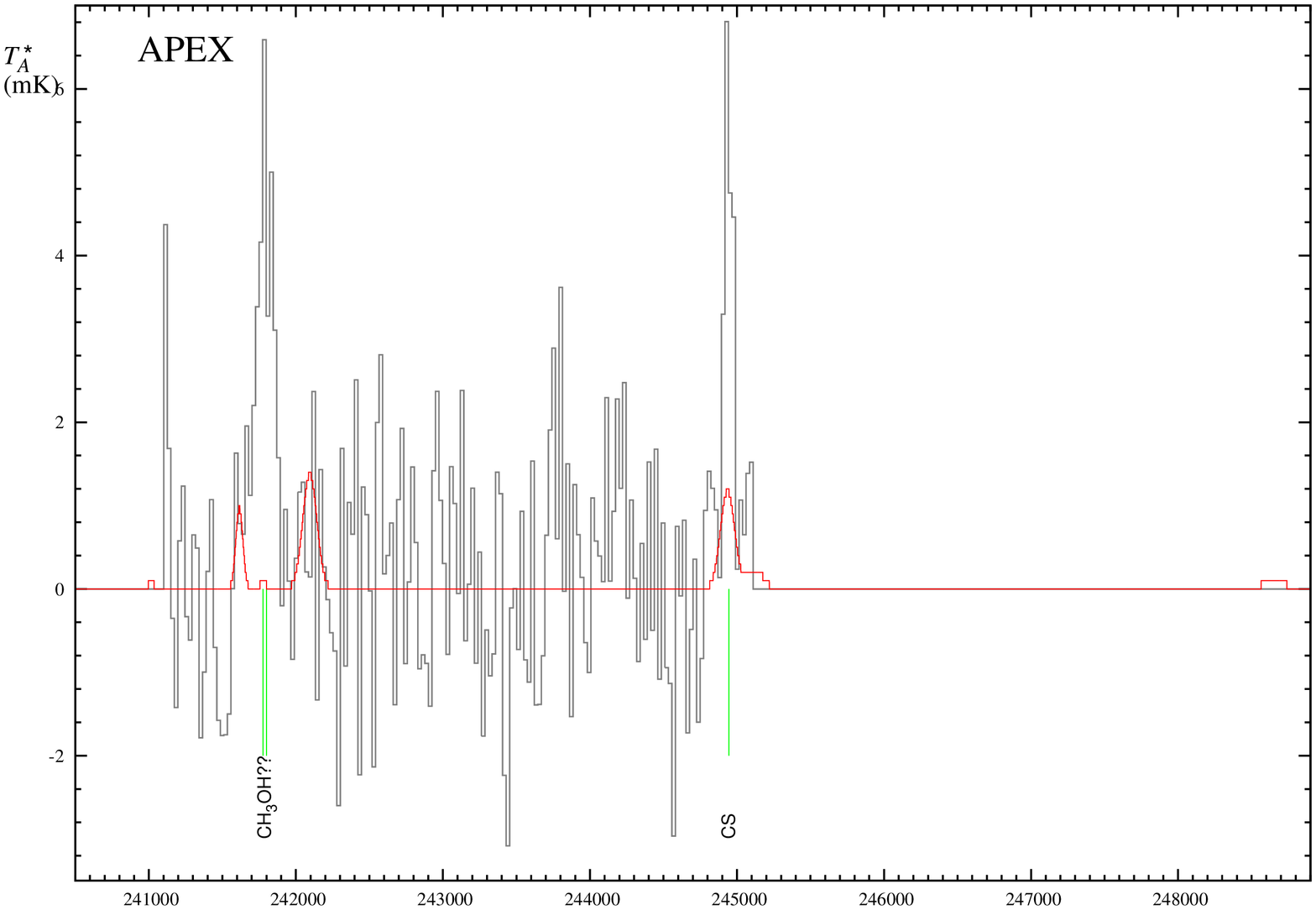}
\caption{Continued.}
\end{figure*}

  \setcounter{figure}{1}%

\begin{figure*} [tbh]
\centering
\includegraphics[angle=270,width=0.85\textwidth]{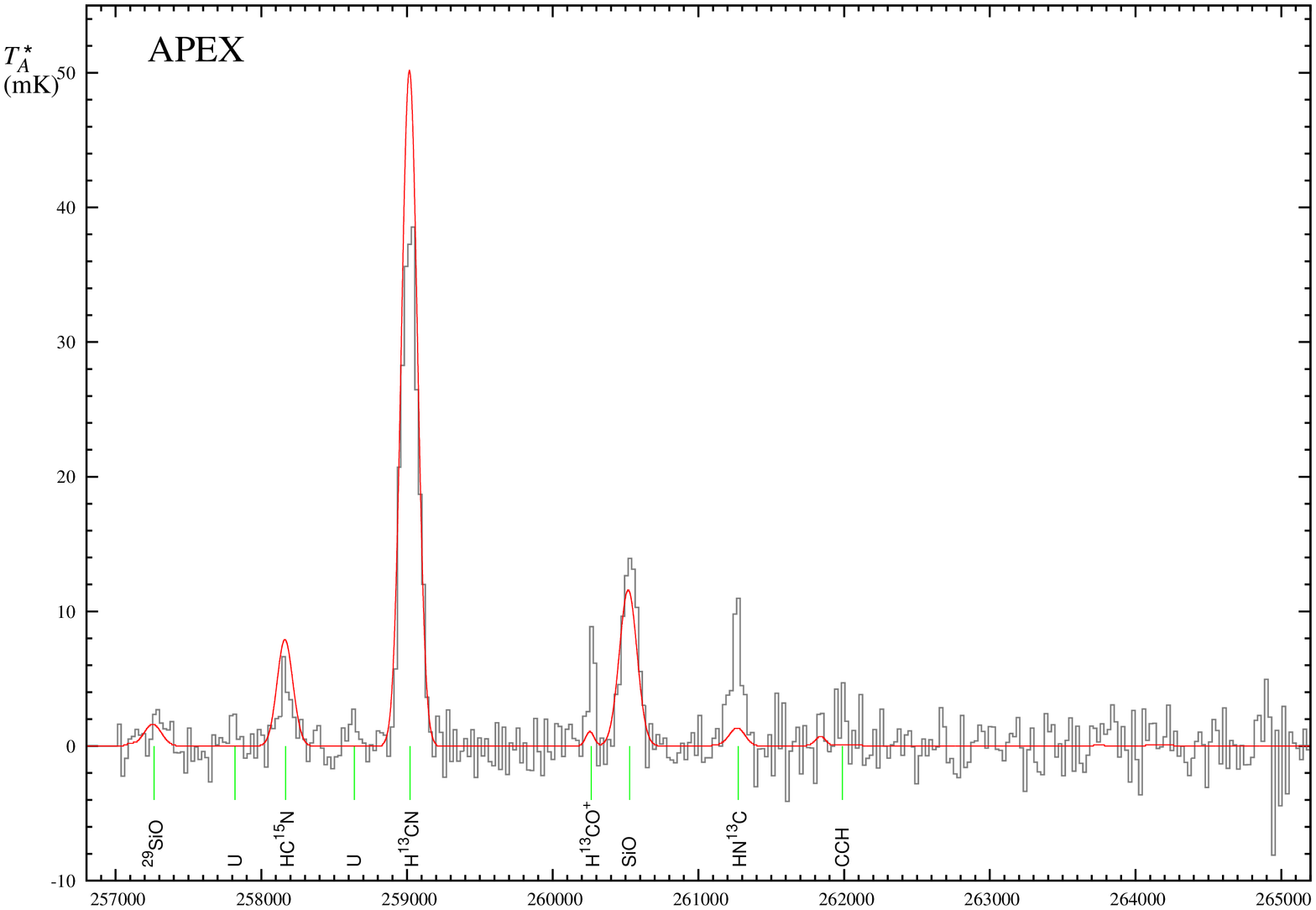}
\includegraphics[angle=270,width=0.85\textwidth]{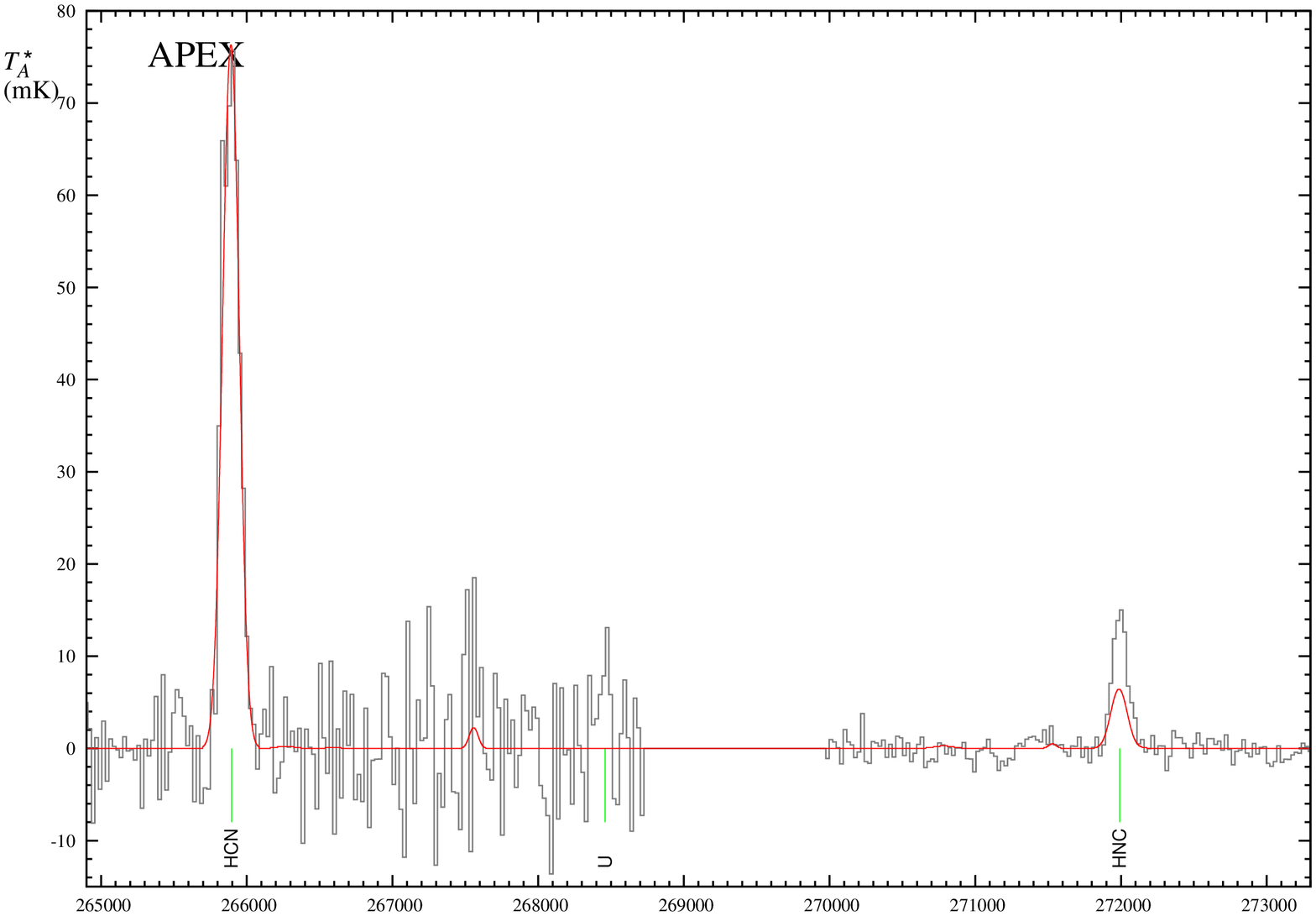}
\caption{Continued.}
\end{figure*}

  \setcounter{figure}{1}%

\begin{figure*} [tbh]
\centering
\includegraphics[angle=270,width=0.85\textwidth]{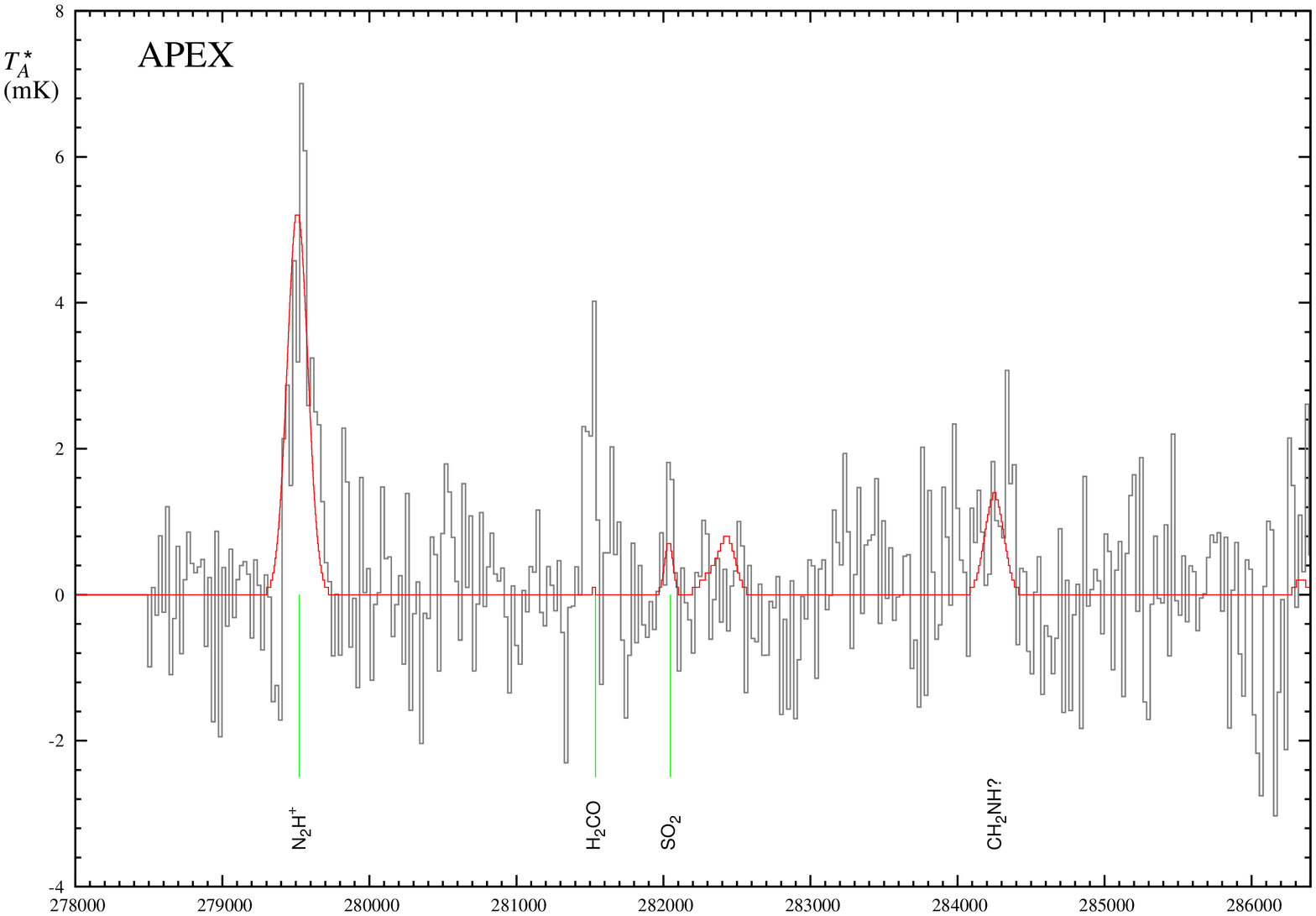}
\includegraphics[angle=270,width=0.85\textwidth]{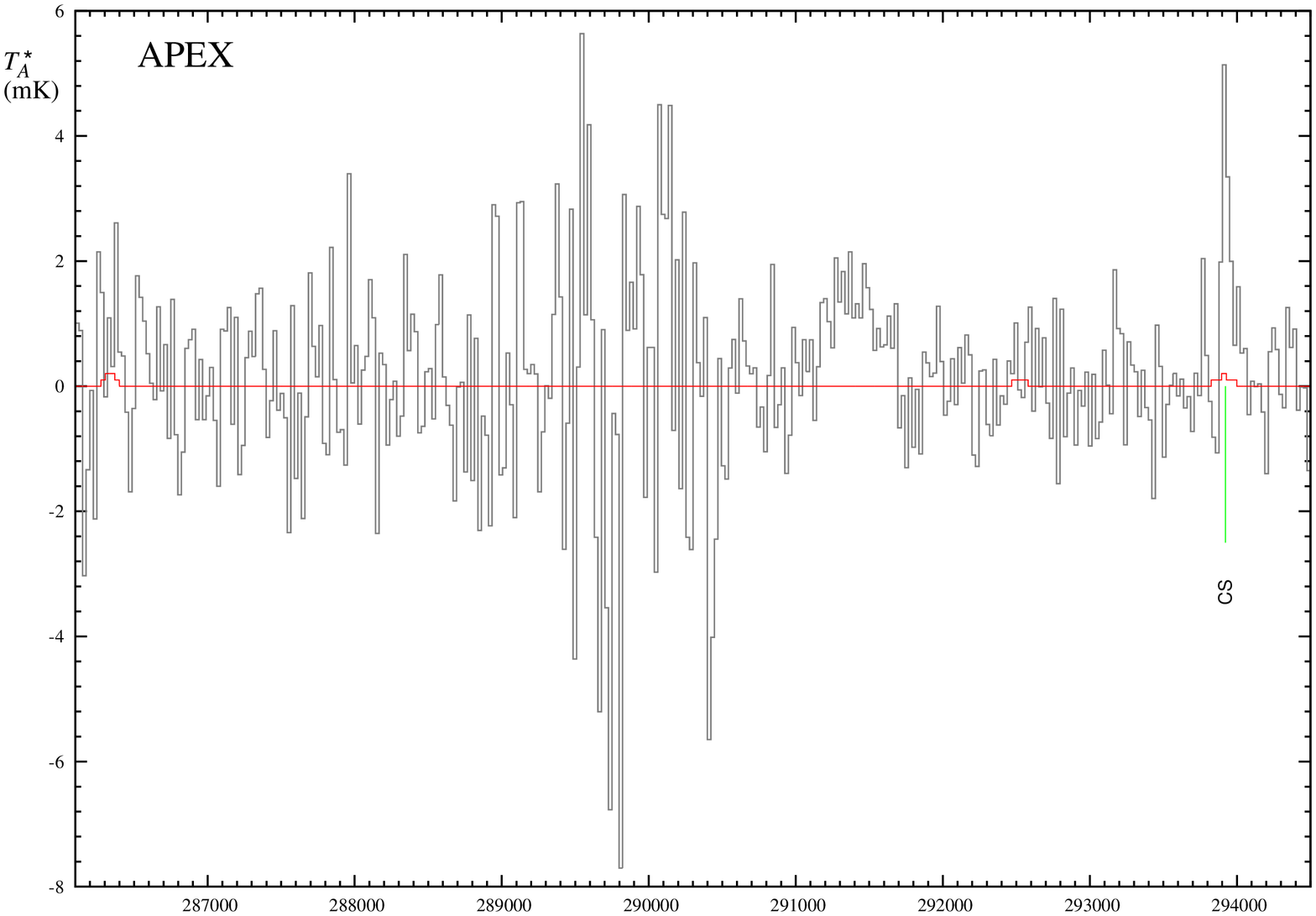}
\caption{Continued.}
\end{figure*}

  \setcounter{figure}{1}%

\begin{figure*} [tbh]
\centering
\includegraphics[angle=270,width=0.85\textwidth]{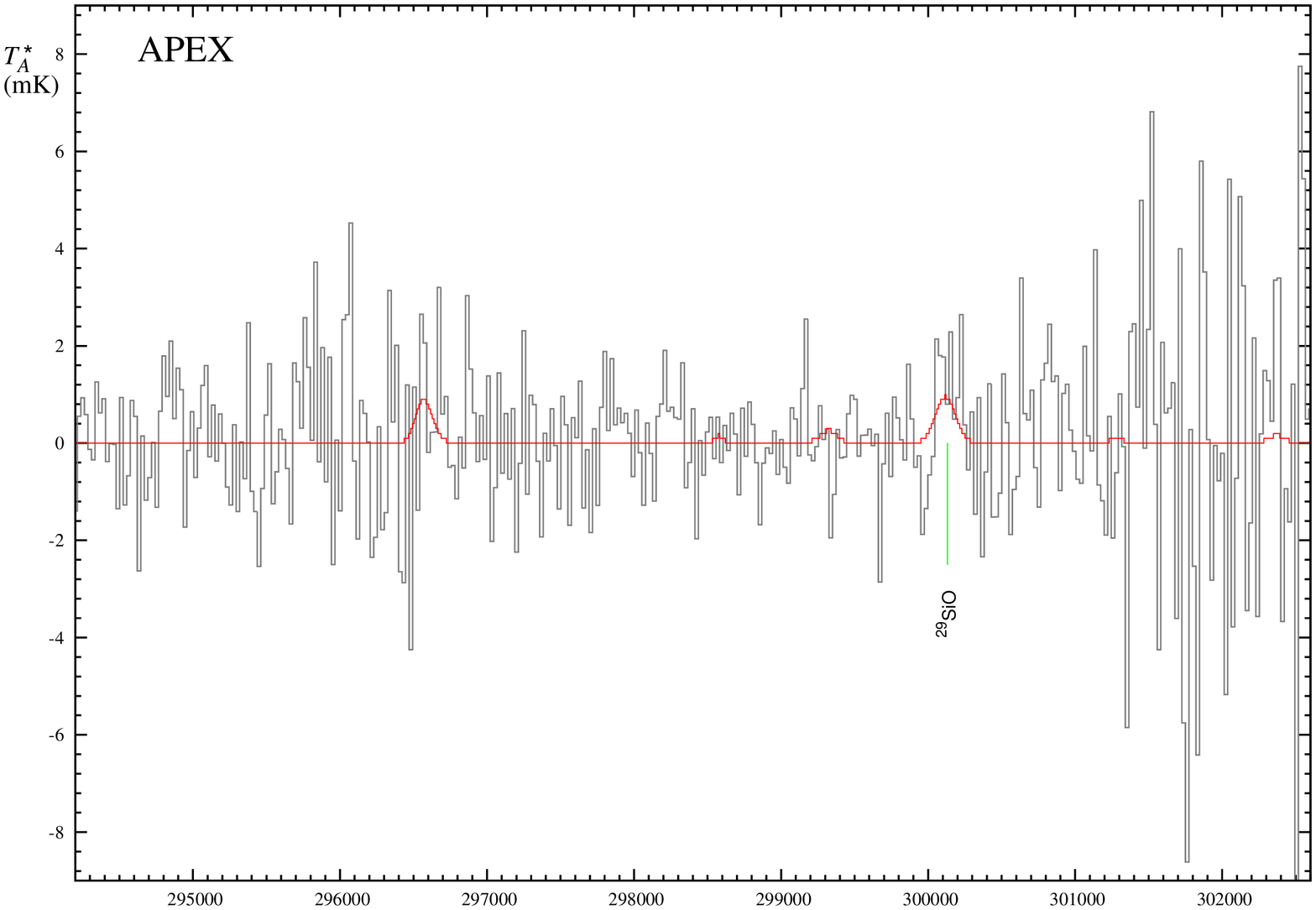}
\includegraphics[angle=270,width=0.85\textwidth]{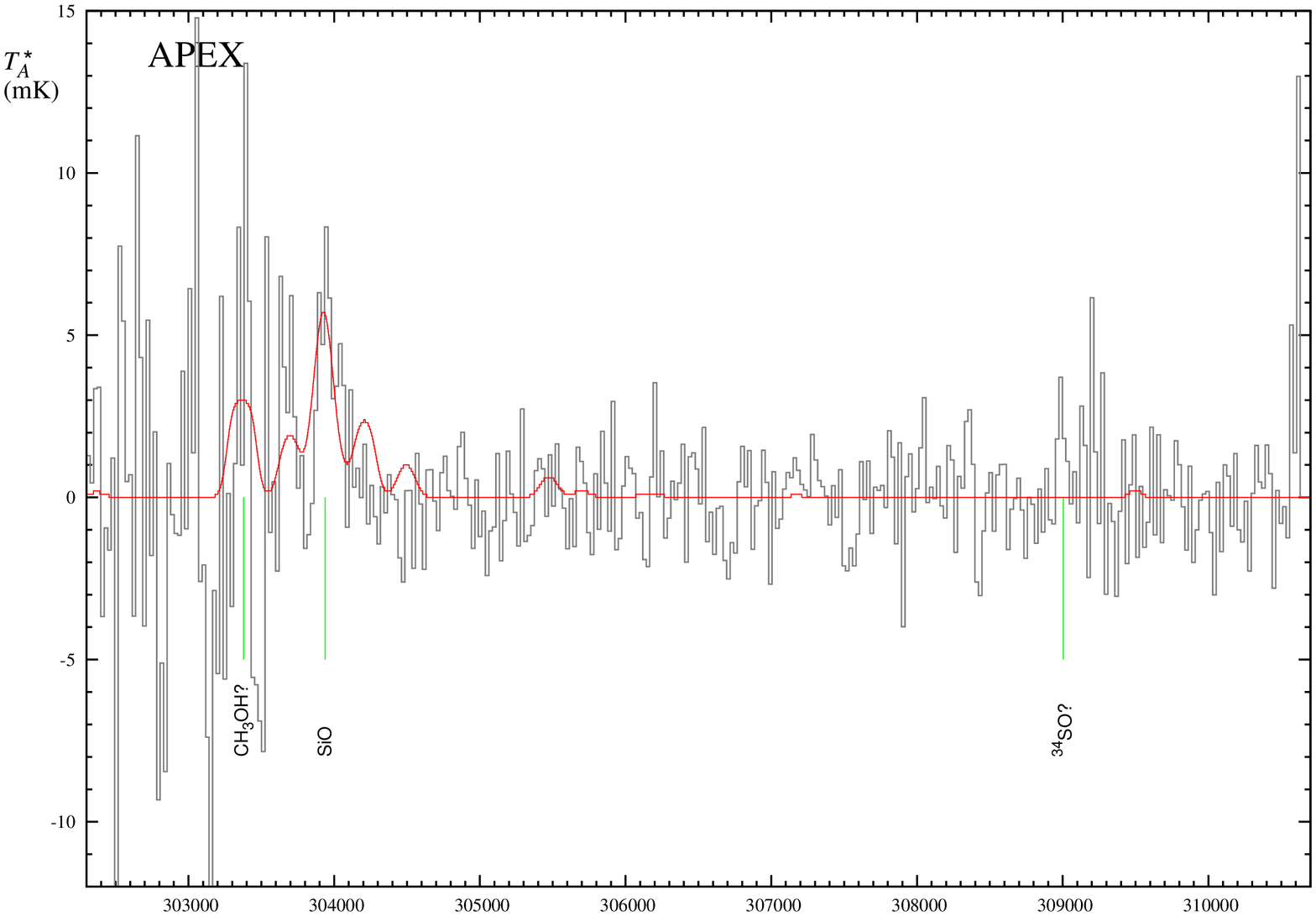}
\caption{Continued.}
\end{figure*}

  \setcounter{figure}{1}%

\begin{figure*} [tbh]
\centering
\includegraphics[angle=270,width=0.85\textwidth]{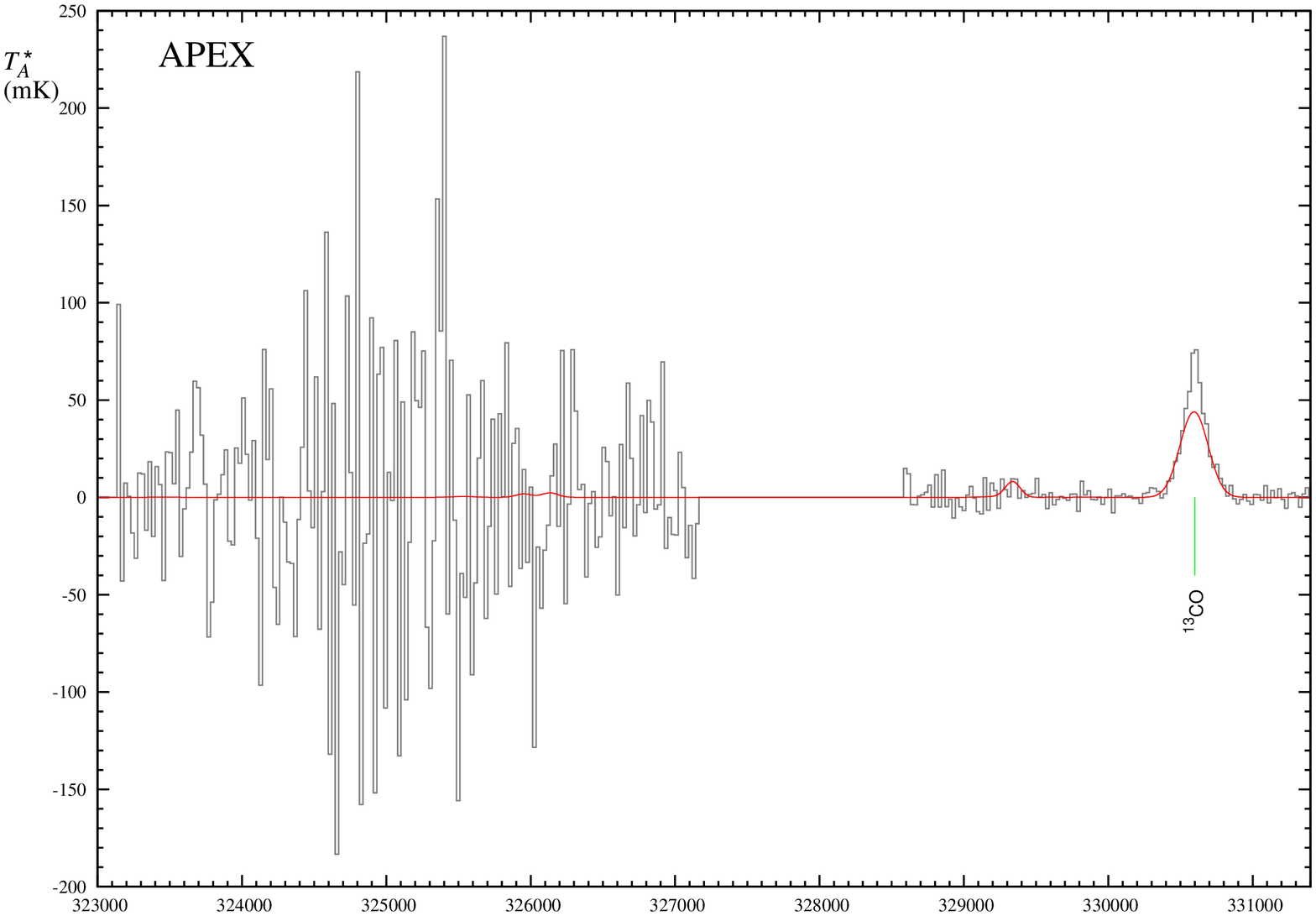}
\includegraphics[angle=270,width=0.85\textwidth]{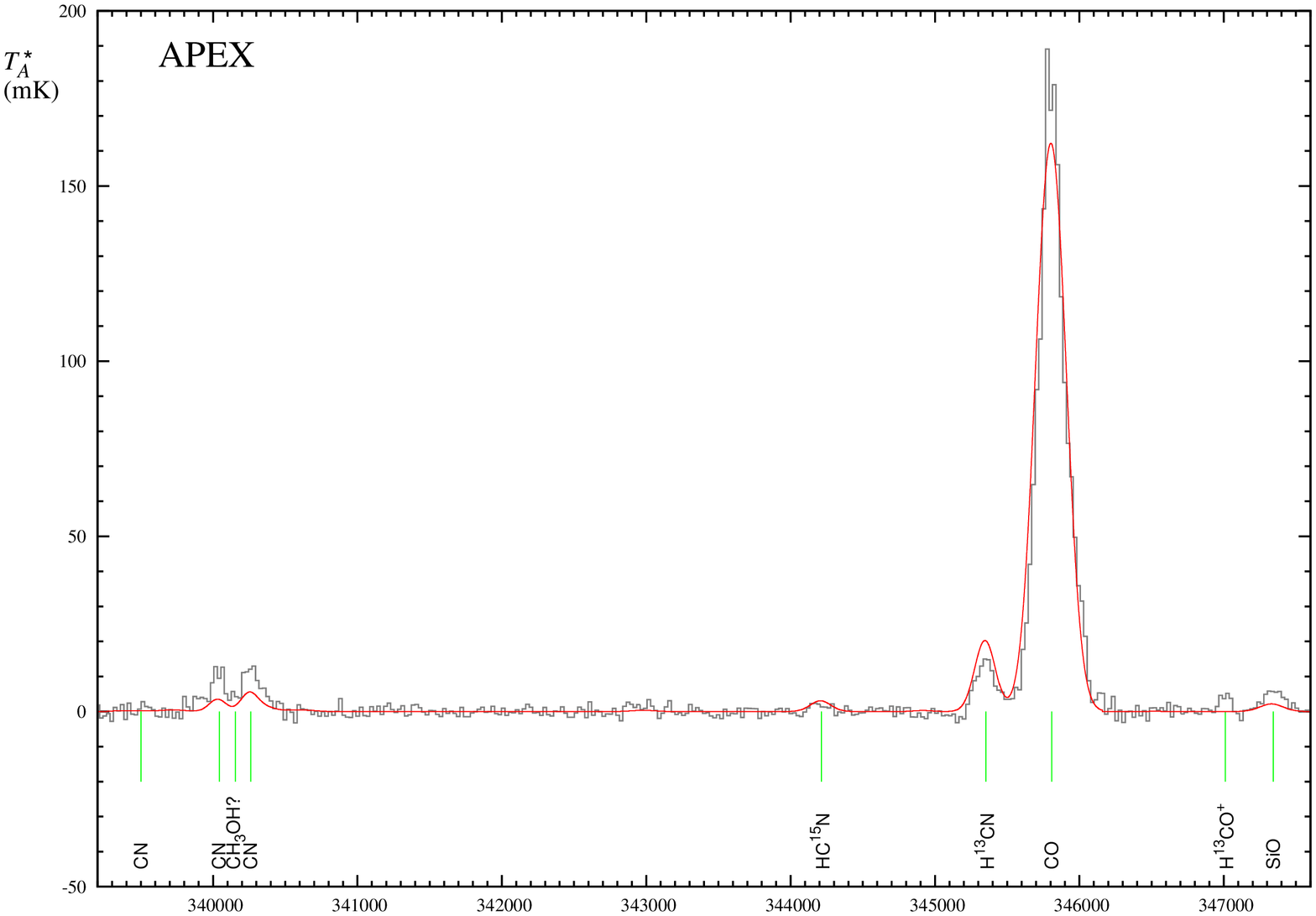}
\caption{Continued.}
\end{figure*}

  \setcounter{figure}{1}%

\begin{figure*} [tbh]
\centering
\includegraphics[angle=270,width=0.85\textwidth]{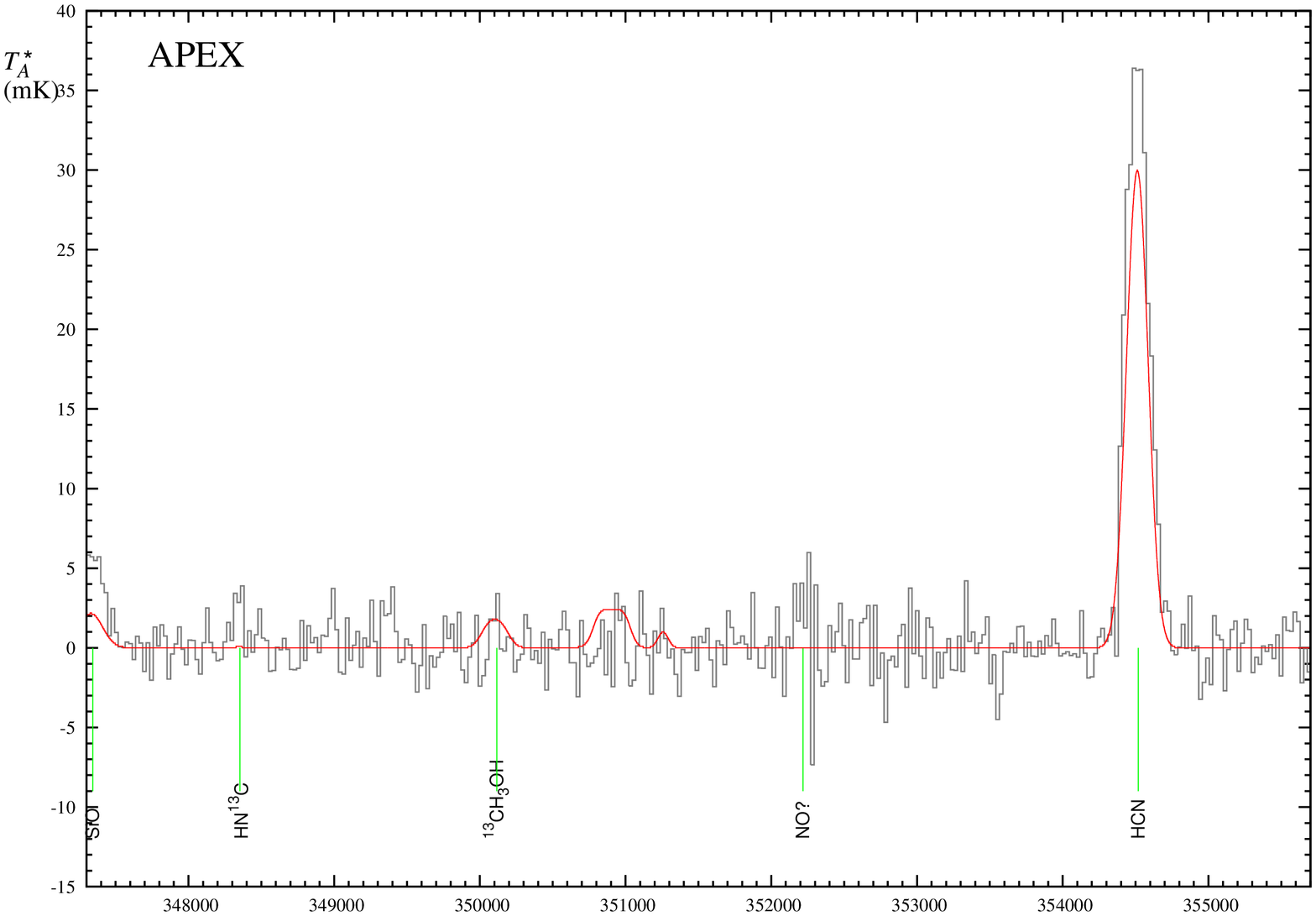}
\includegraphics[angle=270,width=0.85\textwidth]{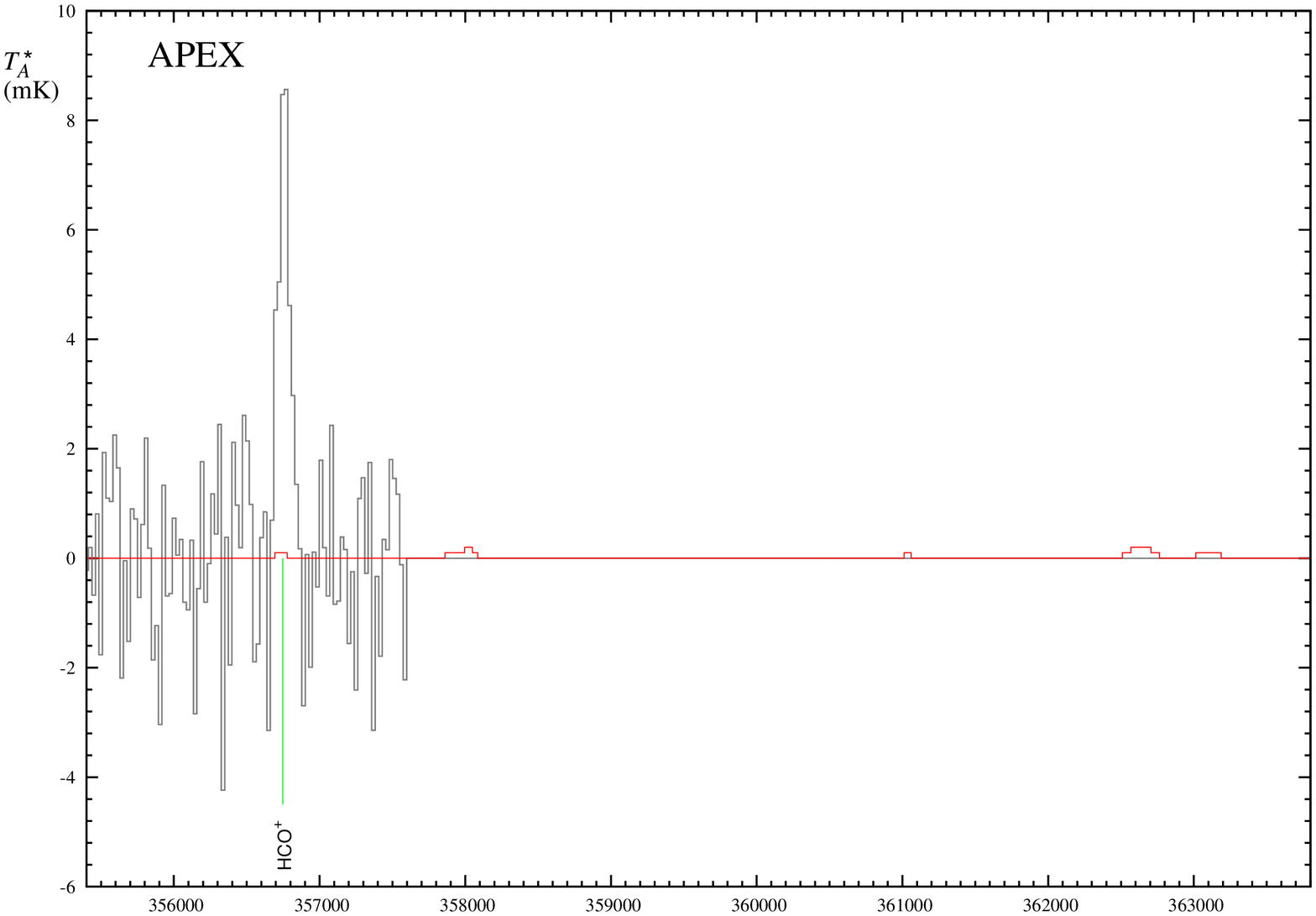}
\caption{Continued.}
\end{figure*}

  \setcounter{figure}{1}%

\begin{figure*} [tbh]
\centering
\includegraphics[angle=270,width=0.85\textwidth]{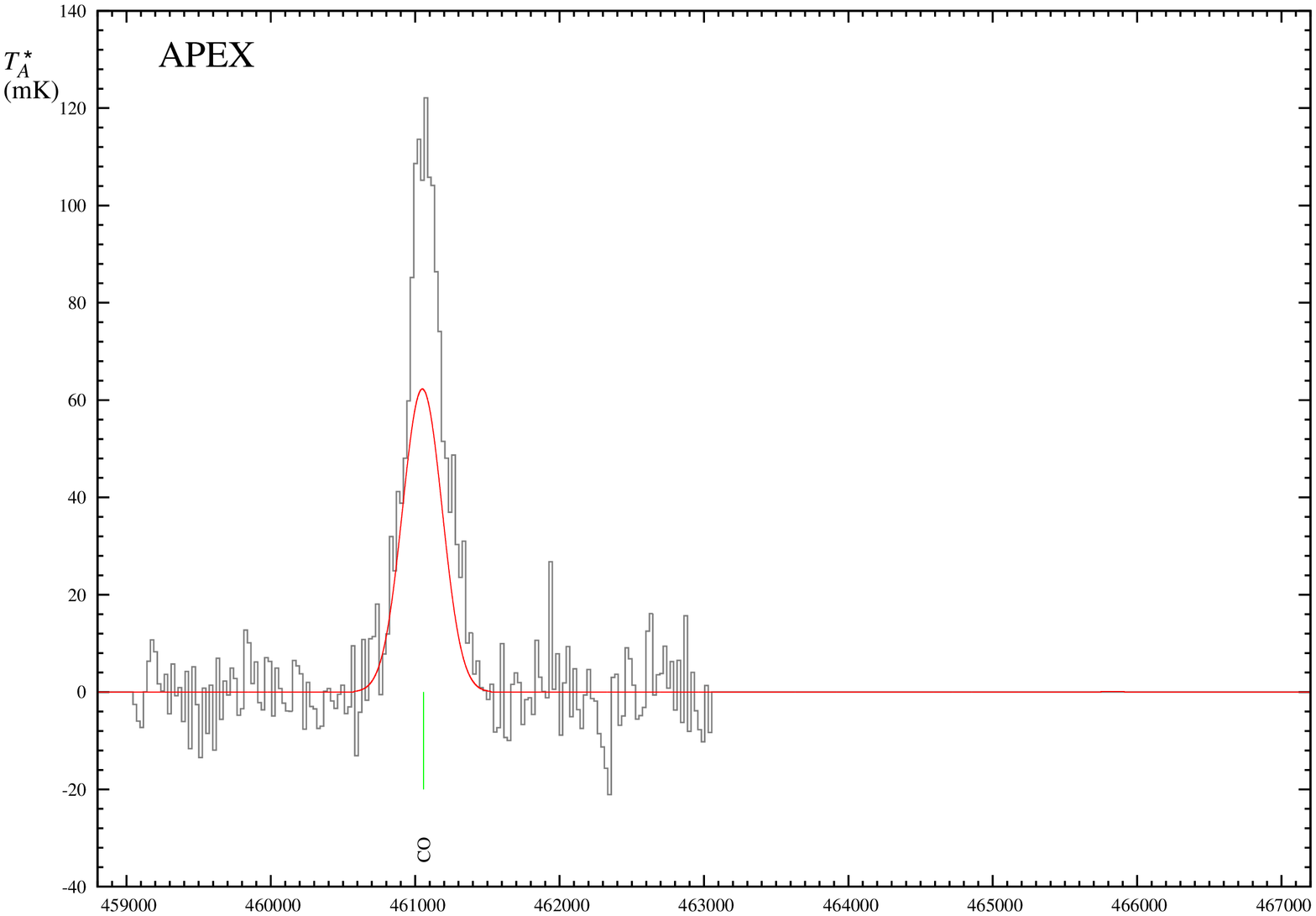}
\includegraphics[angle=270,width=0.85\textwidth]{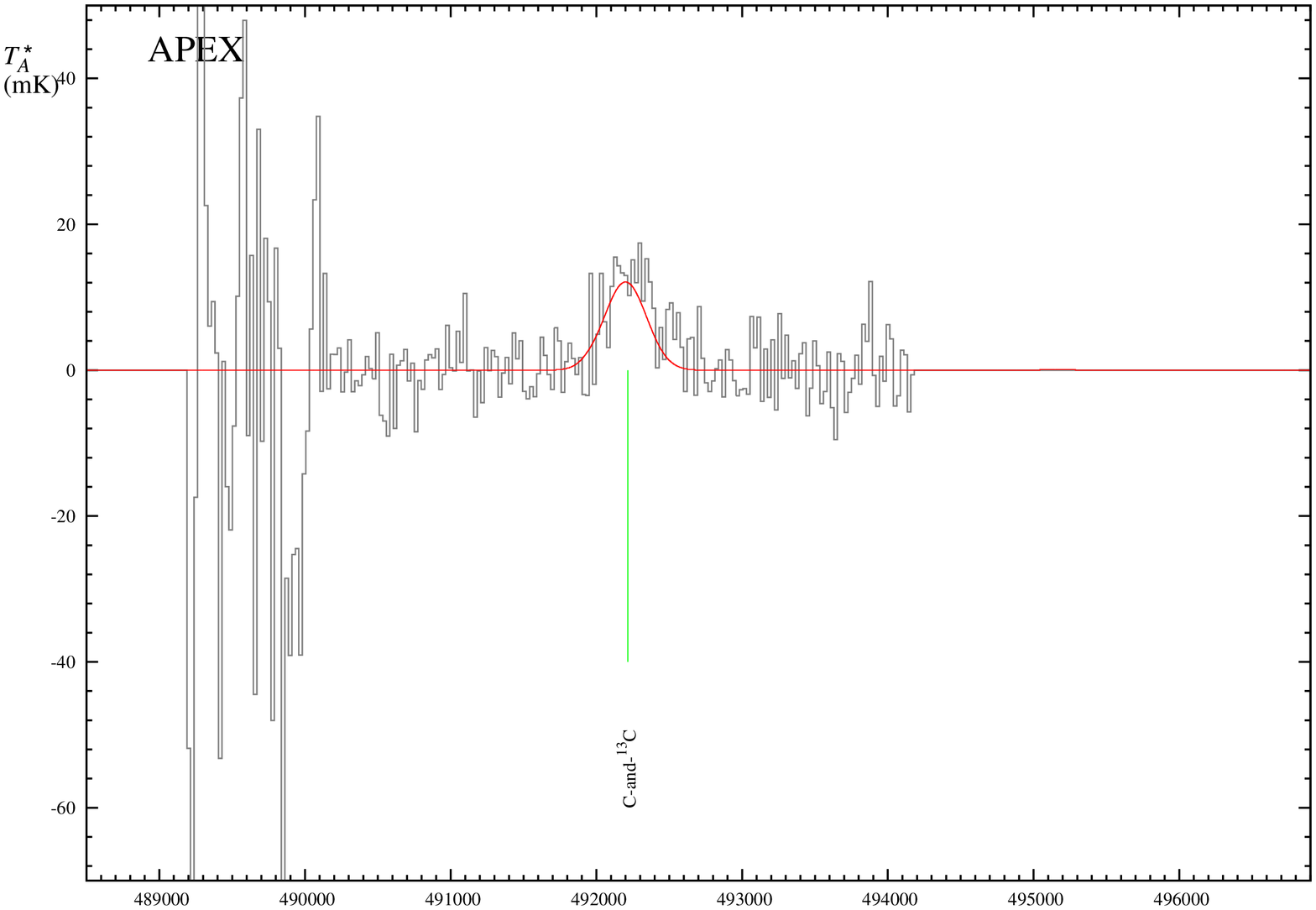}
\caption{Continued.}
\end{figure*}


\begin{thebibliography}{}
\bibitem[Af{\c s}ar \& Bond(2007)]{afsar} Af{\c s}ar, M., \& Bond, H.~E.\ 2007, \aj, 133, 387 
\bibitem[Amari et al.(2001)]{Amari2001} Amari, S., Gao, X., Nittler, L.~R., et al.\ 2001, \apj, 551, 1065 
\bibitem[Arnould et al.(1999)]{arnould1999} Arnould, M., Goriely, S., \& Jorissen, A.\ 1999, \aap, 347, 572 
\bibitem[Billade et al.(2012)]{sepia} Billade, B., Nystrom, O., Meledin, D., et al.\ 2012, IEEE Transactions on Terahertz Sci. and Techn., 2, 208 
\bibitem[Bohigas(2017)]{adiabatic} Bohigas, J.\ 2017, \mnras, 466, 1412 
\bibitem[Bujarrabal et al.(1988)]{bujarrabal88} Bujarrabal, V., Gomez-Gonzalez, J., Bachiller, R., \& Martin-Pintado, J.\ 1988, \aap, 204, 242 
\bibitem[Carter et al.(2012)]{emir} Carter, M., Lazareff, B., Maier, D., et al.\ 2012, \aap, 538, A89 
\bibitem[Castro-Carrizo et al.(2005)]{FrostyLeo} Castro-Carrizo, A., Bujarrabal, V., S{\'a}nchez Contreras, C., Sahai, R., \& Alcolea, J.\ 2005, \aap, 431, 979 
\bibitem[Cherchneff(2006)]{Cherchneff2006} Cherchneff, I.\ 2006, \aap, 456, 1001 
\bibitem[Cernicharo et al.(2000)]{cernicharoAlF} Cernicharo, J., Gu{\'e}lin, M., \& Kahane, C.\ 2000, \aaps, 142, 181 
\bibitem[Clayton et al.(2007)]{Clayton2007} Clayton, G.~C., Geballe, T.~R., Herwig, F., Fryer, C., \& Asplund, M.\ 2007, \apj, 662, 1220 
\bibitem[Cohen(1985)]{cohen85} Cohen, J.~G.\ 1985, \apj, 292, 90 
\bibitem[Coutens et al.(2017)]{coutens} Coutens, A., Rawlings, J.~M.~C., Viti, S., \& Williams, D.~A.\ 2017, \mnras, 467, 737 
\bibitem[Downes et al.(1997)]{SimbadPosition} Downes, R., Webbink, R.~F., \& Shara, M.~M.\ 1997, \pasp, 109, 345 
\bibitem[Evans et al.(2002)]{evans2002} Evans, A., van Loon, J.~T., Zijlstra, A.~A., et al.\ 2002, \mnras, 332, L35 
\bibitem[Evans et al.(2016)]{evans} Evans, A., Gehrz, R.~D., Woodward, C.~E., et al.\ 2016, \mnras, 457, 2871 
\bibitem[Goldsmith \& Langer(1999)]{RD} Goldsmith, P.~F., \& Langer, W.~D.\ 1999, \apj, 517, 209 
\bibitem[G{\"u}sten et al.(2006)]{apex} G{\"u}sten, R., Nyman, L.~{\AA}., Schilke, P., et al.\ 2006, \aap, 454, L13 
\bibitem[Hajduk et al.(2007)]{hajduk2007} Hajduk, M., Zijlstra, A.~A., van Hoof, P.~A.~M., et al.\ 2007, \mnras, 378, 1298 
\bibitem[Hajduk et al.(2013)]{hajduk2013} Hajduk, M., van Hoof, P.~A.~M., \& Zijlstra, A.~A.\ 2013, \mnras, 432, 167 
\bibitem[Harrison(1996)]{harrison96} Harrison, T.~E.\ 1996, \pasp, 108, 1112 
\bibitem[Hartquist et al.(1995)]{hartquist} Hartquist, T.~W., Menten, K.~M., Lepp, S., \& Dalgarno, A.\ 1995, \mnras, 272, 184 
\bibitem[Hema et al.(2012)]{hema} Hema, B.~P., Pandey, G., \& Lambert, D.~L.\ 2012, \apj, 747, 102 
\bibitem[Hevelius(1671)]{hevelius} Hevelius, J.\ 1671, Philosophical Transactions of the Royal Society of London Series I, 6, 2197 
\bibitem[Highberger et al.(2001)]{AlFinPPN} Highberger, J.~L., Savage, C., Bieging, J.~H., \& Ziurys, L.~M.\ 2001, \apj, 562, 790 
\bibitem[Iliadis(2007)]{Iliadis2007} Iliadis, C.\ 2007, Nuclear Physics of Stars,~ISBN 978-3-527-40602-9. Wiley-VCH Verlag, Wenheim, Germany, 2007
\bibitem[Jeffery et al.(2011)]{jeffery} Jeffery, C.~S., Karakas, A.~I., \& Saio, H.\ 2011, \mnras, 414, 3599 
\bibitem[Jos{\'e} \& Hernanz(2007)]{JoseHernanz} Jos{\'e}, J., \& Hernanz, M.\ 2007, Meteoritics and Planetary Science, 42, 1135 
\bibitem[Jos{\'e}(2002)]{JoseRev} Jos{\'e}, J.\ 2002, Classical Nova Explosions, 637, 104
\bibitem[Jos{\'e} et al.(2004)]{jose2004} Jos{\'e}, J., Hernanz, M., Amari, S., Lodders, K., \& Zinner, E.\ 2004, \apj, 612, 414 
\bibitem[J{\o}rgensen et al.(2004)]{shockOutflow} J{\o}rgensen, J.~K., Hogerheijde, M.~R., Blake, G.~A., et al.\ 2004, \aap, 415, 1021 
\bibitem[Jorissen et al.(1992)]{jorissen} Jorissen, A., Smith, V.~V., \& Lambert, D.~L.\ 1992, \aap, 261, 164 
\bibitem[Kami{\'n}ski et al.(2009)]{kamiV838} Kami{\'n}ski, T., Schmidt, M., Tylenda, R., Konacki, M., \& Gromadzki, M.\ 2009, \apjs, 182, 33 
\bibitem[Kami{\'n}ski et al.(2010)]{kamiV4332} Kami{\'n}ski, T., Schmidt, M., \& Tylenda, R.\ 2010, \aap, 522, A75 
\bibitem[Kami{\'n}ski et al.(2011)]{kami_coecho} Kami{\'n}ski, T., Tylenda, R., \& Deguchi, S.\ 2011, \aap, 529, A48 
\bibitem[Kami{\'n}ski et al.(2013b)]{kami_surv} Kami{\'n}ski, T., Gottlieb, C.~A., Young, K.~H., Menten, K.~M., \& Patel, N.~A.\ 2013, \apjs, 209, 38 
\bibitem[Kami{\'n}ski et al.(2015a)]{kamiNat} Kami{\'n}ski, T., Menten, K.~M., Tylenda, R., et al.\ 2015, \nat, 520, 322 
\bibitem[Kami{\'n}ski et al.(2015b)]{kamiV1309} Kami{\'n}ski, T., Mason, E., Tylenda, R., \& Schmidt, M.~R.\ 2015, \aap, 580, A34 
\bibitem[Kasemann et al.(2006)]{champ} Kasemann, C., G{\"u}sten, R., Heyminck, S., et al.\ 2006, \procspie, 6275,  
\bibitem[Kato(2003)]{kato} Kato, T.\ 2003, \aap, 399, 695 
\bibitem[Kipper et al.(2004)]{kipper} Kipper, T., Klochkova, V.~G., Annuk, K., et al.\ 2004, \aap, 416, 1107 
\bibitem[Klein et al.(2012)]{ffts} Klein, B., Hochg{\"u}rtel, S., Kr{\"a}mer, I., et al.\ 2012, \aap, 542, L3 
\bibitem[Klein et al.(2014)]{flash} T. Klein, T., Ciechanowicz, M., Leinz, Ch., et al., IEEE Trans. on Terahertz Science and Technology, Vol.4, No. 5, pp 588 - 596
\bibitem[Liu et al.(2016)]{Liu} Liu, N., Nittler, L.~R., O'D.~Alexander, C.~M., et al.\ 2016, \apj, 820, 140 
\bibitem[Lodders(2003)]{lodders} Lodders, K.\ 2003, \apj, 591, 1220 
\bibitem[Menon et al.(2013)]{menon} Menon, A., Herwig, F., Denissenkov, P.~A., et al.\ 2013, \apj, 772, 59 
\bibitem[Mamon et al.(1987)]{mamon87} Mamon, G.~A., Glassgold, A.~E., \& Omont, A.\ 1987, \apj, 323, 306 
\bibitem[Marka et al.(2015)]{rejection} Marka, C., Kramer, C., Navarro, S., \& John, D., Report on E150 image band rejections V3.0, 2015, \url{http://www.iram.es/IRAMES/mainWiki/EmirforAstronomers#Reports_and_publications}
\bibitem[Marty et al.(2011)]{Marty15N} Marty, B., Chaussidon, M., Wiens, R.~C., Jurewicz, A.~J.~G., \& Burnett, D.~S.\ 2011, Science, 332, 1533 
\bibitem[Miller Bertolami et al.(2011)]{DIN} Miller Bertolami, M.~M., Althaus, L.~G., Olano, C., \& Jim{\'e}nez, N.\ 2011, \mnras, 415, 1396 
\bibitem[Mitchell(1984)]{shocksSchem} Mitchell, G.~F.\ 1984, \apj, 287, 665 
\bibitem[Motiyenko et al.(2016)]{13ch3nh2} Motiyenko, R.~A., Margul{\`e}s, L., Ilyushin, V.~V., et al.\ 2016, \aap, 587, A152 
\bibitem[M{\"u}ller et al.(2001)]{cdms1} M{\"u}ller, H. S. P., Thorwirth, S., Roth, D. A., \& Winnewisser, G. 2001, \aap, 370, L49
\bibitem[M{\"u}ller et al.(2005)]{cdms2} M{\"u}ller, H. S. P., Schl{\"o}der, F., Stutzki, J.,  \& Winnewisser, G. 2005, J. Mol. Struct., 742, 215
\bibitem[Naylor et al.(1992)]{naylor} Naylor, T., Charles, P.~A., Mukai, K., \& Evans, A.\ 1992, \mnras, 258, 449 
\bibitem[Nittler \& Hoppe(2005)]{NH2005} Nittler, L.~R., \& Hoppe, P.\ 2005, \apjl, 631, L89 
\bibitem[Pardo et al.(2001)]{atm} Pardo, J.~R., Cernicharo, J., \& Serabyn, E.\ 2001, IEEE Transactions on Antennas and Propagation, 49, 1683 
\bibitem[Pardo et al.(2007)]{Pardo2007} Pardo, J.~R., Cernicharo, J., Goicoechea, J.~R., Gu{\'e}lin, M., \& Asensio Ramos, A.\ 2007, \apj, 661, 250 
\bibitem[Pearson et al.(1976)]{h15n13c} Pearson, E.~F., Creswell, R.~A., Winnewisser, M., \& Winnewisser, G.\ 1976, Zeitschrift Naturforschung Teil A, 31, 1394 
\bibitem[Pickett et al.(1998)]{jpl} Pickett, H.~M., Poynter, R.~L., Cohen, E.~A., et al.\ 1998, \jqsrt, 60, 883
\bibitem[Plume et al.(1999)]{Plume} Plume, R., Jaffe, D.~T., Tatematsu, K., Evans, N.~J., II, \& Keene, J.\ 1999, \apj, 512, 768 
\bibitem[Pulliam et al.(2011)]{hcoplus} Pulliam, R.~L., Edwards, J.~L., \& Ziurys, L.~M.\ 2011, \apj, 743, 36 
\bibitem[Rao \& Lambert(2008)]{RaoLambert} Rao, N.~K., \& Lambert, D.~L.\ 2008, \mnras, 384, 477 
\bibitem[Quintana-Lacaci et al.(2013)]{NOpaper} Quintana-Lacaci, G., Ag{\'u}ndez, M., Cernicharo, J., et al.\ 2013, \aap, 560, L2 
\bibitem[Rodriguez Kuiper et al.(1977)]{14C} Rodriguez Kuiper, E.~N., Kuiper, T.~B.~H., Kakar, R.~K., \& Zuckerman, B.\ 1977, \apj, 214, 394 
\bibitem[Roueff et al.(2015)]{roueff2015} Roueff, E., Loison, J.~C., \& Hickson, K.~M.\ 2015, \aap, 576, A99 
\bibitem[Rushton et al.(2005)]{rushton} Rushton, M.~T., Geballe, T.~R., Filippenko, A.~V., et al.\ 2005, \mnras, 360, 1281 
\bibitem[Sahai \& Nyman(1997)]{sahai} Sahai, R., \& Nyman, L.-{\AA}.\ 1997, \apjl, 487, L155 
\bibitem[Sch{\"o}ier et al.(2006)]{sio} Sch{\"o}ier, F.~L., Olofsson, H., \& Lundgren, A.~A.\ 2006, \aap, 454, 247 
\bibitem[Shara \& Moffat(1982)]{shara82}Shara, M.~M., \& Moffat, A.~F.~J.\ 1982, \apjl, 258, L41 
\bibitem[Shara et al.(1985)]{shara85} Shara, M.~M., Moffat, A.~F.~J., \& Webbink, R.~F.\ 1985, \apj, 294, 271 
\bibitem[Sievers et al.(2015)]{ghosts} Sievers, A., John, D., Navarro, S., \& Kramer, C. Report on suppression of E150 ghost lines after installation of an LO filter, V1.2, 2015, \url{http://www.iram.es/IRAMES/mainWiki/EmirforAstronomers#Reports_and_publications}
\bibitem[Soker \& Tylenda(2003)]{ST2003} Soker, N., \& Tylenda, R.\ 2003, \apjl, 582, L105 
\bibitem[Spitzer(1978)]{spitzer78} Spitzer, L.\ 1978, Physical processes in the interstellar medium, by Lyman Spitzer.~ New York Wiley-Interscience, 1978
\bibitem[St{\c e}pie{\'n}(2011)]{stepien} St{\c e}pie{\'n}, K.\ 2011, \aap, 531, A18 
\bibitem[S{\'a}nchez Contreras et al.(2015)]{contreras2015} S{\'a}nchez Contreras, C., Velilla Prieto, L., Ag{\'u}ndez, M., et al.\ 2015, \aap, 577, A52 
\bibitem[Tajitsu et al.(2015)]{7Be} Tajitsu, A., Sadakane, K., Naito, H., Arai, A., \& Aoki, W.\ 2015, \nat, 518, 381 
\bibitem[Tenenbaum et al.(2010)]{aro_surv} Tenenbaum, E.~D., Dodd, J.~L., Milam, S.~N., Woolf, N.~J., \& Ziurys, L.~M.\ 2010, \apjs, 190, 348 
\bibitem[Terzieva \& Herbst(2000)]{TH2000} Terzieva, R., \& Herbst, E.\ 2000, \mnras, 317, 563 
\bibitem[Tylenda \& Soker(2006)]{TS2006} Tylenda, R., \& Soker, N.\ 2006, \aap, 451, 223 
\bibitem[Tylenda et al.(2005)]{tyl_v838} Tylenda, R., Soker, N., \& Szczerba, R.\ 2005, \aap, 441, 1099 
\bibitem[Tylenda et al.(2011)]{v1309} Tylenda, R., Hajduk, M., Kami{\'n}ski, T., et al.\ 2011, \aap, 528, A114 
\bibitem[Tylenda(2005)]{tyl838} Tylenda, R.\ 2005, \aap, 436, 1009 
\bibitem[Tylenda et al.(2013)]{tyl-blg360} Tylenda, R., Kami{\'n}ski, T., Udalski, A., et al.\ 2013, \aap, 555, A16 
\bibitem[Vassilev et al.(2008)]{shfi} Vassilev, V., Meledin, D., Lapkin, I., et al.\ 2008, \aap, 490, 1157 
\bibitem[Vastel(2014)]{vastel} Vastel, C., 2014, Formalism for the CASSIS Software,\url{http://cassis.irap.omp.eu/docs/RadiativeTransfer.pdf}
\bibitem[Watanabe \& Kouchi(2002)]{WC2002} Watanabe, N., \& Kouchi, A.\ 2002, \apjl, 571, L173 
\bibitem[Zhang et al.(2008)]{ngc7027} Zhang, Y., Kwok, S., \& Dinh-V-Trung 2008, \apj, 678, 328 
\bibitem[Zhang(2016)]{zhang2016} Zhang, Y.\ 2016, arXiv:1611.03593 
\bibitem[Ziurys et al.(1994)]{ziurysAlF} Ziurys, L.~M., Apponi, A.~J., \& Phillips, T.~G.\ 1994, \apj, 433, 729 
\end{thebibliography}
\end{document}